\documentclass[12pt]{article}
\usepackage{amsmath}
\usepackage{amsfonts}
\usepackage{latexsym}
\usepackage{verbatim}
\usepackage{epsfig}
\usepackage{graphicx}
\usepackage{color}
\newtheorem{theorem}{Theorem}[section]

\newtheorem{lemma}{Lemma}[section]
\newtheorem{definition}{Definition}[section]
\newtheorem{proposition}{Proposition}[section]
\definecolor{darkgreen}{rgb}{0.,0.6,0.2}
\definecolor{aqua}{rgb}{0.0, 1.0, 1.0}
\begin{document}
\newcounter{INDEX}
\setcounter{section}{0}
\setcounter{subsection}{0}
\setcounter{equation}{0}
%

\newcommand{\abov}[2]{\genfrac{}{}{0pt}{}{#1}{#2}}
\newcommand{\eps}{{\varepsilon}}
\newcommand{\BL}{{{\mathcal B}_{\Lambda}}}
\newcommand{\Cd}{{{\mathbb C}^d}}
\newcommand{\Cm}{{{\mathbb C}^m}}
\newcommand{\Cn}{{{\mathbb C}^n}}
\newcommand{\Cp}{{{\mathbb C}^p}}
\newcommand{\cp}{{{\mathbb C}^{p_1}}}
\newcommand{\cpp}{{{\mathbb C}^{p_2}}}
\newcommand{\cppp}{{{\mathbb C}^{p_3}}}
\newcommand{\Cq}{{{\mathbb C}^q}}
\newcommand{\C}{{\mathbb C}}
\newcommand{\D}{{\mathcal D}}
\newcommand{\Di}{{\rm diag}}
\newcommand{\Ed}{{\cal E}_d}
\newcommand{\Edc}{\check{\cal E}_d}
\newcommand{\Edh}{\hat{\cal E}_d}
\newcommand{\fbar}{{\overline{f}}}
\newcommand{\fwig}{{\widetilde{f}}}
\newcommand{\Homeo}{{\rm Homeo}}
\newcommand{\M}{{\mathcal M}}
\newcommand{\N}{{\mathbb N}}
\newcommand{\OL}{{{\Omega}_{\Lambda}}}
\newcommand{\Q}{{\mathbb Q}}
\newcommand{\R}{{\mathbb R}}
\newcommand{\Rd}{{{\mathbb R}^d}}
\newcommand{\Rm}{{{\mathbb R}^m}}
\newcommand{\Rn}{{{\mathbb R}^n}}
\newcommand{\Rp}{{{\mathbb R}^p}}
\newcommand{\rp}{{{\mathbb R}^{p_1}}}
\newcommand{\rpp}{{{\mathbb R}^{p_2}}}
\newcommand{\rppp}{{{\mathbb R}^{p_3}}}
\newcommand{\Rq}{{{\mathbb R}^q}}
\newcommand{\Rt}{{{\mathbb R}^2}}
\newcommand{\Rmo}{{{\mathbb R}^m_0}}
\newcommand{\Smo}{{{\mathbb S}^{m-1}}}
\newcommand{\T}{{\mathbb T}}
\newcommand{\Td}{{{\mathbb T}^d}}
\newcommand{\TL}{{{\mathcal T}_{\Lambda}}}
\newcommand{\Tm}{{{\mathbb T}^m}}
\newcommand{\Z}{{\mathbb Z}}
\newcommand{\Zd}{{\mathbb Z}^d}
\newcommand{\Zm}{{\mathbb Z}^m}
\newcommand{\Zn}{{\mathbb Z}^n}
\newcommand{\Zp}{{\mathbb Z}^p}
\newcommand{\Zq}{{\mathbb Z}^q}
\newcommand{\ZL}{{{\mathcal Z}_{\Lambda}}}
\newcommand\be{\begin{equation}}
\newcommand\ee{\end{equation}}
\newcommand\bea{\begin{eqnarray}}
\newcommand\eea{\end{eqnarray}}

\title{
A detailed and unified treatment of spin-orbit systems using tools 
distilled from the
theory of bundles}

\author{  K.~Heinemann$^b$, D.~P.~Barber$^a$, J.~A. ~Ellison$^b$ and
Mathias Vogt$^a$\\
$^a$ \small{Deutsches~Elektronen--Synchrotron, DESY, ~22607 ~Hamburg,~Germany}
\\
$^b$ \small{Department of Mathematics and Statistics, The University of New
Mexico,} \\
\small{Albuquerque, New Mexico 87131, U.S.A. }}

\date{\today }
\allowdisplaybreaks
\maketitle

\begin{abstract}
We return to our study \cite{BEH} of invariant spin fields and spin tunes
for polarized beams in storage rings
but in contrast to the continuous-time treatment in \cite{BEH}, we
now employ a discrete-time formalism, beginning with the
  $\rm{Poincar\acute{e}}$ maps of the continuous time formalism. 
We then substantially extend our toolset and generalize the notions of invariant
spin field and invariant frame field. We revisit some old theorems and prove
several theorems believed to be new.
In particular we study two transformation rules, one of them known 
and the other new, where the former turns out to be
an $SO(3)$-gauge transformation rule. 
We then apply the theory to the dynamics of
spin-$1/2$ and spin-$1$ particle bunches and their
density matrix functions, describing semiclassically
the particle-spin content of bunches.
Our approach thus unifies the spin-vector dynamics from
the T-BMT equation with the spin-tensor dynamics and other dynamics.
This unifying aspect of our approach relates the examples elegantly and
 uncovers relations between the various underlying dynamical systems in a 
transparent way. 
As in \cite{BEH}, the particle motion
is integrable but we now allow for nonlinear particle motion on each torus.

Since this work is inspired by notions from the theory of bundles,
we also provide
insight into the underlying bundle-theoretic aspects of the well-established
concepts of invariant spin field, spin tune and invariant frame field.
Thus the group theoretical notions hidden
in \cite{BEH} and in its forerunners
\cite{DK73,Yo2} will be exhibited. Since we neglect, as is usual,
the Stern-Gerlach force,
the underlying principal bundle is of product form
so that we can present the theory
in a fashion which does not use bundle theory at all.
Nevertheless we occasionally mention the bundle-theoretic meaning
of our concepts and we also mention the
similarities with the geometrical approach to Yang-Mills Theory.
\end{abstract}
\tableofcontents
%
\setcounter{equation}{0}
\section{Introduction}
\label{1}
In \cite{BEH} we undertook an extensive
study of the concept of spin tune in storage rings on the basis of the
Thomas--Bargmann--Michel--Telegdi (T--BMT) equation \cite{Ja} of spin
precession. This naturally included a discussion of the invariant spin field and
the invariant frame field. 
This work is a sequel to \cite{BEH} and is based largely on
mathematical concepts and ideas in the PhD Thesis \cite{He2} of the first author
(KH), where a method from 
Dynamical-Systems Theory is
exploited to distil some essential features of 
particle-spin motion in
storage rings. As to be seen in Chapter \ref{10} 
this method clarifies and considerably extends the current theory
of \cite{BEH}. In fact it generalizes the concepts of
invariant frame field, spin tune, spin-orbit
resonances, invariant polarization field and
invariant spin field to an arbitrary subgroup $H$ of $SO(3)$ by using
the concept of $H$-normal form and invariant $(E,l)$-field. This
leads us to the Normal Form Theorem and various theorems
which generalize some standard theorems
that are also presented in this work. For short versions of
the present work, see  \cite{HBEV1,HBEV2}.

In \cite{BEH}  we assumed the particle motion to be
independent of the spin, i.e., we neglected the Stern-Gerlach force.
Also, the particle motion was described by an integrable
Hamiltonian system in action-angle variables, $J,\phi\in\Rd$.
We further assumed that the electric and
magnetic fields were of class $C^1$, i.e., continuously
differentiable) both in $\phi$ and
$\theta$. Thus the T--BMT equation became a linear system of ordinary
differential equations (ODE) for the particle-spin-vector
motion with smooth
coefficients depending quasiperiodically on $\theta$.
This quasiperiodic structure led us to a generalization of the Floquet theorem
and a new approach to the spin tune.

Although accelerator physicists tend to concentrate on studying specific
models of particle-spin
motion in real storage rings, many of the issues surrounding the
spin tune and the invariant spin field depend just on
the {\em structure}
of the equations of particle-spin motion 
and can be treated in general terms. This is the strategy to be
adopted here and it clears the way for the focus on purely
mathematical matters and in particular for the exploitation of methods
from Dynamical-Systems Theory and the theory of bundles.

In storage-ring physics there are two main approaches for dealing with
the independent variable in the equations of motion (EOM), namely use
of the flow formalism or the map formalism.  In the flow formalism the
EOM is an ODE, whence the independent
variable is the continuous variable $\theta \in \R$ describing the
distance around the ring.  In the map formalism the independent
variable in the EOM is the discrete variable $n \in \Z$ labelling 
the turn number where ${\mathbb Z}$ denotes the set of integers.
In Dynamical-Systems Theory
it is common practice to refer to
the independent variable in the EOM, such as $\theta$, the ``time''
and that is the convention that we will use here. Thus there is
a continuous-time and a
discrete-time formalism. In \cite{BEH} we used the former, 
here the emphasis is on discrete time. Nevertheless it would 
be possible to present the machinery of this work in the continuous-time 
formalism.

The external electrodynamic fields inside an accelerator's
vacuum chamber are smooth, i.e., of class $C^\infty$.
So the $C^1$ assumption adopted in \cite{BEH}
appears to be  perfectly reasonable.
On the other hand, practical numerical spin--orbit tracking
simulations are usually carried out with fields which cut off sharply
at the ends of magnets and/or with thin-lens approximations.
Thus in \cite{BEH} our formalism involved class $C^1$ in the time variable
$\theta$ although numerical calculations cited there in Sec.\ X  had been
obtained using hard-edged and thin-lens fields.
However, hard-edged and thin-lens ring elements fit naturally into the
discrete-time formalism.
In particular, for this,
we merely require that the fields are
continuous (i.e., of class $C^0$) in the orbital phases and we allow
jump discontinuities
in $\theta$.  Of course, this still allows study of systems with
fields smooth in $\theta$
and/or the orbital phases.
The way that the discrete-time formalism derives from the continuous-time
formalism is explained in
Section \ref{2.1}.

This work is designed so that it can be read independently of \cite{BEH}.
However, we wish to avoid repeating the copious contextual material
contained in \cite{BEH}. We therefore invite the reader to consult the
Introduction and the Summary and Conclusion in \cite{BEH}
in order to acquire a better appreciation of the context.
In this work, as in \cite{BEH}, the particle motion
is integrable
and we allow the number of angle variables, $d$, to be arbitrary (but
$\geq 1$) although for particle-spin motion 
in storage rings, the case $d = 3$
is the most important.  We use the symbols $\phi =
(\phi_1,...,\phi_d)^t$, $J = (J_1,...,J_d)^t$ and $\omega(J) =
(\omega_1(J),...,\omega_d(J))^t $
respectively for the lists of orbital angles, orbital actions
and orbital tunes where $^t$ denotes the transpose and  where with continuous time $d\phi/d\theta = \omega(J)$.
In the continuous-time formalism, the T--BMT equation is written as
$d {\bf {S}}/{d\theta}= {\bf {\Omega}}(\theta,J,\phi(\theta)) \times \bf
{S}$ where the $3$-vector 
$\bf S$ is the spin expectation value (``the
spin vector'') in the rest frame of a particle and ${\bf {\Omega}}$ is the
precession vector obtained as indicated in \cite{BEH} from the
electric and magnetic fields on the particle trajectory.
For the purposes of this
work we don't need to consider the whole $(J,\phi)$ phase space
since it will suffice to confine ourselves to a 
fixed $J$-value, i.e., to particle motion on a single
torus. Thus the actions $J$ are just parameters.
However it is likely that our work can be easily generalized to arbitrary
particle motion if one maintains our condition that
the particle motion is unaffected by the spin motion.

This work, in which we aim to present particle-spin motions
in terms of Dynamical-Systems Theory, is structured as follows.
In Section \ref{2.1} we discuss
the continuous-time formalism
which will motivate, in Section \ref{2.2},
the discrete-time concept of the ``spin-orbit system''
$(j,A)$ which characterizes a given setup by its
$1$-turn particle map $j$ on the torus $\Td$. 
While $j$ characterizes the integrable particle dynamics, 
$A$ is the $1$-turn spin transfer matrix function,
the latter being a continuous function from $\Td$ to $SO(3)$.
In the special case of the torus translation we have
$j={\cal P}_\omega$ where $\omega$ is the orbital tune and 
${\cal P}_\omega$ is the corresponding translation on the torus after one turn.
Thus in Section \ref{2.1} we derive the discrete-time 
$\rm{Poincar\acute{e}}$ map formalism from 
the continuous-time formalism and in Section \ref{2.1b} we
introduce the torus $\Td$ as a topological space.
For the torus the angle variable $\phi$ is
replaced by the angle variable $z$.
Then in Section \ref{2.2} we define the set ${\cal SOS}(d,j)$ of
spin-orbit systems $(j,A)$ to be considered in this work.
and in Section \ref{2.3} we introduce three important tools: the topological group, 
the group action, and the cocycle. These will carry us through the
whole work and will reveal a host of
well-known and less well-known structures underlying spin-orbit systems. 
In Chapter \ref{6} we define polarization field trajectories and these lead
to the definition of the invariant spin field (ISF).
A transformation rule, $(j,A)\mapsto(j,A')$,
is introduced in Chapter \ref{3}. This partitions ${\cal SOS}(d,j)$
into equivalence classes and spin-orbit systems belonging to the 
same equivalence class have similar properties.
For the notions of
partition and equivalence class, see Appendix \ref{A.2}.
It also leads us 
to structure-preserving  transformations of 
particle-spin-vector trajectories
and to structure-preserving  transformations
of polarization-field trajectories. In Chapter \ref{4.3} the partition of 
${\cal SOS}(d,j)$ leads us to several important 
subsets of ${\cal SOS}(d,j)$ which are denoted by
${\cal CB}_H(d,j)$.
Each of these subsets of ${\cal SOS}(d,j)$ is defined in terms of 
a simple form of $A$. In particular a $(j,A)$ in 
${\cal SOS}(d,j)$ belongs to
${\cal CB}_H(d,j)$ iff it can be transformed to a $(j,A')$ 
such that $A'$ is $H$-valued 
where $H$ is a subgroup of $SO(3)$. Then
$(j,A')$ is said to be an ``$H$-normal form'' of $(j,A)$.
The concept of $H$-normal form is also the driving force which leads 
us to the general theory of Chapter \ref{10}. 
In Chapter \ref{4.3} we also formulate and prove a standard theorem, which connects
the notions of ISF and invariant frame field (IFF) and which will
turn out in Chapter \ref{10} as 
the special case $H=SO(2)$ of the Normal Form Theorem.

In Chapter \ref{4} the partition of 
${\cal SOS}(d,j)$ leads us to the important 
subset ${\cal ACB}(d,j)$ of ${\cal SOS}(d,j)$.
This subset ${\cal ACB}(d,j)$ 
of ${\cal SOS}(d,j)$ is defined in terms of 
another simple form of $A$. In particular a $(j,A)$ in 
${\cal SOS}(d,j)$ belongs to
${\cal ACB}(d,j)$ iff it can be transformed to a $(j,A')$ 
such that $A'$ is constant. 
On the other hand spin tunes describe constant
rates of precession in appropriate reference frames so that one needs 
special spin-orbit systems which can be reached by transforming from
the original spin-orbit systems to such frames. Indeed, it is
${\cal ACB}(d,j)$ which leads in Section \ref{4.2}
to the notion of spin tune and
to the notion of spin-orbit resonance. 
Chapter \ref{VII} covers the topic of polarization. In particular
in Section \ref{6.4} we derive various formulas which 
estimate the bunch polarization with special emphasis on the situation
where only two ISF's exist. In Section \ref{8} we state and prove an
important and well-known theorem which provides conditions under
which only two ISF's exist.
Then, in Chapter \ref{10} we revisit and generalize the studies of the
previous chapters using an approach that we call 
the ``Technique of Association''(ToA) by which
the $SO(3)$-spaces $(E,l)$ label the different ``contexts'',
covering all the different spin variables.
The basic features of the ToA are defined in Section \ref{10.2}
and finer details in Sections \ref{10.3} and \ref{10.6} whereas
applications are considered in Sections \ref{10.4}-\ref{10.5} and the
bundle-theoretic origins of the ToA in Section \ref{10.7}.
With the ToA we will see that the
particle-spin-vector motion, i.e., the 
particle-spin-vector trajectories
and the polarization-field trajectories introduced in
Chapters \ref{2}-\ref{VII} turn out to be tied to the special
context, $(E,l)=(\R^3,l_{v})$, of the ToA where the $\R^3$-valued
spin variable is the spin vector $S$ and where $l_{v}(r,S)=rS$.
In \cite{BEH} we didn't go beyond $(\R^3,l_{v})$ but in this work
we do. For example we will study $(E_{t},l_{t})$ (see, e.g., 
Section \ref{10.4})
which encompasses the behavior of the spin tensor needed
for spin-$1$ particles, and we will study other important 
$(E,l)$ as well, in particular those needed for density matrices.
With Chapter \ref{10} it also becomes clear which of the
concepts of Chapters \ref{2}-\ref{VII} are 
$(E,l)$-dependent and which not.
For example, the concepts of spin-orbit system,
particle $1$-turn map, spin transfer matrix function, 
spin tune, spin-orbit resonance, 
invariant frame field, $H$-normal form 
are $(E,l)$-independent since they only depend on 
$(j,A)$. Clearly $(E,l)$-independent
concepts are very general. In contrast the concepts of invariant field
and the two ToA transformation rules are $(E,l)$-dependent.
While the main dynamical themes of
Chapter \ref{10} are the Normal Form Theorem and various invariant-field 
theorems,
a host of other results will be found along the way as well.
In Appendix \ref{A} we introduce the basic analytic notions like
continuous functions and partitions. 
Appendix \ref{11} contains some of our proofs.

Although many of our concepts, and in particular the ToA of Sections 
\ref{10.1}-\ref{10.6}, 
have their origin in bundle theory as outlined 
in Section \ref{10.7} we do not explicitly use bundle theory in 
those sections.
Thus it is appropriate to briefly mention
that the bundle  machinery has many
similarities with the so-called geometrical approach to Yang-Mills Theory.
The hallmark of most bundle approaches is a carefully chosen principal
bundle which allows one to store all data in the
associated bundles of that principal bundle. Of course,
one of the associated bundles is the principal bundle itself.
In our application the underlying principal bundle carries the data from
the particle motion and of the spin transfer matrix functions.
Moreover the associated bundles are labeled by the 
$SO(3)$-spaces $(E,l)$, i.e., they correspond to the above-mentioned
``contexts''. Thus each associated bundle carries a specific spin variable
$x$, e.g., the spin vector $S$ for the T--BMT spin motion, 
or a matrix $M$ for the spin tensor motion needed for spin-$1$
particles. The specific design of our underlying principal bundle
takes advantage of the fact that in polarized beam physics one neglects the
Stern-Gerlach force, thereby allowing us to use
techniques which were developed by R.J.Zimmer, R. Feres and others since
the 1980's to study so-called rigidity problems in Dynamical-Systems Theory
(see \cite{Fe,Zi2,Zi3} and Chapter 9 in \cite{HK1}).
In contrast, in the geometrical approach to Yang-Mills Theory one
picks a principal bundle which carries the data from
the space-time and from the gauge potentials and gauge fields while
the matter fields (leptons, quarks, Higgs particles, magnetic monopoles etc.)
reside in specific associated bundles.
The advantage of the use of bundles is its great flexibility and 
its ability to store and reveal
data and structures. For example in our application we take advantage of 
the cocycle structure of the spin transfer matrix functions, of a
SO(3)-gauge transformation structure connecting different spin-orbit systems,
and of the duality of particle-spin and field motion.
The duality provides  the practically important ability to track polarization
field trajectories, tensor field trajectories etc. 
in terms of the accelerator's particle
trajectories. In the special case $(E,l)=(\R^3,l_{v})$ the above
duality is the duality between particle-spin-vector 
trajectories and
polarization-field trajectories. 
\setcounter{equation}{0}
\section{Spin-orbit systems}
\setcounter{equation}{0}
\label{2}
A central aim of this work is a study of
the $1$-turn particle-spin-vector map ${\cal P}[j,A]$ given by
(\ref{eq:2.3b}), i.e.,
\begin{eqnarray}
&& {\cal P}[j,A](z,S):=  \left( \begin{array}{c}
 j(z) \\ A(z)S
\end{array}\right) \; ,
\nonumber
\end{eqnarray}
where $z$ is the angle variable on the torus and where
$j$ represents the $1$-turn particle map whereas, in the case of
real spin motion, a spin vector $S$ would be mapped to $A(z)S$ after one turn 
according to the $1$-turn spin transfer 
matrix function $A$ derived from the T-BMT equation.
These objects will be defined in
detail in this section and the above map will be generalized   
in Chapter \ref{10}, from spin vectors to other objects related to spin.
In Section \ref{2.1b} we discuss the torus as a topological space.
In Section \ref{2.2} we discuss the basic properties of 
${\cal P}[j,A]$ and in 
Section \ref{2.3} we define some group theoretical
notions underlying ${\cal P}[j,A]$.
\subsection{Deriving the discrete-time particle-spin-vector motion
from the continuous-time particle-spin-vector motion}
\label{2.1}
We begin our study by deriving our discrete-time  
particle-spin-vector motion from a continous-time
initial value problem (IVP) which takes the form
\begin{eqnarray}
&& \frac{d\phi}{d\theta}=\omega \; , \qquad
 \phi(0)=\phi_0 \in {\mathbb R}^d \; ,
\label{eq:2.10} \\
&& \frac{dS}{d\theta}={\cal A}(\theta, \phi)S \; ,  \qquad
 S(0)=S_0 \in {\mathbb R}^3
\; ,
\label{eq:2.12}
\end{eqnarray}
where $\omega\in\Rd$ and where the matrix-valued function
${\cal A}:\R^{d+1}\rightarrow\R^{3\times 3}$
is continuous in $\phi$ and  piecewise continuous in $\theta$.
More precisely, ${\cal A}$ is either continuous or has not more than
finitely many jump discontinuities at $\theta$ values
$\theta_1,...,\theta_N$ such that
${\cal A}$ is continuous on
$(\R\setminus\lbrace \theta_1,...,\theta_N\rbrace)\times\Rd$ and such that
${\cal A}(\theta_1;\cdot),...,{\cal A}(\theta_N;\cdot)$ are continuous.
For the $\cdot$ notation see Appendix \ref{A.1}.
Moreover we assume that ${\cal A}$ is $2\pi$-periodic
in each of its $d+1$ arguments and that it is skew--symmetric, i.e.,
${\cal A}^t(\theta,\phi)=-{\cal A}(\theta,\phi)$.
Without loss of generality and
for simplicity of notation we choose $\theta=0$ as the initial time.
We denote the set of ${\cal A}$, where
${\cal A}$ satisfies the above conditions, by ${\cal BMT}(d)$.

As is clear from the above and the Introduction, the above IVP
and the assumptions on ${\cal A}$ are motivated by our underlying interest 
in particle-spin-vector motion in storage rings.
In the application to particle-spin-vector motion
in storage rings,
$S$ is a column vector of components of the spin $\bf S$
and ${\cal A}(\theta,\phi)\equiv{\cal A}_J(\theta,\phi)$
represents the rotation rate vector ${\bf {\Omega}}(\theta,J,\phi)$
of the T--BMT equation \cite{BEH}.
Here $J,\phi$ are the action-angle variables of an integrable particle
motion. Note that ${\cal A}(\theta,\phi)$ is $2\pi$-periodic
in $\theta$ because we deal with storage rings and
${\cal A}(\theta,\phi)$ is $2\pi$-periodic in the
$d$ components of $\phi$ since the latter are angle variables.
Moreover ${\cal A}$ is skew-symmetric by its origin in the 
T--BMT equation, thus preserving the norm of $S$.
We suppress the $J$, except for a few occasions 
where we need it, since we work mainly on a fixed-$J$ torus.
The set ${\cal BMT}(d)$ includes standard 
particle-spin-vector motion but need not, 
and is only restricted by the above mentioned constraints on ${\cal A}$,
in keeping with our wish to 
investigate
the properties of any system defined by (\ref{eq:2.10}) and (\ref{eq:2.12}).

Since the system (\ref{eq:2.10}),(\ref{eq:2.12}) is periodic in $\theta$ it is
convenient  to study the behavior of solutions in terms of the
$\rm{Poincar\acute{e}}$ map (PM) \cite{AP,HK2}. We now derive a convenient
representation for the PM.
Solving (\ref{eq:2.10}) gives
\begin{eqnarray}
&& \phi(\theta) = \phi_0+\omega\theta \; ,
\label{eq:2.17av}
\end{eqnarray}
whence (\ref{eq:2.12}) reads as
\begin{eqnarray}
&&  \frac{dS}{d\theta}={\cal A}(\theta,\phi_0+\omega\theta)S \; ,  \qquad
 S(0)=S_0 \in {\mathbb R}^3 \; .
\label{eq:2.17aw}
\end{eqnarray}
Since ${\cal A}(\theta;\phi)$
is piecewise continuous in $\theta$ it
can be shown \cite{Cr} that the IVP (\ref{eq:2.17aw}) has a unique solution $S$
in the sense that
\begin{eqnarray}
&&  S(\theta)=S_0 + \int_0^\theta
{\cal A}(t,\phi_0+\omega t) S(t)dt
\; .
\label{eq:2.17awa}
\end{eqnarray}
It follows that
$S(\theta)$ is continuous in $\theta$.
The proof in Cronin \cite{Cr} does not include
the parameter $\phi_0$ but it is easily added.

Since (\ref{eq:2.17aw}) is linear in $S$
the general solution of (\ref{eq:2.17awa}) can be written as
\begin{eqnarray}
&&  S(\theta) =  \Phi_{CT}[\omega,{\cal A}](\theta;\phi_0)S_0 \; ,
\label{eq:2.10ng}
\end{eqnarray}
where,  with (\ref{eq:2.17awa}), the function 
$\Phi_{CT}[\omega,{\cal A}]:\R\times\R^d\rightarrow SO(3)$
satisfies
\begin{eqnarray}
&&  \Phi_{CT}[\omega,{\cal A}](\theta;\phi_0) = I_{3\times 3}
+ \int_0^\theta   {\cal A}(t,\phi_0+\omega t) \Phi_{CT}[\omega,{\cal A}]
(t;\phi_0)dt
\; ,
\label{eq:2.17awb}
\end{eqnarray}
and where $I_{3\times 3}$ is the $3\times 3$ unit matrix and where the subscript
``CT'' indicates ``continuous time''.
Since the values
of $\cal A$ are real skew--symmetric $3 \times 3$ matrices, 
$\Phi_{CT}[\omega,{\cal A}]$ is $SO(3)$-valued where $SO(3)$
is the set of
real $3\times 3$--matrices $R$ for which $R^t R=I_{3\times 3}$ 
and $\det(R)=1$.
By adding the parameters $\phi_0$ and $\omega$
in Cronin's proof, and using the fact
that ${\cal A}(\theta;\phi)$ is continuous in $\phi$, we conclude from
(\ref{eq:2.17awb})
that $\Phi_{CT}[\omega,{\cal A}]\in{\cal C}(\R^{d+1},SO(3))$
where ${\cal C}({\R^{d+1}},SO(3))$
is the set of continuous functions from
$\R^{d+1}$ into $SO(3)$. See Appendix \ref{A.4} too.
Furthermore $\Phi_{CT}[\omega,{\cal A}](\theta,\phi)$ is $2\pi$-periodic in the
components of $\phi$. Using (\ref{eq:2.17av}) and (\ref{eq:2.10ng}),
the solution of the IVP
(\ref{eq:2.10}),(\ref{eq:2.12}) can now be written
\begin{eqnarray}
&& \left( \begin{array}{c}
  \phi(\theta) \\ S(\theta)
\end{array}\right) =
\varphi(\theta;\phi_0,S_0) \; ,
\label{eq:5.84axb}
\end{eqnarray}
where the function $\varphi\in{\cal C}(\R^{d+4},\R^{d+3})$, is defined by
\begin{eqnarray}
&& \varphi(\theta;\phi,S):=\left( \begin{array}{c}
 \phi + \omega\theta \\
\Phi_{CT}[\omega,{\cal A}](\theta,\phi)S
\end{array}\right) \; .
\label{eq:5.84axa}
\end{eqnarray}
The PM on $\R^{d+3}$ is defined by
$\varphi(2\pi;\cdot)$ and it reads as
\begin{eqnarray}
&&
\varphi(2\pi;\phi,S)
=\left( \begin{array}{c}
 \phi + 2\pi\omega \\
\Phi_{CT}[\omega,{\cal A}](2\pi;\phi)S
\end{array}\right) \; .
\label{eq:5.84ax}
\end{eqnarray}
With (\ref{eq:5.84ax}) the $\rm{Poincar\acute{e}}$ map $\varphi(2\pi;\cdot)$
is determined by the parameters $\omega$ and ${\cal A}$.
With this the study of the non-autonomous
continuous-time Dynamical System (DS) of
(\ref{eq:2.10}),(\ref{eq:2.12})
has now been replaced by a study of an autonomous 
discrete-time DS given by
the PM (\ref{eq:5.84ax}).
In the following section we will see how 
the $\rm{Poincar\acute{e}}$ map (\ref{eq:5.84ax}), which is expressed
in terms of the angle variable $\phi$, will be expressed in terms of
the angle variable $z$ on the torus, leading us to the 
$\rm{Poincar\acute{e}}$ map ${\cal P}_{CT}[\omega,{\cal A}]$
in (\ref{eq:5.84axt}) below.
\subsection{The torus $\Td$ as the arena of the particle motion}
\label{2.1b}
Since $\phi$ is an angle variable and since we will have to prove
many analytic properties later, it is very convenient to replace
$\Rd$ by the torus $\Td$. One often defines a torus as the space obtained by
the function from $\phi\in\Rd$ to
$\phi\;{\rm mod\;2\pi}$. However continuity plays a significant role
in our study and so we give a definition suited for defining
continuity in terms of a topology, i.e., in terms of open sets.
For the basic topological notions, see Appendix \ref{A.3}.
A reader familiar with the torus can imagine (\ref{eq:5.84ax}) on the torus
and safely move to Section \ref{2.2}.
We define
\begin{eqnarray}
&& \Td :=  \lbrace \phi+\tilde\Zd :\phi\in\Rd  \rbrace \; ,
\label{eq:2.20}
\end{eqnarray}
where
\begin{eqnarray}
&& \phi +  \tilde\Zd := \lbrace \phi+\tilde{\phi}:
\tilde{\phi}\in \tilde{\Z}^d\rbrace
=\lbrace \phi+2\pi n:n\in\Zd\rbrace \; ,
\label{eq:2.21}
\end{eqnarray}
and where $\tilde\Zd :=\lbrace 2\pi n:n\in\Zd\rbrace$.
So on $\Td$, a chosen $\phi$ is accompanied by a countable infinity
of points separated by $2 \pi$ in each component of $\phi$.
In other words the elements of $\Td$ are countably infinite subsets of
$\Rd$. These subsets form a partition of $\R^d$. For the
definition of partition see Appendix \ref{A.2}.
Each element of $\Td$ can be represented by a unique element of
$[0,2\pi)^d$ and so $[0,2\pi)^d$ is a
``representing set'' of the partition $\Td$ of $\R^d$.
For the definition of representing set of partition 
see Appendix \ref{A.2}.
Clearly $\phi' + \tilde{\Z}^d= \phi + \tilde{\Z}^d$ iff there exists an
$m\in \Z^d$ such that $\phi'-\phi=2\pi m$.
The topology on $\Td$ is defined in a standard way using the topology of $\R^d$.

As we will now show, with this definition of $\Td$,   the PM  
(\ref{eq:5.84ax}) can be written as in (\ref{eq:5.84axt}) as
\begin{eqnarray}
&&
{\cal P}_{CT}[\omega,{\cal A}](z,S):=
\left( \begin{array}{c}
 {\cal P}_\omega(z) \\
A_{CT}[\omega,{\cal A}](z)S
\end{array}\right) \; , 
\nonumber
\end{eqnarray}
using the definitions in (\ref{eq:s2.10t}) and (\ref{eq:5.84ayn}).
The reader who is satisfied with this result might wish to jump to the summary of this subsection
on a first reading. In any case we now  continue with a rigorous justification and proof of the continuity of (\ref{eq:5.84axt}).

For that we now consider the onto function (surjection)
 $\pi_d:\Rd\rightarrow\Td$ where
\begin{eqnarray}
&&  \pi_d(\phi): = \phi + \tilde{\Z}^d =\lbrace \phi+2\pi n:n\in\Zd\rbrace \; ,
\label{eq:2.22}
\end{eqnarray}
then a subset $B\subset\Td$ is said to be open iff 
the inverse image, $\pi_d^{-1}(B)\subset\R^d$, of $B$ under $\pi_d$
is open (for the notion of inverse image see also Appendix \ref{A.1}). 
Thus the function $\pi_d$ and the
natural topology on $\Rd$ are used
to define a topology on $\Td$. It is common to say that
the topology on $\Td$ is ``co-induced by $\pi_d$'' \cite{wiki}. 
See Appendix \ref{A.6} too.
Of course, $\pi_d$ is continuous, i.e.,
$\pi_d\in{\cal C}(\Rd,\Td)$. Furthermore the topology on $\Td$
is the largest for which $\pi_d$ is continuous.
We will see many more
co-induced topologies (on sets different from $\Td$) in this work.
In an older terminology the above topology on $\Td$
is called the ``identification topology'' w.r.t. $\pi_d$ and 
$\pi_d$ is called an ``identification map'' \cite{Du,Hu}.

\vspace{3mm}
\noindent{\bf Remark:}
\begin{itemize}
\item[(1)] Another common definition of the torus is a 
cartesian product of circles, i.e., a subset  
${\hat {\mathbb T} }^d $ 
of $\R^{2d}$. Here the topology is defined in terms of
the Euclidean norm on $\R^{2d}$ giving the Euclidean metric. The 
two topological spaces $ {\hat {\mathbb T} }^d $ and $\Td$ are 
homeomorphic whence
the topology of $\Td$ has an underlying metric and thus
continuity could be defined in terms of the standard
``$\epsilon-\delta$'' approach. For the notion of
``homeomorphic'' see Appendix \ref{A.4}. 
However for the purposes of this work it is
easier to work with the equivalent open-set definition of continuity.
\hfill $\Box$
\end{itemize}
Clearly, by (\ref{eq:2.20}) and (\ref{eq:2.22}) and the remarks after
(\ref{eq:2.21}),
\begin{eqnarray}
&& \Td =  \lbrace \pi_d(\phi):\phi\in[0,2\pi)^d\rbrace = 
 \lbrace \pi_d(\phi):\phi\in\Rd  \rbrace \; .
\label{eq:2.20a}
\end{eqnarray}
We can now prepare for the definition of the
$\rm{Poincar\acute{e}}$ map ${\cal P}_{CT}[\omega,{\cal A}]$ and
the demonstration of  
its continuity. As a first step we use the above topology on $\Td$ 
to obtain the following lemma:
\setcounter{lemma}{0}
\begin{lemma} \label{L0}(Torus Lemma)\\
Let $Y$ be a topological space and
$F:{\mathbb R}^d\rightarrow Y$ be $2\pi$-periodic in each of its $d$
arguments. Then a unique function
$f:{\mathbb T}^d\rightarrow Y$ exists such that 
\begin{eqnarray}
&& F = f\circ \pi_d \; .
\label{eq:2.22na}
\end{eqnarray}
Moreover if $F$ is continuous then $f$ is continuous, i.e.,
$f\in{\cal C}({\mathbb T}^d,Y)$.
\end{lemma}

\begin{figure}[htbp]
\setlength{\unitlength}{1.5cm}
\begin{picture}(5,2) \thicklines
\put(3.5,-.3){$Y$}
\put(6.9,-.3){$\Td$}
\put(6.9,1.4){$\R^d$}
\put(5.4,-.6){$f$}
\put(5.4,1){$F$}
\put(7.5,0.6){$\pi_d$}
\put(6.7,-.2){\vector(-1,0){2.7}} 
\put(7,1.3){\vector(0,-1){1.2}} 
\put(6.7,1.4){\vector(-2,-1){2.7}} 
\end{picture}
\vspace{1cm}
\caption{Commutative diagram for Lemma \ref{L0} }\label{fig:lift0}
\end{figure}
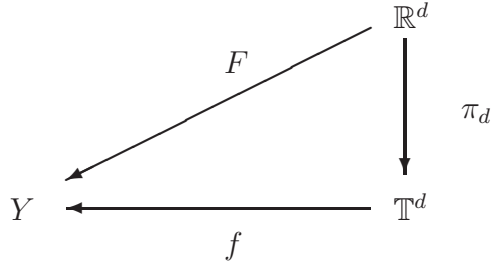

The situation in (\ref{eq:2.22na}) is illustrated
by the commutative diagram in Fig. \ref{fig:lift0}. \\ 

\noindent{\em Proof of Lemma \ref{L0}:}
Given the function $F$ we define the function
$f:\Td \rightarrow Y$ by
\begin{eqnarray}
&& f(z):=F(\phi), \quad \phi\in z \; ,
\label{eq:2.22n}
\end{eqnarray}
where the elements of the set $z$ are defined by (\ref{eq:2.22}) so that,
by (\ref{eq:2.20a}), $\phi\in z$ iff $\pi_d(\phi)=z$.
The function $f$ is well defined since $F(\phi)$ has, by periodicity,
the same value for every choice of $\phi\in z$. 
Clearly $f$ satisfies
(\ref{eq:2.22na}). Furthermore, since $\pi_d$ is a surjection,
$f$ is the only function which satisfies (\ref{eq:2.22na}).
Let $F$ be continuous.
To see that $f$ is continuous we
need to show that the inverse image $f^{-1}(V)$ is open 
for all open subsets $V$ of $Y$.
In fact, by (\ref{eq:2.22na}), we compute, for the inverse images, 
$\pi_d^{-1}(f^{-1}(V))=(f\circ \pi_d)^{-1}(V)=F^{-1}(V)$. Thus if $V$ is 
open, $F^{-1}(V)=\pi_d^{-1}(f^{-1}(V))$ is open
since $F$ is continuous.
Thus indeed $f$ is continuous.
Note that the second part of the proof also follows from
the Continuity Lemma in Appendix \ref{A.6}.
\hfill $\Box$

\vspace{3mm}

\noindent{\bf Remark:}
\begin{itemize}
\item[(2)] We have the following simple corollary to the Torus Lemma.

Let $g\in{\cal C}({\mathbb T}^d,Y)$ and let 
$F\in{\cal C}({\mathbb R}^d,Y)$ be defined by $F:= g\circ \pi_d$.
Then a function
$f:{\mathbb T}^d\rightarrow Y$ exists such that (\ref{eq:2.22na}) holds
and $f=g$.

Proof:
Clearly $F$ is continuous and $2\pi$-periodic in all of its arguments.
Thus we can apply  Lemma \ref{L0} to $F$ giving us a unique function 
$f:{\mathbb T}^d\rightarrow Y$ which satisfies (\ref{eq:2.22na}).
Thus and since $F=g\circ\pi_d$ we have $f=g$.
\hfill $\Box$
\end{itemize}

Since $\pi_d$ is continuous and $2\pi$-periodic in all of its arguments, 
a trivial application of Lemma \ref{L0} is where $F=\pi_d$ and $f=id_\Td$
where $id_\Td$ is the identity function on $\Td$ (for the latter see also
Appendix \ref{A.1}). 
More importantly, with Lemma \ref{L0} we can now rewrite the PM 
(\ref{eq:5.84ax}) in terms of $\Td$.
First we define the function ${\cal P}_\omega:\Td \rightarrow \Td$ by
\begin{eqnarray}
&& {\cal P}_\omega(z):=(\phi+2\pi\omega)+ \tilde{\Z}^d \; , \quad \phi\in z \; .
\label{eq:s2.10t}
\end{eqnarray}
This represents the particle dynamics on $\Td$ and it simply 
is a linear translation on the torus which shifts the
set $z=\phi+ \tilde{\Z}^d$ to the set 
$(\phi+2\pi\omega)+ \tilde{\Z}^d$. Secondly, we define
the function $A_{CT}[\omega,{\cal A}]:\Td\rightarrow SO(3)$ by
\begin{eqnarray}
&& A_{CT}[\omega,{\cal A}](z):= \Phi_{CT}[\omega,{\cal A}](2\pi;\phi) \; , 
\quad \phi\in z \; .
\label{eq:5.84ayn}
\end{eqnarray}
Here the functions $A_{CT}[\omega,{\cal A}]$ resp. 
$\Phi_{CT}[\omega,{\cal A}](2\pi;\cdot)$
correspond to $f$ resp. $F$ of Lemma \ref{L0} and thus 
$A_{CT}[\omega,{\cal A}]$
is continuous. Thus the PM (\ref{eq:5.84ax}) will be rewritten as
the function 
${\cal P}_{CT}[\omega,{\cal A}]:
\Td\times\R^3\rightarrow \Td\times\R^3$ defined by
\begin{eqnarray}
&&
{\cal P}_{CT}[\omega,{\cal A}](z,S):=
\left( \begin{array}{c}
 {\cal P}_\omega(z) \\
A_{CT}[\omega,{\cal A}](z)S
\end{array}\right) \; .
\label{eq:5.84axt}
\end{eqnarray}
We now argue that ${\cal P}_{CT}[\omega,{\cal A}]$ is continuous. 
It is easy to show, by (\ref{eq:s2.10t}),
that $F(\phi):=({\cal P}_\omega\circ\pi_d)(\phi)=
\pi_d(\phi+2\pi\omega)$ whence, and since $\pi_d$ 
is continuous and $2\pi$-periodic in its arguments,
$F$ belongs to ${\cal C}({\mathbb R}^d,\Td)$ and 
is $2\pi$-periodic in all of its arguments so that,
by Lemma \ref{L0}, a unique function
$f:{\mathbb T}^d\rightarrow \Td$ exists such that 
$F = f\circ \pi_d$ and $f$ is continuous. Of course since
$F={\cal P}_\omega\circ\pi_d$ we have $f={\cal P}_\omega$ whence 
${\cal P}_\omega\in{\cal C}(\Td,\Td)$. 
Since ${\cal P}_{-\omega}$ is the inverse of
${\cal P}_\omega$ we can write, due to the discussion after (\ref{eq:5.84axt}),
${\cal P}_\omega\in \Homeo(\Td)$.
Here $\Homeo(Y)$ denotes the set of homeomorphisms on the topological space $Y$
(see also  Appendix \ref{A.4}). 
Since $A_{CT}[\omega,{\cal A}]\in{\cal C}(\Td,SO(3))$, 
(\ref{eq:5.84axt}) implies
that ${\cal P}_{CT}[\omega,{\cal A}]\in{\cal C}(\Td\times\R^3,\Td\times\R^3)$. 

In summary, we have reduced the study of the 
continuous-time non-autonomous DS 
(\ref{eq:2.10}),(\ref{eq:2.12}) to the 
study of the discrete-time autonomous DS given by
the map of (\ref{eq:5.84axt}). This map is determined by
$\omega$ and $A_{CT}[\omega,{\cal A}]$. Thus we define
\begin{eqnarray}
&& {\cal SOS}_{CT}(d,\omega):=
\lbrace ({\cal P}_\omega,A_{CT}[\omega,{\cal A}]):
{\cal A}\in{\cal BMT}(d)\rbrace
\; .
\label{eq:2.3aa}
\end{eqnarray}
In the next section we will generalize (\ref{eq:5.84axt}) to the
maps that we will consider in this work.
\subsection{Introducing the  set ${\cal SOS}(d,j)$
of spin-orbit systems}
\label{2.2}
We now generalize ${\cal SOS}_{CT}(d,\omega)$ to ${\cal SOS}(d,j)$
by generalizing
${\cal P}_\omega$ and $A_{CT}[\omega,{\cal A}]$ to $j$ and $A$ giving us
\begin{eqnarray}
&& {\cal SOS}(d,j):=\lbrace (j,A):
A\in {\cal C}(\Td,SO(3))\rbrace \; ,
\label{eq:2.3a}
\end{eqnarray}
where $j\in \Homeo(\Td)$ and
where the matrix function $A$ is arbitrary in 
${\cal C}(\Td,SO(3))$ and thus is not 
necessarily derived from the $\cal A$ of 
(\ref{eq:2.10}),(\ref{eq:2.12}).  

Since ${\cal P}_\omega\in \Homeo(\Td)$, and since the function
$A_{CT}[\omega,{\cal A}]$ belongs to ${\cal C}(\Td,SO(3))$,
we see from (\ref{eq:2.3aa}) and (\ref{eq:2.3a}) that
\begin{eqnarray}
&& {\cal SOS}_{CT}(d,\omega)\subset 
{\cal SOS}(d,{\cal P}_\omega) \; ,
\label{eq:2.3ab}
\end{eqnarray}
and it will be shown below that the inclusion in (\ref{eq:2.3ab}) is proper, 
i.e., that ${\cal SOS}_{CT}(d,\omega)\neq {\cal SOS}(d,{\cal P}_\omega)$.

We call every pair
$(j,A)$ in ${\cal SOS}(d,j)$ a ``spin-orbit system''.
We call $A$ the 
``$1$-turn spin transfer matrix function'' of a spin-orbit system
$(j,A)$. We call $\omega$ the ``orbital tune vector'' of a spin-orbit system
$({\cal P}_\omega,A)$. 
We denote the union of the ${\cal SOS}(d,j)$ over $j$
by ${\cal SOS}(d)$.

Motivated by (\ref{eq:5.84axt}), we define, for every
$(j,A)$ in ${\cal SOS}(d,j)$, the function
${\cal P}[j,A]:\Td\times\R^3\rightarrow\Td\times\R^3$ by
\begin{eqnarray}
&& {\cal P}[j,A](z,S):=  \left( \begin{array}{c}
 j(z) \\ A(z)S
\end{array}\right) \; ,
\label{eq:2.3b}
\end{eqnarray}
and we call ${\cal P}[j,A]$ the 
``$1$-turn particle-spin-vector map of $(j,A)$''.
The map is invertible with inverse
\begin{eqnarray}
&& {\cal P}[j,A]^{-1}(z,S):=  \left( \begin{array}{c}
 j^{-1}(z) \\ A^t(j^{-1}(z))S
\end{array}\right) \; .
\label{eq:2.3binverse}
\end{eqnarray}
Clearly ${\cal P}[j,A]$ and ${\cal P}[j,A]^{-1}$ belong to
${\cal C}(\Td\times\R^3,\Td\times\R^3)$
whence ${\cal P}[j,A]$ is a homeomorphism and we write
${\cal P}[j,A]\in \Homeo(\Td\times\R^3)$.
In the special case where the spin-orbit system
$(j,A)$ belongs to ${\cal SOS}_{CT}(d,\omega)$
the $1$-turn particle-spin-vector 
map of $(j,A)$ carries the data of the PM, i.e.,
$j={\cal P}_\omega$ and
\begin{eqnarray}
&& {\cal P}_{CT}[\omega,{\cal A}]
={\cal P}[{\cal P}_\omega,A_{CT}[\omega,{\cal A}]]
\; .
\label{eq:2.2.30}
\end{eqnarray}
See also (\ref{eq:5.84ayn}) and (\ref{eq:5.84axt}).
In particular ${\cal P}_{CT}[\omega,{\cal A}]\in \Homeo(\Td\times\R^3)$.

All physical applications we have in mind have 
$j={\cal P}_\omega$ and so in this case $j$ is just a shorthand.
However, since for most notions and results of this work a general
$j$ is perfectly applicable, 
we do not confine ourselves to $j={\cal P}_\omega$.

A central aim of this paper is a study of the DS defined 
by (\ref{eq:2.3b}).
We find it convenient to work in the more general setting of
${\cal SOS}(d,j)$ and (\ref{eq:2.3b}) rather 
than the special setting of
${\cal SOS}_{CT}(d,\omega)$.
However the main physical interest is in a small subset of
${\cal SOS}_{CT}(d,\omega)$.
There is a natural question: given
$({\cal P}_\omega,A)$ in ${\cal SOS}(d,{\cal P}_\omega)$, does it belong to
${\cal SOS}_{CT}(d,\omega)$? 
This is an analogue of the following question
from beam dynamics: given a symplectic map, can it be generated as the
$1$-turn map of a Hamiltonian system? We do not deal with this question.
However, to show that the inclusion
(\ref{eq:2.3ab}) is proper consider $\omega\in\R$ and 
$({\cal P}_\omega,A)\in{\cal SOS}(1,{\cal P}_\omega)$
with $A\in{\cal C}(\T^1,SO(3))$ where $m$ is an integer and
\begin{eqnarray}
&&  A(z):=\left( \begin{array}{ccc} \cos m \phi & -\sin m \phi
& 0 \\ \sin m \phi & \cos m \phi & 0 \\
 0 & 0 & 1 \end{array}\right) \; , \quad \phi\in z \; .
\label{eq:2.50}
\end{eqnarray}
It can be shown with Lemma \ref{L0}
that $A$ in (\ref{eq:2.50}) is well defined and continuous as we did for
$A_{CT}[\omega,{\cal A}]$ after (\ref{eq:5.84ayn}),
thus $({\cal P}_\omega,A)\in{\cal SOS}(1,{\cal P}_\omega)$.
It was shown in \cite[Section 7.2]{He2}, by using simple arguments 
from Homotopy Theory, that $({\cal P}_\omega,A)\in 
{\cal SOS}_{CT}(1,{\cal P}_\omega)$ iff 
$m$ is even. Thus for $m$ odd we have an example showing that the inclusion
(\ref{eq:2.3ab}) is proper. Note also that 
$({\cal P}_\omega,A^2)\in {\cal SOS}_{CT}(1,{\cal P}_\omega)$ 
for every integer $m$.

We now discuss the DS defined by (\ref{eq:2.3b}).
It is a special case of a DS defined by a homeomorphism $f\in {\cal C}(Y,Y)$ on
a topological space $Y$.
The iterates are given by
\begin{eqnarray}
y(n+1)=f(y(n))\; , \quad y(0)=y_0\; , \quad n\in \Z \; ,
\label{eq:2.J10}
\end{eqnarray}
thus $y(1)=f(y_0), y(-1)=f^{-1}(y_0), y(2)=(f\circ f)(y_0), 
y(-2)=(f^{-1}\circ f^{-1})(y_0)$, etc.
The solution of (\ref{eq:2.J10}) can be written as 
\begin{eqnarray}
y(n)=\psi(n;y_0) \; , \quad  \; \psi(0;y_0)=y_0 \; ,
\label{eq:2.J15}
\end{eqnarray}
where the function $\psi:\Z\times Y\rightarrow Y$ satisfies
\begin{eqnarray}
\psi(n+m;y)= \psi(n;\psi(m;y))\; , \quad  \; \psi(0;y_0)=y \; .
\label{eq:2.J20}
\end{eqnarray}
Let $f^n$ be the $n$-fold composition of $f$ with itself.  Then
$\psi(n;y_0)=f^n(y_0)$ and we call $f^n$ the $n$-th iterate of $f$.
We use the standard topology on $\Z$ (see 
Section \ref{2.3}) in which case a function on
${\mathbb Z}\times Y$ is continuous iff it is
continuous in the second argument. Thus  
$\psi\in{\cal C}({\mathbb Z}\times Y,Y)$.
One proves (\ref{eq:2.J20})
by noting that both $H_1(n)=\psi(n+m;y)$ 
and $H_2(n)= \psi(n;\psi(m;y))$ satisfy (\ref{eq:2.J10})
with $H_1(0)=H_2(0)$. Thus by uniqueness they are equal for all $n$.

For our case we have $Y=\Td\times\R^3,f={\cal P}[j,A]$ and
\begin{eqnarray}
&&  \hspace{-1cm}
y(n)=
\left( \begin{array}{c}
z(n) \\ S(n)\end{array}\right)
\label{eq:7.4.10nnt}
\end{eqnarray}
whence, by (\ref{eq:2.3b}),
\begin{eqnarray}
&&  \hspace{-1cm}
\left( \begin{array}{c}
z(n+1) \\ S(n+1)\end{array}\right)
= \left( \begin{array}{c}
j(z(n))\\ A(z(n))S(n)
\end{array}\right)
\label{eq:2.J20a}
\end{eqnarray}
with $z(0)=z_0$ and  $S(0)=S_0$ given. We call $z(\cdot)$ 
a ``particle trajectory of $(j,A)$'' and
$S(\cdot)$ a ``spin-vector trajectory of $(j,A)$''.
Moreover we call a function $(z(\cdot),S(\cdot))$ 
a ``particle-spin-vector trajectory of $(j,A)$''.
The notion of particle-spin-vector 
trajectory will be generalized in Chapter \ref{10}
where we generalize the spin vector to an arbitrary variable 
related to spin.

We now derive a convenient representation for $\psi$ in our case where
$f={\cal P}[j,A]$. We follow the procedure in Section \ref{2.1} going from
(\ref{eq:2.10}) and (\ref{eq:2.12}) to (\ref{eq:5.84axa}).
Clearly $z(n)=j^n(z_0)$. Define $L[j]:\Z\times\Td\rightarrow \Td$ via
\begin{eqnarray}
&& L[j](n;z):= j^n(z) \; ,
\label{eq:3.20}
\end{eqnarray}
then 
\begin{eqnarray}
&&  L[j](n+m,z) = L[j](n;L[j](m;z)) \; ,
\label{eq:5.72b}
\end{eqnarray}
and
$S(n+1)=A(L[j](n;z_0))S(n)$ and $S(-n)=A^t(L[j](-n;z_0))S(-n+1)$ so that
\begin{eqnarray}
&&S(n)=A( L[j](n-1;z))
\cdots A( L[j](1;z))A(z)S_0 \; , \qquad (n=1,2,...) \; \\ \nonumber
&&S(n)=A^t( L[j](n;z))
\cdots A^t( L[j](-1;z))S_0 \; , \qquad (n=-1,-2,...) \; ,
\end{eqnarray}
where we used the fact that $A^t(z)A(z)=I_{3\times 3}$. Thus 
\begin{eqnarray}
&& S(n) := \Psi[j,A](n;z_0)S_0 \; ,
\label{eq:3.1}
\end{eqnarray}
where
\begin{eqnarray}
&& \hspace{-1cm}
\Psi[j,A](0;z)=I_{3\times 3} \; , \nonumber\\
&& \hspace{-1cm}
\Psi[j,A](n;z)=A( L[j](n-1;z))
\cdots A( L[j](1;z))A(z)\; , \qquad (n=1,2,...) \; ,
\nonumber\\
&&\hspace{-1cm}
\Psi[j,A](n;z)=A^t( L[j](n;z))\cdots A^t(L[j](-1;z))
\; ,  \;\; (n=-1,-2,...) \; .
\nonumber\\
\label{eq:3.5}
\end{eqnarray}
We now have the desired representation for $\psi$ given
by the function $L[j,A]:\Z\times\Td\times\R^3\rightarrow
\Td\times\R^3$ defined by the $n$th iteration of ${\cal P}[j,A]$:
\begin{eqnarray}
&& L[j,A](n;z,S):={\cal P}[j,A]^n(z,S)
= \left( \begin{array}{c}
L[j](n;z) \\
\Psi[j,A](n;z)S\end{array}\right) \; .
\label{eq:s2.10}
\end{eqnarray}
With (\ref{eq:2.J15}) or (\ref{eq:s2.10}) the solution of
(\ref{eq:2.J20a}) is
\begin{eqnarray}
&&  \hspace{-1cm}
\left( \begin{array}{c} z(n)\\ S(n)\end{array}\right)
=  L[j,A](n,z_0,S_0) \; .
\label{eq:10.42n}
\end{eqnarray}
Also from (\ref{eq:2.J20}) or (\ref{eq:s2.10}) we get
\begin{eqnarray}
&& \hspace{-1cm}
  L[j,A](n+m,z,S) = L[j,A](n;L[j,A](m;z,S)) \; , \quad
L[j,A](0;z,S) = \left( \begin{array}{c} z \\ S\end{array}\right) \; .
\label{eq:5.72a}
\end{eqnarray}
Inserting (\ref{eq:s2.10}) into (\ref{eq:5.72a}) gives
\begin{eqnarray}
&& \Psi[j,A](n+m;z)=\Psi[j,A](n;L[j](m;z))
\;\Psi[j,A](m;z) \; .
\label{eq:3.25}
\end{eqnarray}

We now introduce some additional terminology which will be useful in the 
following. We call the $n$-th iterate 
${\cal P}[j,A]^n=L[j,A](n;\cdot)$
the ``$n$-turn particle-spin-vector 
map of $(j,A)$'',
we call $\Psi[j,A]$ the ``spin transfer matrix function''
of $(j,A)$ and we call $\Psi[j,A](n;\cdot)$ the
``$n$-turn spin transfer matrix function'' of $(j,A)$. 
Since $\Psi[j,A](n;\cdot)$ is continuous,
every spin transfer matrix function is a continuous function
due to the standard topology on $\Z$.
Clearly
\begin{eqnarray}
&& \hspace{-1cm}
\Psi[j,A](1;z) = A(z) \; ,
\label{eq:3.5nn}
\end{eqnarray}
which justifies calling $A$ the $1$-turn spin transfer matrix
function.

The behavior of the spin-vector trajectories in (\ref{eq:3.1})
depends on the values of
$A$ reached by the particle motion $L[j](n;z_0)$
in its argument, which in turn depends on $j$. 
In the case $j={\cal P}_\omega$ the argument 
$z(n)$ of $A$ in (\ref{eq:2.J20a})
will remain in a confined subset of the torus for some values of $\omega$
and for other values it will cover the torus densely.
To be more precise we define resonance.
We say $\chi\in {\mathbb R}^n$ is {\em resonant} if there exists a non-zero 
integer vector $k \in {\mathbb Z}^n$ such that $k\cdot\chi=0$
and nonresonant if not resonant.
If $j={\cal P}_\omega$ and
$(1,\omega)$ is nonresonant then the argument 
$z(n)$ of $A$ in (\ref{eq:2.J20a}) covers the 
torus densely and since $A$ is continuous all values of $A$ affect the 
spin-vector 
trajectory whereas if $(1,\omega)$ is resonant the values of $A$ are only 
sampled by its values on a sub-torus. 
The spin-orbit system $({\cal P}_\omega,A)$ is said to be
``off orbital resonance'' if $(1,\omega)$ is nonresonant
and ``on orbital resonance'' if $(1,\omega)$ is resonant.
Thus spin-vector trajectories may exhibit significantly different qualitative behaviors on and off orbital resonance.
We will now generalize the notion ``off orbital resonance''.
One says that $j\in \Homeo(\Td)$ is ``topologically transitive'' if
a $z_0\in\Td$ exists such that the set
$B:=\lbrace j^n(z_0):n\in\Z\rbrace$ is dense in $\Td$, i.e., 
$\overline{B}=\Td$ where $\overline{B}$ denotes the topological
closure of $B$, see Appendix A.3. An important special 
case is when $j={\cal P}_\omega$:
then $j$ is topologically transitive iff
$(1,\omega)$ is nonresonant.

It is interesting to relate again the motion defined by 
(\ref{eq:2.10}),(\ref{eq:2.12})
to the motion defined by (\ref{eq:2.3b}).
Let $\phi_0\in\Rd$ and $\omega\in\Rd,{\cal A}\in
{\cal BMT}(d)$ and let $S$ be a solution
of the IVP (\ref{eq:2.17aw}).
Defining the function $\hat{S}:\Z\rightarrow\R^3$
by $\hat{S}(n):=S(2\pi n)$ we observe that $\hat{S}(\cdot)$ is
a spin-vector trajectory of
$({\cal P}_\omega,A_{CT}[\omega,{\cal A}])$, i.e., in addition to
(\ref{eq:2.2.30}) we get
\begin{eqnarray}
&& \Psi[{\cal P}_\omega,A_{CT}[\omega,{\cal A}]](n;\phi_0+ \tilde{\Z}^d) =
\Phi_{CT}[\omega,{\cal A}](2\pi n;\phi_0) \; .
\label{eq:5.84aynnn}
\end{eqnarray}
\subsection{Group actions and cocycles}
\label{2.3}
We  now continue with the DS  defined by (\ref{eq:2.3b}) 
and we will define some group theoretical
notions underlying $\Psi[j,A],L[j]$ and $L[j,A]$ 
which will be crucial for the remainder of this work.
\setcounter{definition}{1}
\begin{definition}
\label{D2.1}
(Group)\\
\noindent A ``group'' is a pair $(G,*)$ where $G$ is a set and
$*$ is a binary operation 
such that
\begin{eqnarray}
&& {\rm (G0)\; (Binary\; operation)} \qquad ~~~~~~~~~~
\forall_{g_1,g_2\in G}\; (g_1*g_2)\in G \; ,
\nonumber\\
&& {\rm (G1)\; (Associativity)} \qquad ~~~~~~~~~~
\forall_{g_1,g_2,g_3\in G}\; (g_1*g_2)*g_3 = g_1*(g_2*g_3) \; ,
\nonumber\\
&& {\rm (G2)\; (Identity\; element}\;e_G) \qquad ~~
\exists_{e_G\in G}\; \forall_{g\in G}\; e_G=e_G*g=g*e_G \; ,
\nonumber\\
&& {\rm (G3)\; (Inverse\; elements)} \qquad ~~~~~~
\forall_{g_1\in G}\;
\exists_{g_2\in G}\; e_G=g_1*g_2=g_2*g_1 \; .
\nonumber
\end{eqnarray}
We will abbreviate $(G,*)$ as $G$ when the operation $*$ is clear from 
the context and we often write $g_1*g_2$ as $g_1g_2$ when
the operation $*$ is clear from the context.
The inverse element of a $g\in G$ is denoted by $g^{-1}$.
If $H$ is a subset of $G$ and if $g,g'\in G$ then we define
$gH g':=\lbrace g h g':h\in H\rbrace$.

A subset $G'$ of $G$ is called a ``subgroup of $G$'' if it is a group w.r.t.
to the restriction of $*$ to $G'$.
Two elements $g',g''$ of a group $G$ are called ``conjugate'' if
$g\in G$ exists such that $g''=g g' g^{-1}$. 
Two subgroups $G',G''$ of a group $G$ are called ``conjugate'' if
$g\in G$ exists such that $G''=gG'g^{-1}$.

A group $(G,*)$ is called ``Abelian'' if
\begin{eqnarray}
&& {\rm (G4)\; (Commutativity)} \qquad ~~~~~~~
\forall_{g_1,g_2\in G}\; g_1*g_2 = g_2*g_1 \; , ~~~~~~~~~~~~~~~~~
\nonumber
\end{eqnarray}
in which case $*$ is often replaced by  $+$.
\hfill $\Box$
\end{definition}
Important examples of groups in Section \ref{2.2} are
$(\Z,+)$,
$(SO(3),*)$ (in the latter case 
the binary operation is matrix multiplication).
%
\setcounter{definition}{2}
\begin{definition}
\label{D2.3}
($G$-action, $G$-set, isotropy group)\\
\noindent Consider a group $G$ and a set $E$.
Then a function $l:G\times E\rightarrow E$ is called a ``$G$-action on
$E$'' if, for $g_1,g_2\in G,x\in E$,
\begin{eqnarray}
&& l(e_G;x) = x \label{eq:nA.1} \\
&& l(g_1g_2;x) =  l(g_1;l(g_2;x)) \; .
\label{eq:nA.2}
\end{eqnarray}
If $l$ is a $G$-action on $E$ then the pair $(E,l)$ is
called a ``$G$-set''. 

Let $x\in E$. Then we denote by $Iso(E,l;x)$ the set of those $g\in G$
for which $x$ is a fixed point of
$l(g;\cdot)$ i.e.,
\begin{eqnarray}
&&  \hspace{-1cm}
Iso(E,l;x):=\lbrace g\in G:l(g;x)=x\rbrace \; .
\label{eq:12.17bnncaa}
\end{eqnarray}
Using (\ref{eq:nA.1}) and (\ref{eq:nA.2}) it is a simple exercise
to show that
$Iso(E,l;x)$ is a subgroup of $G$, and it is 
called the  ``isotropy group'' (or
``stabilizer group'') of $(E,l)$ at $x$. 
\hfill $\Box$
\end{definition}
If $(E,l)$ is a $G$-set then, for each $g\in G$, 
the function $l(g;\cdot):E\rightarrow E$ is onto since, for
every $y\in E$, the equality $l(g;x) = y$ 
is solved by $l(g^{-1};y) = x$.
Moreover $l(g;\cdot)$ is one-one since
the equality $l(g;x) = l(g;y)$ implies that
$x=l(g^{-1};l(g;x)) =l(g^{-1};l(g;y))=y$.
Thus $l(g;\cdot)$ is a bijection with
inverse $l(g^{-1};\cdot)$.
For the definition of ``bijection'', see Appendix \ref{A.1}.

It is clear by (\ref{eq:5.72b})
that $L[j]$ is a $\Z$-action on $\Td$ whence
$(\Td,L[j])$ is a $\Z$-set. 
Analogously it follows from (\ref{eq:5.72a}) that $L[j,A]$
is a $\Z$-action on $\Td\times\R^3$ and that
$(\Td\times\R^3,L[j,A])$ is a $\Z$-set. Apart from this
$\Z$-set we will see many more $\Z$-sets in this work which are
tied with $(j,A)$. In particular in Chapter \ref{10}
we will define an infinite collection of $\Z$-sets tied with $(j,A)$.
\setcounter{definition}{3}
\begin{definition}
\label{D2.3new}
($(E,l)$-orbit)\\
\noindent Let $(E,l)$ be a $G$-set. If $x\in E$ then the set
$\lbrace l(g;x):g\in G\rbrace$ is called the ``$(E,l)$-orbit of $x$''. 
We denote the set of $(E,l)$-orbits by $E/l$ and define the
function $l(G;\cdot):
E\rightarrow E/l$ by 
$l(G;x):=\lbrace l(g;x):g\in G\rbrace$.
Thus $l(G;x)$ is the $(E,l)$-orbit of $x$.

A $G$-set is called ``transitive'' if it has only one orbit, i.e.,
if $E$ is the $(E,l)$-orbit of every $x$ in $E$.
\hfill $\Box$
\end{definition}

The $E/l$ in Definition \ref{D2.3new}
is a partition of $E$ (see Appendix \ref{A.2}) and thus $l(G;\cdot)$ is well
defined and a surjection.
Isotropy groups are important tools for dealing with $G$-sets 
since they allow one to conveniently deal with the $(E,l)$-orbits.
In fact the isotropy groups  
of $SO(3)$-sets will be used in this capacity in
Chapter \ref{10} - in particular they will
give us more insights into 
particle-spin-vector 
motions and polarization-field motions (the latter being defined in Chapter 
\ref{6}). 
\setcounter{definition}{4}
\begin{definition}
\label{D2.2}
(Topological group)\\
\noindent A ``topological group'' is a group $(G,*)$ where $G$ is a
topological space, where  the binary operation
$*$ is continuous and where the function $g\mapsto g^{-1}$ on $G$
is also continuous.
\hfill $\Box$
\end{definition}
The above-mentioned groups $\Z$ and $SO(3)$
in Section \ref{2.2} are topological as we consider them to be
equipped with their standard topologies. Thus
the topology of $\Z$ is discrete, i.e., every subset of $\Z$ is open,
and $SO(3)$ is equipped with the subspace topology from $\R^{3\times 3}$
(for the latter notion, see Appendix A.3).
In this work we are often interested in $G$-sets
where $G$ and $E$ have a topology and $l$ is continuous.
This is formalized in the following definition.
\setcounter{definition}{5}
\begin{definition}
\label{D2.4}
($G$-space)\\
\noindent Let $(E,l)$ be a $G$-set, let $E$ be a topological space,
$G$ be a topological group,
and let $l\in{\cal C}(G\times E,E)$
where $G\times E$ carries the product topology.  
For the latter notion, see Appendix A.5.
Then the pair $(E,l)$ is called a ``$G$-space''.

The definitions of ``transitive'',``isotropy group'', $(E,l)$-orbit and $E/l$ 
are the same as for $G$-sets. Also, we equip 
each $(E,l)$-orbit
with the subspace topology from $E$.
\hfill $\Box$
\end{definition}
If $(E,l)$ is a $G$-space then each $l(g;\cdot)$ is a homeomorphism.
Moreover in the important subcase when the topology of $G$ is discrete
(e.g., when $G={\mathbb Z}$) the condition
that $l$ is continuous is equivalent to $l(g;\cdot)$ being
continuous for all $g\in G$.

Since, by (\ref{eq:3.20}), $L[j](n;\cdot)$ is continuous
it is clear that the $\Z$-set $(\Td,L[j])$ is a $\Z$-space
and  $L[j](n;\cdot)\in \Homeo(\Td)$.
Recalling that  $\Psi[j,A]$ is continuous, it
is equally clear by (\ref{eq:s2.10}) that $L[j,A](n;\cdot)$
is continuous whence the $\Z$-set
$(\Td\times\R^3,L[j,A])$ is also a $\Z$-space 
and $L[j,A](n;\cdot)\in \Homeo(\Td\times\R^3)$.

There are many books which cover $G$-spaces. A useful
book, dedicated exclusively to $G$-sets and $G$-spaces, is \cite{Ka}.
Clearly $G$-sets and $G$-spaces are $2$-tuples $(E,l)$. The use of the
terms set and space in this context simply arise out of the need for
simple names and the fact that $E$ is either a set or a topological space.
The term  $G$-set is synonymous with the term ``transformation group'' 
often used in the physics literature.

\vspace{3mm}

The spin transfer matrix function is an example of a cocycle and
(\ref{eq:3.25}) is the cocycle condition. We thus define:
\setcounter{definition}{6}
\begin{definition}
\label{D2.5}
(Cocycle)\\
\noindent
Let $(E,l)$ be a $G$-space and $K$ be a topological group.
Then a function $f\in{\cal C}(G\times E,K)$ is called a
``$K$ cocycle over the $G$-space $(E,l)$''
if, for $g,g'\in G,x\in E$,
\begin{eqnarray}
&& f(gg';x)=f(g;l(g';x))f(g';x) \; .
\label{eq:nsA.3xbb}
\end{eqnarray}
Here $G\times E$ carries the product topology.
\hfill $\Box$
\end{definition}
For literature on cocycles, see, e.g., \cite{KR,Zi1} and Chapter 1 in
\cite{HK1}.
The reader will easily appreciate the similarity between the structures of
(\ref{eq:3.25}) and (\ref{eq:nsA.3xbb}) and the correspondence
between the functions $\Psi[j,A]\in{\cal C}(\Z\times \Td,SO(3))$ and
$f\in{\cal C}(G\times E,K)$.
Since $(\Td,L[j])$ is a $\Z$-space and $SO(3)$ is a topological
group, the set of $SO(3)$ cocycles over $(\Td,L[j])$ is well defined.
In fact since $\Psi[j,A]\in{\cal C}(\Z\times \Td,SO(3))$ it
follows from (\ref{eq:3.25}) that,
for every $(j,A)\in{\cal SOS}(d,j)$, $\Psi[j,A]$
is a $SO(3)$ cocycle over $(\Td,L[j])$.
Conversely, every $SO(3)$ cocycle $\Psi$ over $(\Td,L[j])$
is the spin transfer matrix function of a
spin-orbit system since,
by defining $A:=\Psi(1;\cdot)$, we have $\Psi[j,A]=\Psi$ so that
$\Psi$ is the spin transfer matrix function of $(j,A)$.
Clearly the cocycle property of the spin transfer matrix 
function $\Psi[j,A]$
is an important structure of the particle-spin-vector motion
of spin-orbit systems and in Section \ref{10.2.2} 
their importance
will be extended to more general spin variables. 
Furthermore cocycles are of key importance in the underlying bundle theory
(see Section \ref{10.7}).
\section{Polarization-field trajectories and invariant 
polarization fields}
\setcounter{equation}{0}
\label{6}
In this chapter we introduce the notions of polarization field,
invariant polarization field, spin field and invariant spin field and we
present their most basic properties.
\subsection{Generalities}
\label{6.1}
High precision measurements of the spin-dependent properties of the 
interactions of colliding particles 
in storage rings depend on the equilibrium
spin polarization being maximized. This, in turn, is facilitated by an understanding of the meaning
of the term equilibrium, both as it applies to the value of the polarization and
to its direction at each point in phase space. We will return to these matters in Section {\ref{6.4}}
but continue now  with a definition and an exploration of the effects of maps.

Suppose therefore that $(j,A)\in{\cal SOS}(d,j)$ and that,
at time $n=0$, a spin vector $S_{z_0}$ has been assigned to
every point $z_0=\phi_0+ \tilde{\Z}^d\in\Td$ of the ``particle torus'' and
consider their evolution according to (\ref{eq:3.1}). Let $S_{z_0}(\cdot)$
denote the spin-vector trajectory with the
initial value $S_0=S_{z_0}(0)$. We
define the field trajectory ${\cal S}={\cal S}(n,z)$ by
${\cal S}(n,j^n(z))=S_z(n)$ where $n$ and $z$ vary over
$\Z$ and $\Td$ respectively. Clearly ${\cal S}(n,\cdot)$ is the
distribution of spins which started at $n=0$ with the assignments
$S_{z_0}$ and evolved under the dynamics of (\ref{eq:3.1}).
Since (\ref{eq:3.1}) gives us
$S_z(n+1)=A(j^n(z))S_z(n)$, we have
\begin{eqnarray}
&& {\cal S}(n+1,z)=A\biggl(j^{-1}(z)\biggr)
{\cal S}\biggl(n,j^{-1}(z)\biggr) \; .
\label{eq:4.1}
\end{eqnarray}
\setcounter{definition}{0}
\begin{definition} \label{D6.1}
(Polarization-field trajectory, invariant polarization field, ISF)\\
\noindent Let $(j,A)\in{\cal SOS}(d,j)$.
We call a function ${\cal S}\in{\cal C}({\mathbb Z}\times \Td,{\mathbb R}^3)$
a ``polarization-field trajectory of $(j,A)$'',
if it satisfies the evolution equation (\ref{eq:4.1}).
Clearly ${\cal S}(n,\cdot)\in {\cal C}(\Td,\R^3)$ and we
call ${\cal S}(0,\cdot)$ the ``initial value of ${\cal S}$''.
A function $f\in {\cal C}(\Td,\R^3)$ is called an ``invariant
polarization field of $(j,A)$'' if it satisfies
\begin{eqnarray}
&& f\circ j =A f
\; .
\label{eq:4.2a}
\end{eqnarray}
A polarization-field trajectory ${\cal S}$ is also called a 
``spin-field trajectory'' if $|{\cal S}|=1$. 
An invariant polarization field $f$ is called an
``invariant spin field (ISF)'' if $|f|=1$.
We denote the set of invariant spin fields of $(j,A)$ by ${\cal ISF}(j,A)$. 
\hfill $\Box$
\end{definition}
At (\ref{eq:2.3b}) we defined the function ${\cal P}[j,A](z,S)$
for transporting particles and their spin vectors.

The IRT, Theorem \ref{P02xxx} in
Section \ref{10.6.1}, will show that the ISF
is a rather deep concept.

\noindent We now define the function 
$\tilde{\cal P}[j,A]:{\cal C}(\Td,{\mathbb R}^3)
\rightarrow  {\cal C}(\Td,{\mathbb R}^3)$ for the field evolution by
\begin{eqnarray}
&& \tilde{\cal P}[j,A](f):=(Af)\circ j^{-1} \; ,
\label{eq:xx13.2n}
\end{eqnarray}
i.e., $(\tilde{\cal P}[j,A](f))(z):=A(j^{-1}(z))f(j^{-1}(z))$
where $f\in{\cal C}(\Td,{\mathbb R}^3)$.
It is an easy exercise to show that, for $(j,A)\in{\cal SOS}(d,j)$ and
$(j',A')\in{\cal SOS}(d,j')$,
\begin{eqnarray}
&& \tilde{\cal P}[j',A']\circ  \tilde{\cal P}[j,A]
=  \tilde{\cal P}[j'\circ j,A''] \; ,
\label{eq:2.55}
\end{eqnarray}
where $A''\in {\cal C}(\Td,SO(3))$ is defined by 
$A'':=(A'\circ j)A$, whence
\begin{eqnarray}
&& \tilde{\cal P}[j,A] =
\tilde{\cal P}[j,A_{d,0}]\circ \tilde{\cal P}[id_\Td,A] \; ,
\label{eq:2.56}
\end{eqnarray}
where $A_{d,0}\in{\cal C}(\Td,SO(3))$ is defined by
$A_{d,0}(z):=I_{3\times 3}$. Then the inverse, $\tilde{\cal P}[j,A]^{-1}$,
of $\tilde{\cal P}[j,A]$ is given by
\begin{eqnarray}
&& \tilde{\cal P}[j,A]^{-1} =
\tilde{\cal P}[id_\Td,A^t] \circ \tilde{\cal P}[j^{-1},A_{d,0}] 
= \tilde{\cal P}[j^{-1},A^t\circ j^{-1} ]
\; .
\label{eq:2.57}
\end{eqnarray}
Thus $\tilde{\cal P}[j,A]$ is a bijection so that the function
$\tilde{L}[j,A]:{\mathbb Z}
\times{\cal C}(\Td,{\mathbb R}^3)
\rightarrow  {\cal C}(\Td,{\mathbb R}^3)$, defined by
\begin{eqnarray}
&& \tilde{L}[j,A](n;\cdot):= \tilde{\cal P}[j,A]^n \; ,
\label{eq:xx13.2}
\end{eqnarray}
is a ${\mathbb Z}$-action on ${\cal C}(\Td,{\mathbb R}^3)$
where $\tilde{\cal P}[j,A]^n$ denotes the
$n$-th iteration of $\tilde{\cal P}[j,A]$. Clearly
$({\cal C}(\Td,{\mathbb R}^3),\tilde{L}[j,A])$ is a ${\mathbb Z}$-set.
Note that, by (\ref{eq:3.5}), (\ref{eq:xx13.2n}) and (\ref{eq:xx13.2}),
\begin{eqnarray}
&& \tilde{L}[j,A](n;f) =
\biggl(\Psi[j,A](n;\cdot)f\biggr)\circ L[j](-n;\cdot)
 \; ,
\label{eq:xx13.2nn}
\end{eqnarray}
i.e., $(\tilde{L}[j,A](n;f))(z)=\Psi[j,A](n;L[j](-n;z))f(L[j](-n;z))$.
Of course, with (\ref{eq:xx13.2n}) the evolution equation
(\ref{eq:4.1}) can be written as
${\cal S}(n+1,\cdot)=\tilde{\cal P}[j,A]({\cal S}(n,\cdot))$
whence, by (\ref{eq:xx13.2}), for every polarization-field trajectory 
${\cal S}$
\begin{eqnarray}
&& {\cal S}(n,\cdot)=\tilde{L}[j,A](n;{\cal S}(0,\cdot)) \; .
\label{eq:x13.3}
\end{eqnarray}
\subsection{Invariant polarization fields}
\label{6.1b}
In this section we take a closer look at invariant polarization fields.

We first recall from (\ref{eq:4.1}) that
if ${\cal S}$ is a polarization-field trajectory 
of $(j,A)$ then
\begin{eqnarray}
&& {\cal S}(n+1,j(z))=A(z){\cal S}(n,z) \; ,
\label{eq:4.2aa}
\end{eqnarray}
whence if ${\cal S}$ is also time-independent 
then
\begin{eqnarray}
&& {\cal S}(n,j(z))=
{\cal S}(n+1,j(z))=A(z){\cal S}(n,z) \; .
\label{eq:4.2ab}
\end{eqnarray}
It follows from (\ref{eq:4.2aa}) and (\ref{eq:4.2ab}) and Definition \ref{D6.1}
and by induction in $n$ that if ${\cal S}$ is a polarization-field trajectory 
of $(j,A)$ then ${\cal S}$ is time-independent iff its initial value
${\cal S}(0,\cdot)$
is an invariant polarization field of $(j,A)$.

Invariant polarization fields play an important role in polarized
beam physics since they can be used to estimate the maximum attainable polarization of a bunch as we explain
in Section \ref{6.4}, 
and since they are closely tied to the
notions of spin tune and spin-orbit resonance 
(see Chapter \ref{4}). 
In fact as indicated  in the Introduction invariant polarization fields are central to this work.
This becomes especially clear when we generalize
the notions of invariant polarization field to the notion 
of invariant $(E,l)$-field in Chapter \ref{10} whereby
(\ref{eq:4.2a}) will turn out to be a so-called stationarity equation.

We now make some comments on the question of the existence of the ISF for 
spin-orbit systems in ${\cal SOS}(d,j)$. 
It should be clear that the constraints involved in the definition of the 
ISF are nontrivial. However,
if a spin-orbit system $(j,A)$ has an ISF $f$ then
$-f$ is also an ISF of $(j,A)$. So since $f\neq -f$, if $(j,A)$ has
a finite number of ISF's, then this number is even.
The important subcase where $(j,A)$ has exactly two ISF's is dealt with in
Chapter \ref{VII}. 

It is also known \cite{BV1} and examined in
Section \ref{10.4b.2}
that spin-orbit systems exist which are
on orbital resonance and which have no continuous ISF of the kind that we treat here. 
At the same time there are some indications, mainly from
numerical computations on ISF's, that practically relevant
spin-orbit systems which have no ISF are ``rare''.
Thus we state the following conjecture, which we call the ``ISF-conjecture'':
If $(j,A)$ is a spin-orbit system such that $j$ is topologically transitive
then $(j,A)$ has an ISF. Note that a special case of this conjecture is:
If a spin-orbit system $({\cal P}_\omega,A)$ is off orbital resonance, 
then it has an ISF.

The ISF-conjecture is, at least to our knowledge, unresolved.
The question of the existence of the ISF is widely considered 
important both as a theoretical matter and as it relates to the
practical matter of deciding whether a beam can have stable, non-vanishing
polarization. 
Chapter \ref{10} presents a new framework for discussing it.

Since the ISF-conjecture deals with topologically transitive $j$
we state the following theorem which considers
this situation.
\setcounter{theorem}{1}
\begin{theorem}
\label{T09t0}
Let $(j,A)\in{\cal SOS}(d,j)$ where $j$ is topologically transitive.
If $f$ is an invariant polarization field 
of $(j,A)$ then $|f|$ is constant, i.e., $|f(z)|$ is independent
of $z$. Also
$(j,A)$ has an ISF iff it has an invariant polarization field 
which is not identically zero.
\end{theorem}
\noindent{\em Proof of Theorem \ref{T09t0}:}
Let $f$ be an invariant polarization field 
of $(j,A)$. Then, by Definition \ref{D6.1}, 
$f\in{\cal C}(\Td,\R^3)$ and
\begin{eqnarray}
&&  \hspace{-1cm}
|f(j(z))| = |f(z)| \; .
\label{eq:er1}
\end{eqnarray}
We pick a $z_0\in\Td$ such that the set $B:=\lbrace j^n(z_0):n\in\Z\rbrace$ 
is dense in $\Td$, i.e., $\overline{B}=\Td$ and we define
$S_0:=f(z_0)$. Since $f$ is an invariant polarization field we have 
$B\subset C:=\lbrace z\in\Td:|f(z)|=|S_0|\rbrace$.
On the other hand, the sphere of radius $|S_0|$, i.e., the set
$\lbrace S\in\R^3:|S|=|S_0|\rbrace$
is a closed subset of $\R^3$ whence, because
$f$ is continuous, $C$ is a closed subset of
$\Td$. Therefore $\Td=\overline{B}\subset\overline{C}=C$ so that  
$\Td=C$. Thus, by the definition of $C$,
we conclude that $|f(z)|=|S_0|$ for all $z\in\Td$.

To prove the second claim, let $f$ be an invariant polarization field 
of $(j,A)$ which is not identically zero.
Clearly by the first claim $|f|$ is constant and takes a nonzero value
because $|f|$ is not identically zero. Thus we define
$g\in{\cal C}(\Td,\R^3)$ by $g:=f/|f|$ whence, by Definition \ref{D6.1}, 
$g$ is an ISF of $(j,A)$. Conversely every ISF of
$(j,A)$ is an invariant polarization field 
of $(j,A)$ which is not identically zero.
\hfill $\Box$

In the special 
case when $j={\cal P}_\omega$ with $(1,\omega)$ nonresonant one can prove
Theorem \ref{T09t0} alternatively
by some simple Fourier Analysis of $f$ \cite{He2}.
With Theorem \ref{T09t0}, the ISF conjecture is equivalent to the 
following statement: If $j$ is topologically transitive
then  $(j,A)$ has an invariant polarization field
which is not identically zero.
Note also that Theorem \ref{T09t0} will be generalized
by Lemma \ref{L10.4}.

A less formal picture surrounding Theorem \ref{T09t0} is as follows. 
When $j$ is topologically transitive, the whole of $\Td$ can 
effectively be reached from any starting position
$z_0$ by repeated application of $j$. Moreover, by a corresponding 
repeated application of $A$, $f(z_0)$   
generates $f(z)$ at effectively all points on $\Td$. 
So the $f(z)$ on $\Td$ are all ``connected''. Also,  since
$A$ is $SO(3)$-valued all the $|f(z)|$ are the same. 
On the other hand, if $j$ is not transitive, 
the $f(z)$ at different $z$ need not be connected. For example the 
$f(z)$ at adjacent $z$ could have opposite signs. 
We will encounter a related situation in Section \ref {10.4b.2} for 
$j={\cal P}_\omega$ with $\omega = 1/2$ and in Remark 12 of
Chapter \ref{10}.
\section{Transforming spin-orbit systems}
\label{3}
\setcounter{equation}{0}
In this chapter we introduce the transformation of any
$(j,A)\in{\cal SOS}(d,j)$ under any $T\in{\cal C}(\Td,SO(3))$ and
we show how this is accompanied by a transformation of 
${\cal P}[j,A]$ and $\tilde{\cal P}[j,A]$ as well as
by a transformation of particle-spin-vector trajectories
and polarization-field trajectories.
\subsection{The transformation rule of spin-orbit systems}
\label{3.1}
We now show how to partition ${\cal SOS}(d,j)$
into subsets within which the dynamics is similar.
Consider  $(j,A)\in{\cal SOS}(d,j)$ and let
$(z(\cdot),S(\cdot))$ be
a particle-spin-vector trajectory of $(j,A)$, i.e., let 
(\ref{eq:2.J20a}) hold so that $S(\cdot)$ is a spin-vector trajectory 
of $(j,A)$ and thus $S(n+1)=A(z(n))S(n)$.
For arbitrary $T\in{\cal C}(\Td,SO(3))$, the function
$S':\Z\rightarrow\R^3$ defined by 
$S'(n) := T^t(z(n))S(n)$ 
satisfies $S'(n+1)=T^t(z(n+1))A(z(n))T(z(n))S'(n)$. So
$S'(\cdot)$ is a spin-vector trajectory of a new spin-orbit system, namely of
$(j,A')\in{\cal SOS}(d,j)$ which is defined by
\begin{eqnarray}
&&  A'(z): =T^t( j(z))A(z)T(z) \; .
\label{eq:5.25}
\end{eqnarray}
Note that (\ref{eq:5.25}) implies $A(z)=T(j(z))A'(z)T^t(z)$. Thus
$(z(\cdot),S'(\cdot))$ is
a particle-spin-vector trajectory of $(j,A')$. 
Recalling from (\ref{eq:2.3b}) that
${\cal P}[j,A](z,S) =  \left( \begin{array}{c}
j(z)\\ A(z)S
\end{array}\right)$ and
${\cal P}[j,A'](z,S) =  \left( \begin{array}{c}
j(z)\\ A'(z)S
\end{array}\right)$
it is easy to show that (\ref{eq:5.25}) holds iff
\begin{eqnarray}
&& {\cal P}[id_\Td,T]^{-1}\circ {\cal P}[j,A]\circ 
{\cal P}[id_\Td,T]
 = {\cal P}[j,A'] \; .
\label{eq:2.3bna}
\end{eqnarray}
Eq. (\ref{eq:5.25}) gives
rise to a partition of ${\cal SOS}(d,j)$ as we formalize
in the next two definitions.
\setcounter{definition}{0}
\begin{definition}
\label{D3a}
(Transformation rule of spin-orbit systems)\\
\noindent
Let $(j,A)$ and $(j,A')$ be in ${\cal SOS}(d,j)$.
Then a $T$ in ${\cal C}(\Td,SO(3))$ is called a
``transfer field from $(j,A)$ to $(j,A')$'' iff
(\ref{eq:5.25}) holds. We also say that
``$(j,A')$ is the transform of $(j,A)$ under $T$''.
We denote the collection of all
transfer fields from $(j,A)$ to $(j,A')$ by
${\cal TF}(A,A';d,j)$.
Note that if $T\in{\cal TF}(A,A';d,j)$ then
$T^t\in{\cal TF}(A',A;d,j)$, i.e., 
$(j,A)$ is the transform of $(j,A')$ under $T^t$.
\hfill $\Box$
\end{definition}
Clearly ${\cal TF}(A,A';d,j)\neq\emptyset$ iff
$(j,A')$ is a transform of $(j,A)$ as in (\ref{eq:5.25}).
Note that in general we don't have transfer fields, i.e.,
${\cal TF}(A,A';d,j)=\emptyset$
(see, e.g., Remark 9 in Chapter \ref{4}).

Following Appendix \ref{A.2} we make the definition:
\setcounter{definition}{1}
\begin{definition}
\label{D3b}
Let $(j,A)$ and $(j,A')$ be in ${\cal SOS}(d,j)$.
Then we write $(j,A)\sim(j,A')$
and say that $(j,A)$ and $(j,A')$ are ``equivalent'' iff
$(j,A')$ is a transform of $(j,A)$ under some
$T\in {\cal C}(\Td,SO(3))$.
Clearly the relation $\sim$ is symmetric, reflexive, and transitive and
thus is an equivalence relation on ${\cal SOS}(d,j)$.
Let $\overline{(j,A)}:=\lbrace (j,A''):(j,A'')\sim(j,A)\rbrace$, i.e.,
the equivalence class of $(j,A)$ under $\sim$.
As outlined in Appendix \ref{A.2}, the sets
$\overline{(j,A)}$ partition ${\cal SOS}(d,j)$.
\hfill $\Box$
\end{definition}
Two spin-orbit systems which are equivalent 
share many important properties, e.g.,
the existence or nonexistence of an ISF (see Remark 3 below).
We will see other properties shared by equivalent
spin-orbit systems throughout this work.
For checking those shared properties it can be convenient to check
them for a ``simple'' element of $\overline{(j,A)}$
(see especially Chapters \ref{4.3} and \ref{4}).

The transformation rule 
$(j,A)\longrightarrow(j,A')$ also gives the following
transformation rule of spin transfer matrix functions:
\begin{eqnarray}
&& \Psi[j,A]\longrightarrow \Psi[j,A'] \; .
\label{eq:120.109tannsfx}
\end{eqnarray}
It follows from (\ref{eq:3.25}) and (\ref{eq:5.25}) and
by induction in $n$ that, if 
$T\in{\cal TF}(A,A';d,j)$,
with $T\in{\cal C}(\Td,SO(3))$ 
then the spin transfer matrix functions
of $(j,A) \in{\cal SOS}(d,j)$ and $(j,A')
\in{\cal SOS}(d,j)$ are related by
\begin{eqnarray}
&& \Psi[j,A'](n;z)= T^t( L[j](n;z))
\Psi[j,A](n;z)T(z) \; .
\label{eq:5.30}
\end{eqnarray}
Recall from Section \ref{2.3} that $\Psi[j,A]$ and 
$\Psi[j,A']$ are cocycles. Then (\ref{eq:5.30}) implies that the cocycles
$\Psi[j,A]$ and $\Psi[j,A']$ are
``cohomologous''. For this notion, see, e.g.,
\cite{He2,KR,Zi1} and Chapter 1 in \cite{HK1}.
\subsection{Transforming particle-spin-vector trajectories and
polarization-field trajectories. Topological $G$-maps of $G$-spaces}
\label{3.2}
With Definition \ref{D3a} we arrive at the following 
transformation rule of $\Z$-actions:
\begin{eqnarray}
&& L[j,A]\longrightarrow L[j,A'] \; ,
\label{eq:120.109tannsf} \\
&& \tilde{L}[j,A]\longrightarrow 
\tilde{L}[j,A']
\; ,
\label{eq:120.109tannpf}
\end{eqnarray}
where $A,A'$ are related by (\ref{eq:5.25}) with
$T\in{\cal C}(\Td,SO(3))$
and $(j,A) \in{\cal SOS}(d,j)$.
It is easy to see how the $\Z$-actions $L[j,A]$
and $L[j,A']$ in the transformation rule 
(\ref{eq:120.109tannsf}) are related.
In fact it follows from (\ref{eq:2.3bna}) that
\begin{eqnarray}
&& {\cal P}[id_\Td,T]^{-1}\circ {\cal P}[j,A]^n\circ 
{\cal P}[id_\Td,T]
 = {\cal P}[j,A']^n \; .
\label{eq:2.3bnc}
\end{eqnarray}
Therefore, by (\ref{eq:s2.10}),
$L[j,A'](n;\cdot)= {\cal P}[id_\Td,T]^{-1}\circ
L[j,A](n;\cdot)\circ {\cal P}[id_\Td,T]$, so that
\begin{eqnarray}
&& {\cal P}[id_\Td,T]^{-1}\circ L[j,A](n;\cdot)
= L[j,A'](n;\cdot)\circ {\cal P}[id_\Td,T]^{-1} \; .
\label{eq:s2.10b}
\end{eqnarray}
Moreover it is easy to see how the $\Z$-actions 
$\tilde{L}[j,A]$ and $\tilde{L}[j,A']$ 
in the transformation rule (\ref{eq:120.109tannpf}) are related.
In fact we conclude from (\ref{eq:2.55}) that
\begin{eqnarray}
&& \tilde{\cal P}[id_\Td,T^t]=\tilde{\cal P}[id_\Td,T]^{-1} \; , 
\label{eq:4.2.20}\\
&&
\tilde{\cal P}[id_\Td,T]^{-1}\circ\tilde{\cal P}[j,A]\circ
\tilde{\cal P}[id_\Td,T] =\tilde{\cal P}[j,A'] \; ,
\label{eq:5.1n}
\end{eqnarray}
whence, by (\ref{eq:xx13.2}),
\begin{eqnarray}
&& \tilde{\cal P}[id_\Td,T]^{-1}\circ \tilde{L}[j,A](n;\cdot)
= \tilde{L}[j,A'](n;\cdot)\circ \tilde{\cal P}[id_\Td,T]^{-1} \; .
\label{eq:5.1na}
\end{eqnarray}
The following definition provides a simple classification of the
relations (\ref{eq:s2.10b}) and (\ref{eq:5.1na}).
Recall that  $G$-sets and  $G$-spaces are defined in
Definitions \ref{D2.3} and \ref{D2.4}.
\setcounter{definition}{2}
\begin{definition}
\label{D3c}
($G$-maps of $G$-sets, topological $G$-maps of $G$-spaces)\\
\noindent
a) Consider $G$-sets $(E_1,l_1)$ and $(E_2,l_2)$. 
A function $f:E_1\rightarrow E_2$
is called a ``$G$-map from $(E_1,l_1)$ to $(E_2,l_2)$'' if, 
for every $g\in G$, $f\circ l_1(g;\cdot)=l_2(g;f(\cdot))$, i.e., if for
every $g\in G,x\in E_1$,
\begin{eqnarray}
&& f(l_1(g;x))=l_2(g;f(x)) \; .
\label{eq:A.55n}
\end{eqnarray}

\noindent
b)  Consider the $G$-spaces
$(E_1,l_1)$ and $(E_2,l_2)$  and let $f\in{\cal C}(E_1,E_2)$.
If $f$ satisfies (\ref{eq:A.55n})
then $f$ is called a ``topological $G$-map from $(E_1,l_1)$ to $(E_2,l_2)$''.
\hfill $\Box$
\end{definition}
If $f$ is a $G$-map from the $G$-set $(E_1,l_1)$ to the $G$-set
$(E_2,l_2)$ and if $f$ is a bijection, 
then $f^{-1}$ is a $G$-map from $(E_2,l_2)$ to $(E_1,l_1)$ and
$(E_2,l_2)$ and $(E_1,l_1)$ are said to be
``isomorphic''.
We then also say that $l_2$ and $l_1$ are ``isomorphic'' and
that $f$ is an ``isomorphism'' from $(E_1,l_1)$ to $(E_2,l_2)$.
Note that isomorphic $G$-sets share many properties.

Analogously, when
$f$ is a topological
$G$-map from the $G$-space $(E_1,l_1)$ to the $G$-space
$(E_2,l_2)$ and if $f$ is a homeomorphism then
$(E_2,l_2)$ and $(E_1,l_1)$ are
said to be ``isomorphic''.
We then also say that $f$ is an 
``isomorphism'' from $(E_1,l_1)$ to $(E_2,l_2)$.
Clearly then $f^{-1}$ is an isomorphism from $(E_2,l_2)$ to $(E_1,l_1)$.
Note that isomorphic $G$-spaces share many properties.

It is clear how the
relations (\ref{eq:s2.10b}) and  (\ref{eq:5.1na}) can be
formulated in terms of Definition \ref{D3c}. 
First, since ${\cal P}[id_\Td,T]^{-1}\in \Homeo(\Td\times\R^3)$
it follows from (\ref{eq:s2.10b}) and Definitions \ref{D3a} and \ref{D3c} 
that if $T$ is a transfer field from $(j,A)$ to
$(j,A')$ then ${\cal P}[id_\Td,T]^{-1}$ is an isomorphism
from the $\Z$-space $(\Td\times\R^3,L[j,A])$ to the $\Z$-space
$(\Td\times\R^3,L[j,A'])$.
In particular, the transformation rule (\ref{eq:120.109tannsf}) 
relates isomorphic $\Z$-spaces. Secondly, since $\tilde{\cal P}[id_\Td,T]^{-1}$
is a bijection it follows from 
(\ref{eq:5.1na}) and Definition \ref{D3c} that $\tilde{\cal P}[id_\Td,T]^{-1}$
is an isomorphism from the $\Z$-set
$({\cal C}(\Td,{\mathbb R}^3),\tilde{L}[j,A])$
to the $\Z$-set
$({\cal C}(\Td,{\mathbb R}^3),\tilde{L}[j,A'])$.

The transformation rules (\ref{eq:120.109tannsf}) and
(\ref{eq:120.109tannpf}) will be generalized in 
Section \ref{10.2.3}.
In particular, as we point out in
Section \ref{10.7}, they derive from an
$SO(3)$-gauge transformation rule.

The transformation rules (\ref{eq:120.109tannsf}) and 
(\ref{eq:120.109tannpf}) transform $\Z$-actions, i.e., they transform dynamics.
We now supplement (\ref{eq:120.109tannsf}) and
(\ref{eq:120.109tannpf}) 
by transformation rules of the underlying histories, i.e., 
transformation rules of particle-spin-vector 
trajectories and polarization-field
trajectories. First of all, 
as mentioned at the beginning of this section,
we arrive at the transformation rule of
particle-spin-vector trajectories:
\begin{eqnarray}
&& (z(\cdot),S(\cdot)) \longrightarrow (z(\cdot),S'(\cdot)) \; , \quad
S'(n) := T^t(z(n))S(n) \; .
\label{eq:5.35}
\end{eqnarray}
Clearly if $(z(\cdot),S(\cdot))$ is a
particle-spin-vector trajectory of $(j,A)$ then
$(z(\cdot),S'(\cdot))$ is a particle-spin-vector trajectory of $(j,A')$.

In parallel to (\ref{eq:5.35}) one can also transform polarization-field
trajectories.
In fact if $f$ is the initial value 
of a polarization-field trajectory ${\cal S}$ of
$(j,A)$ then we can relate it to the polarization-field trajectory 
${\cal S}'$
of $(j,A')$ whose initial value is $\tilde{\cal P}[id_\Td,T]^{-1}(f)=T^tf$.
Thus, by (\ref{eq:x13.3}),
\begin{eqnarray}
&& {\cal S}(n,\cdot)=\tilde{L}[j,A](n;f) \; , \quad
{\cal S}'(n,\cdot)=\tilde{L}[j,A'](n;T^tf) \; ,
\label{eq:x13.3na}
\end{eqnarray}
whence, by (\ref{eq:xx13.2}) and (\ref{eq:5.1n}),
\begin{eqnarray}
&&  {\cal S}'(n,\cdot)=\tilde{L}[j,A'](n;{\cal S}'(0,\cdot))
\nonumber\\
&& \quad =  \tilde{\cal P}[j,A']^n ({\cal S}'(0,\cdot))
=(\tilde{\cal P}[id_\Td,T]^{-1}\circ\tilde{\cal P}[j,A]^n\circ
\tilde{\cal P}[id_\Td,T]) ({\cal S}'(0,\cdot))
\nonumber\\
&&
=(\tilde{\cal P}[id_\Td,T]^{-1}\circ\tilde{\cal P}[j,A]^n\circ
\tilde{\cal P}[id_\Td,T] \circ \tilde{\cal P}[id_\Td,T]^{-1})(f)
=\tilde{\cal P}[id_\Td,T]^{-1}(\tilde{\cal P}[j,A]^n(f))
\nonumber\\
\qquad
&& =\tilde{\cal P}[id_\Td,T]^{-1}(\tilde{L}[j,A](n,f))
=\tilde{\cal P}[id_\Td,T]^{-1}({\cal S}(n,\cdot))
\; ,
\label{eq:5.1nb}
\end{eqnarray}
i.e.,
\begin{eqnarray}
&&  {\cal S}'(n,\cdot)=\tilde{\cal P}[id_\Td,T]^{-1}({\cal S}(n,\cdot))
\; .
\label{eq:5.1nc}
\end{eqnarray}
We conclude from (\ref{eq:5.1nc}) that
if ${\cal S}$ is a polarization-field trajectory of $(j,A)$
then ${\cal S}'$, defined by (\ref{eq:5.1nc}),
is a polarization-field trajectory of $(j,A')$. Thus with
(\ref{eq:5.1nc}) we have a natural transformation rule
of polarization-field trajectories. 
Note that (\ref{eq:xx13.2n}) and
(\ref{eq:5.1nc}) give us 
\begin{eqnarray}
&&  {\cal S} \longrightarrow {\cal S}' \; , \quad
{\cal S}'(n,z): =T^t(z){\cal S}(n,z) \; .
\label{eq:5.1}
\end{eqnarray}
With (\ref{eq:5.1}) and by the remarks after (\ref{eq:4.2ab})
we have the following transformation rule of invariant polarization fields:
\begin{eqnarray}
&&  f \longrightarrow f' \; , \quad
f'(z): =T^t(z)f(z) \; .
\label{eq:5.1j}
\end{eqnarray}
In fact if $f$ is an invariant polarization field of $(j,A)$
then $f'$, defined by (\ref{eq:5.1j}),
is an invariant polarization field of $(j,A')$.
In Section \ref{10.2} we will generalize the notions of 
particle-spin-vector trajectory, 
polarization-field trajectory, and invariant
polarization field. Then the transformation rules 
(\ref{eq:5.35}), (\ref{eq:5.1}) and (\ref{eq:5.1j})
will be generalized accordingly.

\vspace{3mm}
\noindent{\bf Remarks:}
\begin{itemize}
\item[(1)]
The transformation rules 
(\ref{eq:120.109tannsf}), (\ref{eq:120.109tannpf}), (\ref{eq:5.35}),
(\ref{eq:5.1}) and
(\ref{eq:5.1j}) are no strangers to the polarized-beam community. In fact
when researchers deal with the topics of spin tune, spin frequency,
spin resonances,
resonance strengths etc. then they often appeal more or less directly to
these transformation rules. In those applications the aim, typically, 
is to transform $(j,A)$ to a ``simple''  $(j,A')$. 
\item[(2)]
The transformation rule (\ref{eq:5.35}) could be generalized to
\begin{eqnarray}
&& (z(\cdot),S(\cdot)) \longrightarrow (z(\cdot),S'(\cdot)) \; , \quad
S'(n) := R^t(n,z(\cdot))S(n) \; ,
\label{eq:5.35n}
\end{eqnarray}
where $R(n,z(\cdot))$ generalizes $T(z(n))$ by allowing an arbitrary
dependence on the particle trajectory $z(\cdot)$. 
However, as can be easily shown \cite{He2}, if 
$({\cal P}_\omega,A)\in {\cal SOS}(d,{\cal P}_\omega)$ then, in general the
$(z(\cdot),S'(\cdot))$ in (\ref{eq:5.35n}) is not the 
particle-spin-vector
trajectory of {\it any} $({\cal P}_\omega,A')\in {\cal SOS}(d,{\cal P}_\omega)$! 
\item[(3)] 
It is clear that (\ref{eq:5.1}) maps the polarization-field trajectories
of $(j,A)$ bijectively onto the set of polarization-field trajectories of 
$(j,A')$. It is equally clear that (\ref{eq:5.1j}) maps ${\cal ISF}(j,A)$
bijectively onto ${\cal ISF}(j,A')$.
In particular equivalent
spin-orbit systems have the same number of ISF's.
\item[(4)] When transforming $(j,A)$ to a ``simple''  $(j,A')$ 
the polarization-field trajectories of the latter are  ``simple'' too whence
(\ref{eq:5.1}) allows one to compute polarization-field trajectories 
by transforming
 ``simple'' polarization-field trajectories.
For this philosophy see Chapters \ref{4.3} and \ref{4} too.
\hfill $\Box$
\end{itemize}
\subsection{Remarks on conjugate $1$-turn particle-spin-vector 
maps and structure preserving homeomorphisms}
\label{3.2n}
Note that $\Homeo(\Td\times\R^3)$ forms a group, 
where the group multiplication 
is understood to be the composition of functions.
Thus, since ${\cal P}[j,A]\in \Homeo(\Td\times\R^3)$, it follows from
(\ref{eq:2.3bna}) and Definitions \ref{D2.1} and \ref{D3b} that if
$(j,A)\sim(j,A')$
then ${\cal P}[j,A]$ and ${\cal P}[j,A']$
are conjugate elements of the group 
$\Homeo(\Td\times\R^3)$, i.e., a  ${\cal T}\in \Homeo(\Td\times\R^3)$ 
exists such that 
${\cal P}[j,A']={\cal T}^{-1}\circ{\cal P}[j,A]
\circ{\cal T}$.
In fact ${\cal T}={\cal P}[id_\Td,T]$ with
$T\in{\cal TF}(A,A';d,j)$ is an example.
We call a ${\cal T}\in \Homeo(\Td\times\R^3)$ ``structure preserving for 
a $(j,A)\in{\cal SOS}(d,j)$'' if the homeomorphism
${\cal T}^{-1}\circ{\cal P}[j,A]\circ{\cal T}$
in $\Homeo(\Td\times\R^3)$ is of the form 
${\cal P}[j',A']$ for some $(j',A')\in{\cal SOS}(d,j')$.
We call a ${\cal T}\in \Homeo(\Td\times\R^3)$ ``structure preserving'' 
if it is structure preserving for all $(j,A)\in{\cal SOS}(d,j)$.
As we discovered in Section \ref{3.1}, every ${\cal P}[id_\Td,T]$ with
$T\in{\cal C}(\Td,SO(3))$ is structure preserving.

Thus three natural questions arise.
First, what are the structure-preserving 
${\cal T}\in \Homeo(\Td\times\R^3)$ of a given 
$(j,A)\in{\cal SOS}(d,j)$?
Secondly, which ${\cal T}\in \Homeo(\Td\times\R^3)$ are
structure preserving?
Thirdly, which ${\cal TF}(A,A';d,j)$ are nonempty?
While these questions from Dynamical-Systems Theory will not be 
fully addressed in this work we now give a brief glimpse.
Let $(j,A) \in{\cal SOS}(d,j)$ and $(j,A')\in{\cal SOS}(d,j)$
and let ${\cal T}\in \Homeo(\Td\times\R^3)$.
Writing ${\cal T}$ in terms of components ${\cal T}=({\cal T}_{part},
{\cal T}_v)$ we compute
\begin{eqnarray}
&& ({\cal T}\circ{\cal P}[j,A'])(z,S)=
{\cal T}(j(z),A'(z)S) = 
({\cal T}_{part}(j(z),A'(z)S),{\cal T}_{v}
(j(z),A'(z)S)) 
\; , 
\nonumber\\
&&
({\cal P}[j,A]\circ{\cal T})(z,S)=
{\cal P}[j,A]({\cal T}_{part}(z,S),{\cal T}_{v}(z,S))
\nonumber\\
&&\quad = (j({\cal T}_{part}(z,S)),A({\cal T}_{part}(z,S))
{\cal T}_{v}(z,S)) \; , 
\nonumber
\end{eqnarray}
whence ${\cal P}[j,A']={\cal T}^{-1}\circ{\cal P}[j,A]\circ{\cal T}$ iff
\begin{eqnarray}
&& {\cal T}_{part}(j(z),A'(z)S) = 
j({\cal T}_{part}(z,S)) \; , 
\nonumber\\
&&{\cal T}_{v}(j(z),A'(z)S) = 
A({\cal T}_{part}(z,S)){\cal T}_{v}(z,S) \; .
\nonumber\\
\label{eq:A.55nn}
\end{eqnarray}
The system of equations (\ref{eq:A.55nn}) plays a central
role when one addresses the aforementioned questions.
Of course in the special case ${\cal T}={\cal P}[id_\Td,T]$ with
$T\in{\cal C}(\Td,SO(3))$ we see that
${\cal T}_{part}=id_\Td$ and ${\cal T}_{v}(z,S))=T(z)S$ so that in that case
we recover the fact from Section \ref{3.1}
that ${\cal P}[j,A']={\cal T}^{-1}
\circ{\cal P}[j,A]\circ{\cal T}$
iff $T\in{\cal TF}(A,A';d,j)$. We finally mention that 
there are structure preserving
${\cal T}\in \Homeo(\Td\times\R^3)$ which are different from
any ${\cal P}[id_\Td,T]$. For example, defining
${\cal T}\in \Homeo(\Td\times\R^3)$ by 
${\cal T}=({\cal T}_{part},{\cal T}_{v})$ where
${\cal T}_{part}(z)=z,{\cal T}_{v}(z,S)= \left( \begin{array}{ccc}
1 & 0 & 0 \\
0 & 1 & 0 \\
 0 & 0 & -1 \end{array}\right) S$
one easily sees that ${\cal T}$ is structure preserving and is different from
any ${\cal P}[id_\Td,T]$. The latter follows from the fact that  
$\left( \begin{array}{ccc}
1 & 0 & 0 \\
0 & 1 & 0 \\
 0 & 0 & -1 \end{array}\right)$ has determinant $-1$.
\setcounter{equation}{0}
\section{$H$-normal forms and the subsets 
${\cal CB}_H(d,j)$ of ${\cal SOS}(d,j)$}
\label{4.3}
As in Section \ref{3.2n}, we wish to know which
${\cal TF}(A,A';d,j)$ are nonempty, i.e., which spin-orbit systems
in ${\cal SOS}(d,j)$ are equivalent. In fact, by
Remark 9 in Chapter \ref{4}, every ${\cal SOS}(d,j)$
is partitioned into uncountably many equivalence classes
if $j$ is of the form ${\cal P}_\omega$. 
We have already remarked on the advantages of transforming to a ``simple'' 
$(j,A')$.
Now Remark 9 in Chapter \ref{4}
suggests  that to 
gain insight into the partition of ${\cal SOS}(d,j)$ it is
fruitful to find ``simple'' elements $(j,A')$ in an equivalence class
$\overline{(j,A)}$ and to compare different
equivalence classes in terms of their
``simple'' elements. In this chapter we apply this philosophy
by focusing on those ``simple'' elements $(j,A')$ in
$\overline{(j,A)}$ for which $A'$ is $H$-valued where
$H$ is a fixed subgroup of $SO(3)$. Then $(j,A')$ is called an
$H$-normal form of $(j,A)$.
Note that the notion of $H$-normal form is different 
from the usual
definition of normal form for spin \cite{Yo2} but it is inspired
by the $SO(2)$-normal forms studied in \cite{Yo1}.

We will proceed as follows. In Section \ref{V.1} we will define the notion
of $H$-normal form. Then in Section \ref{V.2} we will consider the
important case $H=SO(2)$ where $SO(2)$ is defined by (\ref{eq:6.5})
and we will see that the notion of $SO(2)$-normal form is not new and
is connected with the notion of the ISF via the IFF Theorem,
Theorem \ref{T09t1}b. 
\subsection{Generalities}
\label{V.1}
\setcounter{definition}{0}
\begin{definition}
\label{D6} ($H$-normal form, ${\cal CB}_H(d,j)$)\\
\noindent
Consider a subgroup, $H$, of $SO(3)$ and let
$(j,A)$ be in ${\cal SOS}(d,j)$.
Then we call a $(j,A')$ in ${\cal SOS}(d,j)$
an ``$H$-normal form of $(j,A)$'' if $A'$ is $H$-valued and
$(j,A) \sim (j,A')$, i.e.,
$(j,A')\in\overline{(j,A)}$.
We denote by ${\cal CB}_H(d,j)$ the set of all spin-orbit systems in
${\cal SOS}(d,j)$ which have an $H$-normal form.
Thus $(j,A)\in{\cal CB}_H(d,j)$ iff
$T\in{\cal C}(\Td,SO(3))$ exists such that
\begin{eqnarray}
&& T^t(j(z))A(z)T(z) \in H
\; ,
\label{eq:6.21}
\end{eqnarray}
holds for every $z\in\Td$. The acronym ${\cal CB}$
will be explained in Remark 6 of Chapter \ref{4}.
We also define
\begin{eqnarray}
&&  {\cal TF}_H(j,A):= 
\biggl\lbrace
T\in{\cal C}(\Td,SO(3)):(\forall\;z\in\Td)\; T^t(j(z))A(z) T(z)
\in H \biggr\rbrace\; .
\label{eq:5.25aaa}
\end{eqnarray}
Thus $(j,A)\in {\cal CB}_H(d,j)$
iff ${\cal TF}_H(j,A)$ is nonempty.
Note that the elements of ${\cal TF}_H(j,A)$  are the
transfer fields from $(j,A)$ to those $(j,A')$ for which
$A'$ is $H$-valued.
\hfill $\Box$
\end{definition}
In Chapter \ref{10} we will take a deeper look
into $H$-normal forms for arbitrary subgroups $H$ of $SO(3)$.
See for example the Normal Form Theorem, Theorem \ref{T10.1}, in
Section \ref{10.2.2add}. 

We now make some remarks on Definition \ref{D6}.

\vspace{3mm}
\noindent{\bf Remarks:}
\begin{itemize}
\item[(1)]
Definition \ref{D6} gives us another property shared by 
equivalent
spin-orbit systems since it implies that if $(j,A)$ belongs to
${\cal CB}_H(d,j)$ then every spin-orbit system in $\overline{(j,A)}$
belongs to ${\cal CB}_H(d,j)$.
\item[(2)] 
Let $(j,A)$ be in ${\cal SOS}(d,j)$ and let
$H'$ and $H$ be subgroups of $SO(3)$ such that $H\subset H'$. 
Then, by Definition \ref{D6}, 
${\cal TF}_{H}(j,A)\subset{\cal TF}_{H'}(j,A)$. 
On the other hand, by Definition \ref{D6}, 
if $(j,A)\in{\cal CB}_H(d,j)$ then ${\cal TF}_{H}$ is nonempty whence
${\cal TF}_{H'}$ is nonempty so that, by Definition \ref{D6}, 
$(j,A)\in{\cal CB}_{H'}(d,j)$. Thus
\begin{eqnarray}
&& {\cal CB}_{H}(d,j)\subset{\cal CB}_{H'}(d,j)
\; .
\label{eq:12.17dbtnf}
\end{eqnarray}
This fact 
is even true under more general conditions than $H\subset H'$
as explained after Definition \ref{D5.x}. This fact also implies that
the ``larger $H$'' the more likely it is that a given
$(j,A)$ has an $H$-normal form. For more details on this aspect see
the remarks after the NFT in Section \ref{10.2.2add}.
\item[(3)] 
Let $(j,A)$ be in ${\cal SOS}(d,j)$, let
$H$ be a subgroup of $SO(3)$ and $r\in SO(3)$. Then
it is an easy exercise to show, by Definition \ref{D6}, that
${\cal TF}_{rHr^t}(j,A) = \lbrace Tr^t:T\in {\cal TF}_{H}(j,A)\rbrace$.
\hfill $\Box$
\end{itemize}

To relate $H$-normal forms for different $H$ we make the following definition:
\setcounter{definition}{1}
\begin{definition}
\label{D5.x}
\noindent
Let $H$ and $H'$ be subsets of $SO(3)$.
We write $H\unlhd H'$ if an $r\in SO(3)$ exists such that
$rHr^t\subset H'$. For the notation $rHr^t$ see Definition
\ref{D2.1}. Recalling Appendix \ref{A.2}, 
$\unlhd$ is a relation on the set of subsets of
$SO(3)$ and it is easy to show that $\unlhd$ is reflexive and transitive
but not symmetric.
\hfill $\Box$
\end{definition}
Note that $H\unlhd H'$ if $H\subset H'$.
If $H,H'$ are subgroups of $SO(3)$ then $H\unlhd H'$ iff
$H$ is conjugate to a subgroup of $H'$.
It is an easy exercise to show, by Remarks 2 and 3,
that (\ref{eq:12.17dbtnf}) holds 
if $H\unlhd H'$ (this strengthens Remark 2).
Thus via $\unlhd$ one can order spin-orbit tori in terms of their
normal forms.
It is also a simple exercise to show that 
if $H$ and $H'$ are conjugate subgroups of $SO(3)$ then
$H'\unlhd H$ and $H\unlhd H'$ whence, by (\ref{eq:12.17dbtnf}),
\begin{eqnarray}
&& {\cal CB}_{H'}(d,j)={\cal CB}_H(d,j)
\; .
\label{eq:12.17dbtnfa}
\end{eqnarray}
The relation $\unlhd$ is well-known in Mathematics and
will become an important tool in Chapter \ref{10}.
\subsection{$SO(2)$-normal forms. The IFF
Theorem}
\label{V.2}
In this section we consider $H$-normal forms in the special case
$H=SO(2)$ where the subgroup $SO(2)$ of $SO(3)$
is defined by
\begin{eqnarray}
&& \hspace{-1cm}
SO(2):= \biggl\lbrace \left( \begin{array}{ccc}
\cos(x) & -\sin(x) & 0 \\
\sin(x) & \cos(x) & 0 \\
 0 & 0 & 1 \end{array}\right):x\in{\mathbb R}\biggr\rbrace
\nonumber\\
&&=\lbrace \exp(x{\cal J}):x\in{\mathbb R}\rbrace
=\lbrace \exp(x{\cal J}):x\in[0,2\pi)\rbrace
\; ,
\label{eq:6.5}
\end{eqnarray}
with
\begin{eqnarray}
&& {\cal J}:= \left( \begin{array}{ccc} 0 & -1 & 0 \\
 1 & 0 & 0 \\
 0 & 0 & 0 \end{array}\right) \; .
\label{eq:6.5a}
\end{eqnarray}
For reasons that will become clear below, we now come to:
\setcounter{definition}{2}
\begin{definition}
\label{DIFF}
(Invariant frame field)\\
\noindent Let $(j,A)\in{\cal SOS}(d,j)$.
We call every element of ${\cal TF}_{SO(2)}(j,A)$ 
an ``Invariant Frame Field (IFF) of
$(j,A)$''.  Clearly, by Definition \ref{D6},
${\cal TF}_{SO(2)}(j,A)$ is nonempty iff
$(j,A)\in {\cal CB}_{SO(2)}(d,j)$. 
\hfill $\Box$
\end{definition}
Moreover, for any subgroup $H\neq SO(2)$ of $SO(3)$, we will view the elements
of ${\cal TF}_{H}(j,A)$ as generalized IFF's of
$(j,A)$.
Definition \ref{DIFF} sets the stage for:
\setcounter{theorem}{3}
\begin{theorem}
\label{T09t1} 
a) ($SO(2)$-Lemma)
A matrix $r$ in $SO(3)$ belongs to $SO(2)$ iff 
$r(0,0,1)^t=(0,0,1)^t$.\\

\noindent
b)
(IFF Theorem)
Let $(j,A)\in{\cal SOS}(d,j)$. Then $T$ is an IFF of $(j,A)$
iff $T\in{\cal C}(\Td,SO(3))$ and 
the third column of $T$ is an ISF of $(j,A)$.
In other words, a $T\in{\cal C}(\Td,SO(3))$ belongs to 
${\cal TF}_{SO(2)}(j,A)$ iff
$f(z):=T(z)(0,0,1)^t$ satisfies (\ref{eq:4.2a}).
\end{theorem}
\noindent{\em Proof of Theorem \ref{T09t1}a:}
The claim follows immediately from (\ref{eq:6.5}).
\hfill $\Box$

\noindent{\em Proof of Theorem \ref{T09t1}b:}
``$\Rightarrow$'': Let $T\in{\cal TF}_{SO(2)}(j,A)$.
Then, by Definition \ref{D6}, \\
$T^t(j(z))A(z)T(z) \in SO(2)$ whence, by 
Theorem \ref{T09t1}a,
$T^t(j(z))A(z)T(z)(0,0,1)^t = (0,0,1)^t$
so that $A(z)T(z)(0,0,1)^t =T(j(z))(0,0,1)^t$
whence, by Definition \ref{D6.1}, $T(0,0,1)^t$ is an ISF 
of $(j,A)$.

``$\Leftarrow$'': Let $T\in{\cal C}(\Td,SO(3))$ and let 
$T(0,0,1)^t$ be an ISF 
of $(j,A)$ whence, by Definition \ref{D6.1}, 
$A(z)T(z)(0,0,1)^t =T(j(z))(0,0,1)^t$ so that
$T^t(j(z))A(z)T(z)(0,0,1)^t = (0,0,1)^t$.
Thus, by Theorem \ref{T09t1}a,
$T^t(j(z))A(z)T(z) \in SO(2)$.
It now follows from Definition \ref{D6} that 
$T\in{\cal TF}_{SO(2)}(j,A)$. 
\hfill $\Box$

\vspace{3mm}

By Theorem \ref{T09t1}b, IFF's are those continuous $T$'s whose third
columns are ISF's. In fact this is to be expected given the 
definition  of the IFF in the continuous-time formalism in \cite{BEH}.
There, we begin with the ISF at each point in phase space, and then construct the IFF by attaching 
two unit vectors to the ISF at each point so as to form a local orthonormal coordinate
system for spin at each point in phase space. Spin vector motion 
within the IFF is then a simple precession 
around the ISF. Here, in constrast, we come from the opposite direction by noting that by definition 
spin vector motion w.r.t.
an element of $T\in{\cal TF}_{SO(2)}(j,A)$ as obtained by a transformation of the kind in 
(\ref{eq:5.35}) (say), is a simple precession around the third axis. 
We then discover that the third axis must be an ISF. In this way we prepare to 
state and prove two general theorems in Chapter \ref{10}.
In fact the Normal Form Theorem, Theorem \ref{T10.1} in
Section \ref{10.2.2add}, will
generalize Theorem \ref{T09t1}b.
Most importantly Theorem \ref{T09t1}b 
connects the concepts of normal form and invariant field and the Normal Form
Theorem will generalize this connection from $SO(2)$ to
an arbitrary subgroup $H$ of $SO(3)$ using a
generalization of the notion of invariant polarization field. 
In fact it will turn out that $SO(2)$ is an isotropy group closely related 
to ISF's and IFF's.
The second general theorem to be mentioned is the Cross Section Theorem,
Theorem \ref{P12.10}  in
Section \ref{9.4b.4} which will show that the IFF
is a rather deep concept. See Remark 30 of
Chapter \ref{10} too.

\vspace{3mm}
\noindent{\bf Remarks:}
\begin{itemize}
\item[(4)] A question closely related to
Theorem \ref{T09t1}b is: if
$f\in{\cal C}(\Td,\R^3)$ with $|f|=1$, is there  a 
$T\in{\cal C}(\Td,SO(3))$ such that $f$ is the third column of $T$?
In fact it is well-known, as pointed out in Section \ref{9.4b.4}, that
in general such a $T$ does not exist.
The above question will also be generalized in
Chapter \ref{10} - see Section \ref{10.2.2add}.
\item[(5)] 
One can show, e.g., as in Appendix C in \cite{He2},
that if $A\in{\cal C}(\Td,SO(3))$ is $SO(2)$-valued 
then a constant $N\in{\mathbb Z}^d$ and an
$a\in{\cal C}(\Td,{\mathbb R})$ exist such that
\begin{eqnarray}
&&  A(z)=\exp({\cal J}[N\cdot\phi+2\pi a(z)]) \; ,
\label{eq:5.25aibb}
\end{eqnarray}
where $\phi\in z$.
The fact that $a$ is continuous is the only nontrivial detail of 
(\ref{eq:5.25aibb}).
\item[(6)] If $(j,A)$ belongs to
${\cal CB}_{SO(2)}(d,j)$ then
$(j,A)\sim(j,A')$ where 
$A'$ is $SO(2)$-valued whence, by Remark 5,
a constant $N'\in{\mathbb Z}^d$ and an
$a'\in{\cal C}(\Td,{\mathbb R})$ exist such that
\begin{eqnarray}
&&  A'(z)=\exp({\cal J}[N'\cdot\phi+2\pi a'(z)]) \; ,
\label{eq:5.25aibbp}
\end{eqnarray}
where $\phi\in z$.
It is noteworthy that the constant $N'$ in (\ref{eq:5.25aibbp})
carries interesting information about
$A'$. For example, as shown in Section 7.2 of
\cite{He2} by using simple arguments from Homotopy Theory,
for $({\cal P}_\omega,A')$ to belong to
${\cal SOS}_{CT}(d,\omega)$ it is necessary that
all $d$ components of $N'$ are even integers.

If $(z(\cdot),S'(\cdot))$ is a particle-spin-vector trajectory
of $(j,A')$ then, by 
(\ref{eq:2.J20a}) and (\ref{eq:5.25aibbp}), $S'$  
evolves simply as:
\begin{eqnarray}
&& S'(n+1) = \exp\biggl( {\cal J}[ N'\cdot\phi_1
+ 2\pi a'(L[j](n;z(0)))]\biggr)S'(n) \; ,
\label{eq:2.5n}
\end{eqnarray}
where $\phi_1\in L[j](n;z(0))$. Note
that the spin vector motion in (\ref{eq:2.5n}) is planar, i.e.,
the points $S'(n)$ lie in a plane parallel to the 
$1$-$2$-plane.

If $T\in{\cal TF}(A,A';d,j)$ and if
$(z(\cdot),S(\cdot))$ is a particle-spin-vector trajectory of 
$(j,A)$ then, by the transformation rule (\ref{eq:5.35}),
$(z(\cdot),S(\cdot))$ transforms into the 
particle-spin-vector trajectory 
$(z(\cdot),S'(\cdot))$ of $(j,A')$ where
$S'(n) := T^t(z(n))S(n)$. Thus $S'(\cdot)$ obeys (\ref{eq:2.5n}).
\hfill $\Box$
\end{itemize}
\section{${\cal ACB}(d,j)$ and the notions of 
spin tune and spin-orbit resonance}
\setcounter{equation}{0}
\label{4}
As mentioned at the beginning of Chapter \ref{4.3}, it is
natural to ask which
$(j,A)$ in ${\cal SOS}(d,j)$ are equivalent and it is
fruitful to find ``simple'' elements $(j,A')$ in an equivalence class
$\overline{(j,A)}$ and to compare different
equivalence classes in terms of their
``simple'' elements. In this chapter we apply this philosophy
by focusing on those ``simple'' elements $(j,A')$ in
$\overline{(j,A)}$ for which $A'$ is constant, i.e., for which  $A'(z)$
is independent of $z$. This leads us in Section \ref{4.1} to the 
definition and main basic properties of the
subset ${\cal ACB}(d,j)$ of ${\cal SOS}(d,j)$.

In Section \ref{4.2} this approach will enable us
to associate tunes in addition to $\omega$, namely spin tunes,
with our spin-orbit systems. 
As in other dynamical systems, tunes can lead to the recognition of 
resonances and consequent
instabilities. Here, spin tunes will lead to recognition of 
spin-orbit resonances. In the case
of real spin vector 
motion, where spins are subject to the electric and magnetic 
fields on synchro-betatron
trajectories, the definition of spin-orbit resonance allows us to predict at which orbital tunes
spin vector motion might be particularly unstable. The definition of spin tune 
is also closely related
with the concept of $H$-normal form as Theorem \ref{T2} will reveal. 
\subsection{The  subset ${\cal ACB}(d,j)$ 
of ${\cal SOS}(d,j)$}
\label{4.1}
We first define:
\setcounter{definition}{0}
\begin{definition}
\label{D2}
(${\cal ACB}(d,j)$)\\
\noindent
We denote by ${\cal ACB}(d,j)$ the set of those
$(j,A)\in{\cal SOS}(d,j)$ for which
$\overline{(j,A)}$ contains a $(j,A')$
such that $A'$ is constant, i.e., such that  $A'(z)$
is independent of $z$. 
\hfill $\Box$
\end{definition}
The set ${\cal ACB}(d,j)$ contains the
most important spin-orbit systems in ${\cal SOS}(d,j)$ when it comes
to applications. See the remarks after Definition \ref{D4} too. 
However, as explained in  Section 7.6 in \cite{He2},
it is easy to artificially construct 
$(j,A)\in{\cal SOS}(d,j)$
which are not in ${\cal ACB}(d,j)$
(for an example, see Section \ref{10.4b.2}).
The problem of deciding whether a given
spin-orbit system is in ${\cal ACB}(d,j)$,
is fruitful both theoretically and practically.

\vspace{3mm}
\noindent{\bf Remarks:}
\begin{itemize}
\item[(1)] Definition \ref{D2} gives us another property shared by equivalent
spin-orbit systems since it implies that if $(j,A)$ belongs to
${\cal ACB}(d,j)$ then every spin-orbit system in 
$\overline{(j,A)}$
belongs to ${\cal ACB}(d,j)$.
\item[(2)]
Let $(j,A)\in{\cal SOS}(d,j)$ such that $A$ is constant.
Then, by (\ref{eq:3.5}), 
\begin{eqnarray}
&& \hspace{-1cm}
\Psi[j,A](n;z)=A^n \; ,
\label{eq:6.1l}
\end{eqnarray}
whence every $n$-turn spin transfer matrix function of $(j,A)$
is a constant function so that, by Definition \ref{D2}, 
a spin-orbit system belongs
to some ${\cal ACB}(d,j)$ iff it 
is equivalent
to a spin-orbit system for which
every $n$-turn spin transfer matrix function 
is a constant function. 
This motivates our acronym
${\cal ACB}$ in Definition \ref{D2} since the 
spin transfer matrix functions of the
spin-orbit systems in any ${\cal ACB}(d,j)$ are so-called
``almost coboundaries'' (see, e.g., \cite{KR}).
\item[(3)]
If $({\cal P}_\omega,A)\in{\cal SOS}(d,{\cal P}_\omega)$ such that 
$A$ is constant, 
then $({\cal P}_\omega,A)\in{\cal SOS}_{CT}(d,\omega)$ since
one easily shows that a function
${\cal A}:\R^{d+1}\rightarrow\R^{3\times 3}$ exists which is constant
and whose constant value is a skew-symmetric matrix and such that
$A=\exp(2\pi{\cal A})$.
Then we see, by (\ref{eq:5.84ayn}), that
\begin{eqnarray}
&&  A_{CT}[\omega,{\cal A}] = A \; ,
\label{eq:6.1fa}
\end{eqnarray}
whence, by (\ref{eq:2.3aa}), indeed 
$({\cal P}_\omega,A)\in{\cal SOS}_{CT}(d,\omega)$.
\hfill $\Box$
\end{itemize}
The following theorem gives us insights into 
${\cal ACB}(d,j)$ and for that purpose we need some notation.
We begin by defining, for every $\nu\in[0,1)$ and every positive integer $d$, 
the constant function
$A_{d,\nu}\in{\cal C}(\Td,SO(3))$ as
\begin{eqnarray}
&& 
A_{d,\nu}(z):=
\exp(2\pi\nu{\cal J}) =
\left( \begin{array}{ccc} \cos(2\pi \nu) & -\sin(2\pi \nu)
& 0 \\
\sin(2\pi \nu) & \cos(2\pi \nu) & 0 \\
 0 & 0 & 1 \end{array}\right)\; .
\label{eq:6.1}
\end{eqnarray}
Clearly, for every $j\in \Homeo(\Td)$, the spin-orbit system $(j,A_{d,\nu})$
belongs to ${\cal ACB}(d,j)$ since $A_{d,\nu}$ is a constant function.
Secondly, for every $\nu\in[0,1)$ we define 
\begin{eqnarray}
&& G_\nu:=\lbrace \exp(2\pi n\nu{\cal J}):n\in\Z\rbrace 
=\lbrace \exp(2\pi (n\nu+m){\cal J}):m,n\in\Z\rbrace
\; , 
\label{eq:6.1da}
\end{eqnarray}
where the, trivial, second equality highlights the fact that $G_\nu$ consists of
matrices $\exp(2\pi \mu{\cal J})$ where $\mu\in[0,1)$. It is clear by
(\ref{eq:6.5}) that $G_\nu$ is a subgroup of $SO(2)$.
Due to (\ref{eq:6.1}) every $A_{d,\nu}$ is $G_\nu$-valued and 
$G_0=\{ I_{3 \times 3} \}$ is the trivial subgroup of $SO(3)$.
Finally,  for every $(j,A)\in{\cal SOS}(d,j)$ we define the set
\begin{eqnarray}
&& \hspace{-1cm}
\Xi(j,A):=\lbrace\nu\in[0,1):
(j,A_{d,\nu})\in\overline{(j,A)}\rbrace \; .
\label{eq:2.3aad}
\end{eqnarray}

The following theorem shows some relationships between these concepts. Note that the main purpose of 
part c) is to help prove part d).

\setcounter{theorem}{1}
\begin{theorem}
\label{T2} \noindent
a) Let $(j,A)\in{\cal SOS}(d,j)$. Then
$(j,A)\in{\cal ACB}(d,j)$ iff a
$\nu\in[0,1)$ exists such that $(j,A_{d,\nu})$ belongs to
$\overline{(j,A)}$. \\

\noindent b) Let $(j,A)\in{\cal SOS}(d,j)$. Then 
$(j,A)\in{\cal ACB}(d,j)$ iff 
$\Xi(j,A)$ is nonempty.\\

\noindent c) Let
$\nu\in[0,1)$ and
$A\in{\cal C}(\Td,SO(3))$ be $G_\nu$-valued. Then $A$ is a constant function.\\

\noindent d) Let $j\in \Homeo(\Td)$. Then
\begin{eqnarray}
&& {\cal ACB}(d,j) = \bigcup_{\nu\in[0,1)}\; {\cal CB}_{G_\nu}(d,j) \; .
\label{eq:6.1gaan}
\end{eqnarray}
\noindent e) Let $T\in{\cal C}(\Td,SO(3))$ and let
$(j,A')\in{\cal SOS}(d,j)$ be the
transform of $(j,A)\in{\cal SOS}(d,j)$ under $T$, i.e.,
$T\in{\cal TF}(A,A';d,j)$. Then $T$ belongs to
$\bigcup_{\nu\in[0,1)}\; {\cal TF}_{G_\nu}(j,A)$ iff 
$T$ is an IFF of $(j,A)$ and $A'$ is a constant function.\\

\noindent
f) Let $({\cal P}_\omega,A)\in{\cal SOS}(d,{\cal P}_\omega)$.
If $\nu\in\Xi({\cal P}_\omega,A)$ then
\begin{eqnarray}
&& 
\Xi({\cal P}_\omega,A)=[0,1)\cap
\biggl\lbrace \varepsilon\nu + m\cdot \omega+n:
\varepsilon\in\lbrace 1,-1\rbrace,m\in{\mathbb Z}^d,n\in{\mathbb Z}
\biggr\rbrace \;  .
\label{eq:6.15}
\end{eqnarray}
\end{theorem}
\noindent{\em Proof of Theorem \ref{T2}a:}
If $\nu\in[0,1)$ exists such that $(j,A_{d,\nu})$ belongs to
$\overline{(j,A)}$ then, by Definition \ref{D2},
$(j,A)\in{\cal ACB}(d,j)$ since $A_{d,\nu}$ is constant. 
To prove the converse,  let 
$(j,A)\in{\cal ACB}(d,j)$.
Then, by Definition \ref{D2},
$\overline{(j,A)}$ contains a $(j,A')$
such that $A'$ is constant with constant value, say $r$.
By some simple Linear Algebra, a $\nu\in[0,1)$ and a $W\in SO(3)$ 
can be found such that
\begin{eqnarray}
&& 
r = W \exp(2\pi\nu{\cal J}) W^t \; .
\label{eq:6.1a}
\end{eqnarray}
See, e.g., Lemma 2.1 of \cite{BEH}. Thus, defining the constant function
$T\in{\cal C}(\Td,SO(3))$ by $T(z):=W$ we observe by (\ref{eq:6.1a})
and Definition \ref{D3a} that $T\in{\cal TF}(A',A_{d,\nu},d,j)$ 
whence $(j,A')\sim (j,A_{d,\nu})$ so that
$(j,A)\sim (j,A_{d,\nu})$ which implies that $(j,A_{d,\nu})$ belongs to
$\overline{(j,A)}$. 
\hfill $\Box$

\noindent{\em Proof of Theorem \ref{T2}b:}
The claim is a simple consequence of 
(\ref{eq:2.3aad}) and Theorem \ref{T2}a.
\hfill $\Box$

\noindent{\em Proof of Theorem \ref{T2}c:}
Since $A$ is $G_\nu$-valued it follows from (\ref{eq:6.1da})
that a function $\tilde{n}:\Td\rightarrow\Z$ exists such that
$A(z)=\exp({\cal J}2\pi \nu \tilde{n}(z))$ whence
\begin{eqnarray}
&&  A(z)
=\exp({\cal J}2\pi \nu \tilde{n}(z)) \; .
\label{eq:5.25aibbn}
\end{eqnarray}
Clearly $A$ is $SO(2)$-valued whence, by Remark 5 in Chapter \ref{4.3},
a constant $N\in{\mathbb Z}^d$ and an
$a\in{\cal C}(\Td,{\mathbb R})$ exist such that
\begin{eqnarray}
&&  
\exp({\cal J}2\pi \nu \tilde{n}(z))
= A(z)
=\exp({\cal J}[N\cdot\phi+2\pi a(z)]) \; ,
\label{eq:5.25aibbna}
\end{eqnarray}
where $\phi\in z$ and where
in the first equality we used (\ref{eq:5.25aibbn}).
It follows from (\ref{eq:6.5a}) and (\ref{eq:5.25aibbna}) that 
a function $k:\Td\rightarrow\Z$ exists such that
\begin{eqnarray}
&& 2\pi \nu \tilde{n}(\phi+ \tilde{\Z}^d) 
+ 2\pi k(\phi+ \tilde{\Z}^d)
= N\cdot\phi+2\pi a(\phi+ \tilde{\Z}^d)\; .
\label{eq:5.25aibbnb}
\end{eqnarray}
Since $\tilde{n}(\phi+ \tilde{\Z}^d)$, $k(\phi+ \tilde{\Z}^d)$
and $a(\phi+ \tilde{\Z}^d)$ are $2\pi$-periodic
in the components of $\phi$ it follows from (\ref{eq:5.25aibbnb}) that
$N\cdot\phi$ is $2\pi$-periodic
in the components of $\phi$ whence
$N=0$ so that, by (\ref{eq:5.25aibbnb}), for all $z\in\Td$
\begin{eqnarray}
&& \nu \tilde{n}(z) + k(z) = a(z)  \; .
\label{eq:5.25aibbnc}
\end{eqnarray}
Since $\tilde{n}$ and $k$ are $\Z$-valued, the function $\nu \tilde{n} + k$
can take at most countably many values whence, by (\ref{eq:5.25aibbnc}),
the function $a$ can take at most countably many values. 
On the other hand since 
$a$ is continuous and since its domain, $\Td$, is a path-connected
topological space, the range of $a$ is a path-connected
subset of $\R$ , i.e., an interval, say $I$
(for the notion of range see also Appendix \ref{A.1}). 
However since 
$a$ takes at most countably many values, $I$ contains at most only 
countably many points whence, being an interval, $I$ contains just one
point which implies that $a$ is constant.
Since $a$ is constant and $N=0$ it follows
from (\ref{eq:5.25aibbna}) that 
$A$ is constant.
\hfill $\Box$

\noindent{\em Proof of Theorem \ref{T2}d:}
``$\subset $'': Let $(j,A)\in{\cal ACB}(d,j)$. Then, by 
Theorem \ref{T2}a, a $\nu\in[0,1)$ exists such that 
$(j,A_{d,\nu})$ belongs to
$\overline{(j,A)}$. By a remark after
(\ref{eq:6.1da}), $A_{d,\nu}$ is $G_\nu$-valued whence, by
Definition \ref{D6}, $(j,A)\in{\cal CB}_{G_\nu}(d,j)$.\\
\noindent
``$\supset $'': Let $\nu\in[0,1)$ and 
$(j,A)\in{\cal CB}_{G_\nu}(d,j)$ whence,
by Definition \ref{D6}, ${\cal TF}_{G_\nu}(j,A)$ is nonempty. So  
pick a $T\in{\cal TF}_{G_\nu}(j,A)$. Then, by
Definitions \ref{D3a} and \ref{D6},
$T\in{\cal TF}(A,A';d,j)$ where $A'$ is $G_\nu$-valued.
Since $A'$ is $G_\nu$-valued it follows from Theorem \ref{T2}c that
$A'$ is constant which implies, by Definition \ref{D2},
that $(j,A)\in{\cal ACB}(d,j)$.
\hfill $\Box$

\noindent{\em Proof of Theorem \ref{T2}e:}
``$\Rightarrow$'': Let $T\in{\cal TF}_{G_\nu}(j,A)$.
Since $G_\nu$ is a subgroup of $SO(2)$ we conclude from Remark 2
in Chapter \ref{4.3} that $T\in{\cal TF}_{SO(2)}(j,A)$ so that $T$ is an
IFF of $(j,A)$. 
Also, $A'$ is $G_\nu$-valued whence, by Theorem \ref{T2}c,
$A'$ is constant.
\\
\noindent
``$\Leftarrow$'': Let $T$ be an IFF of $(j,A)$ and let $A'$ be constant.
Clearly, by  Definition \ref{DIFF}, $A'$ is $SO(2)$-valued whence 
$\nu\in[0,1)$ exists such that $A'=A_{d,\nu}$ which implies that
$A'$ is $G_\nu$-valued so that
$T\in{\cal TF}_{G_\nu}(j,A)$.
\hfill $\Box$

\noindent{\em Proof of Theorem \ref{T2}f:} 
The claim
is proved in Chapter 8 of \cite{He2}
by using the tool of quasiperiodic functions \cite{He2}.
\hfill $\Box$

\vspace{3mm}

Theorem \ref{T2}d provides the insight that every
${\cal ACB}(d,j)$ can be understood in terms of $H$-normal forms,
a fact that is not obvious by Definition \ref{D2}.
The purpose of Theorem \ref{T2}e is to  lead us to the
definition of the uniform IFF in Remark 5 below.
As shown in Remark 9 below, Theorem \ref{T2}f gives an insight into
how the ${\cal SOS}(d,{\cal P}_\omega)$ are partitioned w.r.t. to the equivalence
relation $\sim$ and it will also 
play a role for the notion of spin-orbit resonance
in Section \ref{4.2}. 
Theorem \ref{T2}a will be used in Remark 5 
and was used for proving
Theorems \ref{T2}b and \ref{T2}d while
Theorem \ref{T2}b will be used in Remark 5 and in Section \ref{4.2}.

\vspace{3mm}
\noindent{\bf Remarks:}
\begin{itemize}
\item[(4)] Let $j\in \Homeo(\Td)$ and
$\nu\in[0,1)$. By a remark after (\ref{eq:6.1da})
we have $G_\nu\subset SO(2)$
whence, by (\ref{eq:12.17dbtnf}),
\begin{eqnarray}
&& {\cal CB}_{G_\nu}(d,j) 
\subset {\cal CB}_{SO(2)}(d,j) \; .
\label{eq:5.25aibbndn}
\end{eqnarray}
It follows from (\ref{eq:5.25aibbndn}) and
Theorem \ref{T2}d that
\begin{eqnarray}
&&  {\cal ACB}(d,j)\subset {\cal CB}_{SO(2)}(d,j) \; .
\label{eq:5.25aibba}
\end{eqnarray}
Then, by Definition \ref{DIFF},
$(j,A)\in{\cal ACB}(d,j)$ has an IFF whence, by Theorem \ref{T09t1}b, 
$(j,A)$ has an ISF, say $f$. In fact, recalling Section \ref{6.1b},
$(j,A)$ has at least the two ISF's $f$ and $-f$.
For example, the constant functions on $\Td$ 
with values $(0,0,1)^t$ and $(0,0,-1)^t$ are ISF's of every spin-orbit
system of the form $(j,A_{d,\nu})$.
\item[(5)] Let $(j,A)\in{\cal SOS}(d,j)$.
A $T\in{\cal C}(\Td,SO(3))$ is called a 
``uniform IFF of $(j,A)$'' iff
it is an IFF of $(j,A)$ and $A'$, defined by (\ref{eq:5.25}),
is a constant function. It is clear that
the uniform IFF's are the discrete-time analogues of
the so-called uniform invariant frame fields 
introduced in the continuous-time formalism of
\cite{BEH}. Also, by Theorem \ref{T2}e, the
elements of $\bigcup_{\nu\in[0,1)}\; {\cal TF}_{G_\nu}(j,A)$ are the
uniform IFF's of  $(j,A)$.
This implies, by Theorem \ref{T2}c,
that $A'$, defined by (\ref{eq:5.25}),
is of the form $A_{d,\nu}$.
It thus follows that the set of 
uniform IFF's of  $(j,A)$ reads as
\begin{eqnarray}
&& \hspace{-1cm}
\bigcup_{\nu\in[0,1)}\; {\cal TF}_{G_\nu}(j,A)
=\bigcup_{\nu\in[0,1)}\;
{\cal TF}(A,A_{d,\nu};d,j)
=\bigcup_{\nu\in \Xi(j,A)}\;{\cal TF}(A,A_{d,\nu};d,j) \; ,
\label{eq:6.10}
\end{eqnarray}
where in the second equality we used (\ref{eq:2.3aad}).
Using (\ref{eq:6.10}) and Theorem \ref{T2}b
we also observe that a $(j,A)$
has a uniform IFF iff $(j,A)\in{\cal ACB}(d,j)$. 

Note that if $(j,A)$ has a uniform IFF, say $T$, then
$TA_{d,\mu}$ is also a uniform IFF of $(j,A)$ where $\mu\in[0,1)$
whence every $(j,A)\in{\cal ACB}(d,j)$ has uncountably many 
uniform IFF's.
Note also, by (\ref{eq:6.10}), that $(j,A)$ has at least as
many uniform IFF's as there are elements in $\Xi(j,A)$.
This is especially evident in the case when $j={\cal P}_\omega$ 
and $(j,A)\in{\cal ACB}(d,j)$ since then,
by (\ref{eq:6.15}), $\Xi(j,A)$ has at most countably many elements
so here the uniform IFF's considerably outnumber the elements of
$\Xi(j,A)$. On the other hand, if $({\cal P}_\omega,A)\in{\cal ACB}(d,j)$ 
it rarely happens that 
$\Xi({\cal P}_\omega,A)$ has finitely many elements since one can show 
\cite{He1} that this only happens iff
all $d$ components of $\omega$ are rational numbers.
\item[(6)]
Let $(j,A)\in{\cal CB}_{G_0}(d,j)$. Thus, by
Definition \ref{D6}, ${\cal TF}_{G_0}(j,A)$ is nonempty and every
$T$ in ${\cal TF}_{G_0}(j,A)$ satisfies
$(T^t\circ j)A T = A_{d,0}$ so that, by (\ref{eq:5.30}),
\begin{eqnarray}
&& \Psi[j,A](n;z)= T(L[j](n;z))T^t(z) \; .
\label{eq:5.30b}
\end{eqnarray}
Eq. (\ref{eq:5.30b}) motivates the acronym
${\cal CB}$ in Definition \ref{D6} since the spin transfer matrix function
$\Psi[j,A]$ in (\ref{eq:5.30b}) belongs to that class of cocycles which are
called ``coboundaries'' (see \cite{HK2} and Chapter 1 in
\cite{HK1}).
\item[(7)] Let $({\cal P}_\omega,A)\in{\cal SOS}(d,{\cal P}_\omega)$. 
It can be easily shown, by using (\ref{eq:2.3aad}), that
$({\cal P}_\omega,A)\in {\cal CB}_{G_0}(d,{\cal P}_\omega)$ iff 
$0\in\Xi({\cal P}_\omega,A)$.
\item[(8)] Let $(j,A)\in{\cal SOS}(d,j)$. It is easy to show, by 
(\ref{eq:6.1}), (\ref{eq:6.1da}) and (\ref{eq:2.3aad}) and
for every $\nu\in[0,1)$, that either
${\cal TF}_{G_\nu}(j,A)$ is empty or there exist integers $m,n$ such that
$(m\nu+n)\in\Xi(j,A)$ 
(note that $n$ ensures that $(m\nu+n)\in[0,1)$).
This implies, by Theorem \ref{T2}f,
that if $j$ is of the form ${\cal P}_\omega$ then a subset $B$
of $[0,1)$ exists which has at most countably many elements and such that
$\bigcup_{\nu\in[0,1)}\; {\cal TF}_{G_\nu}({\cal P}_\omega,A)=
\bigcup_{\nu\in B}\; {\cal TF}_{G_\nu}({\cal P}_\omega,A)$.
\item[(9)] 
As mentioned in Section \ref{3.2n}, in this work we do not fully address
how the ${\cal SOS}(d,j)$ are partitioned w.r.t. to the equivalence
relation $\sim$. Thus it may come as a surprise that Theorem \ref{T2}f
sheds light on this issue. In fact if $j$ is of the form
${\cal P}_\omega$ then ${\cal SOS}(d,j)$ 
contains uncountably many equivalence classes as
follows. 

To prove this claim we first of all note that ${\cal SOS}(d,{\cal P}_\omega)$ has
uncountably many elements since $\nu$ is a continuous parameter whence
there are uncountably many $A_{d,\nu}$, i.e., the spin-orbit systems
$({\cal P}_\omega,A_{d,\nu})$ form
an uncountable subset, say $B$, of ${\cal SOS}(d,{\cal P}_\omega)$
(note that $\omega$ is fixed but $\nu$ varies over $[0,1)$).
Note also 
that both $B$ and $\overline{({\cal P}_\omega,A_{d,\nu})}$ have uncountably
many elements but, as will be shown below,  
$B\cap \overline{({\cal P}_\omega,A_{d,\nu})}$ has at most countably many
elements. In fact in
our proof the sets $B\cap \overline{({\cal P}_\omega,A_{d,\nu})}$ 
for each $\nu$ 
will play a key role and we already note here that they form a
partition of $B$ since the $\overline{({\cal P}_\omega,A_{d,\nu})}$, being
equivalence classes, are mutually disjoint.
In particular, if $\overline{({\cal P}_\omega,A_{d,\nu})}$
and $\overline{({\cal P}_\omega,A_{d,\mu})}$ are different then they are
disjoint and belong to different equivalence classes of the equivalence 
relation $\sim$. The crucial question now is: how
many of the sets $B\cap 
\overline{({\cal P}_\omega,A_{d,\nu})}$ are different? In other words
how common is it that two spin-orbit systems in $B$
are equivalent?  
This is where Theorem \ref{T2} engages.
In fact, by (\ref{eq:6.15}),
each set $\Xi({\cal P}_\omega,A_{d,\nu})$ contains at most
countably many elements. On the other hand if $\nu,\mu\in[0,1)$ then,
by (\ref{eq:2.3aad}), $({\cal P}_\omega,A_{d,\mu})\in
\overline{({\cal P}_\omega,A_{d,\nu})}$ iff $\mu\in\Xi({\cal P}_\omega,A_{d,\nu})$.
Thus every set of the form $B\cap
\overline{({\cal P}_\omega,A_{d,\nu})}$ contains at most countably many 
elements of $B$. Thus we need uncountably many of the
sets $B\cap\overline{({\cal P}_\omega,A_{d,\nu})}$ to overlap $B$ whence the 
$B\cap\overline{({\cal P}_\omega,A_{d,\nu})}$ form an uncountable partition of
$B$. Since different $B\cap\overline{({\cal P}_\omega,A_{d,\nu})}$
are contained in different equivalence classes we thus have shown
that there are uncountably many
equivalence classes of the form $\overline{({\cal P}_\omega,A_{d,\nu})}$.
Thus, as was to be shown, 
${\cal SOS}(d,{\cal P}_\omega)$ is partitioned into uncountably many 
equivalence classes w.r.t. to the equivalence
relation $\sim$.
\hfill $\Box$
\end{itemize}
\subsection{Spin tunes and spin-orbit resonances}
\label{4.2}
Definition \ref{D2}
and Theorem \ref{T2} lead us naturally to the notions of spin tune
and spin-orbit resonance.
A $\nu\in[0,1)$ is said to be a spin tune for
$(j,A)\in{\cal SOS}(d,j)$ if
$(j,A)$ is equivalent to $(j,A')$ with $A'(z)=
\exp(2\pi\nu{\cal J})$, i.e., if
$(j,A_{d,\nu})$ belongs to
$\overline{(j,A)}$.
We thus arrive at the following definition:
\setcounter{definition}{2}
\begin{definition}
\label{D4}
(Spin tune, spin-orbit resonance)\\
\noindent
We call the elements of $\Xi(j,A)$ the spin tunes of $(j,A)$.
We say that $({\cal P}_\omega,A)$ is ``on spin-orbit resonance (SOR)'' 
if $({\cal P}_\omega,A)\in{\cal ACB}(d,{\cal P}_\omega)$ and if for every
$\nu\in\Xi({\cal P}_\omega,A)$ we can find
$m\in{\mathbb Z}^d,n\in{\mathbb Z}$ such that
\begin{eqnarray}
\nu = m\cdot \omega + n  \; .
\label{eq:6.20}
\end{eqnarray}
We say that $({\cal P}_\omega,A)$ is ``off spin-orbit resonance''
iff $({\cal P}_\omega,A)\in{\cal ACB}(d,{\cal P}_\omega)$ and if 
$({\cal P}_\omega,A)$ is not on
spin-orbit resonance.
Note that a spin-orbit system which has no spin tunes
is neither on nor off spin-orbit resonance. This happens in particular when
$j$ is not a torus translation.
\hfill $\Box$
\end{definition}

It follows from Definition \ref{D4} and
Theorem \ref{T2}b that a $(j,A)\in{\cal SOS}(d,j)$ 
has spin tunes iff $(j,A)\in{\cal ACB}(d,j)$.
This has the implication that, by Theorem \ref{T2}d, spin tunes
can be understood in terms of normal forms.
Furthermore it has the implication, by Remark 5, that 
$(j,A)$ has a spin tune iff it has a uniform IFF.

In \cite{BEH} spin-orbit systems with spin tunes belong to the class 
of ``well tuned'' systems and most of the systems
with no spin tunes are said to be ``ill-tuned''.

In \cite{He2} the spin tune and spin-orbit resonances defined here
are called  spin tune of the first kind and spin-orbit 
resonances of the first kind respectively since \cite{He2} finds it convenient to distinguish between
two kind of spin tune. That distinction is not needed here.

If one considers a family $(j_J,A_J)_{J\in\Lambda}$ of
spin-orbit systems (see the Introduction and Chapter \ref{VII}) and if
every $(j_J,A_J)$ has a spin tune, say $\nu_J$, then $\nu_J$ is
called an amplitude dependent spin tune (ADST).  
Recall from Remark 5 that if $T_J$ is a uniform IFF of
$(j_J,A_J)$ then $T_J^t( j(z))A_J(z)T_J(z)=A_{d,\nu_J}(z)
=\exp(2\pi\nu_J{\cal J})$.

As stated at the beginning of this chapter spin-orbit resonance
can lead to a large angular spread of the ISF and that can lead to unacceptably
low equilibrium polarization as explained in Chapter \ref{VII}.
The large angular spread also means that if a particle beam occupies a large volume of
phase space at injection while the spins all point in roughly the same direction, the polarization of the beam 
can be very unstable while the spin precess around their individual ISF's.
See \cite{Ho} for an example of this.
See \cite{Ho},\cite{Mane},\cite{mv2000},\cite{Yo1} for formalisms and
calculations which have demonstrated the potential for a large spread
of the ISF near spin-orbit
resonances. For detailed further comments see Section X in \cite{BEH}.

Moreover, since the ADST can vary with orbital amplitude $J$, particles at
one amplitude can be close to spin-orbit resonance while particles
as nearby amplitudes need not be.  Manifestations of this are
beautifully demonstrated in \cite{Ho,mv2000,BHV00,HV} 
where the value of a rigorous
definition of spin tune is made crystal clear.  Note that as shown in
those works, spin-orbit resonances
tend to be rather repelling than attractive.
The rigorous definition of spin tune and of spin-orbit resonance also will 
lead us in Chapter \ref{VII} 
to the Uniqueness Theorem for the ISF \cite{Yo1,DK73}.
In summary, a rigorous definition, as in
Definition {\ref{D4}}, is very important for a detailed  
understanding of real spin vector motion.

As explained in Section X of \cite{BEH} and in \cite{BV1}, as well
as in other literature, a real spin-orbit system 
$({\cal P}_\omega,A)$ on orbital resonance
normally has  no spin tune. 
One exception is the so-called single resonance model 
underlying the 
model with two Siberian snakes in Section \ref{10.4b.2}.
Nevertheless, such a system
can, but need not, have an ISF of the continuous kind defined
here. An example of a spin-orbit system on orbital resonance
which has no ISF, and thus no spin tune, is studied in Section
\ref{10.4b.2}. If the $d$ components of $\omega$ are rational numbers
then it is easy to calculate an ISF $f$
by finding the real eigenvector $f(z)$ of
the matrix $\Psi[{\cal P}_\omega,A](n;z)$ for the number of turns $n$ for
which the particle returns to its starting position $z$. The
discontinuous ``ISF'' of \cite{BV1} can also be calculated in this
way (and this is also done in our example
in Section \ref{10.4b.2}).
Recall also from the ISF conjecture in Chapter \ref{6} that
we expect an ISF to exist off orbital resonance.

The ISF and the ADST for real spin vector motion off orbital resonance 
in storage rings can be computed in a number of ways 
\cite{B},\cite{F},\cite{HH96},\cite{Ho},\cite{Mane},\cite{mv2000},\cite{Yo3}.
Here we describe two of them and we start with 
a method of computing the ADST, implemented
in the computer code SPRINT \cite{He2,Ho,mv2000}
(as an alternative
method, SPRINT offers an implementation of the SODOM-2 algorithm).
The calculations proceed in
two steps \cite{BEH00,BHV98,Ho,mv2000}. 
For simplicity we consider a fixed but arbitrary action value $J$ and
assume that the spin-orbit system belongs to
${\cal ACB}(d,{\cal P}_\omega)$ and is off orbital resonance and
off spin-orbit resonance.
As we will see in Chapter \ref{VII},
by the
Uniqueness Theorem, Theorem \ref{T7.1}b, the given spin-orbit 
system $({\cal P}_\omega,A)$ has only
two ISF's, say $f$ and $-f$. Of course $f$ and $-f$ in general 
are unknown and in fact
one only attempts to compute a discretization of them. In
the first step, $f$ is computed at some point $z$ 
on the torus at some point $\theta$ 
on a ring using stroboscopic averaging \cite{EH,HH96} giving us $f(z)$.
By Remark 4 an IFF, say $T$, exists and, due to
Theorem \ref{T09t1}b, the third column of $T$ is either $f$ or $-f$ and
here $T(z)$ is constructed by
a simple orthonormalization procedure in which $f(z)$ is the
third column is $T(z)$. The axis represented by the second column of 
$T(z)$ could, for example, be chosen
so as to have no component along the direction of the beam.
In the next step the spin value $f(z)$ is
tracked forwards turn by turn, according to (\ref{eq:4.2a}),
resulting in the discretization $f(z),f({\cal P}_\omega(z)),
f({\cal P}_\omega^2(z)),...,f({\cal P}_\omega^N(z))$ of $f$
for some large integer $N$. Accordingly
$T(z),T({\cal P}_\omega(z)),T({\cal P}_\omega^2(z)),...,
T({\cal P}_\omega^N(z))$ are constructed at the
end of each turn according to the chosen prescription. 
Then, the average spin precession angle around the
ISF w.r.t. this IFF is computed for a very large number of
turns $N$. In fact since
$T$ is an IFF, by Remark 6 in Chapter \ref{4.3}, 
an $N\in{\mathbb Z}^d$ and an
$a\in{\cal C}(\Td,{\mathbb R})$ exist such that
\begin{eqnarray}
&& T^t\Biggl( \biggl(\phi+2\pi (n+1)\omega\biggr)+ 
\tilde{\Z}^d\Biggr)A\Biggl(\biggl(\phi+2\pi n\omega\biggr)+\tilde{\Z}^d\Biggr)
T\Biggl(\biggl(\phi+2\pi n\omega \biggr)+ \tilde{\Z}^d\Biggr)
\nonumber\\
&&\quad =
\exp\Biggl({\cal J}[N\cdot(\phi+2\pi n\omega)+
2\pi a\biggl((\phi+2\pi n\omega) 
+ \tilde{\Z}^d\biggr)]\Biggr) \; ,
\label{eq:6.40}
\end{eqnarray}
where $\phi\in z$ and $n=0,...,N$.
One can show \cite{He2,mv2000}
that the average $<a>$ of $a$, given by
\begin{eqnarray}
&& <a>:=\frac{1}{(2\pi)^d}\int_{[0,2\pi]^d}\;a(\pi_d(\phi))d\phi \; ,
\label{eq:6.41}
\end{eqnarray}
is a spin tune of $({\cal P}_\omega,A)$. On the other hand, (\ref{eq:6.40}),
gives us $a(z),a({\cal P}_\omega(z)),a({\cal P}_\omega^2(z)),...,
a({\cal P}_\omega^N(z))$ which allows one to approximate
the average of $a$.
This delivers an ADST for the given $J$ 
but the member of the set $\Xi
({\cal P}_\omega, A)$ that emerges will depend on the convention used to choose
the first and second axes of $T$.

Another practical way to compute
spin tunes  is by using the spectrum of the spin vector motion 
as follows.
For simplicity we consider a fixed but arbitrary action value $J$ and
assume that the spin-orbit system belongs to
${\cal ACB}(d,{\cal P}_\omega)$.
Then let $({\cal P}_\omega,A)$ have a spin-vector 
trajectory $S(\cdot)$.
The discrete Fourier transform (DFT) 
of $S(0),...,S(N)$ is defined by $\hat{S}$ where
\begin{eqnarray}
\hat{S}(k):=\frac{1}{N+1}
\sum_{n=0}^N\; S(n)\exp(-2\pi i n k/(N+1)) \; ,
\label{eq:8.20d}
\end{eqnarray}
and where $k=0,...,N$. 
We define, for $\lambda\in[0,1)$ and nonnegative integer $N$, 
\begin{eqnarray}
a_N(S,\lambda):=
(N+1)^{-1}\sum_{n=0}^N\; S(n)\exp(-2\pi i n\lambda) \ .
\label{eq:8.20e}
\end{eqnarray}
It can be easily shown \cite{He2} that $a_N(S,\lambda)$ converges as
$N\rightarrow\infty$ and we
denote the limit of $a_N(S,\lambda)$
by $a(S,\lambda)$ and we define the ``spectrum $\Lambda(S)$
of $S$'' by $\Lambda(S):=\lbrace \lambda\in[0,1):
a(S,\lambda)\neq 0\rbrace$. From 
(\ref{eq:8.20d}) and (\ref{eq:8.20e}) it is clear that  $a(S,\lambda)$ can be
approximated by using standard DFT software.
Then, as can be easily shown \cite{He2}, spin tunes are contained in the spectrum
since 
\begin{eqnarray}
&& 
\Lambda(S)\subset \Xi({\cal P}_\omega,A) \cup 
\lbrace l\cdot\omega+n:l\in{\mathbb Z}^d,n\in{\mathbb Z}
\rbrace \; .
\label{eq:8.10a}
\end{eqnarray}
Moreover, the spectrum can contain many of the spin tunes in 
$\Xi({\cal P}_\omega,A)$. 
Theorem 9.1c 
in the continuous-time formalism of \cite{BEH} reaches the same 
conclusions.
With this we have a direct relationship between the set  $\Xi({\cal P}_\omega,A)$ appearing in Theorem \ref{T2}
and a ``measureable'' quantity, namely the spectrum. This way of getting ADST's has been essential for 
interpreting spin vector motion near to resonance with oscillating external magnetic fields 
\cite{Ba}.

\vspace{3mm}
\noindent{\bf Remark:}
\begin{itemize}
\item[(10)]
By (\ref{eq:6.15}) and Definition \ref{D4}
an $({\cal P}_\omega,A)\in{\cal ACB}(d,{\cal P}_\omega)$
is on spin-orbit resonance iff (\ref{eq:6.20}) holds for just one choice of
$m\in{\mathbb Z}^d,n\in{\mathbb Z},\nu\in\Xi({\cal P}_\omega,A)$.
Thus a single spin tune $\nu$ of $({\cal P}_\omega,A)$ is sufficient to determine if
$({\cal P}_\omega,A)$ is on spin-orbit resonance.
Note also, by (\ref{eq:6.15}) and
Definition \ref{D4}, that a spin-orbit system
$({\cal P}_\omega,A)\in{\cal SOS}(d,{\cal P}_\omega)$
is on spin-orbit resonance iff
$0\in\Xi({\cal P}_\omega,A)$. 
Thus, by Remark 7, 
$({\cal P}_\omega,A)$ is on SOR iff $({\cal P}_\omega,A)\in
{\cal CB}_{G_0}(d,{\cal P}_\omega)$.
\hfill $\Box$
\end{itemize}

In analogy with Theorem \ref{T09t1}b we now state:
\setcounter{theorem}{3}
\begin{theorem}
\label{T09t2}
a) 
Let $(j,A)\in{\cal SOS}(d,j)$ and
$T\in{\cal C}(\Td,SO(3))$. Then $T$ satisfies
\begin{eqnarray}
&& T \circ j = A T
\; ,
\label{eq:5.30cn}
\end{eqnarray}
iff it belongs to ${\cal TF}_{G_0}(j,A)$. \\

\noindent b) (SOR Theorem)
Let $({\cal P}_\omega,A)\in{\cal SOS}(d,{\cal P}_\omega)$. 
Then $({\cal P}_\omega,A)$ is on SOR iff 
${\cal TF}_{G_0}({\cal P}_\omega,A)$ is nonempty, i.e., iff $({\cal P}_\omega,A)\in
{\cal CB}_{G_0}(d,{\cal P}_\omega)$.
\end{theorem}
\noindent{\em Proof of Theorem \ref{T09t2}a:}
By Definition \ref{D6}, $T\in{\cal TF}_{G_0}(j,A)$ iff
$T^t(j(z))A(z)T(z) \in G_0$ whence, by (\ref{eq:6.1da}),
$T\in{\cal TF}_{G_0}(j,A)$ iff $T^t(j(z))A(z)T(z)=I_{3\times 3}$
which proves the claim.
\hfill $\Box$

\noindent{\em Proof of Theorem \ref{T09t2}b:}
By Remark 10, $({\cal P}_\omega,A)$ is on SOR iff $0\in\Xi({\cal P}_\omega,A)$
iff $({\cal P}_\omega,A)\in
{\cal CB}_{G_0}(d,{\cal P}_\omega)$. The claim now follows from
Definition \ref{D6}.
\hfill $\Box$

We will use Theorem \ref{T09t2} in the proof of the
Uniqueness Theorem, Theorem \ref{T7.1}b.
Moreover the Normal Form Theorem, Theorem \ref{T10.1} in
Section \ref{10.2.2add}, will
generalize Theorem \ref{T09t2}a from $G_0$ to
an arbitrary subgroup $H$ of $SO(3)$.
It will thereby turn out
that (\ref{eq:5.30cn})
is an example of a so-called stationarity equation. 
The second general theorem to be mentioned is the Cross Section Theorem,
Theorem \ref{P12.10} in
Section \ref{9.4b.4} which will show that the SOR
is a rather deep concept.
Spin-orbit resonances will be further studied
in the following chapter \ref{VII}.
\setcounter{equation}{0}
\section{Polarization}
\label{VII}
In this chapter we 
tie together the concepts of polarization field and polarization.
\subsection{Estimating the polarization}
\label{6.4}
Consider a family $(j_J,A_J)_{J\in\Lambda}$ of spin-orbit systems
where $(j_J,A_J)\in{\cal SOS}(d,j_J)$
and $\Lambda\subset\Rd$ is the set of action values.

We note (see also \cite{BH,BV1}) that,
for every $J\in\Lambda$, we have a so-called
``local polarization'',
say ${\cal S}_{loc,J}$, which by definition is a
polarization-field trajectory of $(j_J,A_J)$ satisfying
\begin{eqnarray}
&& |{\cal S}_{loc,J}|\leq 1 \; .
\label{eq:4.5an}
\end{eqnarray}
The associated polarization on the torus $J$ at time $n$
is then given by
\begin{eqnarray}
&& P_J(n):=(\frac{1}{2\pi})^d
\bigg| \int_{[0,2\pi]^d} \;d\phi {\cal S}_{loc,J}(n,\pi_d(\phi))\bigg| \; .
\label{eq:4.5l}
\end{eqnarray}
We will see below how $P_J$ can be estimated 
by (\ref{eq:4.7l}) which makes $P_J$ a convenient tool
for analyzing the bunch polarization.
In the so-called ``spin equilibrium'' the polarization-field trajectory
${\cal S}_{loc,J}$ is, by the definition of the spin equilibrium,
time-independent for every $J$ 
whence its initial value, ${\cal S}_{loc,J}(0,\cdot)$ 
is an invariant polarization field of $(j_J,A_J)$.
Thus for the spin equilibrium we get
\begin{eqnarray}
&& P_J(n)=P_J(0)=(\frac{1}{2\pi})^d \bigg|
\int_{[0,2\pi]^d}  \;d\phi {\cal S}_{loc,J}(0,\pi_d(\phi))\bigg| \; .
\label{eq:4.5nl}
\end{eqnarray}
Let $j_J$ be topologically transitive. Then, by Theorem \ref{T09t0},
$|{\cal S}_{loc,J}(0,z)|$ is independent of $z$ and, if
${\cal S}_{loc,J}(0,\cdot)$ is not the zero function, then
$|{\cal S}_{loc,J}(0,z)|>0$ and
${\cal S}_{loc,J}(0,\cdot)/|{\cal S}_{loc,J}(0,\cdot)|$ is an ISF of  
$(j_J,A_J)$ whence, by (\ref{eq:4.5an}),(\ref{eq:4.5nl}),
\begin{eqnarray}
&& P_J(n)=P_J(0)=(\frac{1}{2\pi})^d 
\bigg|\int_{[0,2\pi]^d}  \;d\phi 
|{\cal S}_{loc,J}(0,\pi_d(\phi))|\;\frac{{\cal S}_{loc,J}(0,\pi_d(\phi))}
{|{\cal S}_{loc,J}(0,\pi_d(\phi))|}\bigg| 
\nonumber\\
&& \quad \leq 
(\frac{1}{2\pi})^d 
\bigg|
\int_{[0,2\pi]^d}  \;d\phi \frac{{\cal S}_{loc,J}(0,\pi_d(\phi))}
{|{\cal S}_{loc,J}(0,\pi_d(\phi))|}\bigg| 
\; ,
\label{eq:4.5nla}
\end{eqnarray}
so that
\begin{eqnarray}
&& 
P_J(n) = P_J(0)\leq  P_{J,max} \; ,
\label{eq:4.7l}
\end{eqnarray}
where
\begin{eqnarray}
&&  \hspace{-15mm}
P_{J,max}:=(\frac{1}{2\pi})^d
\sup\biggl\lbrace
\bigg|\int_{[0,2\pi]^d}  \;d\phi  f(\pi_d(\phi))\bigg|:
 f\in{\cal ISF}(j_J,A_J)\biggr\rbrace \; .
\label{eq:4.8l}
\end{eqnarray}
Of course (\ref{eq:4.7l}) also holds if 
${\cal S}_{loc,J}(0,\cdot)$ is the zero function because in that case
$P_J(n) =P_J(0)=0$.
Thus (\ref{eq:4.7l}) holds for the spin equilibrium if
$j_J$ is topologically transitive and $(j_J,A_J)$ has an ISF.
We conclude from  (\ref{eq:4.7l}) that
the ISF's provide an upper bound for $P_J$ and this
is one reason why they are so important in practice.
One can simplify (\ref{eq:4.8l}) in the important case where
the spin-orbit system $(j_J,A_J)$ in (\ref{eq:4.8l})
has exactly two ISF's, say $f_J,-f_J$. 
Then (\ref{eq:4.8l}) simplifies to
\begin{eqnarray}
&& P_{J,max}=(\frac{1}{2\pi})^d
\bigg|\int_{[0,2\pi]^d}  \;d\phi f_J(\pi_d(\phi))\bigg| \; .
\label{eq:7.10l}
\end{eqnarray}
Clearly $P_{J,max}$ is small if the range of $f_J$ is spread out.
In Section \ref{8} we will see how the Uniqueness Theorem leads 
to the situation underlying (\ref{eq:7.10l}).

Of course $P_J$ can also be used for an estimation of the
bunch polarization which is given by
\begin{eqnarray}
&& P(n)=(\frac{1}{2\pi})^d
\bigg|\int_\Lambda  \;dJ \rho_{eq}(J)
\int_{[0,2\pi]^d} \;d\phi {\cal S}_{loc,J}(n,\pi_d(\phi))\bigg| \; ,
\label{eq:4.5}
\end{eqnarray}
where $(\frac{1}{2\pi})^d\rho_{eq}$ is the 
equilibrium particle phase-space density
(for more details underlying (\ref{eq:4.5}) see Section
\ref{10.5.1}).
Thus the bunch polarization for the combined beam equilibrium and
spin equilibrium reads as
\begin{eqnarray}
&& P(n)=P(0)=(\frac{1}{2\pi})^d
\bigg|\int_\Lambda  \;dJ \rho_{eq}(J)
\int_{[0,2\pi]^d}  \;d\phi {\cal S}_{loc,J}(0,\pi_d(\phi))\bigg| \; .
\label{eq:4.5n}
\end{eqnarray}
Let the conditions underlying (\ref{eq:4.7l}) hold for almost
all $J$, i.e., let a set $M\subset\Lambda$ exist which has Lebesgue measure
zero and such that, for every $J\in (\Lambda\setminus M)$,
the spin-orbit system $(j_J,A_J)$ has an ISF 
and $j_J$ is topologically transitive. Then, 
by (\ref{eq:4.5nl}),(\ref{eq:4.7l}),(\ref{eq:4.5n}), we have 
for the spin equilibrium
\begin{eqnarray}
&& P(n)=P(0)\leq (\frac{1}{2\pi})^d\int_\Lambda  \;dJ \rho_{eq}(J)
\bigg|\int_{[0,2\pi]^d}  \;d\phi 
{\cal S}_{loc,J}(0,\pi_d(\phi))\bigg| 
\nonumber\\
&&\quad = \int_\Lambda  \;dJ \rho_{eq}(J) P_J(0) \leq
\int_\Lambda  \;dJ \rho_{eq}(J) P_{J,max} 
\; .
\label{eq:4.6}
\end{eqnarray}
Note that we assume that $\rho_{eq}(J)$ and $P_{J,max}$
depend on $J$ regularly enough to ensure that 
the integrals in (\ref{eq:4.5}), (\ref{eq:4.5n}) and
(\ref{eq:4.6}) are meaningful. 
Using (\ref{eq:7.10l}) one
can simplify (\ref{eq:4.6}) in the case where, 
for every $J\in (\Lambda\setminus M)$,
the spin-orbit system $(j_J,A_J)$ has two ISF's 
$f_J,-f_J$ and no others. Then (\ref{eq:4.6}) simplifies, thanks to
(\ref{eq:7.10l}), to
\begin{eqnarray}
&& P(n) = P(0) \leq (\frac{1}{2\pi})^d
\int_\Lambda \;dJ \rho_{eq}(J)
\bigg|\int_{[0,2\pi]^d}  \;d\phi f_J(\pi_d(\phi))\bigg| \; ,
\label{eq:7.10}
\end{eqnarray}
where we also assume that the functional dependences of  $\rho_{eq}(J)$
and $f_J$ on $J$ are
regular enough to ensure that the integrals in (\ref{eq:7.10}) are meaningful.
For more details on estimating the bunch polarization,
also for non-equilibrium spin fields, see \cite{Ho,mv2000}.
\subsection{The Uniqueness Theorem of invariant spin fields}
\label{8}
We saw in (\ref{eq:4.7l}) and (\ref{eq:7.10l}),
how in a situation where only two ISF's exist,
the invariant
spin fields govern the estimation of $P_J$.
In this section we will see that this situation is very common
off spin-orbit resonance.

Let $(j,A)\in{\cal ACB}(d,j)$. 
Then, by Remark 4 in Chapter \ref{4}, $(j,A)$ has an ISF and so 
it natural to ask about the impact of the set
$\Xi(j,A)$ on ${\cal ISF}(j,A)$. In fact, if $j={\cal P}_\omega$ and
$({\cal P}_\omega,A)$ is off orbital resonance,
this question is partially answered by part b) of the following theorem.
\setcounter{theorem}{0}
\begin{theorem} \label{T7.1} 
\noindent 
a) Let $(j,A)\in{\cal SOS}(d,j)$ and let $f$ and $g$ be invariant 
polarization fields of $(j,A)$. Then $h\in {\cal C}(\Td,\R^3)$, defined by
$h(z):=f(z)\times g(z)$, is an invariant 
polarization field of $(j,A)$ where $\times$ denotes the vector product.\\

\noindent b)
(The Uniqueness Theorem)
\noindent
Let $({\cal P}_\omega,A)\in{\cal ACB}(d,{\cal P}_\omega)$ 
be off orbital resonance, i.e.,
let $(1,\omega)$ be nonresonant. Also, let $({\cal P}_\omega,A)$ be 
off spin-orbit
resonance. Then $({\cal P}_\omega,A)$ has an ISF, say $F$, and
$F$ and $-F$ are the only ISF's of $({\cal P}_\omega,A)$.
\end{theorem}

\noindent {\em Proof of Theorem \ref{T7.1}a:} 
Since $f$ and $g$ are invariant 
polarization fields of $(j,A)$ it follows from Definition \ref{D6.1}
that $f\circ j = Af$ and $g\circ j = Ag$ whence
\begin{eqnarray}
&&  \hspace{-5mm}
h(j(z)) = (f(j(z))\times g(j(z)))= (A(z)f(z)\times A(z)g(z))
=  A(z)(f(z)\times g(z)) = A(z)h(z) \; ,
\nonumber
\end{eqnarray}
so that, by Definition \ref{D6.1}, $h$ is an invariant 
polarization field of $(j,A)$.
\hfill $\Box$  

\noindent {\em Proof of Theorem \ref{T7.1}b:} 
Let $({\cal P}_\omega,A)\in{\cal ACB}(d,{\cal P}_\omega)$ 
be off orbital resonance. The claim to be proved is equivalent
to its contrapositive which is
the following claim:
If the total number of ISF's of $({\cal P}_\omega,A)$ is not $2$, then
$({\cal P}_\omega,A)$ is not off spin-orbit resonance. 
Now, we know from
Remark 4 in Chapter \ref{4} that $({\cal P}_\omega,A)$
has at least two ISF's so that if the number of ISF's differs from $2$, 
there are more than two ISF's. Also,
since $({\cal P}_\omega,A)\in{\cal ACB}(d,{\cal P}_\omega)$ we know from
a remark after Definition \ref{D4} 
that $({\cal P}_\omega,A)$ has spin tunes. Then if the system is  
not off spin-orbit resonance, it must be on spin-orbit resonance.
Thus the above claim we have to prove is equivalent
to the following claim:
If the total number of ISF's of $({\cal P}_\omega,A)$ is larger than two, then
$({\cal P}_\omega,A)$ is on spin-orbit resonance.

In fact we will now prove the latter claim.
So let $({\cal P}_\omega,A)$ 
have more than two ISF's. Recalling Section \ref{6.1b}, we then conclude
that $({\cal P}_\omega,A)$ has ISF's, say $f$ and $g$, such
that $g\neq f$ and $g\neq -f$.  Note that $f,-f$ and $g$ are three different
ISF's of $({\cal P}_\omega,A)$.  We define
$h\in {\cal C}(\Td,\R^3)$ by $h(z):=f(z)\times g(z)$ and observe, by 
Theorem \ref{T7.1}a, that $h$ is an invariant 
polarization field of $({\cal P}_\omega,A)$.
On the other hand, since $({\cal P}_\omega,A)$ is 
off orbital resonance, ${\cal P}_\omega$ is topologically transitive
whence, by Theorem \ref{T09t0},
$|h|$ is constant, i.e., $|h(z)|=:\lambda$ is independent of $z$.
We first consider the case where
$\lambda=0$, i.e., where $f\times g$ is the zero function. Then
a function $\tilde{h}:\Td\rightarrow\R$ exists such that
$g=\tilde{h}f$ whence $g\cdot f=\tilde{h}|f|^2=\tilde{h}$ which implies that
$\tilde{h}$ is continuous. On the other hand 
$1=|g|=|\tilde{h}f|=|\tilde{h}|$ whence $\tilde{h}$ 
can take values only in $\lbrace 1,-1\rbrace$
whence, since $\tilde{h}$ is continuous and
$\Td$ is pathwise connected, $\tilde{h}$ is constant. Thus either
$g=f$ or $g=-f$ which is a contradiction. So the case where $\lambda=0$
cannot occur. Thus $\lambda>0$.
Since $h$ is an invariant polarization field of $({\cal P}_\omega,A)$ 
and since the real number $\lambda$ is positive we define
$k\in {\cal C}(\Td,\R^3)$ by $k(z):=h(z)/\lambda=h(z)/|h(z)|$ 
and observe, by using
Definition \ref{D6.1}, that $k$ is an invariant polarization field of 
$({\cal P}_\omega,A)$. Of course $|k(z)|=|h(z)|/|h(z)|=1$ whence  
$k$ is an ISF of $({\cal P}_\omega,A)$.
We also define $l\in {\cal C}(\Td,\R^3)$ by $l(z):=k(z)\times f(z)$ 
and observe, by 
Theorem \ref{T7.1}a, that $l$ is an invariant 
polarization field of $({\cal P}_\omega,A)$.
Of course $f(z)\cdot k(z)=(f(z)\cdot h(z))/|h(z)|
=f(z)\cdot(f(z)\times g(z))/\lambda=0$ 
whence, for every $z\in \Td$,
\begin{eqnarray}
&& 0=l(z)\cdot k(z) =l(z)\cdot f(z) =f(z)\cdot k(z) \; .
\label{eq:7.5}
\end{eqnarray}
Clearly $|l(z)|=|k(z)\times f(z)|=\sqrt{ |k(z)|^2\;|f(z)|^2-(k(z)\cdot f(z))^2}
=\sqrt{1-(k(z)\cdot f(z))^2}=1$ 
which implies that $l$ is an ISF of $({\cal P}_\omega,A)$ and that
\begin{eqnarray}
&& 1=|l(z)|=|k(z)|=|f(z)| \; .
\label{eq:7.6}
\end{eqnarray}
It follows from (\ref{eq:7.5}) and (\ref{eq:7.6}) that 
\begin{eqnarray}
&& [l(z),k(z),f(z)]^t[l(z),k(z),f(z)]=I_{3\times 3} \; .
\label{eq:7.7}
\end{eqnarray}
Moreover, by (\ref{eq:7.6}),
$\det([l(z),k(z),f(z)])
=l(z)\cdot(k(z)\times f(z))=|l(z)|^2=1$ whence, by (\ref{eq:7.7}), 
for every $z\in \Td$, the $3\times 3$-matrix
$[l(z),k(z),f(z)]$ belongs to $SO(3)$.
We thus define $T\in {\cal C}(\Td,SO(3))$ by
$T(z):=[l(z),k(z),f(z)]$. Since all three columns of $T$ are 
invariant polarization fields of
$({\cal P}_\omega,A)$ we have, by Definition \ref{D6.1},
\begin{eqnarray}
&& A(z)T(z)=A(z)[l(z),k(z),f(z)]=[A(z)l(z),A(z)k(z),A(z)f(z)]
\nonumber\\
&&\quad =[l({\cal P}_\omega(z)),k({\cal P}_\omega(z)),f({\cal P}_\omega(z))]
=T({\cal P}_\omega(z)) \; ,
\nonumber
\end{eqnarray}
whence $T \circ {\cal P}_\omega = A T$ so that, by Theorem \ref{T09t2}a,
$T$ belongs to ${\cal TF}_{G_0}(j,A)$. 
Thus, by Theorem \ref{T09t2}b,
$({\cal P}_\omega,A)$ is on spin-orbit resonance as was to
be shown.
\hfill $\Box$ 

\vspace{3mm}

The claim of Theorem \ref{T7.1}b that $({\cal P}_\omega,A)$
has an ISF is trivial because of Remark 4 in Chapter \ref{4}. Thus the essence 
of the claim of Theorem \ref{T7.1}b is that $({\cal P}_\omega,A)$
has only two ISF's. Recall also from Chapter \ref{6}
that the set of ISF's of a spin-orbit system is either infinite
or contains an even number of elements.
Note that in this work the term 
``finite number'' includes the
case of zero. Indeed if a spin-orbit system has no ISF then its
number of ISF's is zero, an even number!
\section{Unified treatment of spin-orbit systems by using the 
Technique of Association (ToA)}
\label{10}
\setcounter{equation}{0}
\subsection{Orientation}
\label{10.1}
We now come to our generalization of the notions of particle-spin-vector
motion and polarization-field motion by introducing our
``Technique of Association'' (ToA).
With this we will see that while the
spin-orbit systems are still the same, 
they can be exploited further to generate
and encompass new perspectives. We thereby see 
that while the spin-orbit systems do not change, their scope
widens. 
By the ToA the well established notions of invariant polarization field and
invariant spin field are generalized to invariant
$(E,l)$-fields where $(E,l)$ is an $SO(3)$-space.
The origin of the ToA is an underlying principal bundle and its associated
bundles, hence the terminology. 
However since the principal bundle is of product form, we can 
easily present the theory
in a fashion which does not use bundle theory.
For a short account of the bundle aspect see Section \ref{10.7}
where we also briefly mention the relation to Yang-Mills Theory.
Several major theorems are presented, among them 
the Normal Form Theorem which ties invariant fields with the notion
of normal form, the Decomposition Theorem,
which allows one to compare different invariant fields,
the Invariant Reduction Theorem, which gives
new insights into the question of existence of 
invariant fields (and in particular invariant spin fields), 
and 
the Cross Section Theorem which supplements the Invariant Reduction Theorem.
It thus turns out that the well established notions of
invariant frame field, spin tune, and spin-orbit
resonance are generalized by the normal form concept whereas
the well established notions of invariant polarization field and
invariant spin field are generalized to invariant $(E,l)$-fields.
In particular we see that the $SO(3)$-space $(\R^3,l_{v})$ 
has been implicitly used in Chapters \ref{2}-\ref{VII}.
With the flexibility in the choice of $(E,l)$
we also have a unified way to study the dynamics of spin-$1/2$ 
and spin-$1$ particles.
Accordingly the special cases $(E,l)=(\R^3,l_{v})$ (for spin vectors)
and $(E,l)=(E_{t},l_{t})$ (for spin tensors) 
are discussed in some detail.
\subsection{Defining the ToA}
\label{10.2}
\subsubsection{The maps}
\label{10.2.1}
For given $(E,l)$ each
particle carries, in addition to its position $z$ on the torus $\Td$,
an $E$-valued quantity $x$ that we call spin.
Depending on the choice of $(E,l)$, $x$ can be the spin vector $S$
or another quantity related to spin motion.
We consider the autonomous DS given by the $1$-turn particle-spin map
\begin{eqnarray}
&&  \hspace{-1cm}
\left( \begin{array}{c} z\\ x\end{array}\right)
\mapsto
\left( \begin{array}{c} z'\\ x'\end{array}\right)
=\left( \begin{array}{c} j(z)\\
l(A(z);x)\end{array}\right) =: {\cal P}[E,l,j,A](z,x)
 \; ,
\label{eq:10.15}
\end{eqnarray}
where $z,z'\in\Td,x,x'\in E$. 
In our formalism, (\ref{eq:10.15}) is the most
general description of particle-spin dynamics
and the choice of $(E,l)$ depends on the situation, e.g.,
$(E,l)=(\R^3,l_{v})$ for spin-$1/2$ particles - see below
(in that case $x$ is the spin vector $S$).
Note that the function
${\cal P}[E,l,j,A]:\Td\times E\rightarrow\Td\times E$, defined by
(\ref{eq:10.15}), belongs to $\Homeo(\Td\times E)$ since its inverse
is ${\cal P}[E,l,j^{-1},A^t\circ j^{-1}]$.

\noindent 
For an $E$-valued field
$f:\Td\rightarrow E$ set 
$x=f(z)$ in (\ref{eq:10.15}) so that the particle motion
moves $z$ to $j(z)$
and the field value at $j(z)$ becomes $l(A(z);f(z))$. Thus the field
$f$ evolves into the field 
$f':\Td\rightarrow E$ where
$f'(z):=l(A(j^{-1}(z));f(j^{-1}(z)))$.
Therefore we have obtained a map of fields, i.e., 
the autonomous discrete-time DS given by the
$1$-turn field map
\begin{eqnarray}
&&  \hspace{-1cm}
f\mapsto f':= l\biggl(A\circ j^{-1};f\circ j^{-1}\biggr) 
 =: \tilde{\cal P}[E,l,j,A](f)
\; .
\label{eq:10.17}
\end{eqnarray}
We are only interested in continuous fields, i.e., $f\in {\cal C}(\Td,E)$
so that   $f'\in {\cal C}(\Td,E)$) too.
Note that the function
$\tilde{\cal P}[E,l,j,A]:{\cal C}(\Td,E)\rightarrow
{\cal C}(\Td,E)$, defined by (\ref{eq:10.17}), is a bijection
since its inverse is $\tilde{\cal P}[E,l,j^{-1},A^t\circ j^{-1}]$.
We call an $f\in {\cal C}(\Td,E)$
an ``invariant $(E,l)$-field of $(j,A)$'' if it is mapped by 
(\ref{eq:10.17}) into itself, i.e., if
\begin{eqnarray}
&&  \hspace{-1cm}
f\circ j = l(A;f) \; .
\label{eq:10.18}
\end{eqnarray}
We call (\ref{eq:10.18}) the ``$(E,l)$-stationarity
equation of $(j,A)$''. Clearly an $f\in {\cal C}(\Td,E)$
is an invariant $(E,l)$-field of $(j,A)$ iff 
$\tilde{\cal P}[E,l,j,A](f)=f$.
We call an $f\in {\cal C}(\Td,E)$
an ``invariant $n$-turn $(E,l)$-field of $(j,A)$'' iff 
$\tilde{\cal P}[E,l,j,A]^n(f)=f$ where $n$ is a positive integer.
Of course the notions of invariant $(E,l)$-field 
and invariant $1$-turn $(E,l)$-field are identical. 
Invariant $n$-turn $(E,l)$-fields with $n>1$ 
play a role for spin-orbit systems
$j={\cal P}_\omega$ with $(1,\omega)$ nonresonant (see Remark 12 and
Section \ref{10.4b.2}).

\vspace{3mm}
\noindent{\bf Remark:}
\begin{itemize}
\item[(1)] 
Consider the special case $(E,l)=(\R^3,l_{v})$ where we define the function
$l_{v}:SO(3)\times\R^3\rightarrow\R^3$ by
\begin{eqnarray}
&&  \hspace{-1cm}
l_{v}(r,S):=rS \; .
\label{eq:10.19}
\end{eqnarray}
Then the above particle-spin and field maps of
(\ref{eq:10.15}) and (\ref{eq:10.17}) 
become the particle-spin-vector and polarization 
field maps we know from Chapters 
\ref{2}-\ref{VII}. 
In fact it is a simple exercise to show that
$(\R^3,l_{v})$ is an $SO(3)$-space and that, by (\ref{eq:2.3b}), 
the map (\ref{eq:10.15}) becomes
\begin{eqnarray}
&&  \hspace{-1cm}
\left( \begin{array}{c} z\\ S\end{array}\right)
\mapsto\left( \begin{array}{c} 
j(z)\\
l_{v}(A(z);S)\end{array}\right) 
=\left( \begin{array}{c} 
j(z)\\
A(z)S\end{array}\right) 
= {\cal P}[j,A](z,S) \; ,
\label{eq:10.22}
\end{eqnarray}
i.e., ${\cal P}[\R^3,l_{v},j,A]={\cal P}[j,A]$
and, by (\ref{eq:xx13.2n}), the map (\ref{eq:10.17}) becomes
\begin{eqnarray}
&&  \hspace{-1cm}
f\mapsto f' =l_{v}\biggl(A\circ j^{-1};f\circ j^{-1}
\biggr) 
=(Af)\circ j^{-1}=\tilde{\cal P}[j,A](f) \; ,
\label{eq:10.23}
\end{eqnarray}
i.e., $\tilde{\cal P}[\R^3,l_{v},j,A]=\tilde{\cal P}[j,A]$.

With (\ref{eq:10.22}) and (\ref{eq:10.23}) we see
that it is the special case $(E,l)=(\R^3,l_{v})$
that underlies the particle-spin-vector 
and polarization field motion of Chapters 
\ref{2}-\ref{VII}. Thus the ToA is a generalization of
the particle-spin-vector and polarization 
field motion of Chapters 
\ref{2}-\ref{VII} to arbitrary $(E,l)$. 

We also recover the notion of invariant polarization field.
In fact an $f\in {\cal C}(\Td,E)$ is an 
invariant $(\R^3,l_{v})$-field of $(j,A)$ iff it satisfies
the $(\R^3,l_{v})$-stationarity
equation of $(j,A)$, $f\circ j= l_{v}(A;f)$, i.e.,
iff $f\circ j= Af$. Thus, by Definition \ref{D6.1},
the notion of invariant polarization
field is identical with the notion of
 ``invariant $(\R^3,l_{v})$-field''.
Since we use the terminology ``field'' so often, it
is important to mention that the notions of 
IFF and uniform IFF are different from the notion
of invariant $(E,l)$-field.
\hfill $\Box$
\end{itemize}
The $SO(3)$-spaces $(E,l)$ can take a variety of forms. For example
in addition to $(\R^3,l_{v})$ we will consider 
$(E_{t},l_{t})$ where $E_{t}\subset\R^{3\times 3}$ is defined
in Section \ref{10.4.1}. This $SO(3)$-space is used for 
studies of polarized beams of spin-$1$ particles like deuterons.
\subsubsection{The trajectories}
\label{10.2.2}
Iterating the particle-spin 
map (\ref{eq:10.15}), the particle-spin trajectories 
$(z(\cdot),
x(\cdot))$ are defined by
\begin{eqnarray}
&&  \hspace{-1cm}
\left( \begin{array}{c} z(n+1)\\ x(n+1)\end{array}\right)
=\left( \begin{array}{c} j(z(n))\\
l(A(z(n));x(n))\end{array}\right) \; ,
\label{eq:10.25}
\end{eqnarray}
with $z(0)=z_0,x(0)=x_0$ whence
\begin{eqnarray}
&&  \hspace{-1cm}
\left( \begin{array}{c} z(n)\\ x(n)\end{array}\right)
=\left( \begin{array}{c} j^n(z_0)\\
l\biggl(A(j^{n-1}(z_0))A(j^{n-2}(z_0))\cdots A(z_0));x_0)
\end{array}\right) \; .
\label{eq:10.30}
\end{eqnarray}
It is convenient to introduce the corresponding $\Z$-action which is
the function $L[E,l,j,A]:\Z\times \Td\times E\rightarrow \Td\times E$ 
defined by
\begin{eqnarray}
&& L[E,l,j,A](n;z,x):= {\cal P}[E,l,j,A]^n(z,x)=
\left( \begin{array}{c}
L[j](n;z) \\ l( \Psi[j,A](n;z);x)
\end{array}\right) \; .
\label{eq:10.36}
\end{eqnarray}
It is easy to show that $(\Td\times E,L[E,l,j,A])$ is a $\Z$-space.
In the study of this $\Z$-space, which will not be 
fully addressed in this work, it is
of key importance that $\Psi[j,A]$ is a cocycle (recall Definition
\ref{D2.5}).
With (\ref{eq:3.20}) and $L[E,l,j,A]$
the solution (\ref{eq:10.30}) of (\ref{eq:10.25}) can be written as
\begin{eqnarray}
&&  \hspace{-1cm}
\left( \begin{array}{c} z(n)\\ x(n)\end{array}\right)
=  L[E,l,j,A](n,z_0,x_0) \; .
\label{eq:10.42}
\end{eqnarray}
For the record, we call a function
$(z(\cdot),x(\cdot)):\Z\rightarrow \Td\times E$ 
an $(E,l)$-trajectory (or, just 
``particle-spin trajectory'') of $(j,A)$
if (\ref{eq:10.42}) holds for all
$n\in\Z$ (i.e., if (\ref{eq:10.25})  holds for all $n\in\Z$).

On iteration of the field map (\ref{eq:10.17}),
the field trajectories $F$ emerge in terms of the equation of motion
\begin{eqnarray}
&& F(n+1,z)=
l\biggl(A(j^{-1}(z));F(n,j^{-1}(z))\biggr) \; ,
\label{eq:10.43}
\end{eqnarray}
whence
\begin{eqnarray}
&&  \hspace{-1cm}
F(n,z)=l\biggl(  \Psi[j,A](n;L[j](-n;z));F(0,L[j](-n;z))\biggr) \; ,
\label{eq:10.44}
\end{eqnarray}
where, as always, $\Psi[j,A](n;\cdot)$ is the 
spin transfer matrix function of $(j,A)$. 
For the record, we call a continuous function
$F:\Z\times\Td\rightarrow E$ 
an $(E,l)$-field trajectory of $(j,A)$ if it satisfies (\ref{eq:10.43}) 
or, equivalently, (\ref{eq:10.44})). 
The terminology ``trajectory'' is justified since the function
$n\mapsto F(n,\cdot)$ is a ``trajectory'' of fields belonging to
${\cal C}(\Td,E)$. Clearly an $(E,l)$-field trajectory is
time-independent iff its initial value 
$F(0,\cdot)\in {\cal C}(\Td,E)$ is an invariant $(E,l)$-field.
The notion of $(E,l)$-field trajectory generalizes the notion of
polarization-field trajectory as pointed out in Remark 2 below.

We define the $\Z$-action
$\tilde{L}[E,l,j,A]:\Z\times {\cal C}(\Td,E)
\rightarrow {\cal C}(\Td,E)$ by
\begin{eqnarray}
&& \hspace{-1cm}
\tilde{L}[E,l,j,A](n;f):= g \; , \quad
g(z):=l\biggl( \Psi[j,A]\biggl(n;L[j](-n;z)\biggr);f(L[j](-n;z))\biggr) \; .
\label{eq:10.45}
\end{eqnarray}
With (\ref{eq:10.45}) we can write (\ref{eq:10.44}) as
\begin{eqnarray}
&&  \hspace{-1cm}
F(n,\cdot) =  \tilde{L}[E,l,j,A](n,F(0,\cdot)) \; .
\label{eq:10.46}
\end{eqnarray}
Clearly a continuous function
$F:\Z\times\Td\rightarrow E$ is
an $(E,l)$-field trajectory of $(j,A)$ iff it satisfies (\ref{eq:10.46}).

If $F$ is an $(E,l)$-field trajectory and $z(\cdot)$ 
is an particle trajectory of $(j,A)$ then by (\ref{eq:10.44})
the function $n\mapsto (z(n),F(n,z(n))$ is an
$(E,l)$-trajectory of $(j,A)$. Thus the particle-spin motion can be viewed
as a characteristic motion of the field motion. Of course in the
special case $(E,l)=(\R^3,l_{v})$ this is well-known and is the basic 
fact underlying all spin vector tracking methods.

\vspace{3mm}
\noindent{\bf Remark:}
\begin{itemize}
\item[(2)] In the special case where
$(E,l)=(\R^3,l_{v})$ the
above trajectories become trajectories from Chapters 
\ref{2}-\ref{VII}. In fact, by (\ref{eq:s2.10}), we have
$L[\R^3,l_{v},j,A]=L[j,A]$.
This implies that the notion of ``$(\R^3,l_{v})$-trajectory 
is identical to the notion of
 ``particle-spin-vector trajectory''.

Moreover, by (\ref{eq:xx13.2}), we have
$\tilde{L}[\R^3,l_{v},j,A]=\tilde{L}[j,A]$.
This implies that the notion of ``$(\R^3,l_{v})$-field trajectory 
is identical to the notion of
 ``polarization-field trajectory''.
\hfill $\Box$
\end{itemize}
Since we work in the framework of topological dynamical systems, $A,j,l$ are
continuous functions and we therefore require our fields 
to be continuous, in particular the invariant $(E,l)$-fields.
Thus every $(E,l)$-field trajectory $F$ 
fulfills two different conditions:
the ``dynamical'' condition (\ref{eq:10.43})
and the ``regularity'' condition that $F(0,\cdot)$ is continuous.
However, in contrast to the dynamical condition, the
regularity condition is a matter of choice.
While in this work, and in \cite{He2}, we choose continuity
as the regularity property, this  property
can basically vary between the extremes
``Borel measurable'' and ``of class $C^\infty$''.
\subsubsection{The First ToA Transformation Rule}
\label{10.2.3}
The First ToA Transformation Rule generalizes the transformation rule given
by (\ref{eq:2.3bna}) and (\ref{eq:5.1n}) to
(\ref{eq:10.55}) and (\ref{eq:10.55a}), i.e., it is a generalization
from $(\R^3,l_{v})$ to $(E,l)$ whence it
is closely related to the concept of
the $H$ normal form.
We aim to understand the dependence of particle-spin
and field motions on $A$ in
the general ToA setting just as we did in the
$(E,l)=(\R^3,l_{v})$ setting of Chapters \ref{2}-\ref{VII}. 
By the transformation rule (\ref{eq:5.35}),
the map (\ref{eq:10.22}), i.e.,
\begin{eqnarray}
&&  \hspace{-1cm}
\left( \begin{array}{c} z\\ S\end{array}\right)
\mapsto\left( \begin{array}{c} 
j(z)\\
A(z)S\end{array}\right) \; ,
\label{eq:10.50}
\end{eqnarray}
is transformed into the map
\begin{eqnarray}
&&  \hspace{-1cm}
\left( \begin{array}{c} \zeta \\ \sigma\end{array}\right)
\mapsto
\left( \begin{array}{c} \zeta'\\ \sigma'\end{array}\right)
=\left( \begin{array}{c} 
j(\zeta)\\
A'(\zeta)\sigma\end{array}\right) \; ,
\label{eq:10.51}
\end{eqnarray}
where
\begin{eqnarray}
&&  \hspace{-1cm}
\left( \begin{array}{c} \zeta \\ \sigma\end{array}\right)
:=\left( \begin{array}{c} z \\
T^t(z) S\end{array}\right) \; ,
\label{eq:10.52}
\end{eqnarray}
with $T\in {\cal C}(\Td,SO(3))$ and $A'\in {\cal C}(\Td,SO(3))$ defined by
(\ref{eq:5.25}), i.e., \\
$A'(z): =T^t(j(z))A(z)T(z)$.
It is easy to generalize the transformation rule 
(\ref{eq:10.50}) and (\ref{eq:10.51}) by replacing
$(\R^3,l_{v})$ with $(E,l)$.
Thus the map (\ref{eq:10.15}), i.e.,
\begin{eqnarray}
&&  \hspace{-1cm}
\left( \begin{array}{c} z\\ x\end{array}\right)
\mapsto\left( \begin{array}{c} 
j(z)\\
l(A(z);x)\end{array}\right) \; ,
\label{eq:10.50b}
\end{eqnarray}
is transformed into the map
\begin{eqnarray}
&&  \hspace{-1cm}
\left( \begin{array}{c} \zeta \\ \xi \end{array}\right)
\mapsto
\left( \begin{array}{c} \zeta'\\ \xi'\end{array}\right)
=\left( \begin{array}{c} 
j(\zeta)\\
l(A'(\zeta);\xi)\end{array}\right) \; ,
\label{eq:10.51b}
\end{eqnarray}
where
\begin{eqnarray}
&&  \hspace{-1cm}
\left( \begin{array}{c} \zeta \\ \xi\end{array}\right)
:=\left( \begin{array}{c} z \\
l(T^t(z);x)\end{array}\right) \; ,
\label{eq:10.52ba}
\end{eqnarray}
with $T\in {\cal C}(\Td,SO(3))$ and $A'\in {\cal C}(\Td,SO(3))$ defined by
(\ref{eq:5.25}), i.e., $A'(z): =T^t(j(z))A(z)T(z)$.
\vspace{3mm}
\noindent{\bf Remark:}
\begin{itemize}
\item[(3)] 
Using the notation of (\ref{eq:10.15}) one sees that
${\cal P}[E,l,j,A]$ is the map 
(\ref{eq:10.50b}) and that
${\cal P}[E,l,j,A']$ is the map (\ref{eq:10.51b}) whence
${\cal P}[E,l,j,A]$ is transformed into ${\cal P}[E,l,j,A']$.
Moreover it is a simple exercise to show that
\begin{eqnarray}
&&  \hspace{-1cm}
{\cal P}[E,l,j,A']= {\cal P}[E,l,id_\Td,T]^{-1}
\circ {\cal P}[E,l,j,A]
\circ {\cal P}[E,l,id_\Td,T] \; .
\label{eq:10.55}
\end{eqnarray}

In fact by 
defining ${\cal T}:= {\cal P}[E,l,id_\Td,T]^{-1}$ we 
first note that
\begin{eqnarray}
&&  \hspace{-1cm}
{\cal T}^{-1}(z,x)= 
\left( \begin{array}{c} z \\
l(T(z);x)\end{array}\right) \; ,
\label{eq:10.60}
\end{eqnarray}
whence, by (\ref{eq:3.20}),(\ref{eq:5.25}) and (\ref{eq:10.15}),
\begin{eqnarray}
&&  \hspace{-1cm}
\biggl( {\cal T}\circ {\cal P}[E,l,j,A]
\circ {\cal T}^{-1}\biggr)(z,x) =
\biggl({\cal T}\circ {\cal P}[E,l,j,A]\biggr)
\left( \begin{array}{c} z \\
l(T(z);x)\end{array}\right)
\nonumber\\
&& \quad
=  {\cal T}\left( \begin{array}{c}
j(z) \\ l( A(z);l(T(z);x))
\end{array}\right) 
=  {\cal T}\left( \begin{array}{c}
j(z) \\ l( A(z)T(z);x)
\end{array}\right) 
\nonumber\\
&&= \left( \begin{array}{c}
j(z) \\ l\biggl(T^t(j(z));l(A(z)T(z);x)\biggr)
\end{array}\right) 
= \left( \begin{array}{c}
j(z) \\ l\biggl(T^t(j(z))A(z)T(z);x\biggr)
\end{array}\right)
\nonumber\\
&& \quad
= {\cal P}[E,l,j,A'](z,x) \; ,
\nonumber
\end{eqnarray}
as was to be shown.
\hfill $\Box$
\end{itemize}
By iteration of the maps, the First ToA 
Transformation Rule, (\ref{eq:10.50b}) and (\ref{eq:10.51b}), delivers the following
transformation rule of trajectories:
\begin{eqnarray}
&& (z(\cdot),x(\cdot)) \longrightarrow (z(\cdot),x'(\cdot)) \; , \quad
x'(n) := l(T^t(z(n));x(n)) \; ,
\label{eq:10.60a}
\end{eqnarray}
and we observe that if $(z(\cdot),x(\cdot))$ is an
$(E,l)$-trajectory of $(j,A)$ then  $(z(\cdot),x'(\cdot))$ is a
$(E,l)$-trajectory of $(j,A')$.
In the special case where $(E,l)=(\R^3,l_{v})$
the transformation rule (\ref{eq:10.60a}) becomes (\ref{eq:5.35}).

\vspace{3mm}
\noindent{\bf Remark:}
\begin{itemize}
\item[(4)] By iterating (\ref{eq:10.55}) it is an easy exercise to show that
${\cal P}[E,l,id_\Td,T]^{-1}$ is an isomorphism from the $\Z$-space
$(\Td\times E,L[E,l,j,A])$ to the $\Z$-space
$(\Td\times E,L[E,l,j,A'])$. Thus the particle-spin motion
of $(j,A')$ is redundant since it can be covered by
the particle-spin motion of $(j,A)$.
\hfill $\Box$
\end{itemize}
With fields we proceed analogously.
In fact the map (\ref{eq:10.17}), i.e.,
\begin{eqnarray}
&&  \hspace{-1cm}
f\mapsto f':= l\biggl(A\circ j^{-1};f\circ j^{-1}\biggr) \; ,
\label{eq:10.65}
\end{eqnarray}
is transformed into the map 
\begin{eqnarray}
&&  \hspace{-1cm}
g\mapsto g':= l\biggl(A'\circ j^{-1};g\circ j^{-1}\biggr) \; ,
\label{eq:10.66}
\end{eqnarray}
where
\begin{eqnarray}
&&  \hspace{-1cm}
g:=l(T^t;f) \; ,
\label{eq:10.67}
\end{eqnarray}
with $T\in {\cal C}(\Td,SO(3))$ and $A'\in {\cal C}(\Td,SO(3))$ defined by
(\ref{eq:5.25}). 

\vspace{3mm}
\noindent{\bf Remark:}
\begin{itemize}
\item[(5)] 
Using the notation of 
(\ref{eq:10.17}) one sees that
$\tilde{\cal P}[E,l,j,A]$ is the map (\ref{eq:10.65}) and that
$\tilde{\cal P}[E,l,j,A']$ is the map (\ref{eq:10.66}) whence
$\tilde{\cal P}[E,l,j,A]$ is transformed into 
$\tilde{\cal P}[E,l,j,A']$.
Moreover it is a simple exercise to show that
\begin{eqnarray}
&&  \hspace{-1cm}
\tilde{\cal P}[E,l,j,A'] = \tilde{\cal P}[E,l,id_\Td,T]^{-1}
\circ \tilde{\cal P}[E,l,j,A]
\circ \tilde{\cal P}[E,l,id_\Td,T] \; .
\label{eq:10.55a}
\end{eqnarray}
\hfill $\Box$
\end{itemize}
By iteration of  maps the First ToA 
Transformation Rule, (\ref{eq:10.65}) and (\ref{eq:10.66}), delivers the following
transformation rule of $(E,l)$-field trajectories:
\begin{eqnarray}
&&   F \longrightarrow F' \; , \quad
F'(n,z): =l(T^t(z);F(n,z)) \; ,
\label{eq:10.68}
\end{eqnarray}
and we observe that if $F$ is an
$(E,l)$-field trajectory of $(j,A)$ then $F'$ is a
$(E,l)$-field trajectory of $(j,A')$
with $T\in {\cal C}(\Td,SO(3))$ and $A'\in {\cal C}(\Td,SO(3))$ defined by
(\ref{eq:5.25}).
In the special case where $(E,l)=(\R^3,l_{v})$ 
the transformation rule (\ref{eq:10.68}) becomes (\ref{eq:5.1}).

\vspace{3mm}
\noindent{\bf Remark:}
\begin{itemize}
\item[(6)] By iterating (\ref{eq:10.55a}) it is an easy exercise to show that
$\tilde{\cal P}[E,l,id_\Td,T]^{-1}$
is an isomorphism from the $\Z$-set
$({\cal C}(\Td,E),\tilde{L}[E,l,j,A])$ to the $\Z$-set
$({\cal C}(\Td,E),\tilde{L}[E,l,j,A'])$. 
So the field motion of $(j,A')$
is redundant since it can be covered by 
the field motion of $(j,A)$.
\hfill $\Box$
\end{itemize}
While in the First ToA Transformation Rule $(E,l)$ and $j$ are held
fixed and $A$ is transformed, we will also introduce, in Section \ref{10.2.4},
the Second ToA Transformation Rule where $j$ and
$A$ are held fixed and $(E,l)$ is transformed.
\subsubsection{The Normal Form Theorem (NFT)}
\label{10.2.2add}
In this section we generalize the IFF Theorem, Theorem \ref{T09t1}b, and
Theorem \ref{T09t2}a into the NFT. With this we expect the latter 
to have implications for the concepts of ISF, IFF and SOR. In fact its
importance goes even beyond this as will become evident later. 
Recall that  Theorems \ref{T09t1}b and
\ref{T09t2}a are linked to the groups $SO(2)$ and $G_0$ respectively.
In fact we will show in the examples in Remarks 8 and 9 below that in
Theorems \ref{T09t1}b and Theorem \ref{T09t2}a 
the groups $SO(2)$ and $G_0$ play the role of isotropy groups.
The NFT links the notions of invariant $(E,l)$-field and
$H$-normal form for arbitrary subgroups $H$ of $SO(3)$.

\setcounter{theorem}{0}
\begin{theorem} \label{T10.1} (NFT)
Let $(E,l)$ be an $SO(3)$-space, fix $x\in E$ and define\\
$H(x):=Iso(E,l;x)=\lbrace r\in SO(3):l(r;x)=x\rbrace$.
Moreover, let $T\in{\cal C}(\Td,SO(3))$,
$(j,A)\in{\cal SOS}(d,j)$ and define the function
$f\in{\cal C}(\Td,E)$ by 
\begin{eqnarray}
&& f(z):=l(T(z);x) \; .
\label{eq:10.152a}
\end{eqnarray}
Then, for all $z\in\Td$,
\begin{eqnarray}
&& T^t(j(z))A(z)T(z) \in H(x) \; ,
\label{eq:10.130a}
\end{eqnarray}
iff, for all $z\in\Td$,
\begin{eqnarray}
&& f(j(z)) = l(A(z);f(z))
\; .
\label{eq:10.146a}
\end{eqnarray}
In other words, $T\in{\cal TF}_{H(x)}(j,A)$ iff $f$ 
is an invariant $(E,l)$-field of $(j,A)$.
\end{theorem}

\noindent{\em Proof of Theorem \ref{T10.1}:}
``$\Rightarrow$'': 
Let $T$ satisfy (\ref{eq:10.130a}), i.e., 
let $l(T^t(j(z))A(z)T(z);x)=x$. We compute
\begin{eqnarray}
&& f(j(z)) = l(T(j(z));x)=  l\biggl(T(j(z));
l(T^t(j(z))A(z)T(z);x)\biggr) 
\nonumber\\
&&\quad =  l(A(z)T(z);x) 
= l(A(z);l(T(z);x)) 
= l(A(z);f(z)) 
\; ,
\nonumber
\end{eqnarray}
whence $f$ is an invariant $(E,l)$-field of $(j,A)$.\\
\noindent 
``$\Leftarrow$'': Let $f$ satisfy
the $(E,l)$-stationarity equation (\ref{eq:10.146a}) of $(j,A)$. Thus, by 
(\ref{eq:10.152a}), 
\begin{eqnarray}
&&  \hspace{-1cm} l(T(j(z));x) =  f(j(z)) 
= l(A(z);f(z))
\nonumber\\
&& 
=l\biggl(A(z);l(T(z);x)\biggr) 
=l(A(z)T(z);x) \; ,
\nonumber
\end{eqnarray}
whence $x = l\biggl(T^t(j(z));l(A(z)T(z);x)\biggr)
= l\biggl(T^t(j(z))A(z)T(z);x\biggr)$ which implies (\ref{eq:10.130a}) by
(\ref{eq:12.17bnncaa}).
\hfill $\Box$

\vspace{1cm}

The moral of the NFT is that
$f$ and $T$ are effectively equivalent, 
i.e., that one can view invariant fields
from two different perspectives: 
the perspective of Definition \ref{D6.1}
and the perspective of $H$-normal form.
To shed further light on this we note that
$f$ in (\ref{eq:10.152a}) takes values in only one
$(E,l)$-orbit. In fact if $(E,l)$ is an $SO(3)$-space and 
$x\in E$ and if $f\in{\cal C}(\Td,E)$ takes values only in
$l(SO(3);x)$ then we call $T\in{\cal C}(\Td,SO(3))$ an
``$(E,l)$-lift of $f$'' if (\ref{eq:10.152a}) holds.
Thus the NFT says that if an invariant $(E,l)$-field has an 
$(E,l)$-lift, say $T$, then $T\in{\cal TF}_{H(x)}(j,A)$ where
$H(x):=Iso(E,l;x)$ and $x\in E$ has the property that 
$f$ takes only values in $l(SO(3);x)$.
Thus the notion of $(E,l)$-lift will give us insight into invariant
fields via isotropy groups.
Note that, by the IFF Theorem,
Theorem \ref{T09t1}b, for an ISF the notions of IFF and
$(\R^3,l_v)$-lift are identical.
Experience with ISF's lets us believe that in practice an ISF has
an $(\R^3,l_v)$-lift. Thus, by the ISF conjecture in Chapter \ref{6},
we expect that an IFF exists in practice if $j$ is topologically transitive,
i.e., that $(j,A)$ has an $H$-normal form with $H\unlhd SO(2)$.
However it is well-known, as pointed out in Section \ref{9.4b.4}, that
$f\in{\cal C}(\Td,\R^3)$ exist such that 
$f$ takes values only in $l_{v}(SO(3);(0,0,1)^t)$
and which have no $(\R^3,l_v)$-lift.
In any case, the NFT gives insight into invariant $(E,l)$-fields which 
have an $(E,l)$-lift. In fact if an $f\in{\cal C}(\Td,E)$
has an $(E,l)$-lift and then, by the NFT, $f$ can only be
an invariant $(E,l)$-field of a spin-orbit system 
$(j,A)$ if $(j,A)\in{\cal CB}_{H(x)}(d,j)$ where
$H(x):=Iso(E,l;x)$. Thus if $H(x)$ is ``large enough'', e.g.,
$SO(2)\unlhd H(x)$ then chances are that
$f$ is an invariant $(E,l)$-field of $(j,A)$ and if $H(x)$ is small
then chances are that $f$ is not an invariant $(E,l)$-field of $(j,A)$.
We will come back to this point of view in Section \ref{10.4.1}
where we will deal with large and small isotropy groups.
We will also consider lifts in more detail in the CST
in Section \ref{9.4b.4}.

It is also clear that the NFT deals with arbitrary
$SO(3)$-spaces $(E,l)$. Moreover it can also be easily seen that it
deals with arbitrary subgroups $H$ of $SO(3)$ since every subgroup
$H$ of $SO(3)$ is an isotropy group of some $SO(3)$-space $(E,l)$ 
\cite{He1}. 

The following remark compares the NFT for $x,x'$ belonging
to the same $(E,l)$-orbit and at the same time it provides us with
the useful formula (\ref{eq:10.110ln}).

\vspace{3mm}
\noindent{\bf Remark:}
\begin{itemize}
\item[(7)] 
Let $(E,l)$ be an $SO(3)$-space, let $x\in E$ and let $x'\in l(SO(3);x)$, i.e.,
let $r\in SO(3)$ such that $x'=l(r;x)$. We define
$H:=Iso(E,l;x),H':=Iso(E,l;x')$. Then, by (\ref{eq:12.17bnncaa}),
\begin{eqnarray}
&&  \hspace{-1cm}
H'=Iso(E,l;x')=Iso(E,l;l(r;x))=\lbrace r'\in SO(3):l(r';l(r;x))=l(r;x)\rbrace
\nonumber\\
&& 
=\lbrace r'\in SO(3):l(r'r;x)=l(r;x)\rbrace
=\lbrace r'\in SO(3):l(r^t;l(r'r;x))=x\rbrace
\nonumber\\
&& 
=\lbrace r'\in SO(3):l(r^tr'r;x)=x\rbrace
=\lbrace rr'r^t:r'\in SO(3),l(r';x)=x\rbrace
\nonumber\\
&& 
=r\lbrace r'\in SO(3):l(r';x)=x\rbrace r^t
=rIso(E,l;x)r^t =rHr^t
 \; ,
\label{eq:10.110ln}
\end{eqnarray}
whence isotropy groups on the same orbit are conjugate.

We now consider the NFT.  So let $(j,A)\in{\cal SOS}(d,j)$ and
$T\in{\cal TF}_{H}(j,A)$ whence, by the NFT, the function 
$f\in{\cal C}(\Td,E)$ defined by
$f:=l(\cdot;x)\circ T$ is an invariant $(E,l)$-field of $(j,A)$.

Because of (\ref{eq:10.110ln}) and Remark 3 in Chapter \ref{4.3} the function
$T'\in{\cal C}(\Td,SO(3))$ defined by
$T'(z):=T(z)r^t$ belongs to ${\cal TF}_{H'}(j,A)$.
Thus, by the NFT, the function
$f'\in{\cal C}(\Td,E)$, defined by
$f'(z):=l(T'(z);x')$ is an invariant $(E,l)$-field of $(j,A)$.
However $f'(z)=l(T'(z);x')=l(T(z)r^t;x')=
l(T(z);l(r^t;x'))=l(T(z);x)=f(z)$ whence, as perhaps expected,
the function $f$ is independent of the chosen ``reference point'' $x$.
\hfill $\Box$
\end{itemize}

The following remark shows that Theorem \ref{T09t1}b 
is a special case of the NFT.

\vspace{3mm}
\noindent{\bf Remark:}
\begin{itemize}
\item[(8)] 
We now show that Theorem \ref{T09t1}b is a special case 
of the NFT when $(E,l)=(\R^3,l_{v})$. 
We first define, for $\lambda\in[0,\infty)$,
$S_\lambda:=\lambda(0,0,1)^t$. Let
$x=S_\lambda=\lambda(0,0,1)^t$ where $\lambda>0$.
First, if
$T\in{\cal C}(\Td,SO(3))$ then $f$ in (\ref{eq:10.152a}) is the function
$f\in{\cal C}(\Td,\R^3)$ defined by $f(z):=l_{v}(T(z),S_\lambda)
=T(z)S_\lambda$. Secondly, by (\ref{eq:12.17bnncaa})
and (\ref{eq:10.19}),
\begin{eqnarray}
&&  \hspace{-1cm}
Iso(\R^3,l_{v};S_\lambda) =
\lbrace r\in SO(3):l_{v}(r;\lambda(0,0,1)^t)=\lambda(0,0,1)^t\rbrace
\nonumber\\
&& =
\lbrace r\in SO(3):r(0,0,1)^t=(0,0,1)^t\rbrace = SO(2) \; ,
\label{eq:10.186}
\end{eqnarray}
where in the third equality we used the $SO(2)$-Lemma, Theorem \ref{T09t1}a.
We conclude by Remark 2
that in the case, where $(E,l)=(\R^3,l_{v})$ and $x=S_\lambda$ with $\lambda>0$,
the NFT reads as follows: 
If $T\in{\cal C}(\Td,SO(3))$ and 
$(j,A)\in{\cal SOS}(d,j)$ then
$T\in{\cal TF}_{SO(2)}(j,A)$ iff $TS_\lambda$ 
is an invariant polarization field of $(j,A)$.
Thus, in the special case $\lambda=1$, the NFT reads as follows: 
If $T\in{\cal C}(\Td,SO(3))$ and 
$(j,A)\in{\cal SOS}(d,j)$ then
$T\in{\cal TF}_{SO(2)}(j,A)$ iff the third column of $T$ 
is an ISF of $(j,A)$. Thus indeed
Theorem \ref{T09t1}b is the NFT in the special case 
where $(E,l)=(\R^3,l_{v})$ and $x=(0,0,1)^t$.
We also see that $l(T(z);x)$ generalizes the concept
of ``third column of $T(z)$'' from $\R^3,l_{v},(0,0,1)^t$ to $E,l,x$.
For more details on the isotropy groups of $(\R^3,l_{v})$ see
Section \ref{10.3.3}.
\hfill $\Box$
\end{itemize}
The following remark shows that Theorem \ref{T09t2}a 
is a special cases of the NFT.

\vspace{3mm}
\noindent{\bf Remark:}
\begin{itemize}
\item[(9)] We now show that Theorem \ref{T09t2}a is the special case 
of the NFT for which 
$(E,l)=(SO(3),l_{SOR})$ and $x=I_{3\times 3}$ where
the $SO(3)$-space $(SO(3),l_{SOR})$ is defined in terms of the
function $l_{SOR}:SO(3)\times SO(3)\rightarrow SO(3)$ given by
\begin{eqnarray}
&& l_{SOR}(r';r) := r'r\; ,
\label{eq:10.187}
\end{eqnarray}
where $r,r'\in SO(3)$. It is a simple exercise to show that
$(SO(3),l_{SOR})$ is an $SO(3)$-space.

To make our point we first note that, if
$T\in{\cal C}(\Td,SO(3))$, then $f$ in (\ref{eq:10.152a}) is the function
$f\in{\cal C}(\Td,SO(3))$ defined by $f(z):=l_{SOR}(T(z),I_{3\times 3})
=T(z)$, i.e., $f=T$. Secondly, by (\ref{eq:12.17bnncaa}),
(\ref{eq:6.1da}) and (\ref{eq:10.187}),
\begin{eqnarray}
&&  \hspace{-1cm}
Iso(SO(3),l_{SOR};I_{3\times 3}) =
\lbrace r\in SO(3):l_{SOR}(r;I_{3\times 3})=I_{3\times 3}\rbrace 
\nonumber\\
&& 
=
\lbrace r\in SO(3):r=I_{3\times 3}\rbrace = G_0 \; .
\label{eq:10.188}
\end{eqnarray}
We conclude that in the case where $(E,l)=(SO(3),l_{SOR})$ and $x=I_{3\times 3}$ 
the NFT reads as follows: If $T\in{\cal C}(\Td,SO(3))$ and 
$(j,A)\in{\cal SOS}(d,j)$ then
$T\in{\cal TF}_{G_0}(j,A)$ iff $T$ 
is an invariant $(SO(3),l_{SOR})$-field of $(j,A)$.
On the other hand, by (\ref{eq:10.18}),
the $(SO(3),l_{SOR})$-stationarity
equation of $(j,A)$ reads as $T\circ j= AT$ so that indeed
Theorem \ref{T09t2}a is the NFT in the special case 
where $(E,l)=(SO(3),l_{SOR})$ and $x=I_{3\times 3}$.

We also conclude from Theorem \ref{T09t2}a and Remark 5 in
Chapter \ref{4} that every continuous solution of the 
$(SO(3),l_{SOR})$-stationarity equation of $(j,A)$ is a uniform IFF of
$(j,A)$.

For  $(E,l)=(SO(3),l_{SOR})$ and arbitrary $x=r_0\in SO(3)$ it is
an easy exercise to show that 
the NFT reads as follows: If $T\in{\cal C}(\Td,SO(3))$ and 
$(j,A)\in{\cal SOS}(d,j)$ then
$T\in{\cal TF}_{G_0}(j,A)$ iff $T$ satisfies, for all $z\in\Td$,
$T(j(z))r_0 = A(z) T(z)r_0$. Of course since $r_0$ cancels out,
this is equivalent to the case $r_0=I_{3\times 3}$, i.e., equivalent to \\
Theorem \ref{T09t2}a.
\hfill $\Box$
\end{itemize}
Remarks 8 and 9 illustrate how, by feeding in appropriate objects, the NFT
is capable of covering seemingly disparate aspects of the dynamics.
In fact with the ISF and SOR, two kinds of invariance are covered by the NFT
by suitable choice of $(E,l)$.
\subsubsection{The decomposition method.
Invariant sets for $(E,l)$ particle-spin dynamics}
\label{10.2.5}
Recall from Definition \ref{D2.4}
that $l(SO(3);x)$ is the $(E,l)$-orbit of $x$ and that the 
$(E,l)$-orbits form a partition $E/l$ of $E$. 
Since a particle-spin trajectory satisfies
\begin{eqnarray}
&&  \hspace{-1cm}
\left( \begin{array}{c} z(n)\\ x(n)\end{array}\right)
=\left( \begin{array}{c} j^n(z(0))\\
l\biggl(\Psi[j,A](n;z(0));x(0)\biggr)
\end{array}\right) \; ,
\label{eq:10.70}
\end{eqnarray}
we see that $(z(n),x(n))$ belongs to $\Td\times l(SO(3);x(0))$ for all
$n\in\Z$. Thus, for every $x\in E$,
the set $\Td\times l(SO(3);x)$ is invariant under the 
particle-spin motion.
This implies, as the following lemma shows, 
that for the description of the particle-spin motion we can 
replace $L[E,l,j,A]$ by the $L[l(SO(3);x),l_{dec}[x],j,A]$ where, 
for arbitrary $x\in E$, 
the function $l_{dec}[x]:SO(3)\times l(SO(3);x)\rightarrow l(SO(3);x)$ 
is defined as a restriction of the function $l$, i.e.,
\begin{eqnarray}
&&  \hspace{-1cm}
l_{dec}[x](r;y):=l(r;y) \; ,
\label{eq:10.69}
\end{eqnarray}
where $y\in l(SO(3);x),r\in SO(3)$. It is easy 
to show that $l_{dec}[x]$ is a group action of the group
$SO(3)$ on the subset $l(SO(3);x)$ of $E$.
In fact since $l_{dec}[x]$ is a restriction of $l$, it
is even easy to show that
$(l(SO(3);x),l_{dec}[x])$ is an $SO(3)$-space (recall
Definition \ref{D2.4}). We now summarize this in a lemma for further
reference.
\setcounter{lemma}{1}
\begin{lemma} \label{L10.2}
Let $(E,l)$ be an $SO(3)$-space. For every $x\in E$, 
$(l(SO(3);x),l_{dec}[x])$ is a transitive $SO(3)$-space.
Let $(j,A)\in {\cal SOS}(d,j)$ 
and let
$(z(\cdot),x(\cdot))$ be an $(E,l)$-trajectory of $(j,A)$.
Then $(z(\cdot),x(\cdot))$ is also an 
$\biggl(l(SO(3);x(0)),l_{dec}[x(0)]\biggr)$-trajectory of $(j,A)$.
\hfill $\Box$ 
\end{lemma}

Each set $\Td\times l(SO(3);x)$ is invariant under the 
particle-spin motion of (\ref{eq:10.15})
so that  the ``decomposition method'' decomposes $\Td\times E$ 
into the  $\Td\times l(SO(3);x)$.
Thus for  particle-spin motion,
the $SO(3)$-space $(E,l)$ can be replaced by the $SO(3)$-spaces
$(l(SO(3);l_{dec}[x])$ where $x$ ranges over $E$.

The following remark considers the special case where $(E,l)=(\R^3,l_{v})$.

\vspace{3mm}
\noindent{\bf Remark:}
\begin{itemize}
\item[(10)]
In the special case where $(E,l)=(\R^3,l_{v})$ the
$(E,l)$-orbits are spheres. In fact it follows from (\ref{eq:10.19})
and Definition \ref{D2.3new} that the $(\R^3,l_{v})$-orbit of an arbitrary 
$S\in\R^3$ reads as
\begin{eqnarray}
&&  \hspace{-1cm}
l_{v}(SO(3),S)=\lbrace S'\in\R^3:|S'|=|S|\rbrace 
= {\mathbb S}^2_{|S|} \; ,
\label{eq:10.78}
\end{eqnarray}
where ${\mathbb S}^2_\lambda:=\lbrace S \in {\mathbb R}^{3}:|S|=
\lambda\rbrace$ with $\lambda\in[0,\infty)$.
Thus the $(\R^3,l_{v})$-orbits are the 
spheres ${\mathbb S}^2_\lambda$ 
of radius $\lambda\in[0,\infty)$ around $(0,0,0)^t$. 
Moreover $R_{v}:=\lbrace S_\lambda:\lambda\in[0,\infty)\rbrace$ 
is a representing set of the partition $\R^3/l_{v}$ of $\R^3$
(recall from Remark 8 that $S_\lambda=\lambda(0,0,1)^t$). 
Note that all values of $\lambda$ are of physical importance
as can be seen for example in Section \ref{10.5} where
spin vectors 
appear as coefficients in the
density matrix functions of spin-$1/2$ and spin-$1$ particles.
\hfill $\Box$
\end{itemize}

For the field dynamics the situation is analogous.
For the following it is useful to keep in mind from Section \ref{10.2.1}
that if $f\in{\cal C}(\Td,E)$ and $z\in\Td$ then $(z,f(z))$ is mapped into
$(j(z),f'(z))$ where the field
$f$ evolves into the field $f'$ where
$f'(z):=l(A(j^{-1}(z));f(j^{-1}(z)))$ and that $f'=f$ iff $f$ is invariant.

Let $f\in{\cal C}(\Td,E)$ and let us define for every $x\in E$ the 
inverse image of $l(SO(3);x)$ under the function $f$, i.e., the set
\begin{eqnarray}
&& \Sigma_x[E,l,f]:=f^{-1}(l(SO(3);x)) := 
\lbrace z\in\Td: f(z)\in l(SO(3);x)\rbrace
\; ,
\label{eq:10.73}
\end{eqnarray}
where the second equality is just the definition of the inverse image.
Then the nonempty among the $\Sigma_x[E,l,f]$ form a partition of $\Td$ 
which tells us how the values of $f$ are distributed over the 
various $(E,l)$-orbits.   
Note that $\Sigma_x[E,l,f]$ is nonempty iff $f$ takes a
value in $l(SO(3);x)$. For each $x$ such that $\Sigma_x[E,l,f]$ is nonempty, we
define the function 
$f_x:\Sigma_x[E,l,f]\rightarrow l(SO(3);x)$ by
\begin{eqnarray}
&& f_x(z):=f(z)
\; .
\label{eq:10.72}
\end{eqnarray}
\vspace{3mm}
\noindent{\bf Remark:}
\begin{itemize}
\item[(11)] 
Let us illustrate (\ref{eq:10.73}) and (\ref{eq:10.72})
in the special case, where $(E,l)=(\R^3,l_{v})$.
Due to (\ref{eq:10.72}) and
for every $f$ in ${\cal C}(\T^d,\R^3)$ and every
$S\in\R^3$ the values of $f_S$ lie on the sphere
${\mathbb S}^2_{|S|}$. To give a concrete example
we consider the function $f:\T^1\rightarrow \R^3$ defined by
$f(z)\equiv f(\phi + \tilde{\Z}):=\cos(\phi)(0,0,1)^t$ where $\phi\in z$.
It follows from (\ref{eq:10.73}), for every $S\in\R^3$, that
\begin{eqnarray}
&&  \Sigma_S[\R^3,l_{v},f]= \lbrace z\in\T^1: f(z)\in l_{v}(SO(3);S)\rbrace
 = \lbrace z\in\T^1: |f(z)|=|S|\rbrace
\nonumber\\
&&\quad = \lbrace \pi_1(\phi):\phi\in\R, |f(\pi_1(\phi))|=|S|\rbrace
= \lbrace \pi_1(\phi):\phi\in\R, |\cos(\phi)|=|S|\rbrace \; ,
\nonumber
\end{eqnarray}
where in the second equality we used Remark 10. Clearly 
$\Sigma_S[\R^3,l_{v},f]$ is
nonempty iff $|S|\leq 1$ whence $f$ takes values in the infinitely many
$(\R^3,l_{v})$-orbits ${\mathbb S}^2_\lambda$ ($0\leq \lambda \leq 1$).
Note also, by (\ref{eq:2.22}),
that $f(\pi_d(\phi))=\cos(\phi)(0,0,1)^t$ whence $f\circ\pi_d$ is
continuous so that, by the Torus Lemma, Lemma \ref{L0}, $f$ is continuous.
\hfill $\Box$
\end{itemize}
The following lemma shows us how
the time evolution of $f$ changes the $\Sigma_x[E,l,f]$.

\setcounter{lemma}{2}
\begin{lemma} \label{L10.0}
Let $(E,l)$ and $x\in E$ and let $(j,A)\in {\cal SOS}(d,j)$.
Let us map $f\in{\cal C}(\Td,E)$ under $(j,A)$ 
into $f'\in{\cal C}(\Td,E)$ which is 
given by (\ref{eq:10.17}). Then 
$\Sigma_x[E,l,f']$ is the image of $\Sigma_x[E,l,f]$ under $j$, i.e.,
$\Sigma_x[E,l,f']=j(\Sigma_x[E,l,f])$ (for the notion of
``image'', see also Appendix \ref{A.1}).
Moreover if $f$ is an invariant $(E,l)$-field of $(j,A)$ then
$\Sigma_x[E,l,f]=j(\Sigma_x[E,l,f])$.
\end{lemma}

\noindent{\em Proof of Lemma \ref{L10.0}:}
By (\ref{eq:10.17}) and (\ref{eq:10.73})
\begin{eqnarray}
&&  \hspace{-1cm}
\Sigma_x[E,l,f'] = \lbrace z\in\Td: f'(z)\in l(SO(3);x)\rbrace
\nonumber\\
&& 
= \lbrace z\in\Td: l(A(j^{-1}(z));f(j^{-1}(z))) \in l(SO(3);x)\rbrace
\nonumber\\
&& \quad
= \lbrace z\in\Td: f(j^{-1}(z)) \in l(SO(3);x)\rbrace
= \lbrace j(z'):z'\in\Td,f(z')\in l(SO(3);x)\rbrace
\nonumber\\
&& = j(\lbrace z'\in\Td: f(z')\in l(SO(3);x)\rbrace)
=j(\Sigma_x[E,l,f])
\; ,
\label{eq:10.73n}
\end{eqnarray}
where in the third equality we used the fact that $l$ is a group action 
and where in the fifth and sixth equalities we dealt
with images under $j$.

If $f$ is, in addition, an invariant $(E,l)$-field of $(j,A)$ then
$f'=f$ whence (\ref{eq:10.73n}) implies $\Sigma_x[E,l,f]=j(\Sigma_x[E,l,f])$
which proves the second claim.
\hfill $\Box$

With Lemma \ref{L10.0} we see that if $f$
is an invariant $(E,l)$-field of $(j,A)$ then, at least locally,
$f_x$ behaves like an invariant $(E,l)$-field since, for all
$z\in\Sigma_x[E,l,f]$, we have $f_x(j(z)) = l(A(z);f_x(z))$
and, by (\ref{eq:10.69}),
\begin{eqnarray}
&&  \hspace{-1cm}
f_x(j(z)) = l_{dec}[x](A(z);f_x(z)) \; .
\label{eq:10.73na}
\end{eqnarray}
With (\ref{eq:10.73na}) it is thus natural to generalize the notion
of invariant field to a notion of a ``localized'' invariant 
field whose domain is $\Sigma_x[E,l,f]$.
This also implies that if $f\in{\cal C}(\Td,E)$ takes only values in
one $(E,l)$-orbit, say $l(SO(3);x)$, then
$f_x$ is an invariant $(l(SO(3);x),l_{dec}[x])$-field of $(j,A)$ on
$\Td$ iff $f$ is an invariant 
$(E,l)$-field of $(j,A)$.
Indeed our focus in this work is on invariant fields which take values in just
one orbit.
This restriction of ours is for brevity and because 
(see Lemma \ref{L10.4} below) 
the most important invariant fields have this property.

Thus for field motion,
the $SO(3)$-space $(E,l)$ can be replaced by the $SO(3)$-spaces
$(l(SO(3);l_{dec}[x])$ where $x$ ranges over $E$.
The benefit of this is, as we will see further below,
that the $(l(SO(3);x),l_{dec}[x])$-stationarity equations
are easier to handle than the $(E,l)$-stationarity equations
and allow us to use methods which are not available for
$(E,l)$ when $(E,l)$ is not transitive
(we use these methods in the
Decomposition Theorem, Theorem \ref{T10.6}, of
Section \ref{10.3}).

We now will show that invariant $(E,l)$-fields 
often take values in only
one $(E,l)$-orbit.
In fact the following lemma is a straightforward generalization of
Theorem \ref{T09t0} (see also Remark 13 below).
\setcounter{lemma}{3}
\begin{lemma} \label{L10.4}
Let $(E,l)$ be an $SO(3)$-space and let $E$ be Hausdorff
(for the notion of ``Hausdorff'' see Appendix \ref{A.5}).
Let also $(j,A)\in {\cal SOS}(d,j)$ such that
$j$ is topologically transitive.
Then every invariant $(E,l)$-field of $(j,A)$ 
takes values in only one $(E,l)$-orbit.
\end{lemma}
\noindent{\em Proof of Lemma \ref{L10.4}:}
Let $f$ be an
invariant $(E,l)$-field of $(j,A)$.
We pick a $z_0\in\Td$ such that the set
$B:=\lbrace j^n(z_0):n\in\Z\rbrace$ is dense in $\Td$.
Because the nonempty among the $\Sigma_x[E,l,f]$ form a partition of $\Td$ 
we can pick an $x\in E$ such that $z_0\in\Sigma_x[E,l,f]$
whence, by Lemma \ref{L10.0} and since $f$ is invariant, 
$B\subset \Sigma_x[E,l,f]$.  On the other hand,
the continuous function $l(\cdot;x):SO(3)\rightarrow E$ 
has the range $l(SO(3);x)$ 
whence, since $SO(3)$ is compact, the range $l(SO(3);x)$
of this function is a compact subset of $E$ \cite{Mu}.
Because $E$ is Hausdorff the compact subset $l(SO(3);x)$ of $E$
is a closed subset of $E$ \cite{Mu} whence, since
$f$ is continuous, it follows
from (\ref{eq:10.73}) that $\Sigma_x[E,l,f]$ is a closed subset of
$\Td$. Because $B$ is a dense subset of
$\Td$, we get $\Td=\overline{B}\subset\overline{\Sigma_x[E,l,f]}$ so that,
since $\Sigma_x[E,l,f]$ is a closed subset of $\Td$, we
conclude that $\Td\subset\Sigma_x[E,l,f]$ which implies that  
$\Td=\Sigma_x[E,l,f]$. Thus, by the definition of $\Sigma_x[E,l,f]$,
we conclude that $f$ takes values only in $l(SO(3);x)$.
\hfill $\Box$

\vspace{3mm}

The above lemma tells us that 
invariant fields, which take values in only one orbit, are of major
importance and thus in the sequel most of our theorems are stated for that
situation.
Note also that the contexts of the NFT and of 
Lemma \ref{L10.4} overlap since the NFT makes statements about
invariant fields which take values in only one $(E,l)$-orbit
(see (\ref{eq:10.152a})).

\vspace{3mm}
\noindent{\bf Remark:}
\begin{itemize}
\item[(12)] 
Consider  the $SO(3)$-space  $(E,l)$ where $E$ is Hausdorff.
Clearly one can apply
Lemma \ref{L10.4} when $j={\cal P}_\omega$ with $(1,\omega)$ nonresonant.
Perhaps surprisingly one can even use
Lemma \ref{L10.4} to get analogous results when
$j={\cal P}_\omega$ and $(1,\omega)$ is resonant.
In fact in such an approach invariant $n$-turn fields
and tori with dimension smaller than
$d$ play a role \cite{He1}.
\hfill $\Box$
\end{itemize}
Since $\R^3$ is Hausdorff, 
we can apply Lemma \ref{L10.4} to $(E,l)=(\R^3,l_{v})$ as the 
following remarks shows.

\vspace{3mm}
\noindent{\bf Remark:}
\begin{itemize}
\item[(13)] 
Let $(j,A)\in{\cal SOS}(d,j)$
with $j$ topologically transitive and let $f$ 
be an invariant polarization field of $(j,A)$, i.e., by Remark 1,
an invariant $(\R^3,l_{v})$-field of $(j,A)$.
Since $\R^3$ is Hausdorff we can apply 
Lemma \ref{L10.4} and conclude that
$f$ takes values in only one $(\R^3,l_{v})$-orbit, say 
${\mathbb S}^2_{|S|}$ whence $|f(z)|=|S|$.
Note that we already proved that unknowingly in 
Theorem \ref{T09t0}. Of course $\Sigma_S[\R^3,l_{v},f]=\Td$.
\hfill $\Box$
\end{itemize}

By iterating the above procedure which led  to $\Sigma_x[E,l,f]$ and
$f_x$ one arrives at the function 
$F_x:\bigcup_{n\in\Z}\;(\lbrace n\rbrace\times j^n(\Sigma_x[E,l,f]))
\rightarrow l(SO(3);x)$ defined by
\begin{eqnarray}
&& F_x(n,z):=F(n,z) \; ,
\label{eq:10.71}
\end{eqnarray}
where $F:\Z\times\Td\rightarrow E$ is the
$(E,l)$-field trajectory of $(j,A)$ with initial value
$F(0,\cdot)=f(\cdot)$.
Since $F$ is a field trajectory we find, by
(\ref{eq:10.44}) and (\ref{eq:10.71}), that
\begin{eqnarray}
&&  \hspace{-1cm}
F_x(n,z)=l\biggl(  \Psi[j,A](n;L[j](-n;z));f_x(L[j](-n;z))\biggr) \; .
\label{eq:10.75}
\end{eqnarray}
Clearly a necessary condition for $F_x$ to be time-independent is 
that, for all $x$, $j(\Sigma_x[E,l,f])=\Sigma_x[E,l,f]$.
\subsubsection{The Second ToA Transformation Rule}
\label{10.2.4}
The Second ToA Transformation Rule tells us how the motions of different
$SO(3)$-spaces $(E,l)$ are related. This point of view will render the
decomposition method into a useful tool in Sections \ref{10.3}-\ref{10.4b}.
While in the First ToA Transformation Rule $(E,l)$ and $j$ are held
fixed and $A$ is transformed, in the Second ToA Transformation Rule $j$ and
$A$ are held fixed and $(E,l)$ is transformed into another
$SO(3)$-space.

Let $(E_1,l_1)$ and $(E_2,l_2)$ be $SO(3)$-spaces and suppose there exists 
a topological $SO(3)$-map $\beta$ from $(E_1,l_1)$ to $(E_2,l_2)$, i.e.,
$\beta\in{\cal C}(E_1,E_2)$ and 
$\beta(l_1(r;x))=l_2(r;\beta(x))$. Let also $(j,A)\in{\cal SOS}(d,j)$
be fixed but arbitrary.

Consider the mappings (\ref{eq:10.15}) for these $SO(3)$-spaces, i.e.,
\begin{eqnarray}
&&  \hspace{-1cm}
\left( \begin{array}{c} z\\ x\end{array}\right)
\mapsto
\left( \begin{array}{c} 
z' \\
x'\end{array}\right) 
=
\left( \begin{array}{c} 
j(z)\\
l_1(A(z);x)\end{array}\right) 
\; ,
\label{eq:10.50bn}
\end{eqnarray}
\begin{eqnarray}
&&  \hspace{-1cm}
\left( \begin{array}{c} \zeta \\ \xi \end{array}\right)
\mapsto
\left( \begin{array}{c} \zeta'\\ \xi'\end{array}\right)
=\left( \begin{array}{c} 
j(\zeta)\\
l_2(A(\zeta);\xi)
\end{array}\right) \; .
\label{eq:10.51bn}
\end{eqnarray}
From (\ref{eq:10.50bn})
$\beta(x')=\beta(l_1(A(z);x))=l_2(A(z);\beta(x))$. Thus if $\xi$ in
(\ref{eq:10.51bn}) is $\beta(x)$ then 
$\xi'=l_2(A(z);\beta(x))=\beta(x')$.
It follows that if $(z(n),x(n))$ is the solution  
of the IVP of (\ref{eq:10.50bn}) with $(z(0),x(0))=(z_0,x_0)$
then $(\zeta(n),\xi(n))=(z(n),\beta(x(n)))$
is the solution  of the IVP of (\ref{eq:10.51bn}) with 
$(\zeta(0),\xi(0))=(z(0),\beta(x(0)))=(z_0,\beta(x_0))$.

If $\beta$ is not one-one, two IVP's 
for (\ref{eq:10.50bn}) can give rise to the same
IVP of (\ref{eq:10.51bn}). 
If $\beta$ is not onto, then some IVP's of (\ref{eq:10.51bn}) 
are not related to any IVP of (\ref{eq:10.50bn}), e.g., pick
$\xi(0)$ not in the range of $\beta$.
The most interesting case is if
$\beta$ is a surjection, then every solution of 
the IVP of (\ref{eq:10.51bn})  is related to
a solution of an IVP of (\ref{eq:10.50bn}). 
Thus the $(E_1,l_1)$-particle-spin motion
gives insights into all $(E_2,l_2)$-particle-spin motions.
If $\beta$ is a homeomorphism then every 
IVP of (\ref{eq:10.51bn}) can be written in terms of
an IVP of (\ref{eq:10.50bn}) (and vice versa).

\vspace{3mm}
\noindent{\bf Remark:}
\begin{itemize}
\item[(14)] 
We will identify and study several important examples of the function 
$\beta$ in later sections.
We already mention 
two examples of the function $\beta$ here. First of all the function 
$\beta$ defined by (\ref{eq:10.120la}) and (\ref{eq:10.120lb}) 
maps ${\mathbb S}_\lambda^2$ onto ${\mathbb S}_\mu^2$ and demonstrates
how $\beta$ can provide a way to connect and relate
motions of the same spin variable. Moreover the function $\beta$
defined by (\ref{eq:10.175na}) 
in Section \ref{10.4b.1} maps
${\mathbb S}_\lambda^2$ into $\R^{3\times 3}$ and is a striking
example of how $\beta$ can connect and relate
very different spin variables.
The Isotropy-Conjugacy Lemma in Section \ref{10.3.1}
will show how examples like these are enabled by certain subgroups of $SO(3)$. 
\hfill $\Box$
\end{itemize}
The following remark summarizes the above Second ToA Transformation Rule 
for particle-spin motion in a concise way:

\vspace{3mm}
\noindent{\bf Remark:}
\begin{itemize}
\item[(15)] 
Consider the $SO(3)$-spaces $(E_1,l_1)$ and $(E_2,l_2)$ and let
$\beta$ be a topological $SO(3)$-map from $(E_1,l_1)$ to $(E_2,l_2)$.
Moreover let  $(j,A)\in{\cal SOS}(d,j)$.
Using Definition \ref{D3c} it
is an easy exercise to show that
\begin{eqnarray}
&&  \hspace{-1cm}
{\cal P}[E_2,l_2,j,A]\circ \beta_{tot} = \beta_{tot}\circ {\cal P}[E_1,l_1,j,A] \; ,
\label{eq:10.55nn}
\end{eqnarray}
where the function 
$\beta_{tot}\in{\cal C}(\Td\times E_1,\Td\times E_2)$ is 
defined, for $z\in\Td,x\in E_1$, by $\beta_{tot}(z,x):=(z,\beta(x))$. 
If $\beta$ is a homeomorphism then 
$\beta_{tot}\in\Homeo(\Td\times E_1,\Td\times E_2)$ and
\begin{eqnarray}
&&  \hspace{-1cm}
\beta_{tot}^{-1}\circ
{\cal P}[E_2,l_2,j,A] = {\cal P}[E_1,l_1,j,A] \circ \beta_{tot}^{-1}
\; .
\label{eq:10.55nna}
\end{eqnarray}
\hfill $\Box$
\end{itemize}

With fields we proceed analogously.
Consider the mappings (\ref{eq:10.17}) for the $SO(3)$-spaces
$(E_1,l_1)$ and $(E_2,l_2)$:
\begin{eqnarray}
&&  \hspace{-1cm}
f\mapsto f':= l_1\biggl(A\circ j^{-1};f\circ j^{-1}\biggr) \; ,
\label{eq:10.65n}
\end{eqnarray}
\begin{eqnarray}
&&  \hspace{-1cm}
g\mapsto g':= l_2\biggl(A\circ j^{-1};g\circ j^{-1}\biggr)
 \; .
\label{eq:10.66n}
\end{eqnarray}
From (\ref{eq:10.65n})
$\beta(f'(z))=\beta(l_1\biggl(A(j^{-1}(z));f(j^{-1}(z))\biggr))
=l_2\biggl(A(j^{-1}(z));\beta(f(j(z)))\biggr)$.
Thus, if
\begin{eqnarray}
&&  \hspace{-1cm}
g=\beta\circ f \; ,
\label{eq:10.67n}
\end{eqnarray}
then $g'=\beta\circ f'$.

We now formulate this and some consequences as a theorem which
will become important for the
decomposition method (see Section \ref{10.3}).

\setcounter{theorem}{4}
\begin{theorem} \label{TT2}
Let $(E_1,l_1)$ and $(E_2,l_2)$ be $SO(3)$-spaces 
and suppose there exists 
a topological $SO(3)$-map $\beta$ from $(E_1,l_1)$ to $(E_2,l_2)$.
If $f\in{\cal C}(\Td,E_1)$ and $g\in{\cal C}(\Td,E_2)$
then the field mappings (\ref{eq:10.65n}),(\ref{eq:10.66n}) satisfy:\\

\noindent a) If $g=\beta\circ f$ then $g'=\beta\circ f'$.\\

\noindent b) If $g=\beta\circ f$ and $f$ is an invariant $(E_1,l_1)$-field 
of $(j,A)$
then $g$ is an invariant $(E_2,l_2)$-field of $(j,A)$.\\

\noindent c) If $\beta$ is a homeomorphism and
$g=\beta\circ f$ is an invariant $(E_2,l_2)$-field of $(j,A)$
then $f$ is an invariant $(E_1,l_1)$-field of $(j,A)$.\\

\noindent d) If $F$ is the solution  
of the IVP of (\ref{eq:10.65n}) with $F(0,z)=F_0(z)$
then $G$ given by $G(n,z)=\beta(F(n,z))$  
is the solution of the IVP of (\ref{eq:10.66n}) with 
$G(0,z)=\beta(F(0,z))=\beta(F_0(z))$. 
\hfill $\Box$
\end{theorem}

\vspace{3mm}

If $\beta$ is not onto then some IVP's of (\ref{eq:10.66n}) 
have solutions which are not related to any IVP of (\ref{eq:10.65n}). 
If $\beta$ is not one-one, then two IVP's 
of (\ref{eq:10.65n}) can give rise to the same
IVP of (\ref{eq:10.66n}).

The following remark summarizes the above Second ToA Transformation Rule 
for the field motion in a concise way:

\vspace{3mm}
\noindent{\bf Remark:}
\begin{itemize}
\item[(16)] 
Let $(E_1,l_1)$ and $(E_2,l_2)$ be $SO(3)$-spaces and let
$\beta$ be a topological $SO(3)$-map from $(E_1,l_1)$ to $(E_2,l_2)$.
Let also $(j,A)\in{\cal SOS}(d,j)$.
Using Definition \ref{D3c} it
is a simple exercise to show that
\begin{eqnarray}
&&  \hspace{-1cm}
\tilde{\cal P}[E_2,l_2,j,A]\circ \tilde{\beta}
= \tilde{\beta}\circ \tilde{\cal P}[E_1,l_1,j,A] \; ,
\label{eq:10.55ann}
\end{eqnarray}
where the function 
$\tilde{\beta}:{\cal C}(\Td,E_1)\rightarrow{\cal C}(\Td,E_2)$ is defined,  
for $f\in{\cal C}(\Td,E_1)$, by $\tilde{\beta}(f):=\beta\circ f$.
If $\beta$ is a homeomorphism then $\tilde{\beta}$ is a
bijection and
\begin{eqnarray}
&&  \hspace{-1cm}
\tilde{\beta}^{-1}\circ  \tilde{\cal P}[E_2,l_2,j,A]
= \tilde{\cal P}[E_1,l_1,j,A]\circ \tilde{\beta}^{-1}  \; .
\label{eq:10.55anna}
\end{eqnarray}
\hfill $\Box$
\end{itemize}
Two questions naturally arise: when do topological $SO(3)$-maps
exist and what form do they take? 
This is the subject of Section \ref{10.3}.
\subsection{The Isotropy-Conjugacy Lemma (ICL) and
the Decomposition Theorem (DT)}
\label{10.3}
In this section we address the two questions from the
end of Section \ref{10.2.4} for the important case when 
$(E_1,l_1)$ and $(E_2,l_2)$ both are transitive, i.e., have only one orbit.
Most importantly we
also address these questions in the case of the
decomposition of any given $SO(3)$-spaces $(E,l)$ and $(E',l')$ 
where $(E_1,l_1)=(l(SO(3);x),l_{dec}[x])$
and $(E_2,l_2)=(l'(SO(3);x'),l_{dec}'[x'])$.
All this is achieved by the ICL which we then apply to the field
dynamics, leading us to the DT.
\subsubsection{The Isotropy-Conjugacy Lemma}
\label{10.3.1}
The first question from the
end of Section \ref{10.2.4} is answered by the following proposition
which relates topological $SO(3)$-maps with isotropy groups.
\setcounter{proposition}{5}
\begin{proposition} \label{T10.5}
Let $(E_1,l_1)$ and $(E_2,l_2)$ be transitive 
$SO(3)$-spaces and let $E_1,E_2$ be Hausdorff. For arbitrary
$x_1\in E_1$ and $x_2\in E_2$ the following hold.\\

\noindent a) A topological $SO(3)$-map from $(E_1,l_1)$ to $(E_2,l_2)$
exists iff $Iso(E_1,l_1;x_1)\unlhd Iso(E_2,l_2;x_2)$.\\

\noindent b) The $SO(3)$-spaces $(E_1,l_1)$ and $(E_2,l_2)$ are isomorphic
iff $Iso(E_1,l_1;x_1),Iso(E_2,l_2;x_2)$ are conjugate.
\hfill $\Box$ 
\end{proposition}
The reader finds the proof of this proposition at the end of this section. 
In fact this proposition is a simple corollary of the ICL.

In our applications we start with  $SO(3)$-spaces $(E,l)$ and $(E',l')$ 
which are not transitive (for example, $(\R^3,l_v)$) and we work 
with the decompositions $(E_1,l_1)=(l(SO(3);x),l_{dec}[x])$
and $(E_2,l_2)=(l'(SO(3);x'),l_{dec}'[x'])$.
To formulate the ICL we make the following definition:
\setcounter{definition}{6}
\begin{definition}
\label{D8.x}
\noindent
Let $(E,l)$ and $(E',l')$ be $SO(3)$-spaces and let 
$x\in E$ and $x'\in E'$. We denote by $B(E,l,E',l';x,x')$ the set of all
topological $SO(3)$-maps from 
$(l(SO(3);x),l_{dec}[x])$ to $(l'(SO(3),x'),l_{dec}'[x'])$.
In the case where $(E',l')=(E,l)$ 
we abbreviate $B(E,l,E',l';x,x')$ by $B(E,l;x,x')$. 

If $H$ and $H'$ are subsets of $SO(3)$ then we define
\begin{eqnarray}
&& N(H,H'):=\lbrace r\in SO(3):
rHr^t\subset H' \rbrace \; .
\label{eq:12.17dbt}
\end{eqnarray}
Note that, by Definition \ref{D5.x}, $N(H,H')$ is nonempty iff
$H\unlhd H'$.

If $Iso(E,l;x)\unlhd Iso(E',l';x')$, i.e., 
if $N(Iso(E,l;x),Iso(E',l';x'))$ is nonempty
then we can pick 
$r_0\in N(Iso(E,l;x),Iso(E',l';x'))$
and so we get $r_0 Iso(E,l;x)r_0^t \subset Iso(E',l';x')$.
Then we define the function $\hat{\beta}[E,l,E',l';x,x',r_0]:
l(SO(3);x)\rightarrow l'(SO(3);x')$ by
\begin{eqnarray}
&&  \hspace{-1cm}
\hat{\beta}[E,l,E',l';x,x',r_0](l(r_1;x)):=l'(r_1r_0^t;x') \; .
\label{eq:10.83}
\end{eqnarray}
That $\hat{\beta}[E,l,E',l';x,x',r_0]$ is a function, i.e., is single-valued, 
is shown below. Clearly \\
$\hat{\beta}[E,l,E',l';x,x',r_0]$ is a surjection.
In the case where $(E',l')=(E,l)$ we abbreviate\\ 
$\hat{\beta}[E,l,E',l';x,x',r_0]$ by $\hat{\beta}[E,l;x,x',r_0]$.
The ICL will show us that all elements of\\
$B(E,l,E',l';x,x')$ are of the form  $\hat{\beta}[E,l,E',l';x,x',r_0]$
if $E$ and $E'$ are Hausdorff.
\hfill $\Box$
\end{definition}
To show that
$\hat{\beta}[E,l,E',l';x,x',r_0]$ is a function, i.e., is single-valued, let\\
$r_0\in N(Iso(E,l;x),Iso(E',l';x'))$.
If $l(r_1;x)=l(r_2;x)$ then $h:=r_1^t r_2\in Iso(E,l;x)$ whence
we get, by (\ref{eq:10.83}),
\begin{eqnarray}
&&  \hspace{-1cm}
\hat{\beta}[E,l,E',l';x,x',r_0](l(r_2;x))=l'(r_2r_0^t;x')
=l'(r_1 h r_0^t;x')=l'(r_1 r_0^tr_0 h r_0^t;x')
\nonumber\\
&&\quad =l'(r_1 r_0^t;l'(r_0 h r_0^t;x'))=l'(r_1 r_0^t;x')
=\hat{\beta}[E,l,E',l';x,x',r_0](l(r_1;x)) \; ,
\nonumber
\end{eqnarray}
where in the fifth equality we used that 
$r_0 Iso(E,l;x)r_0^t \subset Iso(E',l';x')$. Thus indeed
$\hat{\beta}[E,l,E',l';x,x',r_0]$ is a function. 

If $H$ and $H'$ are nonempty subsets of $SO(3)$ then it 
is an easy exercise to show that
\begin{eqnarray}
&& N(H,H')= \bigcap_{h\in H}\; \bigcup_{h'\in H'}\;
 N(\lbrace h\rbrace,\lbrace h' \rbrace)
= \bigcap_{h\in H}\; N(\lbrace h\rbrace,H') \; .
\label{eq:12.17dbtn}
\end{eqnarray}
The sets $N(H,H')$ are well-known and
will become convenient below.

\setcounter{lemma}{7}
\begin{lemma} \label{L10.5}
(ICL)\\
Let $(E,l)$ and $(E',l')$ be 
$SO(3)$-spaces and let $E,E'$ be Hausdorff. Let also
$x\in E$ and $x'\in E'$. Then the following hold.\\

\noindent a) $B(E,l,E',l';x,x')$ is nonempty iff
$Iso(E,l;x)\unlhd Iso(E',l';x')$. Moreover
\begin{eqnarray}
&&  \hspace{-15mm}
B(E,l,E',l';x,x') = \Biggl\lbrace
\hat{\beta}[E,l,E',l';x,x',r_0]:r_0\in N\biggl(Iso(E,l;x),Iso(E',l';x')\biggr)
\Biggr\rbrace \; .
\label{eq:10.83a}
\end{eqnarray}
\noindent b) Let $Iso(E,l;x)\unlhd Iso(E',l';x')$
and pick a $r_0\in N(Iso(E,l;x),Iso(E',l';x'))$.
Let also $y\in l(SO(3);x),y'\in l'(SO(3),x')$, i.e., 
$r_1,r_2\in SO(3)$ exist such that $y=l(r_1;x)$ and $y'=l'(r_2;x')$.
Then $(r_2 r_0 r_1^t)\in N(Iso(E,l;y),Iso(E',l';y'))$ and
$Iso(E,l;y)\unlhd Iso(E',l';y')$ as well as
$\hat{\beta}[E,l,E',l';y,y',r_2r_0r_1^t]=\hat{\beta}[E,l,E',l';x,x',r_0]$.
Moreover one can choose $y,y'$ such that
$Iso(E,l;y)\subset Iso(E',l';y')$. \\
\noindent Remark: Since $B(E,l,E',l';x,x')=B(E,l,E',l';y,y')$,
the choice of $y,y'$ such that\\
$Iso(E,l;y)\subset Iso(E',l';y')$ can be helpful for the computation of
$B(E,l,E',l';x,x')$ since in that case $I_{3\times 3}
\in N(Iso(E,l;y),Iso(E',l';y'))$ and since
$\hat{\beta}[E,l,E',l';y,y',I_{3\times 3}]$ is easy to handle. In fact in all our
applications we make use of this choice of $y,y'$.\\

\noindent c) $Iso(E,l;x),Iso(E',l';x')$ 
are conjugate iff the $SO(3)$-spaces 
$(l'(SO(3),x'),l'_{dec}[x'])$, \\
$(l(SO(3);x),l_{dec}[x])$ are isomorphic.
Also, for every $r_0\in SO(3)$ such that 
$r_0 Iso(E,l;x)r_0^t= Iso(E',l';x')$, 
$\hat{\beta}[E,l,E',l';x,x',r_0]$ is an isomorphism from 
$(l(SO(3);x),l_{dec}[x])$ to\\ 
$(l'(SO(3),x'),l_{dec}'[x'])$.
Moreover if $Iso(E,l;x),Iso(E',l';x')$ are conjugate 
then one can choose $y\in l(SO(3);x),y'\in l'(SO(3),x')$ such that
$Iso(E,l;y)=Iso(E',l';y')$.\\
\noindent Remark: In all our
applications we make use of this choice of $y,y'$.
\end{lemma}
\noindent{\em Proof of Lemma \ref{L10.5}:} 
See Appendix \ref{11.1}.
The Hausdorff property of $E,E'$ is needed
for proving (\ref{eq:10.83a}) and, as in the proof of
Lemma \ref{L10.4}, the compactness of $SO(3)$ is used as well.
\hfill $\Box$ 

\vspace{3mm}

The following remark mentions some interesting facts which are not addressed
by Lemma \ref{L10.5} (in order to keep its proof short) and 
which will be confirmed
by our examples.

\vspace{3mm}
\noindent{\bf Remark:}
\begin{itemize}
\item[(17)] 
Let $(E,l)$ be an $SO(3)$-space, let $E$ be Hausdorff and $x\in E$. 
We first mention the trivial fact that $Iso(E,l;x)$ 
is closed (this follows from the fact that the singleton
$\lbrace x\rbrace$ is a closed subset
of the Hausdorff space $E$ 
and that $Iso(E,l;x)$ is the inverse image of $\lbrace x\rbrace$
under the continuous function $l(\cdot;x)$).
Let also $(E',l')$ be an $SO(3)$-space, let $E'$ be Hausdorff and $x'\in E'$.
Since $Iso(E,l;x),Iso(E',l';x')$ are closed and $SO(3)$ is compact
it follows \cite[1.71,1.72]{Ka} that 
either all elements of $B(E,l,E',l';x,x')$ are isomorphisms
or none of them (by Lemma \ref{L10.5}c
the former case occurs iff  $Iso(E,l;x),Iso(E',l';x')$ 
are conjugate and then $B(E,l,E',l';x,x')$ is the set of isomorphisms
from $(l(SO(3);x),l_{dec}[x])$ to $(l'(SO(3),x'),l_{dec}'[x'])$).
Moreover 
since $Iso(E,l;x)$ is closed and $SO(3)$ is compact
it follows  \cite[1.70]{Ka} that 
\begin{eqnarray}
&&  \hspace{-15mm}
N(Iso(E,l;x),Iso(E,l;x))=\lbrace r_0\in SO(3):r_0Iso(E,l;x)r_0^t=Iso(E,l;x)
\rbrace \; ,
\nonumber
\end{eqnarray}
which implies that $N(Iso(E,l;x),Iso(E,l;x))$ is a 
subgroup of $SO(3)$ and that\\
$Iso(E,l;x),Iso(E',l';x')$ are conjugate iff
$Iso(E,l;x)\unlhd Iso(E',l';x')$ and \\
$Iso(E',l';x')\unlhd Iso(E,l;x)$.
The latter fact implies that if
$Iso(E,l;x),Iso(E',l';x')$ are not conjugate then either
$B(E,l,E',l';x,x')$ or $B(E',l',E,l;x',x)$ is empty (or both).
\hfill $\Box$
\end{itemize}
\noindent{\em Proof of Proposition \ref{T10.5}:}
The claims follow by setting $(E,l)=(E_1,l_1)$ and $(E',l')=(E_2,l_2)$
in Lemma \ref{L10.5}a and \ref{L10.5}c and using
Definition \ref{D8.x}.
\hfill $\Box$ 
\subsubsection{The Decomposition Theorem}
\label{10.3.2}
In this section we first
state the DT which is the main corollary to 
Lemma \ref{L10.5}. Then we show how Lemma \ref{L10.5} and
the DT turn
the decomposition method into 
a useful instrument of classifying field motions. 

\setcounter{theorem}{8}
\begin{theorem} \label{T10.6}
(DT)\\
Let $(E,l)$ and $(E',l')$ be a $SO(3)$-spaces where
the topological spaces $E,E'$ are Hausdorff and
let $x,x'\in E$. Moreover let $(j,A)\in{\cal SOS}(d,j)$.
Then the following hold.\\

\noindent a) Let $Iso(E,l;x)\unlhd Iso(E',l';x')$
and pick $r_0\in N(Iso(E,l;x),Iso(E',l';x'))$.
Let $f\in{\cal C}(\Td,E)$ take values
only in the $(E,l)$-orbit $l(SO(3);x)$ of $x$ and let
$f'\in{\cal C}(\Td,E)$ be defined by
$f':=\tilde{\cal P}[E,l,j,A](f)$. Let the functions
$g,g'\in{\cal C}(\Td,E')$ be defined by
$g(z):=\hat{\beta}[E,l,E',l';x,x',r_0](f(z))$
and $g':=\tilde{\cal P}[E',l',j,A](g)$. Then
$g'(z)=\hat{\beta}[E,l,E',l';x,x',r_0](f'(z))$.\\
\noindent Remark: $f'$ takes values
only in $l(SO(3);x)$. Also, $g,g'$ take values
only in $l'(SO(3);x')$.
Moreover if
$f=f'$ then $g=g'$, i.e., if $f$ is an invariant $(E,l)$-field of $(j,A)$
then $g$ is an invariant $(E',l')$-field of $(j,A)$.
\\

\noindent b) Let $Iso(E',l';x'),Iso(E,l;x)$ be conjugate, i.e., 
$r_0\in SO(3)$ exists such that \\
$r_0 Iso(E,l;x)r_0^t = Iso(E',l';x')$.
Let $f\in{\cal C}(\Td,E)$ be a function which takes values
only in the $(E,l)$-orbit of $x$. Let the function
$g\in{\cal C}(\Td,E')$ be defined by
$g(z):=\hat{\beta}[E,l,E',l';x,x',r_0](f(z))$.
Then $f$ is an invariant $(E,l)$-field of $(j,A)$ 
iff $g$ is an invariant $(E',l')$-field of $(j,A)$. \\
\noindent Remark: Thus the
invariant $(E',l')$-fields which take values only in $l'(SO(3);x')$
are redundant since they can be
referred to invariant $(E,l)$-fields.
\end{theorem}

\noindent{\em Proof of Theorem \ref{T10.6}:} See 
Appendix \ref{11.2}.
\hfill $\Box$

\vspace{3mm}

Of course since Theorem \ref{T10.6} deals with functions which take values
in only one orbit, it is naturally applied in the situation 
when $j$ is topologically transitive.
While the claims of Theorem \ref{T10.6} are focused on fields,
it is easy to see how the corresponding statements for the 
particle-spin trajectories would look like. 

If $f$ in Theorem \ref{T10.6}a is an invariant field then, as the theorem
tells us, this is a sufficient condition for $g$ to be invariant, too.
However this is not a necessary condition
as we will see by the example of the $2$-snake model in
Section \ref{10.4b.2}. In fact
Theorem \ref{T10.6}a will play an active role for
the $2$-snake model for which we will apply it in the situation of
an $f\in{\cal C}(\T^1,\R^3)$ which is not an ISF (in fact the $2$-snake model
does not have an ISF).

\vspace{3mm}
\noindent{\bf Remark:}
\begin{itemize}
\item[(18)] 
Clearly the central task when applying the DT 
to $(E,l)$ and $(E',l')$ is to determine for every $x\in E,x'\in E'$ whether
$Iso(E,l;x)\unlhd Iso(E',l';x')$, i.e., whether 
$N\biggl(Iso(E,l;x),Iso(E',l';x')\biggr)$ is nonempty.
If $Iso(E,l;x),Iso(E',l';x')$ are conjugate, 
$l'(SO(3);x')$-valued $(E',l')$-fields are redundant whence,
in this situation,
the elements of $B(E,l,E',l';x,x')$ are of no great 
importance in the present work 
(note also that by Remark 17 that in this case all
elements of $B(E,l,E',l';x,x')$ are isomorphisms).
Moreover, by Remark 7,
isotropy groups on the same orbit are conjugate whence
the above strategy amounts to restricting ourselves to those
$x\in E$ which belong a representing set
of the partition $E/l$ of $E$ and to those
$x'\in E'$ which belong a representing set
of the partition $E'/l'$ of $E'$.
We will apply this strategy in the following to several important
choices of $(E,l),(E',l')$.
\hfill $\Box$
\end{itemize}

Our first application of the DT and of the strategy of Remark 18 is the
case of spin-orbit resonance, i.e., when
$(E,l)=(SO(3),l_{SOR})$ and where $(E',l')$ is kept arbitrary, i.e.,
$(E',l')$ is an $SO(3)$-space and $E'$ is Hausdorff.
This case will show that, on spin-orbit resonance, 
invariant $(E',l')$-fields always exist.
Note that $SO(3)$ is Hausdorff whence indeed we can apply the DT.
It is clear, by (\ref{eq:10.187}), that the $SO(3)$-space
$(SO(3),l_{SOR})$ is transitive whence we only have one orbit
and so we choose, as in Remark 9, $x=I_{3\times 3}$ and recall
from (\ref{eq:10.188}) that
$Iso(SO(3),l_{SOR};I_{3\times 3})=G_0$. Thus
$Iso(SO(3),l_{SOR};I_{3\times 3})=G_0\subset Iso(E',l';x')$
whence $Iso(SO(3),l_{SOR};I_{3\times 3})\unlhd Iso(E',l';x')$.
To compute $B(SO(3),l_{SOR},E',l';I_{3\times 3},x')$
we first note, by (\ref{eq:12.17dbt}), that
\begin{eqnarray}
&&  \hspace{-1cm} 
N(Iso(SO(3),l_{SOR},I_{3\times 3}),Iso(E',l';x'))
=N(G_0,Iso(E',l';x'))=SO(3) \; ,
\label{eq:8.3.50} 
\end{eqnarray}
whence, by Lemma \ref{L10.5}a,
\begin{eqnarray}
&&  \hspace{-1cm}
B(SO(3),l_{SOR},E',l';I_{3\times 3},x') = 
\lbrace
\hat{\beta}[SO(3),l_{SOR},E',l';I_{3\times 3},x',r_0]: r_0\in SO(3) 
\rbrace \; .
\label{eq:8.3.51}
\end{eqnarray}
If $r_0,r\in SO(3)$ then, by (\ref{eq:10.187}),(\ref{eq:10.83}),
\begin{eqnarray}
&& \hat{\beta}[SO(3),l_{SOR},E',l';I_{3\times 3},x',r_0](r)
= \hat{\beta}[SO(3),l_{SOR},E',l';I_{3\times 3},x',r_0](l_{SOR}(r;I_{3\times 3}))
\nonumber\\
&&\quad =l'(r r_0^t;x') \; .
\label{eq:8.3.52}
\end{eqnarray}
Thus if $r_0,r_1\in SO(3)$ then 
$\hat{\beta}[SO(3),l_{SOR},E',l';I_{3\times 3},x',r_0]
=\hat{\beta}[SO(3),l_{SOR},E',l';I_{3\times 3},x',r_1]$ iff, for all
$r\in SO(3)$, 
$l'(r r_0^t;x')=l'(r r_1^t;x')$, i.e., iff
$r_0r_1^t\in Iso(E',l';x')$. Thus\\
$\hat{\beta}[SO(3),l_{SOR},E',l';I_{3\times 3},x',r_0]
=\hat{\beta}[SO(3),l_{SOR},E',l';I_{3\times 3},x',r_1]$ iff
$r_1^t\in r_0^t Iso(E',l';x')$. In other words, 
$B(SO(3),l_{SOR},E',l';I_{3\times 3},x')$ has as
many elements as there are left cosets $rIso(E',l';x')$.
To apply the DT let $(j,A)\in{\cal SOS}(d,j)$, let
$f\in{\cal C}(\Td,SO(3))$ and let $g\in{\cal C}(\Td,E')$ be defined by
$g(z):=\hat{\beta}[SO(3),l_{SOR},E',l';I_{3\times 3},x',r_0](f(z))$ for
fixed but arbitrary $r_0\in SO(3)$. Since 
$(SO(3),l_{SOR})$ is transitive, all values of $f$ are in the 
$(SO(3),l_{SOR})$-orbit of $I_{3\times 3}$.
Note also that $g$ takes values only in the $(E',l')$-orbit 
of $x'$. Let $f$ be an invariant $(SO(3),l_{SOR})$-field of
$(j,A)$ (note, by Remark 9, such an $f$ is a uniform IFF of 
$(j,A)$, i.e., it exists iff
$(j,A)$ is on spin-orbit resonance).
It follows from Theorem \ref{T10.6}a that
$g$ is an invariant $(E',l')$-field of $(j,A)$. 
Thus on spin-orbit resonance invariant $(E',l')$-fields always exist
if $E'$ is Hausdorff. In the subcase 
$(E',l')=(\R^3,l_v)$ this result is not surprising because if we pick 
$x'=(0,0,1)^t$ then, by the IFF Theorem,
the third column of the uniform IFF $f$ is an ISF!
We finally look if we can apply Theorem \ref{T10.6}b as well.
In fact, for every $r\in SO(3)$, 
$rI_{3\times 3}r^t=I_{3\times 3}$ whence $G_0$ is conjugate to
$Iso(E',l';x')$ only in the exceptional case when 
$Iso(E',l';x')=G_0$ (e.g., if $(E',l')=(SO(3),l_{SOR})$).
\subsubsection{Applying the Decomposition Theorem 
in the case \\$(E,l)=(E',l')=(\R^3,l_{v})$}
\label{10.3.3}
In this section we apply the DT to the case $(E,l)=(E',l')=(\R^3,l_{v})$. 
This case serves to illustrate the basic technique because it is
more involved than our first example above. However, unlike the above example, 
it does not add to our basic knowledge about spin motion.
Thus the reader who is interested in the more important applications of the
DT in Sections \ref{10.4}-\ref{10.4b}
may go straight to those sections.

We will see that the representing set
$R_{v}=\lbrace S_\lambda:
\lambda\in[0,\infty)\rbrace =\lbrace \lambda(0,0,1)^t:
\lambda\in[0,\infty)\rbrace$ of the partition $\R^3/l_{v}$ of $\R^3$ 
is very convenient (see also Remark 10).
Choosing $R_{v}$
we must determine for given $\lambda,\mu\in [0,\infty)$
whether $Iso(\R^3,l_{v};S_\lambda)\unlhd Iso(\R^3,l_{v};S_\mu)$ or whether
$Iso(\R^3,l_{v};S_\lambda),Iso(\R^3,l_{v};S_\mu)$ are even conjugate.

To compute the isotropy groups we use
(\ref{eq:12.17bnncaa}) and (\ref{eq:10.19}) to
get, for $\lambda\in[0,\infty)$,
\begin{eqnarray}
&&  \hspace{-1cm}
Iso(\R^3,l_{v};S_\lambda) = 
\lbrace r\in SO(3):l_{v}(r;S_\lambda)=S_\lambda\rbrace
\nonumber\\
&& \hspace{-5mm}
 =
\lbrace r\in SO(3):\lambda r(0,0,1)^t = \lambda  (0,0,1)^t\rbrace
= \left\{ \begin{array}{ll}  SO(2)
& \;\; {\rm if\;} \lambda > 0  \\
SO(3) & \;\; {\rm if\;} \lambda = 0   \; , \end{array}
                  \right.
\label{eq:10.119}
\end{eqnarray}
where again we used the $SO(2)$-Lemma, Theorem \ref{T09t1}a.
Since the group $SO(2)$ is convenient to deal with, we see
by (\ref{eq:10.119}) that it was prescient to have 
chosen the $S_\lambda$ to be the elements of $R_{v}$.
With (\ref{eq:10.119}) we are led to consider the following 
four separate cases: $\lambda>0,\mu>0$; $\lambda=0,\mu=0$;
$\lambda=0,\mu> 0$; $\lambda>0,\mu=0$.

We first consider the case when $\lambda,\mu>0$
and we will find that
$B(\R^3,l_{v};S_\lambda,S_\mu)$ has only two elements.
Thus, by (\ref{eq:10.119}),
$Iso(\R^3,l_{v};S_\lambda)=SO(2)=Iso(\R^3,l_{v};S_\mu)$
whence $I_{3\times 3}Iso(\R^3,l_{v};S_\lambda)I_{3\times 3}^t
=Iso(\R^3,l_{v};S_\mu)$ so that\\
$Iso(\R^3,l_{v};S_\lambda)$ and $Iso(\R^3,l_{v};S_\mu)$ are conjugate. 
To compute $B(\R^3,l_{v};S_\lambda,S_\mu)$
we first note, by (\ref{eq:10.119}), that
\begin{eqnarray}
&&  \hspace{-1cm} 
N(Iso[\R^3,l_{v};S_\lambda],Iso[\R^3,l_{v};S_\mu])=N(SO(2),SO(2)) \; .
\label{eq:10.162xxc} 
\end{eqnarray}
To compute $N(SO(2),SO(2))$ let
$r_0\in N(SO(2),SO(2))$ so that, 
by (\ref{eq:12.17dbt}), $r_0 SO(2) r_0^t\subset SO(2)$ whence, by
(\ref{eq:6.5}), for
every $\nu\in\R$ there exists an $\nu'\in\R$ such that
$r_0 \exp(2\pi\nu{\cal J})r_0^t=\exp(2\pi\nu'{\cal J})$. Thus 
\begin{eqnarray}
&&  \hspace{-1cm} 
\exp(2\pi\nu{\cal J})r_0^t(0,0,1)^t=r_0^t(0,0,1)^t \; ,
\label{eq:10.162xxcn} 
\end{eqnarray}
and it is clear that if $r_0\in N(SO(2),SO(2))$ then
(\ref{eq:10.162xxcn}) holds for 
every $\nu\in\R$. This implies that if $r_0\in N(SO(2),SO(2))$ then
(\ref{eq:10.162xxcn}) holds for $\nu =1/2$, i.e.,
\begin{eqnarray}
&&  \hspace{-1cm} 
\Di(-1,-1,1)r_0^t(0,0,1)^t=r_0^t(0,0,1)^t \; ,
\label{eq:10.162xx} 
\end{eqnarray}
where, for $y_1,y_2,y_3\in\R$, we use the abbreviation \\
$\Di(y_1,y_2,y_3):=\left( \begin{array}{ccc} y_1 & 0
& 0 \\ 0 & y_2 & 0 \\
 0 & 0 & y_3 \end{array}\right)$.
It follows from (\ref{eq:10.162xx}) that $r_0^t(0,0,1)^t$ is a normalized
eigenvector of $\Di(-1,-1,1)$ with eigenvalue $1$ whence
$r_0^t(0,0,1)^t=\pm (0,0,1)^t$.
If $r_0^t(0,0,1)^t=(0,0,1)^t$ then, by the $SO(2)$-Lemma, $r_0\in SO(2)$
and if $r_0^t(0,0,1)^t=-(0,0,1)^t$ then, by the $SO(2)$-Lemma, 
$r_0\in {\cal K}_1SO(2)$ where
\begin{eqnarray}
&& \hspace{-1cm}
{\cal K}_0:=I_{3\times 3} \; , \;\;
{\cal K}_1:=\Di(1,-1,-1) \; , \;\; {\cal K}_2:=\Di(-1,-1,1) \; , \;\;
{\cal K}_3:=\Di(-1,1,-1) \; ,
\nonumber\\
\label{eq:10.162} 
\end{eqnarray}
and where ${\cal K}_0,{\cal K}_2,{\cal K}_3$ 
will come into play later.
Thus we have shown that\\ 
$N(SO(2),SO(2))\subset (SO(2) \bowtie \Z_2)$ where
\begin{eqnarray}
&&  \hspace{-1cm}
SO(2) \bowtie \Z_2:=\lbrace rr':r\in\Z_2,r'\in SO(2)\rbrace\; ,
\quad \Z_2:=\lbrace {\cal K}_0,{\cal K}_1 \rbrace \; .
\label{eq:10.161} 
\end{eqnarray}
It is a simple exercise to show that $SO(2)\bowtie \Z_2$ 
is a subgroup of $SO(3)$, the so-called
``knit product'' (or ``Zappa-Szep product'')
of the subgroups $SO(2)$ and $\Z_2$ of $SO(3)$.
It is also an easy exercise to show that
$N(SO(2),SO(2))\supset (SO(2) \bowtie \Z_2)$ 
whence, by (\ref{eq:10.162xxc}),
\begin{eqnarray}
&&  \hspace{-1cm} 
N(Iso[\R^3,l_{v};S_\lambda],Iso[\R^3,l_{v};S_\mu])=
N(SO(2),SO(2)) =
SO(2) \bowtie \Z_2 \; .
\label{eq:10.162xxa} 
\end{eqnarray}
Thus $N(Iso[\R^3,l_{v};S_\lambda],Iso[\R^3,l_{v};S_\mu])$ 
is a group. 
Since $Iso(\R^3,l_{v};S_\lambda)$ and 
$Iso(\R^3,l_{v};S_\mu)$ are conjugate, this group property 
is no surprise due to Remark 17. It follows from (\ref{eq:10.161}),
(\ref{eq:10.162xxa}) and Lemma \ref{L10.5}a that
\begin{eqnarray}
&&  \hspace{-1cm}
B(\R^3,l_{v};S_\lambda,S_\mu) = 
B(\R^3,l_{v},\R^3,l_{v};S_\lambda,S_\mu) = 
\lbrace
\hat{\beta}[\R^3,l_{v};S_\lambda,S_\mu,r_0]: r_0\in (SO(2) \bowtie \Z_2)
\rbrace 
\nonumber\\
&&  = \lbrace \hat{\beta}[\R^3,l_{v};S_\lambda,S_\mu,r_0]: r_0\in SO(2) \rbrace 
\cup \lbrace \hat{\beta}[\R^3,l_{v};S_\lambda,S_\mu,r_0]: r_0\in {\cal K}_1
SO(2) \rbrace 
\; .
\label{eq:10.162xxb}
\end{eqnarray}
We now show that both sets on the rhs of (\ref{eq:10.162xxb}) are singletons.
If $r_0\in SO(2)$ and $r\in SO(3)$ then, by (\ref{eq:10.83}),
\begin{eqnarray}
&& \hat{\beta}[\R^3,l_{v};S_\lambda,S_\mu,r_0](l_{v}(r;S_\lambda))
=l_{v}(r r_0^t;S_\mu)=l_{v}(r;l_{v}(r_0^t;S_\mu))
\nonumber\\
&&\quad
=l_{v}(r;S_\mu)=\hat{\beta}[\R^3,l_{v};S_\lambda,S_\mu,I_{3\times 3}]
(l_{v}(r;S_\lambda)) \; ,
\label{eq:12.17dbtntps}
\end{eqnarray}
i.e., 
\begin{eqnarray}
&&  \hspace{-15mm}
\lbrace
\hat{\beta}[\R^3,l_{v};S_\lambda,S_\mu,r_0]:r_0\in SO(2) \rbrace  =
\lbrace \hat{\beta}[\R^3,l_{v};S_\lambda,S_\mu,I_{3\times 3}] \rbrace\; ,
\label{eq:12.17dbtntqs}
\end{eqnarray}
where in the third equality of (\ref{eq:12.17dbtntps}) we used the 
relation $SO(2) =Iso(\R^3,l_{v};S_\mu)$
from
(\ref{eq:10.119}).
In analogy to (\ref{eq:12.17dbtntps}),
if $r_0\in{\cal K}_1 SO(2)$, i.e.,  $r_0={\cal K}_1 r_1$
with $r_1\in SO(2)$ and if $r\in SO(3)$ then, by
(\ref{eq:10.83}) and (\ref{eq:10.162}),
\begin{eqnarray}
&&  \hspace{-1cm}
\hat{\beta}[\R^3,l_{v};S_\lambda,S_\mu,r_0](l_{v}(r;S_\lambda))
=l_{v}(r r_0^t;S_\mu)=l_{v}(r r_1^t{\cal K}_1;S_\mu)
\nonumber\\
&& 
=l_{v}(r r_1^t;l_{v}({\cal K}_1;S_\mu))
=-l_{v}(r r_1^t;S_\mu)
=- l_{v}(r;l_{v}(r_1^t;S_\mu))
=- l_{v}(r;S_\mu)
\nonumber\\
&&\quad =l_{v}(r;l_{v}({\cal K}_1;S_\mu))
=l_{v}(r{\cal K}_1;S_\mu)
= \hat{\beta}[\R^3,l_{v};S_\lambda,S_\mu,{\cal K}_1](l_{v}(r;S_\lambda)) \; ,
\nonumber\\
\label{eq:12.17dbtntrs}
\end{eqnarray}
i.e., 
\begin{eqnarray}
&&  \hspace{-17mm}
\lbrace
\hat{\beta}[\R^3,l_{v};S_\lambda,S_\mu,r_0]:r_0\in {\cal K}_1 SO(2) \rbrace  =
\lbrace \hat{\beta}[\R^3,l_{v};S_\lambda,S_\mu,{\cal K}_1] \rbrace\; .
\label{eq:12.17dbtntss}
\end{eqnarray}
In the sixth equality of (\ref{eq:12.17dbtntrs})
we again used the relation $SO(2) =Iso(\R^3,l_{v};S_\mu)$ from
(\ref{eq:10.119}).
We conclude from (\ref{eq:10.162xxb}),(\ref{eq:12.17dbtntqs})
and (\ref{eq:12.17dbtntss}) 
that $B(\R^3,l_{v};S_\lambda,S_\mu)$ has only two elements:
\begin{eqnarray}
&&  \hspace{-1cm}
B(\R^3,l_{v};S_\lambda,S_\mu) = 
\lbrace \hat{\beta}[\R^3,l_{v};S_\lambda,S_\mu,{\cal K}_0],
\hat{\beta}[\R^3,l_{v};S_\lambda,S_\mu,{\cal K}_1] \rbrace\; .
\label{eq:10.83atasta}
\end{eqnarray}
We now take a closer look at these two elements and we first note that if
$r_0\in(SO(2) \bowtie \Z_2)$
then, by (\ref{eq:10.19}) and (\ref{eq:10.83}),
\begin{eqnarray}
&&  \hspace{-1cm}
\hat{\beta}[\R^3,l_{v};S_\lambda,S_\mu,r_0](r S_\lambda)=rr^t_0S_\mu 
\; .
\label{eq:10.120l}
\end{eqnarray}
It follows from (\ref{eq:10.120l}) and Remark 10 that, for $r\in SO(3)$,
$\hat{\beta}[\R^3,l_{v};S_\lambda,S_\mu,I_{3\times 3}](\lambda r(0,0,1)^t)
= \mu r (0,0,1)^t$, 
whence, for every $S$ in the domain ${\mathbb S}^2_\lambda$ of
$\hat{\beta}[\R^3,l_{v};S_\lambda,S_\mu,I_{3\times 3}]$,
\begin{eqnarray}
&&  \hspace{-1cm}
\hat{\beta}[\R^3,l_{v};S_\lambda,S_\mu,I_{3\times 3}](S)= \frac{\mu}{\lambda}S \; .
\label{eq:10.120la}
\end{eqnarray}
It also follows from (\ref{eq:10.120l}) and Remark 10 that, for $r\in SO(3)$,\\
$\hat{\beta}[\R^3,l_{v};S_\lambda,S_\mu,{\cal K}_1](\lambda r(0,0,1)^t)=
\mu r {\cal K}_1(0,0,1)^t =-\mu r (0,0,1)^t$
whence, for every $S$ in the domain ${\mathbb S}^2_\lambda$ of
$\hat{\beta}[\R^3,l_{v};S_\lambda,S_\mu,{\cal K}_1]$,
\begin{eqnarray}
&&  \hspace{-1cm}
\hat{\beta}[\R^3,l_{v};S_\lambda,S_\mu,{\cal K}_1](S)= -\frac{\mu}{\lambda}S \; .
\label{eq:10.120lb}
\end{eqnarray}
With (\ref{eq:10.83atasta}),
(\ref{eq:10.120la}) and (\ref{eq:10.120lb}) it is a simple exercise
to show that both elements of $B(\R^3,l_{v};S_\lambda,S_\mu)$ are not only
topological $SO(3)$-maps
but also isomorphisms (and, since
$Iso(\R^3,l_{v};S_\lambda)$ and $Iso(\R^3,l_{v};S_\mu)$ are conjugate, 
this is predicted by Remark 17). To apply
the DT let $(j,A)\in{\cal SOS}(d,j)$, let
$f\in{\cal C}(\Td,\R^d)$ take values
only in the $(\R^3,l_{v})$-orbit ${\mathbb S}^2_\lambda$
of $S_\lambda$ and let $g_0,g_1\in{\cal C}(\Td,\R^d)$ be defined by
$g_0(z):=\hat{\beta}[\R^3,l_{v};S_\lambda,S_\mu,{\cal K}_0](f(z))$
and $g_1(z):=\hat{\beta}[\R^3,l_{v};S_\lambda,S_\mu,{\cal K}_1](f(z))$. Then
$g_0$ and $g_1$ take values only in the $(\R^3,l_{v})$-orbit ${\mathbb S}^2_\mu$
of $S_\mu$. Moreover, by Theorem \ref{T10.6}b,
$f$ is an invariant polarization field of $(j,A)$ 
iff $g_0$ is an invariant polarization field of $(j,A)$ and
$f$ is an invariant polarization field of $(j,A)$ 
iff $g_1$ is an invariant polarization field of $(j,A)$.
This completes our treatment of the first case.

We now consider the case when $\lambda=0,\mu=0$. Then, by (\ref{eq:10.119}),
$Iso(\R^3,l_{v};S_\lambda)=SO(3)=Iso(\R^3,l_{v};S_\mu)$
whence $I_{3\times 3}Iso(\R^3,l_{v};S_\lambda)I_{3\times 3}^t
=Iso(\R^3,l_{v};S_\mu)$ so that\\
$Iso(\R^3,l_{v};S_\lambda)$ and $Iso(\R^3,l_{v};S_\mu)$ are conjugate. 
Since the $(\R^3,l_{v})$-orbit ${\mathbb S}^2_0$
of $S_0=(0,0,0)^t$ only contains $S_0$, the only element of
$B(\R^3,l_{v};S_\lambda,S_\mu)$ is the constant 
$(0,0,0)^t$-valued function. 
Since $Iso(\R^3,l_{v};S_\lambda)$ and $Iso(\R^3,l_{v};S_\mu)$ are conjugate
it is no surprise that the only element of 
$B(\R^3,l_{v};S_\lambda,S_\mu)$ is an isomorphism (see Remark 17).

We now consider the case when $\lambda=0,\mu>0$. Thus, by (\ref{eq:10.119}),
$Iso(\R^3,l_{v};S_\lambda)=SO(3),Iso(\R^3,l_{v};S_\mu)=SO(2)$.
Since, for every $r_0\in SO(3)$, one has
$r_0 SO(3)r_0^t=SO(3)$ we conclude from (\ref{eq:12.17dbt}) that
$N(SO(3),SO(2))$ is empty so that \\
$N(Iso(\R^3,l_{v};S_\lambda),Iso(\R^3,l_{v};S_\mu))$ is empty 
which implies,
by Lemma \ref{L10.5}a, that 
\\$B(\R^3,l_{v};S_\lambda,S_\mu)$ is empty.
Note also that the only subgroup of $SO(3)$ which is conjugate to $SO(3)$ 
is $SO(3)$. Thus $SO(2)$ and $SO(3)$ are not conjugate so that the emptiness
of $N(SO(3),SO(2))$ is predicted by Remark 17.

We finally consider 
the case when $\lambda>0,\mu=0$. Then, by (\ref{eq:10.119}),\\
$Iso(\R^3,l_{v};S_\lambda)=SO(2),Iso(\R^3,l_{v};S_\mu)=SO(3)$
whence, by (\ref{eq:12.17dbt}), \\
$N(Iso(\R^3,l_{v};S_\lambda),Iso(\R^3,l_{v};S_\mu))=SO(3)$.
Recalling from above that $SO(2)$ and $SO(3)$ are not conjugate we observe 
that  
$Iso(\R^3,l_{v};S_\lambda)$ and $Iso(\R^3,l_{v};S_\mu)$ are not conjugate. 
As in the second case,
the only element of $B(\R^3,l_{v};S_\lambda,S_\mu)$ is the constant 
$(0,0,0)^t$-valued function.
Since $Iso(\R^3,l_{v};S_\lambda)\unlhd Iso(\R^3,l_{v};S_\mu)$
and since $Iso(\R^3,l_{v};S_\lambda)$ and $Iso(\R^3,l_{v};S_\mu)$
are not conjugate it is no surprise that the only element of 
$B(\R^3,l_{v};S_\lambda,S_\mu)$ is a
topological $SO(3)$-map which is not an isomorphism (see Remark 17).

We finally mention some further features of the example
$(E,l)=(E',l')=(\R^3,l_{v})$:

\vspace{3mm}
\noindent{\bf Remark:}
\begin{itemize}
\item[(19)] 
It follows from (\ref{eq:10.119}) that for given 
$\lambda,\mu\in [0,\infty)$ either \\
$Iso(\R^3,l_{v};S_\lambda) \unlhd Iso(\R^3,l_{v};S_\mu)$ 
or $Iso(\R^3,l_{v};S_\mu) \unlhd Iso(\R^3,l_{v};S_\lambda)$. 
This is quite remarkable since in general two subgroups
of $SO(3)$ are not related by $\unlhd$.
Thus more
generally, by Lemma \ref{L10.5}b,
for arbitrary $S,S'\in\R^3$, either
$Iso(\R^3,l_{v};S) \unlhd Iso(\R^3,l_{v};S')$ 
or $Iso(\R^3,l_{v};S') \unlhd Iso(\R^3,l_{v};S)$. We also see
by (\ref{eq:10.119}) that if 
$Iso(\R^3,l_{v};S_\lambda) \unlhd Iso(\R^3,l_{v};S_\mu)$ then 
$Iso(\R^3,l_{v};S_\lambda) \subset Iso(\R^3,l_{v};S_\mu)$.
The latter inclusion is another reason why we have
chosen the $S_\lambda$ to be the elements of $R_{v}$
(note also that this inclusion is predicted by Lemma \ref{L10.5}b).
Since the $S_\lambda$ are the elements
of a representing set
of the partition $\R^3/l_{v}$ of $\R^3$
it also follows from (\ref{eq:10.110ln}) and (\ref{eq:10.119}) that 
every isotropy group of $(\R^3,l_{v})$ is either conjugate to
$SO(2)$ or to $SO(3)$. 
\hfill $\Box$
\end{itemize}
We have thus shown in the simple example of the this section
how the DT classifies invariant fields in terms of the isotropy groups 
of the $SO(3)$-spaces $(E,l)$ and $(E',l')$ at hand 
and how Definition \ref{D8.x} and the ICL play a key role.
So, for example, for the case $\lambda>0,\mu >0$, we decomposed 
$(E,l_{v})$ into the two $(E,l_{v})$ orbits $l_{v}(SO(3);S_\lambda)$
and $l_{v}(SO(3);S_\mu)$ (both of them spheres of nonzero radius).
Then we showed how invariant fields with values confined
to these spheres are related and thus classified via the
functions $\hat{\beta}[\R^3,l_{v};S_\lambda,S_\mu,I_{3\times 3}]$ and
$\hat{\beta}[\R^3,l_{v};S_\lambda,S_\mu,{\cal K}_1]$ which in fact are the only
ones there are. A key role was played by
$N(Iso[\R^3,l_{v};S_\lambda],Iso[\R^3,l_{v};S_\mu])$ which turned out to
be the subgroup $SO(2) \bowtie \Z_2$ of $SO(3)$.
This  will also play a major role
in Section \ref{10.4} (but for a different reason).
We also saw that the invariant fields with values confined
to the sphere $S_\mu$ are redundant since both betas are isomorphisms
(and this implied that the subset 
$N(Iso[\R^3,l_{v};S_\lambda],Iso[\R^3,l_{v};S_\mu])$ of $SO(3)$
is a group).
\subsection{Applying the ToA to $(E_{t},l_{t})$}
\label{10.4}
The $SO(3)$-space $(E,l)=(\R^3,l_{v})$ is needed 
for describing polarized beams of arbitrary nonzero spin.
However, it does not always suffice. 
In particular, for spin-$1$ particles like deuterons \cite{BV2}
we need a framework for handling the spin tensor variable $M$ which is
a real, symmetric and traceless $3\times 3$ matrix.
See \cite{BV2} for the dynamics of $M$ 
under the influence of the T-BMT equation. 
This inspires (\ref{eq:10.130}) below which leads to the correct
$1$-turn map in (\ref{eq:10.130b}).
See \cite{BV2} and Section \ref{10.5.2} for the way in which $M$
appears in the spin-$1$ density matrix function.
So in this section we introduce the $SO(3)$-space $(E_{t},l_{t})$ 
to encompass the spin tensor and allow us to use the ToA for 
spin-$1$ particles \cite{BV2}. As in Section \ref{10.3} the focus
is on the field motion.

We will proceed as follows.
In Section \ref{10.4.1}, after defining $(E_{t},l_{t})$, we will
obtain the representing set $R_{t}$ of
the partition $E_{t}/l_{t}$ of $E_{t}$ and compute the
isotropy groups of $(E_{t},l_{t})$ allowing us to apply 
the DT in the case $(E,l)=(E',l')=(E_{t},l_{t})$.
Then in Section \ref{10.4.2}
we apply the NFT to $(E_{t},l_{t})$.
\subsubsection{Basic properties of $(E_{t},l_{t})$}
\label{10.4.1}
We define $E_{t}:=\lbrace M\in\R^{3\times 3}:M^t=M,Tr[M]=0\rbrace$
and equip $E_{t}$ with the subspace topology from
$\R^{3\times 3}$. Thus, and since $\R^{3\times 3}$ with its natural topology is a 
Hausdorff space, $E_{t}$ is a Hausdorff space, too.
We also define the function $l_{t}:SO(3)\times E_{t}\rightarrow E_{t}$ 
by 
\begin{eqnarray}
&&  l_{t}(r;M):=rMr^t \; ,
\label{eq:10.130}
\end{eqnarray}
with $r\in SO(3),M\in E_{t}$.
It is an easy exercise to show that $(E_{t},l_{t})$ is an
$SO(3)$-space. Note that matrices, which belong to the same
$(E_{t},l_{t})$-orbit, are similar, in particular they have the
same number of distinct eigenvalues. 
If $M\in E_{t}$ then we denote by $\#(M)$ the number of its 
distinct eigenvalues.

The $1$-turn map (\ref{eq:10.15}) 
in the present case is ${\cal P}[E_t,l_t,j,A]$, given by
\begin{eqnarray}
&&  {\cal P}[E_t,l_t,j,A](z,M)=
\left( \begin{array}{c} 
j(z)\\
A(z)MA(z) \end{array}\right) \; ,
\label{eq:10.130b}
\end{eqnarray}
and the $1$-turn field map (\ref{eq:10.17}) by
\begin{eqnarray}
&& \hspace{-1cm} \tilde{\cal P}[E_t,l_t,j,A](f)= (AfA^t)\circ j^{-1} \; .
\label{eq:8.4.100}
\end{eqnarray}

In the following remark we compute the $(E_{t},l_{t})$-orbits, 
giving us the partition 
$E_{t}/l_{t}$ of $E_{t}$.

\vspace{3mm}
\noindent{\bf Remark:}
\begin{itemize}
\item[(20)] 
It follows from (\ref{eq:10.130}), Definition \ref{D2.3new} 
and some simple Linear Algebra \cite{He1} that the \\
$(E_{t},l_{t})$-orbit of an arbitrary 
$M\in E_{t}$ reads as
\begin{eqnarray}
&&  \hspace{-15mm}
l_{t}(SO(3);M)=\lbrace M'\in  E_{t}:
(\det(M'),Tr[M'^2])=(\det(M),Tr[M^2]) \rbrace \; .
\label{eq:10.135}
\end{eqnarray}
Note that (\ref{eq:10.135}) follows easily from the fact that the
characteristic polynomial of $M$ is the function 
$\det(M-x I_{3\times 3})=-x^3+\frac{1}{2}Tr[M^2]x + \det(M)$ and that
this polynomial is the same for all elements of the 
$(E_{t},l_{t})$-orbit of $M$.
We now define
\begin{eqnarray}
&&  \hspace{-1cm}
R_{t}:= \biggl\lbrace 
\Di(y_1,y_2,-y_1-y_2):(y_1,y_2)\in (\Lambda_1\cup 
\Lambda_2\cup \Lambda_3) \biggr\rbrace \subset E_t\; ,
\label{eq:10.140}
\end{eqnarray}
where $\Lambda_1:=\lbrace (0,0) \rbrace, \Lambda_2:=
\lbrace (y,y):0\neq y\in\R \rbrace, \Lambda_3:=
\lbrace (y,y'):
y\in(0,\infty),y'\in(-y/2,y)\rbrace$.
The matrices $\Di(y_1,y_2,-y_1-y_2)$ with $(y_1,y_2)\in \Lambda_j$ have
a simple interpretation: they are
those matrices $M$ in $R_{t}$ which have $\#(M)=j$.
This implies, since matrices of the same
$(E_{t},l_{t})$-orbit are similar, that
an arbitrary matrix $M\in E_{t}$ has $\#(M)=j$
if its $(E_{t},l_{t})$-orbit
contains a matrix $M'=\Di(y_1,y_2,-y_1-y_2)$ with $(y_1,y_2)\in \Lambda_j$
\cite{He1} (of course $M'$ is unique).
By (\ref{eq:10.135}), each element of
$R_{t}$ belongs to a different $(E_{t},l_{t})$-orbit.
Moreover, by some simple Linear Algebra, one can show
\cite{He1} that every element $M$ of $E_{t}$ belongs to the 
$(E_{t},l_{t})$-orbit of some $M'\in R_{t}$.
In other words, $R_{t}$ is a representing set of
the partition $E_{t}/l_{t}$ of $E_{t}$ (recall from
Remark 10 that $R_{v}$ 
is a representing set of the partition $\R^3/l_{v}$ of $\R^3$). 
As with $R_{v}$, the choice $R_{t}$ is very convenient as
will become clear below.
\hfill $\Box$
\end{itemize}
Note that the above technique of using $\det(M),Tr[M^2]$ as ``invariants'' of
$M$ is also used sometimes for the emittance matrix in four-dimensional
linear beam optics.

Remark 20
allows us, in the following remark, to parametrize the
elements of $E_{t}$ in terms of normalized vectors.

\vspace{3mm}
\noindent{\bf Remark:}
\begin{itemize}
\item[(21)] By the above, the set of those $M$ in
$E_{t}$ for which $\#(M)=j$ is given by
\begin{eqnarray}
&&  \hspace{-15mm}
\lbrace r \Di(y_1,y_2,-y_1-y_2)r^t:r\in SO(3),(y_1,y_2)\in \Lambda_j \rbrace 
\; ,
\label{eq:10.145}
\end{eqnarray}
and, for $(y_1,y_2)\in\R^2$,
\begin{eqnarray}
&&  \hspace{-1cm}
\Di(y_1,y_2,-y_1-y_2) = y_1 I_{3\times 3} + \Di(0,y_2-y_1,-2y_1-y_2)
\nonumber\\
&& = y_1 I_{3\times 3} + (y_2-y_1)(0,1,0)(0,1,0)^t
 - (2y_1+y_2)(0,0,1)(0,0,1)^t \; .
\label{eq:10.146}
\end{eqnarray}
\hfill $\Box$
\end{itemize}
The following remark shows the impact of Lemma \ref{L10.4} on
invariant $(E_{t},l_{t})$-fields.

\vspace{3mm}
\noindent{\bf Remark:}
\begin{itemize}
\item[(22)] 
Let $(j,A)\in{\cal SOS}(d,j)$ where $j$ is topologically transitive.
Let $f$ be an invariant 
$(E_{t},l_{t})$-field of 
$(j,A)$. Then, by Lemma \ref{L10.4}, 
$f$ takes values in only one $(E_{t},l_{t})$-orbit, say 
$l_{t}(SO(3);M)$. By Remark 20 we can choose $M$ to belong to
$R_{t}$, i.e., 
$M=\Di(y_1,y_2,-y_1-y_2)$ where $(y_1,y_2)\in\Lambda_i$ with 
$\#(M)=i$.
Thus, by Remark 21, a function $T:\Td\rightarrow SO(3)$ exists such that 
$f(z)= l_{t}(T(z);M)$ and
\begin{eqnarray}
&& f(z) = 
y_1 I_{3\times 3} + (y_2-y_1)\tilde{k}(z)\tilde{k}^t(z) 
-(2y_1+y_2)k(z)k^t(z) \; ,
\label{eq:10.210}
\end{eqnarray}
where the functions $k,\tilde{k}:\Td\rightarrow \R^3$ are defined by 
\begin{eqnarray}
&& k(z):=T(z)(0,0,1)^t \; , \quad  \tilde{k}(z):=T(z)(0,1,0)^t \; .
\label{eq:10.177n}
\end{eqnarray}
Of course $\#(f(z))=i$ for all $z\in\Td$.
In the special case where $i=2$, i.e.,
$y_1=y_2=:y$ with $0\neq y\in\R$, 
(\ref{eq:10.210}) reads as
\begin{eqnarray}
&& f(z)= y I_{3\times 3} -3 yk(z)k^t(z) \; ,
\label{eq:10.178n}
\end{eqnarray}
and in the special case where $i=1$, i.e.,
$M=\Di(0,0,0)$, (\ref{eq:10.210}) reads as $f(z)= \Di(0,0,0)$.
\hfill $\Box$
\end{itemize}
Using (\ref{eq:12.17bnncaa}),(\ref{eq:6.5}),(\ref{eq:10.161}),
(\ref{eq:10.130}) and (\ref{eq:10.140}) it is a simple exercise
\cite{He1} to show that for $M\in R_{t}$
\begin{eqnarray}
&&  \hspace{-1cm}
Iso(E_{t},l_{t};M) 
= \left\{ \begin{array}{ll}  SO(3)
& \;\; {\rm if\;}\#(M)=1 \\
SO(2) \bowtie \Z_2 & \;\; {\rm if\;}\#(M)=2 \\
SO_{diag}(3)
& \;\; {\rm if\;}\#(M)=3\; ,
\end{array}
                  \right.
\label{eq:12.17bnncbtt}
\end{eqnarray}
where
\begin{eqnarray}
&& \hspace{-1cm}
SO_{diag}(3):=  SO(3)\cap  \lbrace \Di(y_1,y_2,y_3):y_1,y_2,y_3\in\R \rbrace 
= \lbrace {\cal K}_0,{\cal K}_1,{\cal K}_2,{\cal K}_3\rbrace \; .
\label{eq:10.163}
\end{eqnarray}
Clearly $SO_{diag}(3)$ is a subgroup of $SO(3)$ and it is
the set of diagonal matrices in $SO(3)$.
Note that, by (\ref{eq:10.161}) and (\ref{eq:10.163}),
\begin{eqnarray}
&&  \hspace{-1cm}
SO_{diag}(3)\subset (SO(2) \bowtie \Z_2)\subset SO(3) \; .
\label{eq:10.164}
\end{eqnarray}
Since the groups $SO(2)\bowtie \Z_2$ and $SO_{diag}(3)$ are
conveniently handled, we see
by (\ref{eq:12.17bnncbtt}) that it was prescient to have 
chosen $R_{t}$ as in (\ref{eq:10.140}).

The following remark mentions some implications of
(\ref{eq:12.17bnncbtt}).

\vspace{3mm}
\noindent{\bf Remark:}
\begin{itemize}
\item[(23)] 
We conclude from 
(\ref{eq:12.17bnncbtt}) and (\ref{eq:10.164})
that, if $M,M'\in R_{t}$, then\\
$Iso(E_{t},l_{t};M)\unlhd Iso(E_{t},l_{t};M')$
iff $\#(M)\geq\#(M')$
(and, by Lemma \ref{L10.5}b, 
this holds for arbitrary
$M,M'\in E_{t}$ since $R_{t}$ is a representing set of
$E_{t}/l_{t}$). Thus, quite remarkably
we see that for all $M,M'\in E_{t}$ either
$Iso(E_{t},l_{t};M)\unlhd Iso(E_{t},l_{t};M')$ or
$Iso(E_{t},l_{t};M')\unlhd Iso(E_{t},l_{t};M)$.

We also see, by (\ref{eq:12.17bnncbtt}) and (\ref{eq:10.164}),
that if $M,M'\in R_{t}$ and 
$Iso(E_{t},l_{t};M)\unlhd Iso(E_{t},l_{t};M')$ then 
$Iso(E_{t},l_{t};M)\subset Iso(E_{t},l_{t};M')$.
The latter inclusion is another reason why we have
chosen $R_{t}$ as in (\ref{eq:10.140})
(note also that this inclusion is predicted by Lemma \ref{L10.5}b).
We recall from Section \ref{10.3.3} that
$SO(3)$ is only
conjugate to itself. Also, by Definition \ref{D2.1},
the finite group $SO_{diag}(3)$ is not conjugate to infinite groups and
the infinite group $SO(2)\bowtie \Z_2$ is not conjugate
to finite groups. Thus it follows from 
(\ref{eq:12.17bnncbtt}) that, if
$M,M'\in R_{t}$, then
$Iso(E_{t},l_{t};M)$ and $Iso(E_{t},l_{t};M')$ are conjugate iff
$\#(M)=\#(M')$ (and this holds for arbitrary
$M,M'\in E_{t}$ since $R_{t}$ is a representing set of
$E_{t}/l_{t}$).
In fact for arbitrary $M\in E_{t}$ we have $\#(M)=1$ iff
$Iso(E_{t},l_{t};M)$ is conjugate to $SO(3)$,
we have $\#(M)=2$ iff
$Iso(E_{t},l_{t};M)$ is conjugate to $SO(2) \bowtie \Z_2$, and
we have $\#(M)=3$ iff
$Iso(E_{t},l_{t};M)$ is conjugate to $SO_{diag}(3)$.
\hfill $\Box$
\end{itemize}
Recalling the remarks after the NFT,
the isotropy groups in (\ref{eq:12.17bnncbtt}) will give us insight
into the possibility of invariant $(E_{t},l_{t})$-fields.
In fact if $(j,A)$ has an ISF then, as we will see in
Section \ref{10.4b.1}, $(j,A)$ has an invariant $(E_{t},l_{t})$-field
whose values are matrices $M$ with $\#(M)=2$. This follows 
from the fact that $SO(2)\subset (SO(2) \bowtie \Z_2)$.
Thus, by the remarks after the NFT, we believe that in practice 
invariant $(E_{t},l_{t})$-fields exist
whose values are matrices $M$ with $\#(M)=2$.
On the other hand, by (\ref{eq:12.17bnncbtt}) and the remarks after the NFT, 
an invariant $(E_{t},l_{t})$-field which has an $(E_t,l_t)$-lift 
and whose values are matrices $M$ with $\#(M)=3$ can only
exist if $(j,A)$ has an $SO_{diag}(3)$-normal form (for the notion of ``lift'',
see the remarks after the NFT).
However $SO_{diag}(3)$ is ``small'' since it contains only
four elements thus we expect that $(j,A)$ has an $SO_{diag}(3)$-normal form
only if it has a $G_0$-normal form, i.e., if $(j,A)$ is on spin-orbit 
resonance. Since in practice we expect that the assumption of
an $(E_t,l_t)$-lift is not strong, we expect that in practice
invariant $(E_{t},l_{t})$-fields
whose values are matrices $M$ with $\#(M)=3$ only
exist on spin-orbit resonance. 
On spin-orbit resonance those 
invariant $(E_{t},l_{t})$-fields exist, as is explained after Remark 18.
Note also that the $\Di(0,0,0)$-valued function is
always an invariant $(E_{t},l_{t})$-field whence 
invariant $(E_{t},l_{t})$-fields always exist
whose values are matrices $M$ with $\#(M)=1$.
Note that $\#(M)=1$ is of physical importance
as can be seen for example in Section \ref{10.5} where
spin tensors $M$ appear as coefficients in the
density matrix functions of spin-$1$ particles.

We now apply the DT
to the case $(E,l)=(E',l')=(E_{t},l_{t})$. This is interesting
since it deals with invariant $(E_{t},l_{t})$-fields and since the
latter have been much less studied than invariant polarization fields.
We follow the strategy of Remark 18 and use again the convenient representing
set $R_{t}$ of the partition $E_{t}/l_{t}$. Thus we have to
determine for given $M,M'\in R_{t}$
whether $Iso(E_{t},l_{t};M)\unlhd Iso(E_{t},l_{t};M')$ or whether
$Iso(E_{t},l_{t};M)$ and $Iso(E_{t},l_{t};M')$ are even conjugate.
In fact by Remark 23 we know that
$Iso(E_{t},l_{t};M)\unlhd Iso(E_{t},l_{t};M')$
iff $\#(M)\geq \#(M')$ and that
$Iso(E_{t},l_{t};M)$ and $Iso(E_{t},l_{t};M')$ are conjugate iff $\#(M)=\#(M')$.
Recalling Remark 18, the latter case compares
invariant $(E_{t},l_{t})$-fields which have identical behavior.
To keep the discussion short we confine ourselves to the
case where $i=3,k=2$ and which is interesting since
it involves the two most important isotropy 
groups of $(E_{t},l_{t})$: $SO_{diag}(3)$ and $SO(2) \bowtie \Z_2$.

The computations are analogous to those in Section \ref{10.3.3}.
So we focus here on the results and leave the somewhat lengthy Linear Algebra
\cite{He1} to the reader. In fact, by Definition \ref{D8.x},
one gets
\begin{eqnarray}
&& \hspace{-1cm}
N([Iso(E_{t},l_{t};M),Iso(E_{t},l_{t};M'))
=N(SO_{diag}(3),SO(2) \bowtie \Z_2) 
\nonumber\\
&&
= (SO(2) \bowtie \Z_2) \cup (SO(2) \bowtie \Z_2) {\cal K}_4
\cup  (SO(2) \bowtie \Z_2){\cal K}_5
\; ,
\label{eq:12.17dbtnta23}
\end{eqnarray}
where
\begin{eqnarray}
&& \hspace{-1cm}
{\cal K}_4:= \left( \begin{array}{ccc}
0 & 0 & -1 \\
0 & 1 & 0 \\
 1 & 0 & 0 \end{array}\right) \; , \quad
{\cal K}_5:= \left( \begin{array}{ccc}
1 & 0 & 0 \\
0 & 0 & -1 \\
 0 & 1 & 0 \end{array}\right) \; .
\label{eq:12.17dbtnta}
\end{eqnarray}
Thus, by Lemma \ref{L10.5}a, and some further Linear Algebra one gets
\begin{eqnarray}
&&  \hspace{-1cm}
B(E_{t},l_{t},M,M') = B(E_{t},l_{t},E_{t},l_{t},M,M') = 
\lbrace
\hat{\beta}[E_{t},l_{t};M,M',{\cal K}_0],
\hat{\beta}[E_{t},l_{t};M,M',{\cal K}_4],
\nonumber\\
&&\hat{\beta}[E_{t},l_{t};M,M',{\cal K}_5]
\rbrace \; .
\label{eq:10.83ata}
\end{eqnarray}
Since $SO(2)\subset (SO(2) \bowtie \Z_2)$ it is no surprise by
Lemma \ref{L10.5}a that $B(E_{t},l_{t},M,M')$ is nonempty.
We now can apply the DT to the current case. So let
$f\in{\cal C}(\Td,E_{t})$ take values only in the 
$(E_{t},l_{t})$-orbit of $M$, i.e., let
a function $T:\Td\rightarrow SO(3)$ exist such that
\begin{eqnarray}
&& f(z):=l_{t}(T(z),M) \; .
\label{eq:10.175}
\end{eqnarray}
Then $f$ satisfies (\ref{eq:10.210}), i.e.,
$f(z) = y_1 I_{3\times 3} + (y_2-y_1)\tilde{k}(z)\tilde{k}^t(z) 
-(2y_1+y_2)k(z)k^t(z)$
where the functions $k,\tilde{k}:\Td\rightarrow \R^3$ are defined by 
(\ref{eq:10.177n}).
Also let $g_n\in{\cal C}(\Td,E_{t})$ be defined by $g_n(z):=
\hat{\beta}[E_{t},l_{t};M,M',{\cal K}_n](f(z))$ where $n=0,4,5$.
Then
\begin{eqnarray}
&& g_0(z) = y I_{3\times 3} -3y k(z)k^t(z) \; , \label{eq:10.180x0} \\
&& g_4(z) = -2y I_{3\times 3} + 3y\tilde{k}(z)\tilde{k}^t(z) 
+3yk(z)k^t(z) \; , \label{eq:10.180x4} \\
&& g_5(z) = y I_{3\times 3} - 3y\tilde{k}(z)\tilde{k}^t(z) \; .
\label{eq:10.180x5}
\end{eqnarray}
Moreover if $(j,A)\in{\cal SOS}(d,j)$ and 
$f$ is an invariant $(E_{t},l_{t})$-field of 
$(j,A)$ then $g_0,g_4,g_5$ are invariant $(E_{t},l_{t})$-fields of
$(j,A)$. Note that $g_0,g_4,g_5$ take values only in $l_{t}(SO(3);M')$.
Thus each value of the functions 
$g_0,g_4,g_5$ is a matrix with two distinct eigenvalues. It is thus
easy to show, by part b) of the DT, that $g_0,g_4,g_5$ are equivalent, i.e.,
are related by isomorphisms of $SO(3)$-spaces.
\subsubsection{Applying the Normal Form Theorem to $(E_{t},l_{t})$}
\label{10.4.2}
We have already applied the NFT to the 
cases of $(\R^3,l_v)$ and $(SO(3),l_{SOR})$ in Remarks  
8 and 9. Now with Section \ref{10.4.1} we are equipped to apply the 
Normal Form Theorem to $(E_{t},l_{t})$.

\setcounter{theorem}{9}
\begin{theorem} \label{T10.7} 
Let $M\in E_{t}$ have $\#(M)=i$, i.e., 
let $r\in SO(3)$ exist such that\\
$M=l_{t}(r;\Di(y_1,y_2,-y_1-y_2))$ where
$(y_1,y_2)\in \Lambda_i$. Moreover let $(j,A)\in{\cal SOS}(d,j)$ and 
$T\in{\cal TF}_H(j,A)$ where we define
$H:=Iso(E_{t},l_{t};M)$.
Then $f\in{\cal C}(\Td,E_{t})$, defined by (\ref{eq:10.175}), 
i.e., $f(z):=l_{t}(T(z),M)$,
is an invariant $(E_{t},l_{t})$-field of $(j,A)$.
Moreover $\#(f(z))=i$
and
\begin{eqnarray}
&& f(z)= y_1 I_{3\times 3} + (y_2-y_1)\tilde{l}(z)\tilde{l}^t(z) 
-(2y_1+y_2)l(z)l^t(z) \; ,
\label{eq:10.176}
\end{eqnarray}
where the functions $l,\tilde{l}\in{\cal C}(\Td,\R^3)$ are defined by 
\begin{eqnarray}
&& l(z):=T(z)r(0,0,1)^t \; , \quad  \tilde{l}(z):=T(z)r(0,1,0)^t \; .
\label{eq:10.177}
\end{eqnarray}
\noindent Remark 1: Note that $|l(z)|=|\tilde{l}(z)|=1$.\\
\noindent Remark 2: In the special case where $i=2$, i.e.,
where $y_1=y_2=:y$ with $0\neq y\in\R$, (\ref{eq:10.176}) reads as
\begin{eqnarray}
&& f(z)= y I_{3\times 3} -3 yl(z)l^t(z) \; ,
\label{eq:10.178}
\end{eqnarray}
and in the special case where $i=1$, i.e.,
$M=\Di(0,0,0)$, (\ref{eq:10.176}) reads as $f(z)= \Di(0,0,0)$.
\end{theorem}

\noindent{\em Proof of Theorem \ref{T10.7}}:
That $f$ in (\ref{eq:10.175})
is an invariant $(E_{t},l_{t})$-field of $(j,A)$ follows
from the NFT, Theorem \ref{T10.1}.
Moreover, by (\ref{eq:10.130}) and (\ref{eq:10.175})
\begin{eqnarray}
&& f(z)=l_{t}(T(z),M) =l_{t}(T(z),l_{t}(r;\Di(y_1,y_2,-y_1-y_2)))
\nonumber\\
&&\quad =l_{t}(T(z)r;
\Di(y_1,y_2,-y_1-y_2))= T(z)\:r\;\Di(y_1,y_2,-y_1-y_2)\;r^t\;T(z)
\; ,
\nonumber
\end{eqnarray}
whence (\ref{eq:10.176}) follows from (\ref{eq:10.146}) and (\ref{eq:10.177}).
The remaining claim follows from Remark 21.
\hfill $\Box$

\vspace{3mm}

Since $SO(2)\subset (SO(2) \bowtie \Z_2)$,
the case $i=2$ in Theorem \ref{T10.7} shows the
impact of IFF's on invariant $(E_{t},l_{t})$-fields
as the following remark demonstrates.

\vspace{3mm}
\noindent{\bf Remark:}
\begin{itemize}
\item[(24)]  
Let $M\in E_{t}$ have $\#(M)=2$,
i.e., let $r\in SO(3)$ exist such that
$M=l_{t}(r;\Di(y,y,-2y))$ where $0\neq y\in \R$.
Also let $(j,A)\in{\cal SOS}(d,j)$ and $T$ be an IFF of $(j,A)$, i.e.,
by Section \ref{V.2}, $T\in{\cal TF}_{SO(2)}(j,A)$. 
We define $H:=Iso(E_{t},l_{t};\Di(y,y,-2y))=SO(2)\bowtie \Z_2$ and 
$H':=Iso(E_{t},l_{t};M)$ where we also used
(\ref{eq:12.17bnncbtt}). It follows from (\ref{eq:10.110ln}) that $H'=rHr^t$.
On the other hand, by (\ref{eq:10.161}),
$SO(2)\subset (SO(2) \bowtie \Z_2)$ whence, by
Remark 2 in Chapter \ref{4.3}, 
${\cal TF}_{SO(2)}(j,A)\subset{\cal TF}_H(j,A)$ so that
$T\in{\cal TF}_H(j,A)$. Now define the function
$T'\in{\cal C}(\Td,SO(3))$ by
$T'(z):=T(z)r^t$. Thus, by Remark 7, $T'\in {\cal TF}_{H'}(j,A)$ whence
$T'$ is a generalized IFF of $(j,A)$ and, by
Theorem \ref{T10.7}, the function
$f\in{\cal C}(\Td,E)$, defined by
$f(z):=l_{t}(T'(z),M)$ is an invariant $(E_{t},l_{t})$-field 
of $(j,A)$. 
Note also that $f(z)=l_{t}(T(z),\Di(y,y,-2y))$. 
This demonstrates how IFF's lead to 
$(SO(2) \bowtie \Z_2)$-normal forms and
invariant $(E_{t},l_{t})$-fields.
\hfill $\Box$
\end{itemize}
\subsection{Applying the Decomposition Theorem 
in the case $(E,l)=(\R^3,l_{v})$ and $(E',l')=(E_{t},l_{t})$}
\label{10.4b}
In this section we apply the DT 
to the case $(E,l)=(\R^3,l_{v})$ and $(E',l')=(E_{t},l_{t})$
and thereby illustrate the connection between
invariant $(E_{t},l_{t})$-fields and 
invariant polarization fields. Moreover, we give
insights into  a model with two Siberian snakes (the ``$2$-snake model'').

We proceed as follows. In Section \ref{10.4b.1}
we apply the DT 
to arrive at Theorem \ref{T10.10}.
Then in Section \ref{10.4b.2} we consider
the $2$-snake model which has normalized, piecewise continuous
solutions of the $(\R^3,l_{v})$-stationarity equation but none of them
continuous whence the $2$-snake model
has no ISF. However we will show that it has
a nonzero invariant $(E_{t},l_{t})$-field whose values
are matrices $M$ with $\#(M)=2$
(and the latter will be derived from Theorem \ref{T10.10}).
%
\subsubsection{A corollary to the DT}
\label{10.4b.1}
To apply the DT in the case where $(E,l)=(\R^3,l_{v})$ and
$(E',l')=(E_{t},l_{t})$ we 
follow the strategy of Remark 18 and again use the convenient 
representing sets $R_{v}$ and $R_{t}$  of
$\R^3/l_{v}$ and $E_{t}/l_{t}$ respectively. Thus we have to
determine for given $S_\lambda\in R_{v},M\in R_{t}$
whether $Iso(\R^3,l_{v};S_\lambda)\unlhd Iso(E_{t},l_{t};M)$ or whether
$Iso(\R^3,l_{v};S_\lambda),Iso(E_{t},l_{t};M)$ are conjugate.
In fact, by (\ref{eq:10.119}) and (\ref{eq:12.17bnncbtt}) and 
Remarks 19 and 23, we only have to consider three cases defined as follows.
In the first case 
$\lambda=0$ and $\#(M)=1$,
in the second case $\lambda>0$ and $\#(M)=1$, and
in the third case $\lambda>0$ and $\#(M)=2$.
In the remaining case
$N(Iso(\R^3,l_{v};S_\lambda),Iso(E_{t},l_{t};M))=\emptyset$
whence $B(\R^3,l_{v},E_{t},l_{t};S_0,M)=\emptyset$. 
The following remark considers the first two cases. 

\vspace{3mm}
\noindent{\bf Remark:}
\begin{itemize}
\item[(25)] 
In the first case, where $S_0=(0,0,0)^t,M=\Di(0,0,0)$, we have, by
(\ref{eq:10.119}) and (\ref{eq:12.17bnncbtt}),
$Iso(\R^3,l_{v};S_0)=SO(3)=Iso(E_{t},l_{t};M)$
whence $Iso(\R^3,l_{v};S_0)$ and
$Iso(E_{t},l_{t};M)$ are conjugate
and $B(\R^3,l_{v},E_{t},l_{t};S_0,M)$ is a singleton
containing the constant, $\Di(0,0,0)$-valued function
which, by the DT, results in the constant
$\Di(0,0,0)$-valued invariant $(E_{t},l_{t})$-field.
Since $Iso(E_{t},l_{t};M)$ and $Iso(E_{t},l_{t};M')$ 
are conjugate it is no surprise that the only element of 
$B(\R^3,l_{v},E_{t},l_{t};S_0,M)$ is an isomorphism (see Remark 17).

In the second case where $\lambda>0,M=\Di(0,0,0)$ 
we have, by (\ref{eq:10.119}) and (\ref{eq:12.17bnncbtt}),\\
$Iso(\R^3,l_{v};S_\lambda)=SO(2)\subset SO(3)=Iso(E_{t},l_{t};M)$.
Since $SO(3)$ is only conjugate to itself, $SO(2),SO(3)$ are not 
conjugate whence, according to our strategy, we will compute the elements of
$B(\R^3,l_{v},E_{t},l_{t};S_\lambda,M)$.
By (\ref{eq:12.17dbt}) we have 
$N(SO(2),SO(3))=SO(3)$. If $r_0\in SO(3)$
then, by (\ref{eq:10.19}), (\ref{eq:10.83}) and (\ref{eq:10.130}),
$\hat{\beta}[\R^3,l_{v},E_{t},l_{t};S_\lambda,M,r_0](r S_\lambda)=rr^t_0Mr_0r^t
= \Di(0,0,0)$ whence, for every $S\in {\mathbb S}^2_\lambda$,
\begin{eqnarray}
&&  \hspace{-1cm}
\hat{\beta}[\R^3,l_{v},E_{t},l_{t};S_\lambda,M,r_0)(S)= \Di(0,0,0) \; .
\label{eq:10.120laxtsst}
\end{eqnarray}
Thus $\hat{\beta}[\R^3,l_{v},E_{t},l_{t};S_\lambda,M,r_0)$ 
is independent of the choice of $r_0$ and so\\
$B(\R^3,l_{v},E_{t},l_{t};S_\lambda,M)$ contains only one element:
\begin{eqnarray}
&&  \hspace{-1cm}
B(\R^3,l_{v},E_{t},l_{t};S_\lambda,M) = \lbrace
\hat{\beta}[\R^3,l_{v},E_{t},l_{t};S_\lambda,M,I_{3\times 3}] \rbrace \; .
\label{eq:10.83atasts}
\end{eqnarray}
Thus the only element of $B(\R^3,l_{v},E_{t},l_{t};S_\lambda,M)$ is the constant 
$\Di(0,0,0)$-valued function which, by the DT, results in the constant
$\Di(0,0,0)$-valued invariant $(E_{t},l_{t})$-field.
Since $Iso(\R^3,l_{v};S_\lambda)\unlhd Iso(E_{t},l_{t};M)$
and since $Iso(\R^3,l_{v};S_\lambda)$ and $Iso(E_{t},l_{t};M)$
are not conjugate it is no surprise that the only element of 
$B(\R^3,l_{v},E_{t},l_{t};S_\lambda,M)$ is a
topological $SO(3)$-map which is not an isomorphism
(recall Remark 17).
\hfill $\Box$
\end{itemize}
We finally consider the third case where 
$\lambda>0,\#(M)=2$ and with this case we can fulfill the
above mentioned aims in the situation where
$(E,l)=(\R^3,l_{v})$ and $(E',l')=(E_{t},l_{t})$.
In the present case where $\lambda>0,M=\Di(y,y,-2y)$ 
with $0\neq y\in\R$ we have, by 
(\ref{eq:10.119}), (\ref{eq:12.17bnncbtt}),
$Iso(\R^3,l_{v};S_\lambda)=SO(2)\subset (SO(2) \bowtie \Z_2)
=Iso(E_{t},l_{t};M)$ whence
\begin{eqnarray}
&& \hspace{-1cm}
N(Iso(\R^3,l_{v};S_\lambda),Iso(E_{t},l_{t};M))
=N(SO(2),SO(2) \bowtie \Z_2) \; .
\label{eq:12.17dbtnta23stsaa}
\end{eqnarray}
Since $SO(2)$ is Abelian and $SO(2)\bowtie \Z_2$ is not Abelian, 
$SO(2),SO(2) \bowtie \Z_2$ are not 
conjugate whence, according to our strategy, we will compute the elements of\\
$B(\R^3,l_{v},E_{t},l_{t};S_\lambda,M)$, i.e., by 
Lemma \ref{L10.5}a and (\ref{eq:12.17dbtnta23stsaa}) we have to compute \\
$N(SO(2),SO(2) \bowtie \Z_2)$. 
It is a simple exercise to show, by (\ref{eq:6.5}),(\ref{eq:12.17dbtn})
and (\ref{eq:10.161}), that
\begin{eqnarray}
&& \hspace{-1cm}
N(SO(2),SO(2) \bowtie \Z_2) \supset (SO(2) \bowtie \Z_2) \; ,
\label{eq:12.17dbtntca}
\end{eqnarray}
and we now show that $N(SO(2),SO(2) \bowtie \Z_2)=SO(2) \bowtie \Z_2$, i.e., we
will show that the inclusion, which is the converse to
(\ref{eq:12.17dbtntca}), holds too. For this we use
(\ref{eq:12.17dbtn}) and (\ref{eq:10.161}) to obtain
\begin{eqnarray}
&& N(SO(2),SO(2) \bowtie \Z_2) = \bigcap_{h\in SO(2)}
\; \bigcup_{h'\in (SO(2) \bowtie \Z_2)}
N(\lbrace h\rbrace,\lbrace h' \rbrace) 
\nonumber\\
&&\quad  =  \bigcap_{h\in SO(2)}
\; \bigcup_{h'\in (SO(2)}
\biggl( N(\lbrace h\rbrace,\lbrace h' \rbrace) 
\cup N(\lbrace h\rbrace,\lbrace {\cal K}_1 h' \rbrace) \biggr)
\nonumber\\
&&\quad = 
\bigcap_{h\in SO(2)}
\; 
\biggl( N(\lbrace h\rbrace,SO(2)) 
\cup N(\lbrace h\rbrace, {\cal K}_1 SO(2)) \biggr)
\nonumber\\
&&\quad \subset
\bigcap_{h\in (SO(2)\setminus \lbrace I_{3\times 3},\exp(\pi{\cal J})\rbrace)}
\; 
\biggl( N(\lbrace h\rbrace,SO(2)) 
\cup N(\lbrace h\rbrace, {\cal K}_1 SO(2)) \biggr) \; .
\label{eq:12.17dbtntst}
\end{eqnarray}
It is an easy exercise to show, by (\ref{eq:6.5}),(\ref{eq:12.17dbtn})
and (\ref{eq:10.162}) and the $SO(2)$-Lemma, that if 
$h\in (SO(2)\setminus \lbrace I_{3\times 3},\exp(\pi{\cal J})\rbrace)$ then
$N(\lbrace h\rbrace, {\cal K}_1 SO(2))=\emptyset$
whence, by (\ref{eq:12.17dbtntst}),
\begin{eqnarray}
&& N(SO(2),SO(2) \bowtie \Z_2) \subset
\bigcap_{h\in (SO(2)\setminus \lbrace I_{3\times 3},\exp(\pi{\cal J})\rbrace)}
\; N(\lbrace h\rbrace,SO(2))  \; .
\label{eq:12.17dbtntstb}
\end{eqnarray}
It is also a simple exercise to show, 
by (\ref{eq:6.5}),(\ref{eq:12.17dbtn}) and
the $SO(2)$-Lemma, that if 
$h\in (SO(2)\setminus \lbrace I_{3\times 3},\exp(\pi{\cal J})\rbrace)$ then
$N(\lbrace h\rbrace, SO(2))\subset (SO(2) \bowtie \Z_2)$ whence, by
(\ref{eq:12.17dbtntstb}),
\begin{eqnarray}
&& \hspace{-1cm}
N(SO(2),SO(2) \bowtie \Z_2) \subset (SO(2) \bowtie \Z_2) \; .
\label{eq:12.17dbtntcaa}
\end{eqnarray}
We conclude from (\ref{eq:12.17dbtntca}) and (\ref{eq:12.17dbtntcaa}) that
$N(SO(2),SO(2) \bowtie \Z_2) = SO(2) \bowtie \Z_2$ whence,
by (\ref{eq:12.17dbtnta23stsaa}),
\begin{eqnarray}
&& \hspace{-1cm}
N(Iso(\R^3,l_{v};S_\lambda),Iso(E_{t},l_{t};M))
= SO(2) \bowtie \Z_2 \; ,
\label{eq:12.17dbtntoan}
\end{eqnarray}
so that, by Lemma \ref{L10.5}a,
\begin{eqnarray}
&&  \hspace{-1cm}
B(\R^3,l_{v},E_{t},l_{t};S_\lambda,M) = \lbrace
\hat{\beta}[\R^3,l_{v},E_{t},l_{t};S_\lambda,M,r_0]:
r_0\in SO(2) \bowtie \Z_2)
\rbrace \; .
\label{eq:10.83atan}
\end{eqnarray}
We will now see that $B(\R^3,l_{v},E_{t},l_{t};S_\lambda,M)$
is a singleton.
In fact if $r_0\in(SO(2) \bowtie \Z_2)$ and $r\in SO(3)$ then, by
(\ref{eq:10.83}),
\begin{eqnarray}
&&  \hspace{-1cm}
\hat{\beta}[\R^3,l_{v},E_{t},l_{t};S_\lambda,M,r_0]
(l_{v}(r;S_\lambda))
=l_{t}(r r_0^t;M)=l_{t}(r;l_{t}(r_0^t;M))
\nonumber\\
&&\quad
=l_{t}(r;M)=\hat{\beta}[\R^3,l_{v},E_{t},l_{t};S_\lambda,M,
I_{3\times 3}](l_{v}(r;S_\lambda)) \; ,
\label{eq:12.17dbtntpan}
\end{eqnarray}
whence, by (\ref{eq:10.83atan}),
\begin{eqnarray}
&&  \hspace{-1cm}
B(\R^3,l_{v},E_{t},l_{t};S_\lambda,M) = \lbrace
\hat{\beta}[\R^3,l_{v},E_{t},l_{t};S_\lambda,M,I_{3\times 3}]
\rbrace \; ,
\label{eq:10.83ataan}
\end{eqnarray}
where in the third equality of (\ref{eq:12.17dbtntpan}) we used 
the second case from (\ref{eq:12.17bnncbtt}) where
$SO(2) \bowtie \Z_2=Iso(E_{t},l_{t};M)$.

We now apply the DT to the current case.
\setcounter{theorem}{10}
\begin{theorem} \label{T10.10}
Let $(j,A)\in{\cal SOS}(d,j)$. Let $\lambda\in(0,\infty)$ and let
$M\in R_{t}$ have $\#(M)=2$, i.e.,
$M=\Di(y,y,-2y)$ where $0\neq y\in\R$. 
Then $B(\R^3,l_{v},E_{t},l_{t};S_\lambda,M)$ 
is given by (\ref{eq:10.83ataan}) and 
for every $S$ in its domain
its only element,
$\hat{\beta}[\R^3,l_{v},E_{t},l_{t};S_\lambda,M,I_{3\times 3}]$,
satisfies
\begin{eqnarray}
&&  \hspace{-1cm}
\hat{\beta}[\R^3,l_{v},E_{t},l_{t};S_\lambda,M,I_{3\times 3}](S)
= y I_{3\times 3} - \frac{3y}{\lambda^2} SS^t \; .
\label{eq:10.175na}
\end{eqnarray}
Let $f\in{\cal C}(\Td,\R^3)$ take values only in the 
$(\R^3,l_{v})$-orbit of $S_\lambda$.
Let the function $g\in{\cal C}(\Td,E_{t})$ be defined by $g(z):=
\hat{\beta}[\R^3,l_{v},E_{t},l_{t};S_\lambda,M,I_{3\times 3}](f(z))$.
Then
\begin{eqnarray}
&&  \hspace{-1cm}
 g(z) = y I_{3\times 3} - \frac{3y}{\lambda^2} f(z)f^t(z) \; .
\label{eq:10.180x0n}
\end{eqnarray}
Let us apply the $1$-turn field map, i.e., let 
the function $f'\in{\cal C}(\Td,\R^3)$ be defined by 
$f':=\tilde{\cal P}[j,A](f)$ and the function 
$g'\in{\cal C}(\Td,E_{t})$ be defined by
$g':=\tilde{\cal P}[E_{t},l_{t},j,A](g)$.
Then
\begin{eqnarray}
&&  \hspace{-1cm}
 g'(z) = y I_{3\times 3} - \frac{3y}{\lambda^2} f'(z)f'^t(z) \; ,
\label{eq:10.180x0na}
\end{eqnarray}
and $g'(z)=
\hat{\beta}[\R^3,l_{v},E_{t},l_{t};S_\lambda,M,I_{3\times 3}](f'(z))$.\\
\noindent Remark: If
$f=f'$ then $g=g'$. In other words if $f$ is an invariant polarization field of
$(j,A)$ then $g$ is an invariant $(E_{t},l_{t})$-field of $(j,A)$.
In particular, if $\lambda=1$ and $f$ is an ISF of
$(j,A)$ then $g$ is an invariant $(E_{t},l_{t})$-field of $(j,A)$.
\end{theorem}

\noindent{\em Proof of Theorem \ref{T10.10}:} 
Using (\ref{eq:10.83}) and (\ref{eq:10.83ataan}) and Remark 21
we get (\ref{eq:10.175na}). Moreover (\ref{eq:10.180x0n}) follows from
(\ref{eq:10.175na}). The remaining claims follow from Theorem \ref{T10.6}a.
\hfill $\Box$

\vspace{2mm}

In the special case $\lambda=1,y=1/\sqrt{6}$, 
(\ref{eq:10.180x0n}) is the expression for the invariant tensor field  
in \cite{BV2}. So we have independently reconstructed the invariant tensor 
field of \cite{BV2} by using the DT!
\subsubsection{The $2$-snake model}
\label{10.4b.2}
In this section we consider a model describing the spin-orbit system
of a flat storage ring which has two thin-lens Siberian Snakes with
mutually perpendicular axes of spin rotation placed at $\theta= 0$ and
$\theta= \pi$. With this layout, the spin tune, $\nu_0$, on the design
orbit, of the ring is $1/2$. Here we are interested in the situation
where, in the absence of snakes, the spin motion is dominated by the
effect of a single harmonic in the Fourier expansion of the radial
component of the ${\bf {\Omega}}(\theta,J,\phi(\theta))$, mentioned in
the Introduction, and due to vertical betatron motion. This case is
often called the ``single resonance model''. The combination of
the single resonance model and two snakes considered in this
section has been studied
extensively. See for example \cite{BV1, mv2000} and the references
therein. The interest in this model stems from the effect on the
polarization of the so-called ``snake resonances''. These occur at
vertical betatron tunes of $1/2, 1/6, 5/6, 1/10, 3/10 \ldots $  Note
that the term snake resonance is a misnomer since it does not refer to
the proper definition of spin-orbit resonance given in
(\ref{eq:6.20}). Our main interest here is in the fact that at snake
resonance, there is no ISF of the kind that we define in this paper.
We have already mentioned this situation in Section \ref{4.2}. For
further background material see \cite{BV1}.

Here we focus on the simplest case, namely that with  vertical 
betatron tune, $\omega = 1/2$,
and we denote the resulting spin-orbit system by $({\cal P}_{1/2},A_{2S})$.
Of course a real bunch is not stable at $\omega = 1/2$ but this does not
play a role in the present section.
We prove two claims.
We first show that $({\cal P}_{1/2},A_{2S})$ has a $2$-turn ISF, defined
below, and a normalized, piecewise continuous
solution of the $(\R^3,l_{v})$-stationarity equation but no ISF.
Secondly we apply the DT via
Theorem \ref{T10.10} to construct, out of the two $2$-turn ISF's,
an invariant nonzero $(E_{t},l_{t})$-field of $({\cal P}_{1/2},A_{2S})$.

We first define $({\cal P}_{1/2},A_{2S})$. For this we define 
the function
$A_{2S}\in {\cal C}(\T^1,SO(3))$, for $\epsilon\in(\R\setminus\Z)$ with \cite{BV1, mv2000} by
\begin{eqnarray}
&& A_{2S}(\phi+ \tilde{\Z}):=
\left( \begin{array}{ccc} 1-2c^2(\phi) & 2b(\phi) c(\phi)  
&  2a(\phi) c (\phi)  \\
2b(\phi) c(\phi)  &  1-2b^2(\phi)  & -2a(\phi) b(\phi) \\
 -2a(\phi)c(\phi)  & 2a(\phi)b(\phi)  &  2a^2(\phi) -1
\end{array}\right) \; ,
\label{eq:10.226}
\end{eqnarray}
where the functions $a,b,c\in{\cal C}(\R,\R)$ are defined by
\begin{eqnarray}
&& a(\phi):=-2\sin^2(\pi\epsilon/2)\sin(\phi)\cos(\phi) \; , \quad
b(\phi):= 
-2\sin(\pi\epsilon/2)\cos(\pi\epsilon/2)\cos(\phi) \; , \nonumber\\
&& c(\phi):=2\sin^2(\pi\epsilon/2)\cos^2(\phi)-1 \; .
\label{eq:10.227}
\end{eqnarray}
Note that
\begin{eqnarray}
&& a^2 + b^2 + c^2 = 1 \; ,
\label{eq:10.228}
\end{eqnarray}
and that we exclude $\epsilon$ from being an integer because in that
case $({\cal P}_{1/2},A_{2S})$ would have an ISF \cite{He1}.
Note also that, by the Torus Lemma, Lemma \ref{L0}, $A_{2S}$ is continuous
since the continuous functions $a,b,c$ are $2\pi$-periodic.

Since ${\cal P}_{1/2}^2=id_\Td$ we will
prove both of our claims by computing the so-called
$2$-turn ISF's of $({\cal P}_{1/2},A_{2S})$. We call
an invariant $2$-turn $(\R^3,l_v)$-field of $({\cal P}_{1/2},A_{2S})$
a ``$2$-turn ISF of $({\cal P}_{1/2},A_{2S})$'' if it is normalized.
Thus, noting that with $\omega = 1/2$, a particle returns to the same 
$z$ over two turns,
an $h\in{\cal C}(\T^1,\R^3)$ is a 
$2$-turn ISF of $({\cal P}_{1/2},A_{2S})$ iff 
\begin{eqnarray}
&& \tilde{\cal P}[{\cal P}_{1/2},A_{2S}]^2(h)=h \; ,
\label{eq:10.229} \\
&& |h|= 1 \; .
\label{eq:10.230} 
\end{eqnarray}
In fact we will see that $({\cal P}_{1/2},A_{2S})$ has just two
$2$-turn ISF's namely $h=k$ and
$h=-k$ where $k$ will be defined below. It is clear that every ISF is
a $2$-turn ISF and we
will show that in fact neither $k$ nor $-k$ is an ISF of
$({\cal P}_{1/2},A_{2S})$ which implies that no ISF exists.
We will then apply 
Theorem \ref{T10.10} to $h=k$ and will thereby obtain
an invariant $(E_{t},l_{t})$-field of $({\cal P}_{1/2},A_{2S})$.
The case $h=-k$ will result in the same invariant $(E_{t},l_{t})$-field.

To begin our computations we first rewrite
(\ref{eq:10.229}), by using (\ref{eq:xx13.2}) and (\ref{eq:xx13.2nn}), into
\begin{eqnarray}
&& h(z) = \Psi[{\cal P}_{1/2},A_{2S}](2;z)h(z) \; .
\label{eq:10.232}
\end{eqnarray}
Thus a $h\in {\cal C}(\T^1,\R^3)$ is a $2$-turn ISF of
$({\cal P}_{1/2},A_{2S})$ iff
(\ref{eq:10.230}),(\ref{eq:10.232}) are fullfilled.
Note that, very conveniently, (\ref{eq:10.232}) is an eigenvalue
problem for $h(z)$ and it is here that we used that $\omega=1/2$.
To obtain the $2$--turn spin transfer matrix function
in (\ref{eq:10.232})
we first conclude from (\ref{eq:10.226}) and (\ref{eq:10.227}) that
\begin{eqnarray}
&& A_{2S}((\phi+\pi)+ \tilde{\Z})
=
\left( \begin{array}{ccc} 1-2c^2(\phi) & -2b(\phi)c(\phi) &  
2a(\phi)c(\phi)  \\
-2b(\phi)c(\phi) &  1-2b^2(\phi) & 2a(\phi)b(\phi) \\
 -2a(\phi)c(\phi) & -2a(\phi)b(\phi) &  2a^2(\phi)-1
\end{array}\right)
\; .
\label{eq:11.2}
\end{eqnarray}
We also conclude from (\ref{eq:3.5}), (\ref{eq:10.226}) and (\ref{eq:11.2})
that the $2$--turn spin transfer matrix function reads as
\begin{eqnarray}
&& \hspace{-1cm}
 \Psi[{\cal P}_{1/2},A_{2S}](2;\phi+ \tilde{\Z})=
A_{2S}((\phi+\pi)+ \tilde{\Z})A_{2S}(\phi+ \tilde{\Z})
\nonumber\\
&& \hspace{-0.5cm}
=\left( \begin{array}{ccc} 1-8c^2(\phi)+8c^4(\phi) & 4b(\phi)c(\phi)
(1-2c^2(\phi))  & 4a(\phi)c(\phi)(1-2c^2(\phi)) \\
-4b(\phi)c(\phi)(1-2c^2(\phi))  &   1-8b^2(\phi)c^2(\phi)  
& -8a(\phi)b(\phi)c^2(\phi) \\
-4a(\phi)c(\phi)(1-2c^2(\phi)) & -8a(\phi)b(\phi)c^2(\phi) & 
1-8a^2(\phi) c^2(\phi) \end{array}\right)
\; .\qquad
\label{eq:11.3}
\end{eqnarray}
Since $\epsilon$ is not an integer, 
$|\sin(\pi\epsilon/2)|$ equals neither $0$ or $1$, and so we define the 
$2\pi$-periodic function
$K\in {\cal C}({\mathbb R},\R^3)$ by
\begin{eqnarray}
&& \hspace{-1cm}
K(\phi):=
\frac{\cos(\pi\epsilon/2)}{|\cos(\pi\epsilon/2)|
  \sqrt{1-\sin^2(\pi\epsilon/2)\cos^2(\phi)}}
\biggl(0, \sin(\pi\epsilon/2)\sin(\phi),-\cos(\pi\epsilon/2)\biggr)  \; .\qquad
\label{eq:11.20}
\end{eqnarray}
By the Torus Lemma, Lemma \ref{L0}, a unique function 
$k\in {\cal C}(\T^1,\R^3)$ exists such that
\begin{eqnarray}
&& k = K\circ \pi_1 \; .
\label{eq:10.240}
\end{eqnarray}
%
It is easy to show that 
(\ref{eq:10.230}) and (\ref{eq:10.232}) are fullfilled for $h=k$, i.e.,
\begin{eqnarray}
&& k(z)=\Psi[{\cal P}_{1/2},A_{2S}](2;z)k(z) \; , 
\label{eq:11.165} \\
&& |k(z)|=1
\; .
\label{eq:11.165a}
\end{eqnarray}
Thus indeed $k$ and $-k$ are $2$-turn ISF's of
$({\cal P}_{1/2},A_{2S})$.
Let $h\in {\cal C}(\T^1,\R^3)$
be an arbitrary  $2$-turn ISF of
$({\cal P}_{1/2},A_{2S})$, i.e., let $h$ satisfy
(\ref{eq:10.230}) and (\ref{eq:10.232}).

To show that either $h=k$ or $h=-k$
let $R\neq I_{3\times 3}$ be a matrix in $SO(3)$. Then $R$ has a real eigenvector
$v\in\R^3$ with eigenvalue $1$ and such that $|v|=1$
whence $r\in SO(3)$ exists such that $v=r(0,0,1)^t$.
Thus $r^t Rr(0,0,1)^t=(0,0,1)^t$ whence, by the $SO(2)$-Lemma,
$r^t Rr\in SO(2)$  
so that a $\nu\in[0,1)$ exists such that
$R=r\exp(2\pi\nu{\cal J})r^t$. This implies, since $R\neq I_{3\times 3}$,
that $\nu\neq 0$. Thus if $w,w'\in\R^3$
are real eigenvectors of $r^t Rr$ with the eigenvalue $1$ and $|w|=|w'|=1$
then $|w\cdot w'|=1$ whence 
if $v,v'\in\R^3$
are real eigenvectors of $R$ with the eigenvalue $1$ and $|v|=|v'|=1$
then $|v\cdot v'|=1$.

Defining the set 
\begin{eqnarray}
&& \hspace{-1cm}
M:=\lbrace z\in \T^1:
\Psi[{\cal P}_{1/2},A_{2S}](2;z)=I_{3\times 3}\rbrace \; ,
\label{eq:11.20an}
\end{eqnarray}
we observe that, if $z\in(\T^1\setminus M)$, then
$\Psi[{\cal P}_{1/2},A_{2S}](2;z)\neq I_{3\times 3}$.
Thus, and since
by (\ref{eq:10.230}), (\ref{eq:10.232}), (\ref{eq:11.165}) and
(\ref{eq:11.165a}),
$h(z),k(z)$ are real eigenvectors of 
$\Psi[{\cal P}_{1/2},A_{2S}](2;z)$ with eigenvalue $1$
and $|h(z)|=|k(z)|=1$ we conclude that, if $z\in(\T^1\setminus M)$,
then $\lambda(z)=1$ where the function $\lambda:\T^1\rightarrow\R$ is defined
by $\lambda(z):=|h(z)\cdot k(z)|$. 
To show that $\lambda(z)=1$ for all $z\in\T^1$ 
we only have to show that $\lambda$ is a constant function.
We thus compute, by (\ref{eq:10.227}) and (\ref{eq:11.3}),
\begin{eqnarray}
&& \hspace{-1cm}
M=\lbrace \phi+ \tilde{\Z}:\phi\in {\mathbb R},c(\phi)(c^2(\phi)-1)=0 \rbrace
=\lbrace \phi+ \tilde{\Z}:\phi\in {\mathbb R},
\cos^2(\phi)=\frac{1}{2\sin^2(\pi\epsilon/2)}\rbrace \; ,
\nonumber\\
\label{eq:11.20a}
\end{eqnarray}
whence $M$ consists of only finitely many points.
Since $\lambda(z)=1$ on $\T^1\setminus M$ and since 
$M$ has only finitely many points we conclude that $\lambda$ is
a continuous function with only finitely many values. 
Since $\T^1$ is path-connected and $\lambda$ is
continuous we use the same argument as in the proof of Theorem 
\ref{T7.1}b and conclude that the range of $\lambda$ is an interval
whence
$\lambda$ is constant so that $\lambda(z)=|h(z)\cdot k(z)|=1$ holds 
for {\it every} $z\in\T^1$.
Thus, and since $|h(z)|=|k(z)|=1$,
either $h=k$ or $h=-k$. So we have shown that the only $2$-turn ISF's
are $h=k$ and $h=-k$. 

To show that neither $k$ nor $-k$ is an ISF we
compute, by (\ref{eq:10.226}) and (\ref{eq:11.20}),
\begin{eqnarray}
&& A_{2S}(\phi+ \tilde{\Z})K(\phi) =-K(\phi+\pi) \; ,
\label{eq:10.250}
\end{eqnarray}
whence, by (\ref{eq:s2.10t}) and (\ref{eq:10.240}),
$A_{2S}(z)k(z) = - k({\cal P}_{1/2}(z))$ so that, by
(\ref{eq:xx13.2n}),
\begin{eqnarray}
&& \tilde{\cal P}[{\cal P}_{1/2},A_{2S}](k)=-k \; ,
\label{eq:10.251} 
\end{eqnarray}
which implies, by Definition \ref{D6.1}, that
$k$ is not an ISF of $({\cal P}_{1/2},A_{2S})$.
Thus $-k$ is not an ISF of $({\cal P}_{1/2},A_{2S})$ either
which completes the proof that the two only 
$2$-turn ISF's of $({\cal P}_{1/2},A_{2S})$
are not ISF's of  $({\cal P}_{1/2},A_{2S})$.
We conclude, by our earlier remarks,
that $({\cal P}_{1/2},A_{2S})$ has no ISF. This proves the first claim.

\vspace{3mm}
\noindent{\bf Remark:}
\begin{itemize}
\item[(26)] 
While $({\cal P}_{1/2},A_{2S})$ has no ISF, it is easy to construct
a normalized, piecewise continuous
solution of the $(\R^3,l_{v})$-stationarity equation (see also \cite{BV2}).
In fact defining $\tilde{K}:\R\rightarrow\R^3$ by
\begin{eqnarray}
&& \tilde{K}(\phi):=\left\{ \begin{array}{ll}  K(\phi)
& \;\; {\rm if\;} \phi\in\bigcup_{n\in\Z}[2\pi n,2\pi n+\pi) 
\\
 -K(\phi)
& \;\; {\rm if\;} \phi\in\bigcup_{n\in\Z}[2\pi n+\pi,2\pi n+2\pi) \; , \end{array}
                  \right.
\label{eq:11.20aa} 
\end{eqnarray}
we observe, by the Torus Lemma in Section \ref{2.1b}, that a unique function
$\tilde{k}:\T^1\rightarrow\R^3$ 
exists such that $\tilde{k} = \tilde{K}\circ \pi_1$.
It is a simple exercise to show that $\tilde{k}$ is a normalized
piecewise continuous
solution of the $(\R^3,l_{v})$-stationarity equation of 
$({\cal P}_{1/2},A_{2S})$.
Of course, $\tilde{k}$ is not an ISF of $({\cal P}_{1/2},A_{2S})$ since 
$({\cal P}_{1/2},A_{2S})$ has no ISF. In fact it is an easy exercise to show,
by (\ref{eq:11.20}) and (\ref{eq:11.20aa}), that 
$\tilde{k}$ is discontinuous at $z=\pi_1(0)$ and $z=\pi_1(\pi)$.
This is an example of a consequence of a lack of topological
transitivity of $j$ mentioned
just after Theorem \ref{T09t0}.

As mentioned at the end of Section \ref{10.2.2} since $A,j,l$
are continuous we require that invariant fields be continuous.
However this requirement is a matter of choice. In fact if one
would impose the weaker condition of Borel measurability then
$\tilde{k}$ would be an ISF.
In fact, as mentioned in Section \ref{4.2}, the requirement of continuity
was relaxed in \cite{BV1}.
\hfill $\Box$
\end{itemize}

We now apply Theorem \ref{T10.10} in the case $\lambda=1$ for the
function $f:=k$. In fact, in the notation of Theorem \ref{T10.10},
it follows from (\ref{eq:10.251}) that
\begin{eqnarray}
&& f'=\tilde{\cal P}[{\cal P}_{1/2},A_{2S}](f)
=\tilde{\cal P}[{\cal P}_{1/2},A_{2S}](k)= - k \; ,
\label{eq:10.251a} 
\end{eqnarray}
Using the notation of Theorem \ref{T10.10},
the functions $g,g'\in{\cal C}(\Td,E_{t})$ are given by
\begin{eqnarray}
&&  \hspace{-1cm}
 g(z) = \hat{\beta}[\R^3,l_{v},E_{t},l_{t};S_1,M,I_{3\times 3}](f(z))
\nonumber\\
&& = \hat{\beta}[\R^3,l_{v},E_{t},l_{t};S_1,M,I_{3\times 3}](k(z))
= y I_{3\times 3} - 3y k(z)k^t(z) \; ,
\\
&&  \hspace{-1cm}
 g'(z) = \hat{\beta}[\R^3,l_{v},E_{t},l_{t};S_1,M,I_{3\times 3}](f'(z))
\nonumber\\
&&= \hat{\beta}[\R^3,l_{v},E_{t},l_{t};S_1,M,I_{3\times 3}](-k(z))
= y I_{3\times 3} - 3y k(z)k^t(z) \; ,
\end{eqnarray}
whence $g=g'$ so that $g$ 
is an invariant $(E_{t},l_{t})$-field of $({\cal P}_{1/2},A_{2S})$.
The same holds for $f=-k$ since, by
repeating the above construction of $g$ and replacing $f=k$ by
$f=-k$ we get the same $g$.
This completes the proof of the second claim.

So although $({\cal P}_{1/2},A_{2S})$ has no ISF, there is a nontrivial
invariant tensor field. This is also expected from \cite{BV1} by
noting that the discontinuities at $z=\pi_1(0)$ and $z=\pi_1(\pi)$
involve a simple change of sign. Then since the invariant tensor field
is quadratic in $\tilde{k}$, those discontinuities do not cause
discontinuities in the invariant tensor field.
\subsection{Applying the ToA to density 
matrix functions}
\label{10.5}
\subsubsection{Spin-$1/2$ particles. Applying the ToA to 
$(E_{dens}^{1/2},l_{dens}^{1/2})$}
\label{10.5.1}
In this section we introduce the $SO(3)$-space $(E_{dens}^{1/2},l_{dens}^{1/2})$ 
to enable the use of the ToA for the
study of the spin-$1/2$ density matrix function 
employed for describing
polarized beams of spin-$1/2$
particles \cite{BV2}.
As in Sections \ref{10.3}-\ref{10.4b} the focus
is on the field motion.

We define 
\begin{eqnarray} 
&&  E_{dens}^{1/2}:= \lbrace R \in\C^{2\times 2}:R^\dagger=R,Tr[R]=1\rbrace \; ,
\label{eq:8.6.10s}
\end{eqnarray}
where $R^\dagger$ denotes the hermitian conjugate of the matrix $R$
and we equip $E_{dens}^{1/2}$ with the subspace topology from
$\C^{2\times 2}$. Thus, and since $\C^{2\times 2}$ with its natural topology is a 
Hausdorff space, $E_{dens}^{1/2}$ is a Hausdorff space, too.
Following a standard parametrization we define
the function $\beta_{dens}^{1/2}:\R^3\rightarrow E_{dens}^{1/2}$ 
for $S\in\R^3$ by
\begin{eqnarray} 
&&
  \beta_{dens}^{1/2}(S):= \frac{1}{2} \biggl( I_{2\times 2} 
+ \sum_{i=1}^3\; S_i\sigma_i\biggr) \; ,
\label{eq:8.6.15s}
\end{eqnarray}
where the Pauli matrices $\sigma_1,\sigma_2,\sigma_3$ are
defined by
\begin{eqnarray}
&&  \hspace{-15mm} 
\sigma_1:= \left( \begin{array}{cc} 0 & -i \\
 i & 0 \end{array}\right) \; , \quad
\sigma_2:= \left( \begin{array}{cc} 1 & 0 \\
 0 & -1 \end{array}\right) \; , \quad
\sigma_3:= \left( \begin{array}{cc} 0 & 1 \\
 1 & 0 \end{array}\right) \; ,
\label{eq:8.6.20s}
\end{eqnarray}
and where $S_i$ denotes the $i$-th component of $S$. 
Since every $\sigma_k$ is hermitian, i.e.,
$\sigma_k^\dagger=\sigma_k$, by (\ref{eq:8.6.15s}), $\beta_{dens}^{1/2}(S)$ 
is hermitian too.
Moreover since, by (\ref{eq:8.6.20s}) and for $i,k=1,2,3$,
\begin{eqnarray}
&& Tr[\sigma_i]= 0 \; , \quad  Tr[\sigma_i\sigma_k]= 2\delta_{ik} \; ,
\label{eq:8.6.25s}
\end{eqnarray}
we find from (\ref{eq:8.6.15s}) that $Tr[\beta_{dens}^{1/2}(S)]=1$.
So $\beta_{dens}^{1/2}(S)$ is a hermitian matrix of
trace $1$, i.e., indeed $\beta_{dens}^{1/2}$ is a function into 
$E_{dens}^{1/2}$.
To show that $\beta_{dens}^{1/2}$ is a homeomorphism we first note, by
(\ref{eq:8.6.20s}), that if $R\in\C^{2\times 2}$ is hermitian then
real numbers $S_0',S_1',S_2',S_3'$ exist such that 
$R=S_0'I_{2\times 2} + \sum_{i=1}^3\; S_i'\sigma_i$ whence, by 
(\ref{eq:8.6.10s}) and (\ref{eq:8.6.25s}) and
if $R\in E_{dens}^{1/2}$, we get $S_0'=1/2$ so 
that $R = \beta_{dens}^{1/2}(S)$ for some $S\in\R^3$.
Thus the function $\beta_{dens}^{1/2}$ is onto
$E_{dens}^{1/2}$, i.e.,
\begin{eqnarray} 
&&  E_{dens}^{1/2} = \lbrace \beta_{dens}^{1/2}(S):S\in\R^3\rbrace \; .
\label{eq:8.6.30s}
\end{eqnarray}
It also follows from (\ref{eq:8.6.15s}) and (\ref{eq:8.6.25s})
that, for $S\in\R^3$ and $i=1,2,3$,
\begin{eqnarray}
&& S_i = Tr[\sigma_i \beta_{dens}^{1/2}(S)] \; .
\label{eq:8.6.27s}
\end{eqnarray}
Thus $S$ is uniquely 
determined by $\beta_{dens}^{1/2}(S)$ whence $\beta_{dens}^{1/2}$ is one-one so that
we conclude that $\beta_{dens}^{1/2}$ is a bijection.
Since $\beta_{dens}^{1/2}$ is a bijection it follows from 
(\ref{eq:8.6.15s}) that
its inverse, $(\beta_{dens}^{1/2})^{-1}$, is defined for $R\in E_{dens}^{1/2}$ by
\begin{eqnarray} 
&&  (\beta_{dens}^{1/2})^{-1}(R):= S \; , \quad  S_i:= Tr[\sigma_iR] \; ,
\label{eq:8.6.28s}
\end{eqnarray}
where $S_i$ denotes the $i$-th component of $S$. 
Moreover, by (\ref{eq:8.6.15s}) and (\ref{eq:8.6.28s}),
both $\beta_{dens}^{1/2}$ and $(\beta_{dens}^{1/2})^{-1}$ are 
continuous functions
whence $\beta_{dens}^{1/2}\in \Homeo(\R^3,E_{dens}^{1/2})$, a fact which plays
a key role in this section.

We now define the function $l_{dens}^{1/2}:SO(3)\times E_{dens}^{1/2}\rightarrow 
E_{dens}^{1/2}$ by 
\begin{eqnarray}
&&  l_{dens}^{1/2}(r;R):=\beta_{dens}^{1/2}
\biggl(l_{v}(r;(\beta_{dens}^{1/2})^{-1}(R))\biggr) \; ,
\label{eq:8.6.31as}
\end{eqnarray}
i.e.,
\begin{eqnarray}
&&  l_{dens}^{1/2}(r;\beta_{dens}^{1/2}(S)):=\beta_{dens}^{1/2}(l_{v}(r;S))
=\beta_{dens}^{1/2}(rS) \; ,
\label{eq:8.6.31s}
\end{eqnarray}
with $r\in SO(3),R\in E_{dens}^{1/2}$ and $S\in\R^3$. 
Since $(\R^3,l_{v})$ is an $SO(3)$-space and\\
$\beta_{dens}^{1/2}\in \Homeo(\R^3,E_{dens}^{1/2})$ it follows from
(\ref{eq:8.6.31as}) that 
$(E_{dens}^{1/2},l_{dens}^{1/2})$ is an
$SO(3)$-space and that $\beta_{dens}^{1/2}$ is an isomorphism from the
$SO(3)$-space $(\R^3,l_{v})$ to the $SO(3)$-space 
$(E_{dens}^{1/2},l_{dens}^{1/2})$. 

Due to (\ref{eq:10.15}), the $1$-turn particle-spin map 
${\cal P}[E_{dens}^{1/2},l_{dens}^{1/2},j,A]$ is given by
\begin{eqnarray}
&&  \hspace{-1cm}
{\cal P}[E_{dens}^{1/2},l_{dens}^{1/2},j,A](z,R)
=\left( \begin{array}{c} j(z) \\
l_{dens}^{1/2}(A(z);R)\end{array}\right) \; ,
\label{eq:8.6.32as}
\end{eqnarray}
where $z\in\Td,R\in E_{dens}^{1/2}$. 
Because the $SO(3)$-spaces $(\R^3,l_{v})$ and $(E_{dens}^{1/2},l_{dens}^{1/2})$
are isomorphic, 
we get easy insight into ${\cal P}[E_{dens}^{1/2},l_{dens}^{1/2},j,A]$ by using the
Second ToA Transformation Rule. In fact recalling Section \ref{10.2.4}
we find via (\ref{eq:10.55nn}) that
${\cal P}[E_{dens}^{1/2},l_{dens}^{1/2},j,A]\circ \beta_{dens,tot}^{1/2} 
= \beta_{dens,tot}^{1/2} \circ {\cal P}[\R^3,l_{v},j,A]$
whence and since, by Remark 1, ${\cal P}[\R^3,l_{v},j,A]={\cal P}[j,A]$
we get
\begin{eqnarray}
&&  \hspace{-1cm}
{\cal P}[E_{dens}^{1/2},l_{dens}^{1/2},j,A]\circ \beta_{dens,tot}^{1/2} 
= \beta_{dens,tot}^{1/2} \circ {\cal P}[j,A] \; ,
\label{eq:8.6.32cs}
\end{eqnarray}
i.e., 
\begin{eqnarray}
&&  \hspace{-1cm}
{\cal P}[E_{dens}^{1/2},l_{dens}^{1/2},j,A](z,\beta_{dens}^{1/2}(S))
= {\cal P}[E_{dens}^{1/2},l_{dens}^{1/2},j,A](\beta_{dens,tot}^{1/2}(z,S))
\nonumber\\
&&=\left( \begin{array}{c} j(z) \\
\beta_{dens}^{1/2}(A(z)S)\end{array}\right) 
\; ,
\label{eq:8.6.32s}
\end{eqnarray}
where $z\in\Td,S\in\R^3$ and where the function $\beta_{dens,tot}^{1/2}
\in \Homeo(\Td\times \R^3,\Td\times  E_{dens}^{1/2})$ is 
defined by $\beta_{dens,tot}^{1/2}(z,S):=
(z,\beta_{dens}^{1/2}(S))$. 

We now come to our main focus, the fields, which in the case 
$(E,l)=(E_{dens}^{1/2},l_{dens}^{1/2})$ are also called  spin-$1/2$ 
density matrix functions and which are functions
$\rho:\Td\rightarrow E_{dens}^{1/2}$ whence, by (\ref{eq:8.6.30s}),
$\rho=\beta_{dens}^{1/2}\circ f$ where the function
$f:\Td\rightarrow \R^3$ is 
defined by $f(z):=(\beta_{dens}^{1/2})^{-1}(\rho(z))$. Thus, using
(\ref{eq:8.6.15s}) and (\ref{eq:8.6.28s}), we get
\begin{eqnarray}
&& f_i(z) = Tr[\rho(z)\sigma_i] \; , 
\label{eq:8.6.34s} \\
&&
\rho(z) =\beta_{dens}^{1/2}(f(z))  = \frac{1}{2} \biggl( I_{2\times 2} 
+ \sum_{i=1}^3\; f_i(z)\sigma_i\biggr) \; ,
\label{eq:8.6.15as}
\end{eqnarray}
where $f_i(z)$ denotes the $i$-th component of $f(z)$.
Of course since $\beta_{dens}^{1/2}\in \Homeo(\R^3,E_{dens}^{1/2})$,
$\rho$ is continuous iff $f$ is continuous.
We call an invariant $(E_{dens}^{1/2},l_{dens}^{1/2})$-field 
an ``equilibrium  spin-$1/2$ density matrix function''.
Due to (\ref{eq:10.17}), 
the $1$-turn field map $\tilde{\cal P}[E_{dens}^{1/2},l_{dens}^{1/2},j,A]$ 
is given by
\begin{eqnarray}
&& \tilde{\cal P}[E_{dens}^{1/2},l_{dens}^{1/2},j,A](\rho)=
l_{dens}^{1/2}\biggl(A\circ j^{-1};\rho\circ j^{-1}\biggr) \; .
\label{eq:8.6.35s}
\end{eqnarray}
Because the $SO(3)$-spaces $(\R^3,l_{v})$ and $(E_{dens}^{1/2},l_{dens}^{1/2})$
are isomorphic, 
we get easy insight into $\tilde{\cal P}[E_{dens}^{1/2},l_{dens}^{1/2},j,A]$ 
by using once again the
Second ToA Transformation Rule. In fact recalling Section \ref{10.2.4}
we find via (\ref{eq:10.55ann}) that
$\tilde{\cal P}[E_{dens}^{1/2},l_{dens}^{1/2},j,A]\circ \tilde\beta_{dens}^{1/2} 
= \tilde{\beta}_{dens}^{1/2} \circ \tilde{\cal P}[\R^3,l_{v},j,A]$
whence and since, by Remark 1, 
$\tilde{\cal P}[\R^3,l_{v},j,A]=\tilde{\cal P}[j,A]$ we get
\begin{eqnarray}
&&  \hspace{-1cm}
\tilde{\cal P}[E_{dens}^{1/2},l_{dens}^{1/2},j,A]\circ \tilde{\beta}_{dens}^{1/2} 
= \tilde{\beta}_{dens,tot}^{1/2} \circ \tilde{\cal P}[j,A] \; ,
\label{eq:8.6.32csn}
\end{eqnarray}
where the function $\tilde{\beta}_{dens}^{1/2}:{\cal C}(\Td,\R^3)
\rightarrow{\cal C}(\Td,E_{dens}^{1/2})$ is defined,  
for $f\in{\cal C}(\Td,\R^3)$, by $\tilde{\beta}_{dens}^{1/2}(f):=
\beta_{dens}^{1/2}\circ f$. It thus follows by Remark 16 
that an $f\in{\cal C}(\Td,\R^3)$
is an invariant polarization field of $(j,A)$ iff
$\beta_{dens}^{1/2} \circ f$
is an invariant $(E_{dens}^{1/2},l_{dens}^{1/2})$-field of $(j,A)$.

We thus have proved:
\setcounter{theorem}{11}
\begin{theorem} \label{T10.8cxs}
The function $\beta_{dens}^{1/2}$ belongs to $\Homeo(\R^3,E_{dens}^{1/2})$.
Let $\rho:\Td\rightarrow E_{dens}^{1/2}$. Then a unique function
$f:\Td\rightarrow \R^3$ exists such that
$\rho=\beta_{dens}^{1/2}\circ f$, i.e.,
(\ref{eq:8.6.15as}) holds where $f_i(z)$ denotes the $i$-th component of $f(z)$.
Moreover $\rho$ is continuous iff $f$ is continuous.
Moreover, let $(j,A)\in{\cal SOS}(d,j)$.
Then $\rho$ is an equilibrium spin-$1/2$ density matrix function
of $(j,A)$, i.e.,
is an invariant $(E_{dens}^{1/2},l_{dens}^{1/2})$-field of $(j,A)$ iff
$f$ is an invariant polarization field of $(j,A)$.
\hfill $\Box$
\end{theorem}

Because of Theorem \ref{T10.8cxs} the study
of equilibrium spin-$1/2$ density matrix functions
effectively amounts to
the study of 
invariant polarization fields. See \cite{BV2} too. Since 
invariant polarization fields 
are studied in other parts of this work the remainder of this section can be
brief and so we 
leave the application of the NFT, DT etc. as an exercise to the reader and
conclude this section with Remarks 27 and 28.

Recalling that $E_{dens}^{1/2}$ is Hausdorff we can address topological
transitivity by applying Lemma \ref{L10.4} to 
$(E,l)=(E_{dens}^{1/2},l_{dens}^{1/2})$ in the following remark:

\vspace{3mm}
\noindent{\bf Remark:}
\begin{itemize}
\item[(27)] 
Recalling that
$\beta_{dens}^{1/2}$ is an isomorphism from $(\R^3,l_{v})$ to 
$(E_{dens}^{1/2},l_{dens}^{1/2})$ we note, by Definition \ref{D3c},
that $\beta_{dens}^{1/2}$ maps each $(\R^3,l_{v})$-orbit, i.e., each
sphere ${\mathbb S}^2_\lambda$ onto an 
$(E_{dens}^{1/2},l_{dens}^{1/2})$-orbit and that
$(\beta_{dens}^{1/2})^{-1}$ maps each 
$(E_{dens}^{1/2},l_{dens}^{1/2})$-orbit onto an  $(\R^3,l_{v})$-orbit.
Let $(j,A)\in{\cal SOS}(d,j)$
with $j$ topologically transitive and let $\rho$ 
be an equilibrium density matrix function of $(j,A)$. 
Since $E_{dens}^{1/2}$ is Hausdorff we can apply 
Lemma \ref{L10.4} and we conclude that
$\rho$ takes values in only one $(E_{dens}^{1/2},l_{dens}^{1/2})$-orbit whence
by the above $f:=(\beta_{dens}^{1/2})^{-1}\circ\rho$ takes values in only one
$(\R^3,l_{v})$-orbit. However this is no surprise since
from Theorem \ref{T10.8cxs} we know that 
$f$ is an invariant polarization field of $(j,A)$ whence, 
by applying Lemma \ref{L10.4} to $(E,l)=(\R^3,l_{v})$, we see once again 
that $f$ takes values in only one $(\R^3,l_{v})$-orbit.
\hfill $\Box$
\end{itemize}

The following remark sketches how one uses 
spin-$1/2$ density matrices for the statistical description of
a bunch of spin-$1/2$ particles:

\vspace{3mm}
\noindent{\bf Remark:}
\begin{itemize}
\item[(28)]  
One can describe a bunch of spin-$1/2$ particles
statistically by a function 
$\rho_{tot}:\Z\times\Td\times\Lambda\rightarrow\C^{2\times 2}$ 
of the form
$\rho_{tot}(n,z,J):=(1/2\pi)^d
\rho_{eq}(J)\rho_{spin}(n,z,J)$ where
$(1/2\pi)^d\rho_{eq}$ describes the equilibrium particle
distribution in the bunch and 
where $\rho_{spin}:\Z\times\Td\times\Lambda\rightarrow E_{dens}^{1/2}$ has the
property that each of the functions $\rho_{spin}(n,\cdot,J)$ 
moves as a spin-$1/2$ density matrix function, i.e., moves into
$\tilde{\cal P}[E_{dens}^{1/2},l_{dens}^{1/2},j,A](\rho_{spin}(n,\cdot,J))$ 
after one turn. Clearly we deal here with the Schroedinger picture.
Of course $\rho_{spin}=\beta_{dens}^{1/2}\circ f$, i.e.,
$\rho_{spin}(n,z,J)= \frac{1}{2} ( I_{2\times 2} 
+ \sum_{i=1}^3\; f_i(n,z,J)\sigma_i)$ where $f_i$ is the 
$i$-th component of $f$.
Note that the domain, $\Lambda$, of the action variable $J$ was introduced
in Section \ref{6.4}.

Let ${\cal O}:\Td\times\Lambda\rightarrow\C^{2\times 2}$ be a 
``physical observable'', i.e., let every value ${\cal O}(z,J)$ be a hermitian
$2\times 2$-matrix whence
${\cal O}(z,J)=g_0(z,J)+ \sum_{i=1}^3\; m_i(z,J)\sigma_i$ 
where $m_0,m_1,m_2,m_3:\Td\times\Lambda\rightarrow\R$. Then 
the ``expectation value'' $<{\cal O}>(n)$ of ${\cal O}$ at time $n$ 
is defined by
\begin{eqnarray}
&&  \hspace{-1cm}
 <{\cal O}>(n):= (1/2\pi)^d
\int_{[0,2\pi]^d\times\Lambda}\;Tr\biggl[\rho_{tot}(n,\pi_d(\phi),J)
{\cal O}(\pi_d(\phi),J)\biggr]d\phi dJ 
\nonumber\\
&&  \hspace{-5mm}
=\int_{[0,2\pi]^d\times\Lambda}\;\rho_{eq}(J)
\biggl( m_0(\pi_d(\phi),J) + \sum_{i=1}^3\; m_i(\pi_d(\phi),J)
f_i(n,\pi_d(\phi),J)\biggr) d\phi dJ \; ,
\nonumber\\
\label{eq:8.6.25as}
\end{eqnarray}
where in the second equality we used (\ref{eq:8.6.25s}). 
For example, in case of the spin observable, i.e.,
${\cal O}_i(z,J):=\sigma_i$, we get from (\ref{eq:8.6.25as})
\begin{eqnarray}
&& <{\cal O}_i>(n) =
(1/2\pi)^d\int_{[0,2\pi]^d\times\Lambda}\;\;\rho_{eq}(J)f_i(n,\pi_d(\phi),J)d\phi dJ 
\; ,
\label{eq:8.6.25cs}
\end{eqnarray}
which is the $i$-th component of the polarization vector of the
bunch, i.e., the bunch polarization is
$P(n)=\sqrt{ \sum_{i=1}^3\; (<{\cal O}_i>(n))^2}$ which we used
in Section \ref{6.4} for the definition of $P(n)$ in
eq. (\ref{eq:4.5}). 
At equilibrium, 
$\rho_{spin}(n,\cdot,J)=\rho_{spin}(0,\cdot,J)$ 
and $\rho_{spin}(0,\cdot,J)$ 
is an equilibrium  spin-$1/2$ density matrix function of $(j,A)$.

The choice $(E_{dens}^{1/2},l_{dens}^{1/2})$
and the above theory of
$\rho_{tot}$ follows from the semiclassical treatment
of Dirac's equation in terms of Wigner functions where the
particle-variables $z$ and $J$ are purely classical (see \cite{mont98}
and the references therein).
\hfill $\Box$
\end{itemize}
\subsubsection{Spin-$1$ particles. Applying the ToA to 
$(E_{dens}^1,l_{dens}^1)$}
\label{10.5.2}
In this section we introduce the $SO(3)$-space $(E_{dens}^1,l_{dens}^1)$ 
to enable the use of the ToA for the
study of the density matrix function to be employed for
polarized beams of spin-$1$
particles \cite{BV2}.
As in Sections \ref{10.3}-\ref{10.4b} the focus
is on the field motion.

To accomplish this we first introduce 
the particle-spin motion and field motion 
of spin-$1$ particles which can be described
by the ToA in terms of the $SO(3)$-space $(E_{v\times t},l_{v\times t})$
where
\begin{eqnarray} 
&&  E_{v\times t}:= \R^3\times E_{t} \; ,
\label{eq:8.6.110}
\end{eqnarray}
and where the function $l_{v\times t}:SO(3)\times E_{v\times t}\rightarrow 
E_{v\times t}$ is defined by 
\begin{eqnarray}
&&  l_{v\times t}(r;S,M):= (l_{v}(r;S),l_{t}(r;M)) = (rS,rMr^t) \; ,
\label{eq:8.6.111}
\end{eqnarray}
with $r\in SO(3),S\in\R^3,M\in E_{t}$.
We equip $E_{v\times t}$ with the subspace topology from
$\R^3\times \R^{3\times 3}$. 
Thus, and since $\R^3\times \R^{3\times 3}$ with its natural topology is a 
Hausdorff space, $E_{v\times t}$ is a Hausdorff space, too.
Since $(\R^3,l_{v})$ and $(E_t,l_{t})$ are $SO(3)$-spaces it follows from
(\ref{eq:8.6.110}) and (\ref{eq:8.6.111}) 
that $(E_{v\times t},l_{v\times t})$ is an
$SO(3)$-space. Using (\ref{eq:10.15}) and (\ref{eq:8.6.111}) it is a simple
exercise to show that the $1$-turn particle-spin map satisfies
\begin{eqnarray}
&&  \hspace{-1cm}
{\cal P}[E_{v\times t},l_{v\times t},j,A](r;z,S,M)
=\biggl( j(z),l_{v}(A(z);S),l_{t}(A(z);M)\biggr) \; .
\label{eq:8.6.112}
\end{eqnarray}
Clearly if $F:\Td\rightarrow E_{v\times t}$ is a function, then unique functions
$f_F:\Td\rightarrow \R^3$ and $T_F:\Td\rightarrow E_{t}$ exist such that
\begin{eqnarray}
&&  \hspace{-1cm}
F(z) =(f_F(z),T_F(z)) \; , 
\label{eq:8.6.113}
\end{eqnarray}
and it is an easy exercise to show that $F$ is continuous iff
$f_F$ and $T_F$ are continuous.
Using (\ref{eq:10.17}), (\ref{eq:8.6.111}) and (\ref{eq:8.6.113}) it is 
simple to show that the $1$-turn field map satisfies
\begin{eqnarray}
&& \hspace{-1cm} \tilde{\cal P}[E_{v\times t},l_{v\times t},j,A](F)=
l_{v\times t}\biggl(A\circ j^{-1};F\circ j^{-1}\biggr)
\nonumber\\
&& =\biggl( l_v(A\circ j^{-1};f_F\circ j^{-1}),
l_t(A\circ j^{-1};T_F\circ j^{-1})\biggr)
=\biggl(\tilde{\cal P}[j,A](f_F),\tilde{\cal P}[E_t,l_t,j,A](t_F)\biggr) \; ,
\nonumber\\
\label{eq:8.6.115}
\end{eqnarray}
where $F\in{\cal C}(\Td,\R^3\times E_t)$.
If $f_F$ is an invariant polarization field of $(j,A)$ and
$T_F$ is an invariant $(E_t,l_t)$-field of $(j,A)$ then, by (\ref{eq:8.6.115}),
$\tilde{\cal P}[E_{v\times t},l_{v\times t},j,A](F)=\\
(\tilde{\cal P}[j,A](f_F),\tilde{\cal P}[E_t,l_t,j,A](t_F))
=(f_F,t_F)=F$ whence $F$ is an invariant $(E_{v\times t},l_{v\times t})$-field 
of $(j,A)$. Conversely if $F$ is an invariant $(E_{v\times t},l_{v\times t})$-field 
of $(j,A)$ then, by (\ref{eq:8.6.115}), 
$(f_F,t_F)=F=\tilde{\cal P}[E_{v\times t},l_{v\times t},j,A](F)
=(\tilde{\cal P}[j,A](f_F),\tilde{\cal P}[E_t,l_t,j,A](t_F))$ whence
$f_F$ is an invariant polarization field of $(j,A)$ and
$T_F$ is an invariant $(E_t,l_t)$-field of $(j,A)$.
We thus have proven that $F$ is an invariant $(E_{v\times t},l_{v\times t})$-field 
of $(j,A)$ iff $f_F$ is an invariant polarization field of $(j,A)$ and
$T_F$ is an invariant $(E_t,l_t)$-field of $(j,A)$.
This completes our outline of 
the particle-spin and field motion of spin-$1$ particles
and we can now study spin-$1$ density matrices. 

We define
\begin{eqnarray} 
&&  E_{dens}^1:= \lbrace R \in\C^{3\times 3}:R^\dagger=R,Tr[R]=1\rbrace \; ,
\label{eq:8.6.10}
\end{eqnarray}
where $R^\dagger$ denotes the hermitian conjugate of the matrix $R$
and we equip $E_{dens}^1$ with the subspace topology from
$\C^{3\times 3}$. Thus, and since $\C^{3\times 3}$ with its natural topology is a 
Hausdorff space, $E_{dens}^1$ is a Hausdorff space, too.
Following a standard parametrization \cite{BV2} we define
the function $\beta_{dens}^1:E_{v\times t}\rightarrow E_{dens}^1$ 
for $S\in\R^3, M\in E_{t}$ by
\begin{eqnarray} 
&&
  \beta_{dens}^1(S,M):= \frac{1}{3} \biggl( I_{3\times 3} 
+ \sum_{i=1}^3\; S_i{\mathfrak J}_i 
+\sqrt{ \frac{3}{2} }  \sum_{i,k=1}^3\; M_{ik}({\mathfrak J}_i{\mathfrak J}_k
+ {\mathfrak J}_k{\mathfrak J}_i)\biggl) \; ,
\label{eq:8.6.15}
\end{eqnarray}
where the matrices ${\mathfrak J}_1,{\mathfrak J}_2,{\mathfrak J}_3$ are
defined by
\begin{eqnarray}
&&  \hspace{-15mm} 
{\mathfrak J}_1:= \sqrt{ \frac{1}{2} } 
\left( \begin{array}{ccc} 0 & -i & 0 \\
 i & 0 & -i \\
 0 & i & 0 \end{array}\right) \; , \quad
{\mathfrak J}_2:= \left( \begin{array}{ccc} 1 & 0 & 0 \\
 0 & 0 & 0 \\
 0 & 0 & -1 \end{array}\right) \; , \quad
{\mathfrak J}_3:= \sqrt{ \frac{1}{2} } 
\left( \begin{array}{ccc} 0 & 1 & 0 \\
 1 & 0 & 1 \\
 0 & 1 & 0 \end{array}\right) \; .
\label{eq:8.6.20}
\end{eqnarray}
Clearly every ${\mathfrak J}_k$ is hermitian, i.e.,
${\mathfrak J}_k^\dagger={\mathfrak J}_k$
whence, by (\ref{eq:8.6.15}), $\beta_{dens}^1(S,M)$ is hermitian too.
For $i,k=1,2,3$,
\begin{eqnarray}
&& Tr[{\mathfrak J}_i]= 0 \; , \quad
Tr[{\mathfrak J}_i{\mathfrak J}_k]
= 2\delta_{ik} \; ,
\label{eq:8.6.25}
\end{eqnarray}
whence, by (\ref{eq:8.6.15}) and since $Tr[M]=0$, we have $Tr[\beta_{dens}^1(S,M)]=1$.
So $\beta_{dens}^1(S,M)$  is a hermitian matrix of
trace $1$, i.e., it is  indeed a function into  $E_{dens}^1$.
To show that $\beta_{dens}^1$ is a homeomorphism we first note \cite{BV2} that 
by (\ref{eq:8.6.20}),
and if $R\in\C^{3\times 3}$ is hermitian, then
real numbers $S_0',S_1',S_2',S_3'$ and an $M'\in E_{t}$
exist such that 
$R=S_0'I_{3\times 3} + \sum_{i=1}^3\; S_i'{\mathfrak J}_i
 +\sum_{i,k=1}^3\; M_{ik}'({\mathfrak J}_i{\mathfrak J}_k
+ {\mathfrak J}_k{\mathfrak J}_i)$ whence, by 
(\ref{eq:8.6.10}) and (\ref{eq:8.6.25}) and
if $R\in E_{dens}^1$, we get $S_0'=1/3$ so 
that $R = \beta_{dens}^1(S,M)$ for some $S\in\R^3$ and $M\in E_{t}$.
Thus the function $\beta_{dens}^1$ is onto
$E_{dens}^1$, i.e.,
\begin{eqnarray} 
&&  E_{dens}^1 = \lbrace \beta_{dens}^1(S,M):S\in\R^3,M\in E_{t}\rbrace \; .
\label{eq:8.6.30}
\end{eqnarray}
It also follows \cite{BV2} from (\ref{eq:8.6.15}) and (\ref{eq:8.6.20})
that, for $S\in\R^3,M\in E_{t}$ and $i,k=1,2,3$,
\begin{eqnarray}
&&  \hspace{-10mm} 
S_i = Tr[{\mathfrak J}_i\beta_{dens}^1(S,M)] \; , \quad
M_{ik} =  -\sqrt{ \frac{2}{3} } \delta_{ik}
+ \sqrt{ \frac{3}{8} }
Tr[ ( {\mathfrak J}_i{\mathfrak J}_k + 
{\mathfrak J}_k{\mathfrak J}_i)\beta_{dens}^1(S,M)] \; ,
\label{eq:8.6.27}
\end{eqnarray}
where $S_i$ denotes the $i$-th component of $S$ and where 
$M_{ik}$ denotes the $(ik)$-th matrix element of $M$. Thus $S$ and $M$ 
are uniquely 
determined by $\beta_{dens}^1(S,M)$ whence $\beta_{dens}^1$ is one-one so that
we conclude that $\beta_{dens}^1$ is a bijection.
Since $\beta_{dens}^1$ is a bijection it follows from 
(\ref{eq:8.6.15}) and (\ref{eq:8.6.27}) that
its inverse, $(\beta_{dens}^1)^{-1}$, is defined for $R\in E_{dens}^1$ by
\begin{eqnarray} 
&&  (\beta_{dens}^{1})^{-1}(R):= (S,M) \; , \quad
S_i: = Tr[{\mathfrak J}_iR] \; , \quad
M_{ik} =  -\sqrt{ \frac{2}{3} } \delta_{ik}
+ \sqrt{ \frac{3}{8} }
Tr[ ( {\mathfrak J}_i{\mathfrak J}_k + 
{\mathfrak J}_k{\mathfrak J}_i)R] \; ,
\nonumber\\
\label{eq:8.6.28}
\end{eqnarray}
where $S_i$ denotes the $i$-th component of $S$ and where 
$M_{ik}$ denotes the $(ik)$-th matrix element of $M$.
Moreover, by (\ref{eq:8.6.15}) and (\ref{eq:8.6.28}),
both $\beta_{dens}^1$ and $(\beta_{dens}^1)^{-1}$ are 
continuous functions
whence $\beta_{dens}^1\in \Homeo(E_{v\times t},E_{dens}^1)$, a fact which plays
a key role in this section.

We now define the function $l_{dens}^1:SO(3)\times E_{dens}^1\rightarrow E_{dens}^1$ 
by 
\begin{eqnarray}
&&  l_{dens}^{1}(r;R):=\beta_{dens}^{1}
\biggl(l_{v\times t}(r;(\beta_{dens}^{1})^{-1}(R))\biggr) \; ,
\label{eq:8.6.31}
\end{eqnarray}
i.e., by (\ref{eq:8.6.111}),
\begin{eqnarray}
&&  l_{dens}^1(r;\beta_{dens}^1(S,M))=
\beta_{dens}^1(l_{v\times t}(r;S,M))
=\beta_{dens}^1(rS,rMr^t) \; ,
\label{eq:8.6.31vt}
\end{eqnarray}
with $r\in SO(3),S\in\R^3,M\in E_{t}$.
Recalling that $(E_{v\times t},l_{v\times t})$ is an $SO(3)$-space and that
$\beta_{dens}^{1}\in \Homeo(E_{v\times t},E_{dens}^{1})$ it follows from
(\ref{eq:8.6.31}) that 
$(E_{dens}^{1},l_{dens}^{1})$ is an
$SO(3)$-space and that $\beta_{dens}^{1}$ is an isomorphism from the
$SO(3)$-space $(E_{v\times t},l_{v\times t})$ to the $SO(3)$-space 
$(E_{dens}^{1},l_{dens}^{1})$. 

Due to (\ref{eq:10.15}), the $1$-turn particle-spin map 
${\cal P}[E_{dens}^{1},l_{dens}^{1},j,A]$ is given by
\begin{eqnarray}
&&  \hspace{-1cm}
{\cal P}[E_{dens}^{1},l_{dens}^{1},j,A](z,R)
=\left( \begin{array}{c} j(z) \\
l_{dens}^{1}(A(z);R)\end{array}\right) \; ,
\label{eq:8.6.32as2}
\end{eqnarray}
where $z\in\Td,R\in E_{dens}^{1}$. 
Because the $SO(3)$-spaces $(E_{v\times t},l_{v\times t})$ and 
$(E_{dens}^{1},l_{dens}^{1})$ are isomorphic, 
we get easy insight into ${\cal P}[E_{dens}^{1},l_{dens}^{1},j,A]$ by using the
Second ToA Transformation Rule. In fact recalling Section \ref{10.2.4}
we obtain via (\ref{eq:10.55nn}) that
\begin{eqnarray}
&&  \hspace{-1cm}
{\cal P}[E_{dens}^{1},l_{dens}^{1},j,A]\circ \beta_{dens,tot}^{1} 
= \beta_{dens,tot}^{1} \circ {\cal P}[E_{v\times t},l_{v\times t},j,A] \; ,
\label{eq:8.6.32cs2}
\end{eqnarray}
i.e., 
\begin{eqnarray}
&&  \hspace{-1cm}
{\cal P}[E_{dens}^{1},l_{dens}^{1},j,A](z,\beta_{dens}^{1}(S,M))
= {\cal P}[E_{dens}^{1},l_{dens}^{1},j,A](\beta_{dens,tot}^{1}(z,S,M))
\nonumber\\
&&=\left( \begin{array}{c} j(z) \\
\beta_{dens}^{1}(A(z)S,A(z)M A^t(z))\end{array}\right) 
\; ,
\end{eqnarray}
where $z\in\Td,S\in\R^3,M\in E_t$ and where the function \\
$\beta_{dens,tot}^{1}
\in \Homeo(\Td\times E_{v\times t},\Td\times  E_{dens}^{1})$ is 
defined, for $z\in\Td,S\in \R^3,M\in E_t$, by $\beta_{dens,tot}^{1}(z,S,M):=
(z,\beta_{dens}^{1}(S,M))$. 

We now come to our main focus, the fields, which in the case 
$(E,l)=(E_{dens}^{1},l_{dens}^{1})$ are also called  spin-$1$ 
density matrix functions and which are functions
$\rho:\Td\rightarrow E_{dens}^{1}$ so that
$\rho=\beta_{dens}^{1}\circ F$ where the function
$F:\Td\rightarrow E_{v\times t}$ is 
defined by $F(z):=(\beta_{dens}^{1})^{-1}(\rho(z))$. 
Clearly $F=(f_F,T_F)$ where 
$f_F:\Td\rightarrow \R^3$ and $T_F:\Td\rightarrow E_{t}$ are uniquely
determined by (\ref{eq:8.6.113}).
Thus, by (\ref{eq:8.6.15}), (\ref{eq:8.6.30}) and (\ref{eq:8.6.28}), we get
\begin{eqnarray}
&& (f_F)_i(z) = Tr[\rho(z){\mathfrak J}_i] \; , \quad 
(T_F)_{ik}(z) =  -\sqrt{ \frac{2}{3} } \delta_{ik}
+ \sqrt{ \frac{3}{8} }
Tr[ ( {\mathfrak J}_i{\mathfrak J}_k + 
{\mathfrak J}_k{\mathfrak J}_i)\rho(z)] \; ,
\label{eq:8.6.34} \\
&& \rho(z)=\beta_{dens}^{1}(f_F(z),T_F(z))
\nonumber\\
&&\quad =\frac{1}{3} \biggl( I_{3\times 3} 
+ \sum_{i=1}^3\; (f_F)_i(z){\mathfrak J}_i 
+\sqrt{ \frac{3}{2} }  \sum_{i,k=1}^3\; 
(T_F)_{ik}(z)({\mathfrak J}_i{\mathfrak J}_k
+ {\mathfrak J}_k{\mathfrak J}_i)\biggl) \; ,
\label{eq:8.6.15vt}
\end{eqnarray}
where $(f_F)_i$ denotes the $i$-th component of $f_F$
and where $(T_F)_{ik}$ denotes the $(ik)$-th matrix element of $T_F$.
Of course, since $\beta_{dens}^{1}\in \Homeo(E_{v\times t},E_{dens}^{1})$,
$\rho$ is continuous iff $F$ is continuous, i.e., iff
$f_F$ and $T_F$ are continuous.
We call an invariant $(E_{dens}^{1},l_{dens}^{1})$-field 
an ``equilibrium  spin-$1$ density matrix function''.
Due to (\ref{eq:10.17}), 
the $1$-turn field map $\tilde{\cal P}[E_{dens}^{1},l_{dens}^{1},j,A]$ 
is given by
\begin{eqnarray}
&& \tilde{\cal P}[E_{dens}^{1},l_{dens}^{1},j,A](\rho)=
l_{dens}^{1}\biggl(A\circ j^{-1};\rho\circ j^{-1}\biggr) \; .
\label{eq:8.6.35svt}
\end{eqnarray}
Because the $SO(3)$-spaces $(E_{v\times t},l_{v\times t})$ 
and $(E_{dens}^{1},l_{dens}^{1})$
are isomorphic, 
we get easy insight into $\tilde{\cal P}[E_{dens}^{1},l_{dens}^{1},j,A]$ 
by using once again the
Second ToA Transformation Rule. In fact recalling Section \ref{10.2.4}
we obtain via (\ref{eq:10.55ann}) that
\begin{eqnarray}
&&  \hspace{-1cm}
\tilde{\cal P}[E_{dens}^{1},l_{dens}^{1},j,A]\circ \tilde\beta_{dens}^{1} 
= \tilde{\beta}_{dens}^{1} \circ \tilde{\cal P}[E_{v\times t},l_{v\times t}] \; ,
\label{eq:8.6.32csnvt}
\end{eqnarray}
where the function $\tilde{\beta}_{dens}^{1}:{\cal C}(\Td,E_{v\times t})
\rightarrow{\cal C}(\Td,E_{dens}^{1})$ is defined,  
for $F\in{\cal C}(\Td,E_{v\times t})$, 
by $\tilde{\beta}_{dens}^{1}(F):=
\beta_{dens}^{1}\circ F$. 
It thus follows by Remark 16
that an $F\in{\cal C}(\Td,E_{v\times t})$
is an invariant $(E_{v\times t},l_{v\times t})$-field of $(j,A)$ iff
$\beta_{dens}^{1} \circ F$
is an invariant $(E_{dens}^{1},l_{dens}^{1})$-field of $(j,A)$, that is,
an  equilibrium  spin-$1$ density matrix function of $(j,A)$.
Thus, by the remarks after (\ref{eq:8.6.115}), 
a $\rho\in{\cal C}(\Td,E_{dens}^{1})$
is an  equilibrium  spin-$1$ density matrix function
of $(j,A)$ iff
$f_F$ is an invariant polarization field of $(j,A)$ and $T_F$
is an invariant $(E_t,l_t)$-field of $(j,A)$ where
$F\in{\cal C}(\Td,E_{v\times t})$ is defined by
$F:=(\beta_{dens}^{1})^{-1}\circ\rho$. 

We thus have proven:
\setcounter{theorem}{12}
\begin{theorem} \label{T10.8cx}
The function $\beta_{dens}^{1}$ belongs to $\Homeo(E_{v\times t},E_{dens}^{1})$.
Let $\rho:\Td\rightarrow E_{dens}^{1}$. Then a unique function
$F:\Td\rightarrow E_{v\times t}$ exists such that 
$\rho=\beta_{dens}^{1}\circ F$. Moreover $F=(f_F,T_F)$ where 
$f_F:\Td\rightarrow \R^3$ and $T_F:\Td\rightarrow E_{t}$ are uniquely
determined by $F$ via (\ref{eq:8.6.113}). Also $F$ is continuous iff
$f_F$ and $T_F$ are continuous.
Furthermore
$\rho=\beta_{dens}^{1}\circ (f_F,T_F)$, i.e., (\ref{eq:8.6.15vt}) holds where 
$(f_F)_i$ denotes the $i$-th component of $f_F$
and $(T_F)_{ik}$ denotes the $(ik)$-th matrix element of $T_F$.
Moreover $\rho$ is continuous iff $f_F$ and $T_F$ are continuous.
In addition let $(j,A)\in{\cal SOS}(d,j)$.
Then $\rho$ is an equilibrium  spin-$1$ 
density matrix function of $(j,A)$, i.e.,
is an invariant $(E_{dens}^{1},l_{dens}^{1})$-field of $(j,A)$ iff
$f_F$ is an invariant polarization field of $(j,A)$
and $T_F$ is an invariant $(E_t,l_t)$-field of $(j,A)$.
\hfill $\Box$
\end{theorem}

Because of Theorem \ref{T10.8cx} the study
of equilibrium  spin-$1$ density matrix functions 
effectively amounts to 
to the study of invariant polarization fields and 
invariant $(E_t,l_t)$-fields. Since those invariant fields 
have been studied in other parts of this work this section has been  rather
brief and we 
leave the application of the NFT, DT etc. as an exercise for the reader.

Moreover we leave a remark on how to 
address topological transitivity by applying Lemma \ref{L10.4} to 
$(E,l)=(E_{dens}^{1},l_{dens}^{1})$ to the reader since it would be
analogous to Remark 27. In particular in the case of
topological transitivity the values of the
invariant $(E_t,l_t)$-field $T_F$ are matrices with the same number
of distinct eigenvalues and the case of three eigenvalues is of
interest only on spin-orbit resonance.

Furthermore we leave a remark on how to use
spin-$1$ density matrices for the statistical description of
a bunch of spin-$1$ particles to the reader since it would be
analogous to Remark 28.
\subsection{The topological spaces $\Edh[E,l,x,f]$}
\label{10.6}
We now come to an intriguing feature
of the ToA, namely the topological spaces $\Edh[E,l,x,f]$ 
to be defined below. 
They allow us to study invariant $(E,l)$-fields $f$, and in 
particular each ISF,
in terms of $\Edh[E,l,x,f]$. This leads 
to a significant
avenue for studying the 
question of the existence of $f$
and for studying IFF's and the generalized IFF's.
\subsubsection{Encapsulating invariant $(E,l)$-fields in the
topological spaces $\Edh[E,l,x,f]$. The Invariant Reduction Theorem (IRT)}
\label{10.6.1}
We first need a definition:
\setcounter{definition}{13}
\begin{definition}
\label{D8.7}
\noindent 
Let $(E,l)$ be an $SO(3)$-space and $x,y\in E$ and let
$f\in{\cal C}(\Td,E)$ take values only in $l(SO(3);x)$.
We define ${\cal R}[E,l,x,y]:= \lbrace r\in SO(3):l(r;x)=y \rbrace$.  
We also define $\Ed:=\Td\times SO(3)$ and 
$p_d\in{\cal C}(\Ed,\Td)$ by $p_d(z,r):=z$ as well as
\begin{eqnarray}
&& \Edh[E,l,x,f]\equiv \Edh[f]:=
 \lbrace (z,r)\in \Ed:l(r;x)=f(z) \rbrace 
= \lbrace (z,r)\in \Ed:r\in {\cal R}[E,l,x,f(z)]\rbrace 
\nonumber\\
&&\quad =\bigcup_{z\in\Td} 
\biggl(\lbrace z \rbrace \times {\cal R}[E,l,x,f(z)]
\biggr) \subset \Ed
\; ,
\label{eq:10.310}
\end{eqnarray}
and equip $\Edh[f]$ with the
subspace topology from $\Ed$. Note that $\Ed$ is the compact topological
space equipped with the product topology from $\Td$ and $SO(3)$.
We will use the abbreviation $\Edh[f]$ when $E,l$ and $x$ are clear from
the context.

We also define the function $\hat{\cal P}[j,A]:\Ed\rightarrow \Ed$ by
\begin{eqnarray}
&& \hat{\cal P}[j,A](z,r)
:=\left( \begin{array}{c} 
j(z) \\ A(z)r
\end{array}\right) \; ,
\label{eq:10.135n}
\end{eqnarray}
where $z\in\Td$ and $r\in SO(3)$.
Note that, by (\ref{eq:10.135n}) and Remark 9,\\
$\hat{\cal P}[j,A]={\cal P}[SO(3),l_{SOR};j,A]$ whence, 
recalling Section
\ref{10.2.1}, $\hat{\cal P}[j,A]\in \Homeo({\cal E}_d)$.
\hfill $\Box$
\end{definition}
Note that if $r\in{\cal R}[E,l,x,y]$ then ${\cal R}[E,l,x,y]=r Iso(E,l,x)$, 
i.e., every ${\cal R}[E,l,x,y]$ is a ``copy'' of $Iso(E,l,x)$.
More precisely, ${\cal R}[E,l,x,y]$ is a so-called left coset of
the subgroup $Iso(E,l,x)$ of $SO(3)$ with respect to $r$.
It is also a simple exercise to show, by (\ref{eq:10.310}), that 
if $\Edh[f]=\Edh[g]$ then $f=g$. 

In the following remark we derive another important
property of $\Edh[f]$.

\vspace{3mm}
\noindent{\bf Remark:}
\begin{itemize}
\item[(29)] 
Let $(E,l)$ be an $SO(3)$-space where $E$ is Hausdorff and let 
$\Edh[f]$ be given as in Definition \ref{D8.7}. 
Then the topological space $\Edh[f]$ is compact
as follows. In fact we note, by (\ref{eq:10.310}), that
\begin{eqnarray}
&& \Edh[f]= \lbrace (z,r)\in {\cal E}_d:l(r;x)=f(z) \rbrace 
=\lbrace (z,r)\in {\cal E}_d: x = l(r^t;f(z))\rbrace 
\nonumber\\
&&\quad =\lbrace (z,r)\in {\cal E}_d: l(r^t;f(z))\in\lbrace x\rbrace
\rbrace \subset \Ed
\; ,
\end{eqnarray}
whence $\Edh[f]$ is the inverse image of 
$\lbrace x\rbrace$ under a continuous function.
Since $E$ is Hausdorff, $\lbrace x\rbrace$ is a closed subset
of $E$ whence we conclude that $\Edh[f]$ is
a closed subset of the compact topological space ${\cal E}_d$. 
This implies that $\Edh[f]$ is compact \cite{Mu}.
\hfill $\Box$
\end{itemize}

With Definition \ref{D8.7} we arrive at:
\setcounter{theorem}{14}
\begin{theorem} (IRT)
\label{P02xxx}
Let $(E,l)$ be an $SO(3)$-space and $x\in E$ and let
$f\in{\cal C}(\Td,E)$ take values only in $l(SO(3);x)$.
Then $f$ is an invariant $(E,l)$-field of $(j,A)$ iff \\
$\hat{\cal P}[j,A](\Edh[f])=\Edh[f]$.
\end{theorem}
\noindent{\em Proof of Theorem \ref{P02xxx}:}
We conclude from (\ref{eq:10.310}),(\ref{eq:10.135n}) that
\begin{eqnarray}
&&  \hspace{-1cm}
\hat{\cal P}[j,A](\Edh[f])
= \hat{\cal P}[j,A] ( \lbrace (z,r)\in {\cal E}_d:
l(r;x)=f(z) \rbrace )
\nonumber\\
&&
= \biggl\lbrace (j(z),A(z)r):(z,r)\in \Ed,l(r;x)=f(z) \biggr\rbrace
\nonumber\\
&& \quad
= \biggl\lbrace (z',r')\in {\cal E}_d:
l\biggl(A^t(j^{-1}(z'))r';x\biggr)=f(j^{-1}(z')) \biggr\rbrace
\nonumber\\
&&
= \biggl\lbrace (z',r')\in {\cal E}_d:
l(r';x)=l\biggl(A(j^{-1}(z'));f(j^{-1}(z'))\biggr) 
\biggr\rbrace
\nonumber\\
&& \quad
= \lbrace (z',r')\in {\cal E}_d:l(r';x)=f'(z')\rbrace
=\Edh[f'] \; ,
\label{eq:10.137}
\end{eqnarray}
where $f'\in{\cal C}(\Td,E)$ is defined by
$f':=\tilde{\cal P}[E,l,j,A](f)$, i.e. (recall (\ref{eq:10.17})),\\
$f'(z)=l(A(j^{-1}(z));f(j^{-1}(z)))$.

If $f$ is an invariant $(E,l)$-field of $(j,A)$ 
then, by Section \ref{10.2.1}, $f=f'$ whence, 
by (\ref{eq:10.137}), $\hat{\cal P}[j,A](\Edh[f])=\Edh[f]$.

Conversely, let $\hat{\cal P}[j,A](\Edh[f])=\Edh[f]$.
Then, by (\ref{eq:10.137}),  $\Edh[f']=\Edh[f]$ where
$f':=\tilde{\cal P}[E,l,j,A](f)$. It follows from the remark after
Definition \ref{D8.7} that $f=f'$ whence $f$ is an invariant $(E,l)$-field 
of $(j,A)$.
\hfill $\Box$

\vspace{1cm}

We will see below how
the topological spaces $\Edh[f]$ in Theorem \ref{P02xxx} 
can be used for the question of the existence of invariant $(E,l)$-fields
in particular ISF's (recall that an ISF takes values
only in $l_v(SO(3);(0,0,1)^t)$).

The terminology ``reduction'' refers to ${\cal E}_d$ being ``reduced'' to
the subspace $\Edh[f]$ and the terminology ``invariant''
refers to the invariance condition:
$\hat{\cal P}[j,A](\Edh[f])=\Edh[f]$, i.e.,
$\Edh[f]$ is an invariant subset of $\Ed$ under the function 
$\hat{\cal P}[j,A]$.
For more details and the definition of invariant reductions, see \cite{Fe,Zi2} 
and Chapter 9 in \cite{HK1} as well as
our comments on bundle theory in Section \ref{10.7}.
By the bundle aspect of the IRT the
concept of ISF is rather deep.

With Definition \ref{D8.7} we have encapsulated $f$ into the
topological space $\Edh[E,l,x,f]\equiv\Edh[f]$. Below we will see how one gets
insight into $\Edh[f]$ in terms of the $f$-independent 
topological spaces $\Edc[z,y]$
which we now define:
\setcounter{definition}{15}
\begin{definition}
\label{D8.7b}
\noindent 
Let $(E,l)$ be an $SO(3)$-space and $x,y\in E$.
Let also $(j,A)\in {\cal SOS}(d,j)$ and $z\in\Td$. Then we define
\begin{eqnarray}
&&  \hspace{-15mm}
\Edc[E,l,j,A,z,x,y]\equiv\Edc[z,y]:=
\bigcup_{n\in\Z} 
\hat{\cal P}[j,A]^n
(\lbrace z \rbrace 
\times \lbrace r\in SO(3):l(r;x)=y\rbrace )
\nonumber\\
&& =
\bigcup_{n\in\Z} 
\biggl(\lbrace j^n(z) \rbrace 
\times \lbrace r\in SO(3):l(r;x)=l(\Psi[j,A](n;z);y)\rbrace 
\biggr) 
\nonumber\\
&& \quad =
\bigcup_{n\in\Z} 
\biggl(\lbrace j^n(z) \rbrace 
\times 
 {\cal R}[E,l,x,l(\Psi[j,A](n;z);y)]\biggr) 
\subset {\cal E}_d
\; ,
\label{eq:10.132n}
\end{eqnarray}
where in the third equality we used 
(\ref{eq:10.36}). Clearly
$\Edc[z,y]$ is nonempty iff
$x,y$ belong to the same $(E,l)$-orbit. 
We equip $\Edc[z,y]$ and 
$\overline{\Edc[z,y]}$ with the
subspace topology from $\Ed$. 
We will use the abbreviation $\Edc[z,y]$ when $E,l,j,A$ and $x$ are clear from
the context.
\hfill $\Box$
\end{definition}
The following corollary to the IRT gives us a first glimpse into how one gets
insight into $\Edh[f]$ by the topological spaces $\Edc[z,y]$:

\setcounter{theorem}{16}
\begin{theorem} \label{C12.10a}
\noindent
Let $(E,l)$ be an $SO(3)$-space and $(j,A)\in {\cal SOS}(d,j)$. Also let
\begin{eqnarray}
&&  \hspace{-5mm}
\Edh[f]=\overline{\Edc[z,y]} \; .
\label{eq:10.132nbaa}
\end{eqnarray}
Then $f$ is an invariant $(E,l)$-field of $(j,A)$ and $j$ is
topologically transitive. Moreover if 
$(z',r')\in \overline{\Edc[z,y]}$
then $f(z')=l(r';x)$.
\end{theorem}
\noindent{\em Proof of Theorem \ref{C12.10a}:}
Since the image of a union is a union of images we conclude from
(\ref{eq:10.132n}) that
\begin{eqnarray}
&&  \hspace{-15mm}
\hat{\cal P}[j,A](\Edc[z,y])
=\bigcup_{n\in\Z} 
\hat{\cal P}[j,A]^{n+1}
(\lbrace z \rbrace 
\times \lbrace r\in SO(3):l(r;x)=y\rbrace )
\nonumber\\
&& =\bigcup_{n\in\Z} 
\hat{\cal P}[j,A]^{n}
(\lbrace z \rbrace 
\times \lbrace r\in SO(3):l(r;x)=y\rbrace ) = \Edc[z,y] \; ,
\label{eq:10.132naa}
\end{eqnarray}
i.e., $\Edc[z,y]$ is invariant under $\hat{\cal P}[j,A]$.
Since $\hat{\cal P}[j,A]$ is a homeomorphism we get
from (\ref{eq:10.132naa})
\begin{eqnarray}
&&  \hspace{-5mm}
\overline{\Edc[z,y]} 
=\overline{\hat{\cal P}[j,A](\Edc[z,y])}
=\hat{\cal P}[j,A](\overline{\Edc[z,y]}) 
\; ,
\label{eq:10.132nab}
\end{eqnarray}
where in the second equality we used \cite[Section III.12]{Du}.
By (\ref{eq:10.132nab}), $\overline{\Edc[z,y]}$
is invariant under $\hat{\cal P}[j,A]$ whence, by our assumption
(\ref{eq:10.132nbaa}), $\Edh[f]$ is 
invariant under $\hat{\cal P}[j,A]$ and this implies, by 
the IRT, that $f$ is an invariant $(E,l)$-field of $(j,A)$.
Using (\ref{eq:10.132nbaa}) and Definition \ref{D8.7} we get
\begin{eqnarray}
&&  \hspace{-5mm}
\Td= p_d(\Edh[f])=p_d(\overline{\Edc[z,y]})
\subset \overline{p_d(\Edc[z,y])}
= \overline{ \bigcup_{n\in\Z} \lbrace j^n(z) \rbrace} 
= \overline{ \lbrace j^n(z):n\in\Z\rbrace} \subset \Td\; ,
\nonumber
\end{eqnarray}
where in the inclusion we used \cite[Section III.8]{Du}. Thus
$\Td=\overline{ \lbrace j^n(z):n\in\Z\rbrace}$ 
whence $j$ is topologically transitive (recall Section \ref{2.2}).
Furthermore if $(z',r')\in \overline{\Edc[z,y]}$
then, by (\ref{eq:10.132nbaa}), $(z',r')\in\Edh[f]$ whence, by 
Definition \ref{D8.7},
$f(z')=l(r';x)$.
\hfill $\Box$

Since $\overline{\Edc[z,y]}$ is a closed subset of the
compact topological space $\Ed$ and since it is
equipped with the subspace topology from $\Ed$,
it is a compact topological space \cite{Mu}.
A key motivation of using $\Edc[z,y]$ for the existence problem
of invariant $(E,l)$-fields
is our belief that, unless (\ref{eq:10.132nbaa}) holds,
the topological spaces
$\overline{\Edc[z,y]}$ are in general different from the $\Edh[f]$ 
and this may display itself in different
homology and homotopy groups and other features of these topological spaces. 
Thus in this approach to the existence problem one ``scans''
through all $\overline{\Edc[z,y]}$ by varying $y$ over
$l(SO(3);x)$. However this issue is beyond the scope of this work.

In the following section we will get more insight into the relation
between $\Edh[f]$ and $\Edc[z,y]$.
\subsubsection{Further properties of the
topological spaces $\Edh[E,l,x,f]$. The Cross Section Theorem (CST)}
\label{9.4b.4}
In the situation of the NFT one has an
$SO(3)$-space $(E,l)$ and an $x\in E$ as well as functions
$T\in{\cal C}(\Td,SO(3))$ and $f\in{\cal C}(\Td,E)$
which are related by $f(z)=l(T(z);x)$, i.e., $T$ is an $(E,l)$-lift of $f$.
Note that a necessary condition to satisfy this relation is
that $f$ takes values only in $l(SO(3);x)$.
Perhaps surprisingly, it is not a sufficient condition, i.e.,
there are situations where
$f\in{\cal C}(\Td,E)$ takes only values in $l(SO(3);x)$ and where
nevertheless no $T\in{\cal C}(\Td,SO(3))$ exists such that
$f(z)=l(T(z);x)$. This is quite remarkable since, if
$f$ takes values only in $l(SO(3);x)$,
then there exists for each $z$ a $T(z)\in SO(3)$ such that $f(z)=l(T(z);x)$, 
i.e., there always exists a function $T:\Td\rightarrow SO(3)$ such that
$f(\cdot)=l(T(\cdot);x)$.
Interesting examples arise in the case of Chapters \ref{2}-\ref{VII}
where $(E,l)=(\R^3,l_{v})$. In fact one can
show by using simple arguments from Homotopy Theory
\cite{He2} that, if $d\geq 2$, then
$f\in{\cal C}(\Td,\R^3)$ exist such that 
$f$ takes values only in $l_{v}(SO(3);(0,0,1)^t)$, i.e., such that
$|f(z)|=1$
and no continuous $T:\Td\rightarrow SO(3)$ exists such that
$f(z)=l_{v}(T(z);(0,0,1)^t)$ 
(of course $T$ exists if we relax the continuity condition on $T$). 
One can also show by using simple arguments from Homotopy Theory
\cite{He2} that for every 
$f\in{\cal C}(\T^1,\R^3)$ such that $|f(z)|=1$
an $T\in{\cal C}(\T^1,SO(3))$ exists such that
$f(z)=l_{v}(T(z);(0,0,1)^t)$. 

The central theme of this section is to show how the set 
$\Edh[f]$ gives a simple sufficient and
necessary condition on $f$ to be of the form $f(z)=l(T(z);x)$
with $T\in{\cal C}(\Td,SO(3))$.
In fact this condition, stated in Theorem \ref{P12.10} below,
is the existence of a ``cross section'' of the
function $p_d[E,l,x,f]$ (to be defined below). Thus in the situation of the
NFT a cross section exists. 

We first define $p_d[f]$ and the function $\Upsilon_d[T]$ and its
restriction.
\setcounter{definition}{17}
\begin{definition}
\label{D8.7c}
\noindent 
Let $(E,l)$ be an $SO(3)$-space and $x\in E$ and let
$f\in{\cal C}(\Td,E)$ take values only in $l(SO(3);x)$.
Thus $\Edh[f]$ is well-defined and we define
the function $p_d[E,l,x,f]\in{\cal C}(\Edh[f],\Td)$ by
\begin{eqnarray}
&& p_d[E,l,x, f](z,r)\equiv  p_d[f](z,r):=
z \; , 
\label{eq:120.30} 
\end{eqnarray}
where $z\in\Td,r\in SO(3)$. 
We will use the abbreviation $p_d[f]$ when $E,l$ and $x$ are clear from
the context. Clearly $p_d[f]$ is surjective since,
by Definition \ref{D8.7} and for every $z\in\Td$, 
there is an $r\in SO(3)$ such that
$(z,r)\in\Ed[f]$.

Given $T:\Td\rightarrow SO(3)$, we define the function 
$\Upsilon_d[T]:{\cal E}_d\rightarrow {\cal E}_d$ by
\begin{eqnarray}
&& \Upsilon_d[T](z,r):=(z,T(z)r) \; .
\label{eq:12.17dbnna}
\end{eqnarray}
We also define the surjection
$\hat{\Upsilon}_d[E,l,x,T]:\Td\times Iso(E,l;x)\rightarrow
\hat{\Upsilon}_d[E,l,x,T](\Td\times Iso(E,l;x))$ as
the restriction of $\Upsilon_d[T]$, i.e.,
\begin{eqnarray}
&& \hat{\Upsilon}_d[E,l,x,T](z,r)\equiv
\hat{\Upsilon}_d[T](z,r)
:=\Upsilon_d[T](z,r)
=(z,T(z)r) \; .
\label{eq:12.17dbnn}
\end{eqnarray}
We will use the abbreviation $\hat{\Upsilon}_d[T]$ when $E,l$ and $x$ 
are clear from the context.
\hfill $\Box$
\end{definition}
It is an easy exercise to show that $\Upsilon_d[T]$ is a bijection and
that $\Upsilon_d[T^t]$ is its inverse. Moreover if $T$ is continuous then, by
(\ref{eq:10.135n}) and (\ref{eq:12.17dbnna}),
$\Upsilon_d[T]=\hat{\cal P}[id_\Td,T]$ whence in this case, by 
Definition \ref{D8.7}, $\Upsilon_d[T]\in \Homeo({\cal E}_d)$.
Note also, by (\ref{eq:12.17dbnn}) and Definition \ref{D8.7}, 
that the range of $\hat{\Upsilon}_d[T]$ reads as
\begin{eqnarray}
&& \hat{\Upsilon}_d[E,l,x,T](\Td\times Iso(E,l;x))=\bigcup_{z\in\Td} 
\biggl(\lbrace z \rbrace \times 
T(z)Iso(E,l;x)\biggr) 
\nonumber\\
&&\quad =\bigcup_{z\in\Td} 
\biggl(\lbrace z \rbrace \times 
{\cal R}[E,l,x,l(T(z);x)]\biggr) \; .
\label{eq:12.17dbnnn}
\end{eqnarray}
Moreover since $\Upsilon_d[T]$ is a bijection, so is
$\hat{\Upsilon}_d[T]$ and the latter's inverse 
$\hat{\Upsilon}_d[T]^{-1}$ is defined by
$\hat{\Upsilon}_d[T]^{-1}(z,r):=\Upsilon_d[T^t](z,r)$.

Since $p_d[f]$ is a surjection it has a right-inverse, i.e., a function
$\sigma:\Td\rightarrow\Edh[f]$ 
such that $p_d[f]\circ\sigma=id_\Td$.
Note that, by Definition \ref{D8.7}, $p_d[f]$ 
has more than one right-inverse except for the case when $Iso(E,l;x)=G_0$. 
Furthermore even though
$p_d[f]$ is continuous, it does not always have a continuous right-inverse. 
A cross section of $p_d[f]$ is, by definition, a continuous
right-inverse (see also Appendix \ref{A.4}) and the
following theorem sheds light at the cross sections of $p_d[f]$.

\setcounter{theorem}{18}
\begin{theorem} \label{P12.10}
(CST)\\
Let $(E,l)$ be an $SO(3)$-space and let $x\in E$. 
Abbreviating $H:=Iso(E,l;x)$ the following hold.\\

\noindent
a) Let $f\in{\cal C}(\Td,E)$ take values only in $l(SO(3);x)$. Then 
$p_d[f]$ has a cross section
iff a $T\in{\cal C}(\Td,SO(3))$ exists such that $f(z)=l(T(z);x)$. 
In other words, $p_d[f]$ has a cross section
iff $f$ has an $(E,l)$-lift (see the remarks after the NFT).
\\

\noindent
b)  Let $f\in{\cal C}(\Td,E)$ take values only in
$l(SO(3);x)$ and let us pick a function
$T:\Td\rightarrow SO(3)$ such that $f(z)=l(T(z);x)$.
Then the bijection $\hat{\Upsilon}_d[T]$ is onto
$\Edh[f]$. Also
$T$ is continuous iff $\hat{\Upsilon}_d[T]$ is a homeomorphism, i.e.,
$\hat{\Upsilon}[T]\in \Homeo(\Td\times H,\Edh[f])$.
\\

\noindent c) 
Let $x\in E$ and $(j,A)\in {\cal SOS}(d,j)$. Let 
$f$ be an invariant $(E,l)$-field
of $(j,A)$ which takes values only in $l(SO(3);x)$.
Then $p_d[f]$ has a cross section iff
a $T$ in ${\cal TF}_H(j,A)$ exists such that $f(z)=l(T(z);x)$.
Moreover if $p_d[f]$ has a cross section 
then $(j,A)\in{\cal CB}_H(d,j)$.
\end{theorem}
\noindent{\em Proof of Theorem \ref{P12.10}a:} 
``$\Rightarrow$'': 
Let $\sigma$ be a cross section of $p_d[f]$. Since 
$\sigma\in{\cal C}(\Td,\Edh[f])$ and $\Edh[f]\subset\Ed$
we have $\sigma(z)=(\tau(z),T(z))$ where 
$\tau\in {\cal C}(\Td,\Td)$ and $T\in {\cal C}(\Td,SO(3))$.
We compute $z=id_\Td(z)=p_d[f](\sigma(z))
=p_d[f](\tau(z),T(z))=\tau(z)$ whence $\sigma(z)=(z,T(z))\in \Edh[f]$
so that, by Definition \ref{D8.7}, $f(z)=l(T(z);x)$.\\

\noindent ``$\Leftarrow$'': 
Let $T\in{\cal C}(\Td,SO(3))$ such that $f(z)=l(T(z);x)$.
Thus, by (\ref{eq:10.310}), $(z,T(z))\in \Edh[f]$ 
whence with the function
$\sigma\in{\cal C}(\Td,\Edh[f])$ defined by
$\sigma(z):=(z,T(z))$, we see that $p_d[f](\sigma(z))=z$. Therefore
$p_d[f]\circ\sigma=id_{\Td}$ so that
$\sigma$ is a cross section of $p_d[f]$.
\hfill $\Box$

\noindent{\em Proof of Theorem \ref{P12.10}b:}
It follows from (\ref{eq:12.17dbnnn}) and Definition \ref{D8.7}
that
\begin{eqnarray}
&& \hat{\Upsilon}_d[E,l,x,T](\Td\times H)=\bigcup_{z\in\Td} 
\biggl(\lbrace z \rbrace \times 
{\cal R}[E,l,x,l(T(z);x)]\biggr) 
\nonumber\\
&&\quad =\bigcup_{z\in\Td} 
\biggl(\lbrace z \rbrace \times 
{\cal R}[E,l,x,f(z)]\biggr) =\Edh[E,l,x,f]
 \; ,
\nonumber
\end{eqnarray}
whence $\hat{\Upsilon}_d[T]$ is onto $\Edh[f]$.
To prove the second claim,
first of all, let $T$ be continuous. Thus, by the remarks 
after Definition \ref{D8.7c}, $\Upsilon_d[T]\in \Homeo({\cal E}_d)$.
Then, recalling from the remarks 
after Definition \ref{D8.7c}, that $\hat{\Upsilon}_d[T]$ is a bijection and a 
restriction of $\Upsilon_d[T]$ we conclude that
$\hat{\Upsilon}_d[T]\in \Homeo(\Td\times H,\Edh[f])$.
Secondly let $\hat{\Upsilon}_d[T]\in \Homeo(\Td\times H,\Edh[f])$.
It follows from (\ref{eq:12.17dbnn})
that $(z,T(z))=\hat{\Upsilon}_d[T](z,I_{3\times 3})$ whence, since
$\hat{\Upsilon}_d[T]$ is continuous, so is $T$.
\hfill $\Box$

\noindent{\em Proof of Theorem \ref{P12.10}c:}
Let first of all $p_d[f]$ have a cross section.
Thus, by Theorem \ref{P12.10}a, a $T\in{\cal C}(\Td,SO(3))$ exists
such that $f(z)=l(T(z);x)$.
Since $f$ is an invariant $(E,l)$-field of $(j,A)$
we thus conclude from the NFT,
Theorem \ref{T10.1}, that $T\in {\cal TF}_H(j,A)$.
Let secondly $T\in {\cal TF}_H(j,A)$ such that
$f(z)=l(T(z);x)$. Thus $T$ is continuous whence, by Theorem \ref{P12.10}a,
$p_d[f]$ has a cross section.
This completes the proof of the first claim. 
It follows from the first claim that if 
$p_d[f]$ has a cross section then
${\cal TF}_H(j,A)$ is nonempty whence, by Definition \ref{D6},
$(j,A)\in{\cal CB}_H(d,j)$.
\hfill $\Box$

\vspace{1cm}

The following remark reconsiders IFF's in terms of the CST:

\vspace{3mm}
\noindent{\bf Remark:}
\begin{itemize}
\item[(30)] 
Let $(j,A)\in {\cal SOS}(d,j)$ have an invariant polarization field
$f$ which takes values only in
$l_v(SO(3);S_\lambda)={\mathbb S}^2_\lambda$ where $\lambda>0$, i.e.,
$|f|=\lambda>0$. Then, by Theorem \ref{P12.10}c, 
$p_d[\R^3,l_v,S_\lambda,f]$ has a cross section
iff a $T$ in ${\cal TF}_H(j,A)$ exists such that $f(z)=l_v(T(z);S_\lambda)$,
i.e., iff $f(z)=\lambda T(z)(0,0,1)^t$ where
$H:=Iso(\R^3,l_v,S_\lambda)=SO(2)$. Thus, by Definition \ref{DIFF},
$p_d[\R^3,l_v,S_\lambda,f]$ has a cross section
iff $(j,A)$ has an IFF whose third column is $f/\lambda$.
\hfill $\Box$
\end{itemize}

\noindent A similar remark could be made about SOR by using 
$(E,l)=(SO(3),l_{SOR})$. For the bundle aspect of the CST see Section \ref{10.7}.
By the bundle aspect of the CST the above remarks indicate that the
concepts of IFF and SOR are rather deep.

As mentioned after Theorem \ref{C12.10a}, a 
key motivation of using $\Edc[z,y]$ for the existence problem
of invariant $(E,l)$-fields
is our belief that the topological spaces
$\overline{\Edc[z,y]}$ are in general different from the
$\Edh[f]$ unless (\ref{eq:10.132nbaa}) holds. 
The CST sheds further light on this issue since, by
Theorem \ref{P12.10}a-b, $\Edh[f]$ is homeomorphic to
$\Td\times Iso(E,l;x)$ if $p_d[f]$ has a cross section whence
the key idea is to compare the topological spaces
$\overline{\Edc[z,y]}$ with $\Td\times Iso(E,l;x)$.
Note that in the important case of the existence problem of the ISF
we have $Iso(\R^3,l_v;(0,0,1)^t)=SO(2)$ whence $\Edh[f]$ in this case 
is homeomorphic to ${\mathbb T}^{d+1}$ (note that
$SO(2)$ is homeomorphic to ${\mathbb T}^{1}$).

The following corollary to the CST
is a partial converse of
Theorem \ref{C12.10a} giving us further
insight into the relation
between $\Edh[f]$ and $\Edc[z,y]$.

\setcounter{theorem}{19}
\begin{theorem} \label{C12.10b}
Let $(E,l)$ be an $SO(3)$-space 
where $E$ is Hausdorff. Let also
$(j,A)\in {\cal SOS}(d,j)$ have an invariant $(E,l)$-field $f$ and let
$j$ be topologically 
transitive, i.e., let a $z_0\in\Td$ exist such that 
$\overline{\lbrace j^n(z_0):n\in\Z\rbrace}=\Td$.
By Lemma \ref{L10.4} an $x\in E$ exists such that 
$f$ takes values only in $l(SO(3);x)$.
If $p_d[f]$ has a cross section then
\begin{eqnarray}
&&  \hspace{-5mm}
\Edh[f]=\overline{\Edc[z_0,f(z_0)]} \; .
\label{eq:10.132nbaan}
\end{eqnarray}
\end{theorem}
\noindent{\em Proof of Theorem \ref{C12.10b}:}
Our strategy is to prove that 
$\hat{\Upsilon}_d[T]^{-1}(\Edh[f])=
\hat{\Upsilon}_d[T]^{-1}(\overline{\Edc[z_0,f(z_0)]})$.
Since $f$ is an invariant $(E,l)$-field
of $(j,A)$ it follows from (\ref{eq:10.44}) that,
for every $n\in\Z$,
$f(j^n(z_0))=l(\Psi[j,A](n;z_0);f(z_0))$ whence 
\begin{eqnarray}
&&  \hspace{-15mm}
\Edc[z_0,f(z_0)]
=\bigcup_{n\in\Z} 
\biggl(\lbrace j^n(z_0) \rbrace 
\times \lbrace r\in SO(3):l(r;x)=l(\Psi[j,A](n;z_0);f(z_0))\rbrace 
\biggr) 
\nonumber\\
&& \hspace{-5mm}
=\bigcup_{n\in\Z} 
\biggl(\lbrace j^n(z_0) \rbrace 
\times \lbrace r\in SO(3):l(r;x)=f(j^n(z_0))\rbrace 
\biggr) \subset \Edh[f]
\; ,
\label{eq:10.132na}
\end{eqnarray}
where the first equality in
(\ref{eq:10.132na}) follows from (\ref{eq:10.132n}).
Of course the inclusion in (\ref{eq:10.132na}) follows from the
definition of $\Edh[f]$. 
By Theorem \ref{P12.10}a, 
a $T\in {\cal C}(\Td,SO(3))$ exists such that
$f(z)=l(T(z);x)$ whence, by Theorem \ref{P12.10}b, 
$\hat{\Upsilon}_d[T]\in \Homeo(\Td\times H,\Edh[f])$
where $H:=Iso(E,l;x)$. 
Because of (\ref{eq:10.132na}), $\hat{\Upsilon}_d[T]^{-1}
(\Edc[z_0,f(z_0)])$ is well-defined and so we
compute
\begin{eqnarray}
&&  \hspace{-5mm}
\overline{\Edc[z_0,f(z_0)]} 
=\overline{\hat{\Upsilon}_d[T](\hat{\Upsilon}_d[T]^{-1}
(\Edc[z_0,f(z_0)]))} 
\nonumber\\
&&
=\hat{\Upsilon}_d[T]( \overline{\hat{\Upsilon}_d[T]^{-1}
(\Edc[z_0,f(z_0)])}) 
\; ,
\label{eq:10.132nc}
\end{eqnarray}
where in the second equality we used \cite[Section XII.2]{Du} and the
fact that $\hat{\Upsilon}_d[T]$ is a homeomorphism.
Using (\ref{eq:12.17dbnna}),(\ref{eq:10.132na}) and the
remarks after  Definition \ref{D8.7c}, we compute
\begin{eqnarray}
&&  \hspace{-5mm}
\hat{\Upsilon}_d[T]^{-1}
(\Edc[z_0,f(z_0)])
=\Upsilon_d[T^t](\Edc[z_0,f(z_0)])
\nonumber\\
&&
=\Upsilon_d[T^t](
\bigcup_{n\in\Z} 
\biggl(\lbrace j^n(z_0) \rbrace 
\times \lbrace r\in SO(3):l(r;x)=f(j^n(z_0))\rbrace 
\biggr))
\nonumber\\
&&
=\Upsilon_d[T^t](
\bigcup_{n\in\Z} 
\biggl(\lbrace j^n(z_0) \rbrace 
\times \lbrace r\in SO(3):l(r;x)=l(T(j^n(z_0));x)\rbrace 
\biggr))
\nonumber\\
&&
=\bigcup_{n\in\Z} 
\biggl(\lbrace j^n(z_0) \rbrace 
\times \lbrace  T^t(j^n(z_0))r:r\in SO(3),l(r;x)=l(T(j^n(z_0));x)\rbrace 
\biggr)\Biggr)
\nonumber\\
&&
=\bigcup_{n\in\Z} 
\biggl(\lbrace j^n(z_0) \rbrace 
\times \lbrace  r'\in SO(3):l(T(j^n(z_0))r';x)
=l(T(j^n(z_0));x)\rbrace 
\biggr)\Biggr)
\nonumber\\
&&
=\bigcup_{n\in\Z} 
\biggl(\lbrace j^n(z_0) \rbrace 
\times \lbrace  r'\in SO(3):l(r';x)=x\rbrace 
\biggr)\Biggr) = \biggl( \bigcup_{n\in\Z} \lbrace j^n(z_0) \rbrace\biggr)
\times H
\nonumber\\
&&
=B\times H
\; ,
\nonumber
\end{eqnarray}
where $B:=\lbrace j^n(z_0):n\in\Z\rbrace$
which implies 
\begin{eqnarray}
&&  \hspace{-5mm}
\hat{\Upsilon}_d[T]^{-1}(\Edc[z_0,f(z_0)]) 
= B\times H \; .
\label{eq:10.132nda}
\end{eqnarray}
To prove (\ref{eq:10.132nbaan}),
we conclude from (\ref{eq:10.132nda}) that
\begin{eqnarray}
&&  \hspace{-15mm}
\overline{\hat{\Upsilon}_d[T]^{-1}
(\Edc[z_0,f(z_0)])}
=\overline{B\times H} = \bar{B}\times \bar{H} = \Td\times \bar{H}
= \Td\times H \; ,
\label{eq:10.132ne}
\end{eqnarray}
where in the second equality we used  \cite[Section IV.1]{Du}
and in the fourth equality we used that $H$ is closed
(the latter follows from Remark 17 and the fact that $E$ is Hausdorff).
Inserting  (\ref{eq:10.132ne}) into (\ref{eq:10.132nc}) results in
$\overline{\Edc[z_0,f(z_0)]} 
=\hat{\Upsilon}_d[T](\Td\times H) 
= \Edh[f]$
where in the second equality we used from Theorem \ref{P12.10}b
that $\hat{\Upsilon}_d[T]$ is
onto $\Edh[f]$.
\hfill $\Box$

The following remark discusses Theorem \ref{C12.10b} in the special case
of the ISF.

\vspace{3mm}
\noindent{\bf Remark:}
\begin{itemize}
\item[(31)] 
Let $(j,A)\in {\cal SOS}(d,j)$ where $j$ is topologically 
transitive and so we have a $z_0\in\Td$ such that the set
$\lbrace j^n(z_0):n\in\Z\rbrace$ is dense in $\Td$. 
Also let $(E,l)=(\R^3,l_v)$ and
$f$ be an invariant polarization field
of $(j,A)$ such that $f$ is not the zero field.
By Lemma \ref{L10.4} and Remark 13
we can pick a $\lambda\in (0,\infty)$ such that 
$f$ takes values only in the sphere $l(SO(3);S_\lambda)={\mathbb S}^2_\lambda$.
Let also $p_d[\R^3,l_v,S_\lambda,f]$ have a cross section, i.e., let
by Remark 30 $(j,A)$ have an IFF whose third column is $f/\lambda$. Then, by
Theorem \ref{C12.10b},
\begin{eqnarray}
&&  \hspace{-5mm}
\Edh[\R^3,l_v,S_\lambda,f] =
\overline{\Edc[\R^3,l_v,j,A,z_0,S_\lambda,f(z_0)]} \; .
\label{eq:10.132nbb}
\end{eqnarray}
Note that 
in the special case $d=1$ the condition, that $p_d[\R^3,l_v,S_\lambda,f]$ 
has a cross section, is redundant (see the remarks at the beginning of this
section). Thus for addressing the ISF conjecture, one should perhaps
start with $d=1$.
\hfill $\Box$
\end{itemize}
\subsection{Underlying bundle theory}
\label{10.7}
While bundle aspects were not needed in the present work
it is  worthwhile to mention them  since they are the origin of 
the ToA (see  \cite{Fe,He2,Zi2} and Chapter 9 in \cite{HK1}) and 
therefore supply 
a steady flow of ideas, many of which not even mentioned here (e.g., 
algebraic hull, characteristic class, rigidity). 
The ``unreduced'' principal bundle underlying our formalism
is a product principal $SO(3)$-bundle with base space $\Td$, i.e.,
it can be written as the $4$-tuple $({\cal E}_d,p_d,\Td,L_d)$ where
$\Ed=\Td\times SO(3)$ is the bundle space, $p_d\in{\cal C}({\cal E}_d,\Td)$ the
bundle projection, i.e., $p_d(z,r):=z$, and 
$({\cal E}_d,L_d)$ the underlying $SO(3)$-space where
$L_d:SO(3)\times \Ed\rightarrow \Ed$ defined by
$L_d(r;z,r'):=(z,r'r^t)$.
For every $(j,A)$, bundle theory gives
us, via the so-called automorphism group of the unreduced principal bundle,
a natural particle-spin map on ${\cal E}_d$ which turns out to be 
$\hat{\cal P}[j,A]$ in (\ref{eq:10.135n}). 
Note that here the cocycle property of the
spin transfer matrix function is crucial.
The reductions are those principal $H$-bundles which are
subbundles of the unreduced bundle such that their bundle space
is a closed subset of ${\cal E}_d$ and such that $H$ is a closed subgroup of
$SO(3)$.
By the well-known Reduction Theorem \cite[Chapter 6]{Fe},
\cite[Chapter 6]{Hus}, every
$(\Edh[f],p_d[E,l,x,f],\Td,L)$, for which $E$ is Hausdorff,
is a reduction where
$L$ is the restriction of $L_d$ to $H\times\Edh[f]$
and conversely, every reduction is of this form.
By bundle theory, the natural particle-spin map on 
$\Edh[f]$ for a given $(j,A)$
is that bijection on $\Ed[f]$ which is
a restriction of $\hat{\cal P}[j,A]$.
Clearly this function is a bijection iff
$\Edh[f]$ is 
$\hat{\cal P}[j,A]$-invariant and then the reduction is called
``invariant under $(j,A)$''.
Thus indeed the IRT deals with invariant reductions and it
states that a reduction is invariant under $(j,A)$ iff
$f$ is an invariant field.

The bundle-theoretic aspect of the CST follows from the simple fact that
the cross sections of $p_d[E,l,x,f]$ are the bundle-theoretic cross sections
of the reduction. Thus, by bundle theory,
$p_d[E,l,x,f]$ has a cross section iff the principal 
bundle $(\Edh[f],p_d[f],\Td,L)$ is trivial, i.e.,
is isomorphic to a product principal bundle by the isomorphism 
$\hat{\Upsilon}_d[T]$ from Section \ref{9.4b.4}.

The First ToA Transformation Rule has its bundle counterpart
in a transformation rule under the
$SO(3)$ gauge transformation group \cite[Chapter 9]{Hus}
of the unreduced principal bundle.

Every $(E,l)$ in the formalism uniquely determines 
an ``associated bundle'' (relative to 
the unreduced bundle) which, up to bundle isomorphism, is of the
form $(\Td\times E,p,\Td)$ where $p(z,x):=z$.
Thereby the fields are just the nontrivial data of the cross sections
of $p$. Moreover the automorphism group of the unreduced principal bundle
acts naturally on the cross sections of $p$ and it is this action which
induces the field map $\tilde{\cal P}[E,l,j,A]$ of 
(\ref{eq:10.17}). Thus invariant $(E,l)$-fields are the nontrivial data
of invariant cross sections of associated bundles. 
This is similar to the situation in gauge field theories where
the matter fields carry the data of cross sections of associated bundles.
Note also that the reduced principal bundles correspond
to a certain subclass of the associated bundles hence the cross sections
in the CST correspond to a subclass of the 
cross sections of the associated bundles.

As a side aspect, the above mentioned reductions reveal a relation to
Yang-Mills Theory via the principal connections.
For example, via Remark 30, in the presence of an IFF we 
have an invariant $SO(2)$ 
reduction which has a cross section and describes planar spin motion.
Since this reduction is a smooth principal bundle, it has a well-defined class
 of principal connections leading via path lifting 
to parallel transport motions which,
remarkably, reproduce
 the form of the well-known T--BMT equation, and thus in discrete time give 
us ${\cal P}[j,A]$.
These aspects will be extended to nonplanar spin motion in future work.
\section{Summary and Outlook}
\label{13}
\setcounter{equation}{0}
In this work we have studied the discrete-time spin motion in 
storage rings in terms of the ToA. We thus generalized the notions of
invariant polarization field and invariant frame field and reconsidered
the notions of spin tune and spin-orbit resonance 
within this framework.
We demonstrated its convenience in many ways, among them the ability
to unify the description of spin-$1/2$ and spin-$1$ particles by
exhibiting common properties of the spin vector motion resp.
spin-tensor motion.

For future work there are several natural avenues.
One obvious avenue is the study of the existence 
and uniqueness problems
of invariant polarization fields and invariant polarization-tensor
fields in terms of the IRT and the closely related notions of ``algebraic hull''
and ``rigidity'' \cite[Section 6]{Fe},\cite[Section 9]{HK1}.
The algebraic hull of $(j,A)$ is, roughly speaking, the ``smallest''
subgroup $H$ of $SO(3)$ for which $(j,A)$ has an $H$-normal form.
The study of invariant fields in terms of the IRT will be
focused on the comparison of the topological spaces $\Edh[f]$ and
$\overline{\Edc[z,y]}$ introduced in
Section \ref{10.6.1}.

Rigidity of $(j,A)$ occurs if, roughly speaking,
the behaviour of $(j,A)$ does not change under the extension of the
time group $\Z$ to a larger time group.
Also the underlying principal bundle invites
the study of the path lifting of its principal connections \cite{Na} and
the study the so-called characteristic classes of its reduced principal
bundles. Characteristic classes occur if one studies a principal
bundle in terms of so-called universal bundles \cite{Hus}.
Note that characteristic classes, the so-called Chern classes, play
a key role in Yang-Mills theory. Another avenue is 
the study, in
the case $j={\cal P}_\omega$, of the spin trajectories $x(\cdot)$
in terms
of a Fourier Analysis since then the equations of motion are 
characterized by quasiperiodic functions in time.
In particular a perturbation analysis via averaging techniques
seems feasible. One could also weaken 
the condition that $A,j,l$ etc. are
continuous functions to the condition of Borel measurability.
Moreover the ToA can be easily modified from our
$SO(3)$ formalism to the quaternion formalism and the
spinor formalism where the group $SU(2)$ will take over the role
of $SO(3)$. 

We now give a summary of Chapter \ref{10} including the
relevant material from Chapters \ref{2}-\ref{VII}.
A spin-orbit system is
a pair $(j,A)$ where $j\in \Homeo(\Td)$ is the particle $1$-turn map 
and $A\in {\cal C}(\Td,SO(3))$ with the torus $\Td$ introduced in Section
\ref{2.1b}. In the special case $j={\cal P}_\omega$, 
$\omega$ is the orbital tune and 
${\cal P}_\omega$ is the corresponding translation on the torus after one turn.
The ToA is defined in Section \ref{10.2} and it considers
arbitrary $SO(3)$-spaces $(E,l)$, defined in
Section \ref{2.3}. Various other group-theoretical notions are
defined in Section \ref{2.3}. 
For every $SO(3)$-space $(E,l)$
and every spin-orbit system $(j,A)$ in ${\cal SOS}(d,j)$ a 
$1$-turn particle-spin map ${\cal P}[E,l,j,A]\in\Homeo(\Td\times E)$
is defined by (\ref{eq:10.15}), i.e.,
${\cal P}[E,l,j,A](z,x):=(j(z),l(A(z);x))$. Also 
a $1$-turn field map $\tilde{\cal P}[E,l,j,A]$
is a bijection on ${\cal C}(\Td,E)$
defined by (\ref{eq:10.17}), i.e.,
$\tilde{\cal P}[E,l,j,A](f):=
l\biggl(A\circ j^{-1};f\circ j^{-1}\biggr)$.
The special case $(E,l)=(\R^3,l_v)$ where $x$ is the spin vector $S$
is studied in Chapters \ref{2}-\ref{VII}
and here ${\cal P}[\R^3,l_{v},j,A]={\cal P}[j,A]$
is given by (\ref{eq:2.3b})
and $\tilde{\cal P}[\R^3,l_{v},j,A]=\tilde{\cal P}[j,A]$
is given by (\ref{eq:xx13.2n}), i.e.,
\begin{eqnarray}
&& \tilde{\cal P}[j,A](f):=(Af)\circ j^{-1} \; .
\nonumber
\end{eqnarray}
We note also that the particle-spin maps are just characteristic maps
of the field maps.
If $f\in {\cal C}(\Td,E)$ satisfies $\tilde{\cal P}[E,l,j,A](f)=f$ then
$f$ is called an invariant $(E,l)$-field of $(j,A)$.
In the special case $(E,l)=(\R^3,l_v)$ 
an invariant $(\R^3,l_v)$-field is also called an invariant polarization field
and in the subcase $|f|=1$ it is called an invariant spin field.
A $j\in \Homeo(\Td)$ is called topologically transitive if
a $z_0\in\Td$ exists such that the topological closure
$\overline{\lbrace j^n(z_0):n\in\Z\rbrace}$ of
$\lbrace j^n(z_0):n\in\Z\rbrace$ equals $\Td$.
 The ISF-conjecture states that
a spin-orbit system $(j,A)$ has an ISF if $j$ is topologically transitive. 

Note that a special case of this conjecture is:
If a spin-orbit system $({\cal P}_\omega,A)$ is off orbital resonance, 
then it has an ISF.

If $(j,A)\in{\cal SOS}(d,j)$ and
$T\in{\cal C}(\Td,SO(3))$ then $(j,A')\in{\cal SOS}(d,j)$
is called the transform of $(j,A)$ under $T$ where
$A'$ is defined by (\ref{eq:5.25}), i.e.,
$A'(z): =T^t( j(z))A(z)T(z)$.
The First ToA Transformation Rule in Section
\ref{10.2.3} transforms
$\tilde{\cal P}[E,l,j,A]$ into 
$\tilde{\cal P}[E,l,j,A']$ and one has
$\tilde{\cal P}[E,l,j,A'] = \tilde{\cal P}[E,l,id_\Td,T]^{-1}
\circ \tilde{\cal P}[E,l,j,A]
\circ \tilde{\cal P}[E,l,id_\Td,T]$.
If $H$ is a subgroup of $SO(3)$ and $(j,A)\in{\cal SOS}(d,j)$ then
$(j,A')$ in ${\cal SOS}(d,j)$ is 
an $H$-normal form of $(j,A)$ if $A'$ is $H$-valued and
$(j,A')$ is a transform of $(j,A)$.
If $H$ and $H'$ are subgroups of $SO(3)$ then
we write $H\unlhd H'$ if an $r\in SO(3)$ exists such that
$rHr^t\subset H'$. If $H\unlhd H'$ and $(j,A)$ has an 
$H$-normal form then it also has an $H'$-normal form
whence spin-orbit tori are ordered in terms of their
normal forms.
The NFT in Section \ref{10.2.2add} states that if
$T\in{\cal C}(\Td,SO(3))$ and $f\in{\cal C}(\Td,E)$ are related by 
$f(z)=l(T(z);x)$ then $f$ is an invariant $(E,l)$-field of $(j,A)$
iff $T^t(j(z))A(z)T(z) \in H(x)=Iso(E,l;x)=
\lbrace r\in SO(3):l(r;x)=x\rbrace$.
Thus invariant fields can be studied in terms of isotropy groups via
the notion of normal form. In particular the ``smaller'' a subgroup
$H$ of $SO(3)$ the less likely it is for 
$(j,A)$ to have an $H$-normal form.
Following Chapter \ref{4},
a spin-orbit system $(j,A)$ has a spin tune 
$\nu\in[0,1)$ if $(j,A')$ with $A'(z)=\exp(2\pi\nu{\cal J})$
is a transform of $(j,A)$.
We say that $({\cal P}_\omega,A)$ is on spin-orbit resonance
if it has spin tunes and if for every spin tune $\nu$ we can find
$m\in{\mathbb Z}^d,n\in{\mathbb Z}$ such that
$\nu = m\cdot \omega + n$.
The Uniqueness Theorem, Theorem \ref{T7.1}b,
states that, if $({\cal P}_\omega,A)$ has spin tunes and is off
orbital resonance and off 
spin-orbit resonance,
then it has only two ISF's and they differ only by a sign.
The polarization of a bunch is defined in terms of the
density matrix function in Section
\ref{10.5.1} and its size is estimated
in Section \ref{6.4}.
The decomposition method in Section
\ref{10.2.5} decomposes each
$SO(3)$-space $(E,l)$ into transitive
$SO(3)$-spaces and it predicts that for
topologically transitive $j$ an invariant 
$(E,l)$-field takes only values in one
$(E,l)$-orbit. This allows us to classify, via the DT, invariant
fields in terms of isotropy groups.
In Sections \ref{10.4}-\ref{10.5} we apply the ToA to 
the $SO(3)$-spaces $(\R^3,l_v)$ and
$(E_t,l_t)$ to study spin-$1/2$ and spin-$1$ particles.
In Section \ref{10.6} we study the existence problem of
invariant $(E,l)$-fields in terms of the
topological spaces $\Edh[E,l,x,f]$
and in Section \ref{10.7} we discuss the 
bundle-theoretic aspects of the present work.
We revisit some old theorems and
prove several theorems which we believe to be new.
Among the former we mention
Theorems \ref{T09t0},~\ref{T09t1},~\ref{T09t2} and \ref{T7.1}
and among the latter
Lemmas \ref{L10.4} and \ref{L10.5} and 
Theorems \ref{T2},~\ref{T10.1},~\ref{T10.6},~\ref{T10.7},~\ref{T10.10},~\ref{T10.8cxs},~\ref{T10.8cx},~\ref{P02xxx},~\ref{C12.10a},~\ref{P12.10} and 
\ref{C12.10b}.
\newpage
\section{Table of Notation}
\label{60}
\addcontentsline{toc}{section}{Table of notation}
\bea
&&\hspace{-8mm} 
\begin{array}{ll}  
{\cal A},A_{CT}[\omega,{\cal A}], A, A_{d,\nu}
& \qquad  (\ref{eq:2.12}), (\ref{eq:5.84ayn}),(\ref{eq:2.3a}),
(\ref{eq:6.1})
\\
{\cal ACB}(d,j) & \qquad {\rm Definition\;}
\ref{D2}
\\
B(E,l,E',l';x,x'),B(E,l;x,x')
& \qquad {\rm Section\;}\ref{10.3.1} \\
 \hat{\beta}[E,l,E',l';x,x',r],\hat{\beta}[E,l;x,x',r] 
& \qquad {\rm Section\;}\ref{10.3.1}\\
{\cal C}(X,Y) {\rm\;set\;of\;continuous\;functions\;from\;}X{\rm\;to\;}Y
& \qquad  {\rm Appendix\;\ref{A.4}}\\ 
{\cal CB}_H & \qquad {\rm Definition\;} \ref{D6}\\
(E,L)-{\rm lift} & \qquad {\rm Section\;}\ref{10.2.2add} \\
(E,L)\;(G-{\rm set}) & 
\qquad {\rm Definition\;} \ref{D2.3} \\
(E,L)\;(G-{\rm space}) & 
\qquad {\rm Definition\;} \ref{D2.4} \\
(E,L)-{\rm orbit},E/L & 
\qquad {\rm Definition\;} \ref{D2.3new},{\rm Definition\;} \ref{D2.4}\\
(E_{t},l_{t}),R_{t}
& \qquad {\rm Section\;}\ref{10.4.1} \\
(E_{v\times t},l_{v\times t}),
(E_{dens}^{1/2},l_{dens}^{1/2}) {\rm\;and\;} (E_{dens}^{1},l_{dens}^{1})
& \qquad {\rm Section\;}\ref{10.5} \\
\Ed, \Edh[E,l,x,f],\Edc[E,l,j,A,z,x,y] & \qquad {\rm Section\;}\ref{10.6.1} \\
{\rm Equivalent\;spin-orbit\;systems},
\overline{(j,A)}  & \qquad 
 {\rm Definition\;}\ref{D3b}
\\
{\rm Group,\;conjugate\;subgroups, topological\;group} & \qquad 
{\rm Definition\;}\ref{D2.1},{\rm \;Definition\;}\ref{D2.2}
\\
{\rm H-normal\;Form} & \qquad {\rm Definition\;} \ref{D6} \\
\Homeo(X,Y) {\rm\; (set\;of\;homeomorphisms\;from\;}X{\rm\;to\;}Y)
& \qquad {\rm Appendix\;\ref{A.4}}\\ 
{\rm Invariant\;frame\;field\;(IFF)} & \qquad  
{\rm Definition\;} \ref{DIFF}\\
 {\rm Uniform\;invariant\;frame\;field} & \qquad 
{\rm Remark\;5\;in\;Chapter\;}\ref{4}\\
{\rm Invariant\;spin\;field\;(ISF)},
{\rm invariant\;polarization\;field} & \qquad  {\rm Definition\;} \ref{D6.1} \\
{\rm Invariant\;}(E,l)-{\rm field}, 
{\rm invariant\;}n-{\rm turn\;}(E,l)-{\rm field} 
 & \qquad {\rm Section\;}\ref{10.2.1} \\
Iso(E,L;x) {\rm \;(Isotropy\;group\;of\;}G-{\rm space\;}(E,L){\rm\;at\;}x) 
& \qquad  {\rm Definition\;} \ref{D2.4} \\
j,{\cal J} & \qquad {\rm Section\;\ref{2.2}},
(\ref{eq:6.5a})\\
L[j],L[j,A],\tilde{L}[j,A] & \qquad (\ref{eq:3.20}),(\ref{eq:s2.10}),
(\ref{eq:xx13.2})\\
L[E,l,j,A],\tilde{L}[E,l,j,A] & \qquad (\ref{eq:10.36}),(\ref{eq:10.45})\\
l_{dec}[x] & \qquad (\ref{eq:10.69}) \\
\Lambda_j  & \qquad {\rm Section\;}\ref{10.4.1}\\
N(H,H'),\unlhd & \qquad (\ref{eq:12.17dbt}),{\rm Section\;\ref{V.1}}
\\ 
\Xi(j,A)  & \qquad (\ref{eq:2.3aad})\\
{\cal P}_\omega & \qquad  (\ref{eq:s2.10t})\\
{\cal P}_{CT}[\omega,{\cal A}] & \qquad (\ref{eq:5.84axt})\\
{\cal P}[j,A],\tilde{\cal P}[j,A] & \qquad (\ref{eq:2.3b}),(\ref{eq:xx13.2n})\\
{\cal P}[E,l,j,A],\tilde{\cal P}[E,l,j,A], \hat{\cal P}[j,A]
& \qquad
(\ref{eq:10.15}), (\ref{eq:10.17}),(\ref{eq:10.135n}) \\
p_d, p_d[f] & \qquad {\rm Section\;}\ref{10.6} \\
\pi_d & \qquad (\ref{eq:2.22})\\
{\rm Resonant, nonresonant, orbital\;resonance}  & \qquad 
{\rm Section\;}\ref{2.2} \\
SO(3) & \qquad {\rm Section\;}\ref{2.1} \\
SO(2) & \qquad (\ref{eq:6.5}) \\
SO(2) \bowtie \Z_2,SO_{diag}(3) & \qquad 
(\ref{eq:10.161}),(\ref{eq:10.163}) \\
S_\lambda, {\mathbb S}_\lambda^2 & \qquad 
 {\rm Sections\;\ref{10.2.2add},\ref{10.2.5}}
\end{array} 
\nonumber
\eea
\newpage
\vspace{-2cm} 
\bea
&&\hspace{-8mm} 
\begin{array}{ll}  
{\cal BMT}(d),
{\cal SOS}_{CT}(d,\omega),
& \qquad {\rm Section\;}\ref{2.1},(\ref{eq:2.3aa}),
\\
{\cal SOS}(d,j){\rm\;(set\;of\;spin-orbit\;systems)}& \qquad  (\ref{eq:2.3a})\\
{\rm Spin\;tune, spin-orbit\;resonance\;} 
& \qquad  {\rm Definition\;} \ref{D4}\\
\Sigma_x[E,l,f]
& \qquad  (\ref{eq:10.73})\\
{\rm Topologically\;transitive} & \qquad {\rm Section\;}\ref{2.2}\\
\Td  {\rm \;(d-torus)} & \qquad (\ref{eq:2.20})\\
{\cal TF}(A,A';d,j),{\cal TF}_H(j,A)
& \qquad {\rm Definition\;}\ref{D3a},
{\rm Definition\;}\ref{D6}\\
{\rm Transitive\;}G-{\rm space} & \qquad {\rm Definition\;} \ref{D2.4} \\
{\rm ToA} & \qquad {\rm Technique\;of\;Association} \\
\Upsilon_d[T],\hat{\Upsilon}_d[E,l,x,T] & \qquad  
 {\rm Section\;}\ref{9.4b.4}\\
 \Phi_{CT}[\omega,{\cal A}]  & \qquad (\ref{eq:2.17awb})\\
 \Psi[j,A]  \; ({\rm spin\;transfer\;matrix\;function}) 
 & \qquad (\ref{eq:3.5})\\
 \omega  \;({\rm orbital\;tune}) 
 & \qquad  {\rm Section\;}\ref{2.1} \\
\Z  & \qquad {\rm Set\;of\;integers} \\
 \tilde{\Z}^d,\Z_2 & 
\qquad {\rm Section\;} \ref{2.1b},(\ref{eq:10.161}) 
\end{array} 
\nonumber
\eea
\section*{Appendix}
\addcontentsline{toc}{section}{Appendix}
\renewcommand{\thesection}{\Alph{section}}
\renewcommand{\thelemma}{\Alph{lemma}}
\setcounter{subsection}{0}
\setcounter{section}{0}
\section{Conventions and terminology}
\label{A}
\setcounter{equation}{0}
\subsection{Function, image, inverse image}
\label{A.1}
A ``function'' $f:X\rightarrow Y$ is determined by its graph and its codomain.
The ``graph'' of $f$ is the set $\lbrace (x,f(x)):x\in X\rbrace$
and the ``codomain''
of $f$ is $Y$. The ``domain'' of $f$ is $X$ and the ``range''
of $f$ is the set
$f(X):=\lbrace f(x):x\in X\rbrace$. 
More generally, if $M$ is a subset of $X$ 
then the ``image'' of $M$ under $f$ is the set
$f(M):=\lbrace f(x):x\in M\rbrace$.
If $M$ is a subset of $Y$ then the ``inverse
image'' of $M$ under $f$ is the set $f^{-1}(M):=\lbrace x\in X:f(x)\in
M\rbrace$.

One calls $f$ a ``surjection'' or ``onto'' if its range and
codomain are equal. One calls $f$ ``one-one'' or an ``injection''
if $f(x)=f(x')$ implies that $x=x'$. One calls $f$ a ``bijection''
if it is one-one and a surjection. 

If $f:X\rightarrow Y$ and $g:Y\rightarrow Z$ are functions
then $g\circ f$ is the function $g\circ f:X\rightarrow Z$
defined by $(g\circ f)(x):=g(f(x))$. One calls the operation $\circ$
the ``composition'' of functions.
If $X$ is a set then the function
$id_X:X\rightarrow X$ is defined by $id_X(x):=x$ and is called the
``identity function'' on $X$. If $f:X\rightarrow Y$ is a bijection then
a unique function $f^{-1}:Y\rightarrow X$ exists such that
$f^{-1}\circ f=id_X,f\circ f^{-1}=id_Y$ and it is called the ``inverse'' of $f$.
Clearly $f$ is a bijection iff it has an inverse.

Note that if $f:X\rightarrow Y$ is a bijection then $f^{-1}$ can either
mean the inverse function or the inverse image operation. However it
will always be clear from the context what the meaning is.

If $f:X'\rightarrow Y$ is a function and $X\subset X'$ 
then we define the function
$f|X:X\rightarrow Y$ as a restriction of $f$ to $X$, i.e.,
by $(f|X)(x):=f(x)$. If $f:X\times Y\rightarrow Z$ is a function
and $x\in X,y\in Y$ then the restriction $f|(\lbrace x\rbrace\times Y)$
is denoted also by $f(x,\cdot)$ and the restriction
$f|(X\times \lbrace y\rbrace)$ is denoted also by $f(\cdot,y)$.

If $f:X\rightarrow X$ is a function then $x\in X$ is called a ``fixed point''
of $f$ if $f(x)=x$. 

Note finally that according to our, very common,
definition of a function two functions with the same graph
are different iff they have different codomains.
Thus the alternative, and
equally common, way to define a function in terms of its graph
(i.e., without invoking the codomain) is different from our definition.
\subsection{Partition, representing set, equivalence relation}
\label{A.2}
If $X$ is a set and if $P$ is a set whose elements are disjoint 
nonempty subsets of $X$ whose union is $X$ then one calls $P$ a ``partition of
$X$''. If $P$ is a partition of $X$ then a subset $X'$ of $X$ 
is called a ``representing set of $P$'' if every element of $P$ contains
exactly one element of $X'$.
Note that partitions and their representing sets are used throughout this
work. When needed, we will always find a representing set.
From a more general view point, one knows that a representing set always
exists if one assumes the Axiom of Choice \cite{Dud}.

If $X$ is a set and $B$ a subset of $X\times X$ then $B$ is called a
``relation'' on $X$. The relation $B$ is called ``symmetric'' if
$(x,y)\in B$ implies that $(y,x)\in B$.
The relation $B$ is called ``reflexive'' if $(x,x)\in B$ for all $x\in X$. 
The relation $B$ is called ``transitive'' if
$(x,y)\in B$ and $(y,z)\in B$ implies that $(x,z)\in B$.

A relation on $X$ is called an ``equivalence relation on $X$'' if
it is symmetric,reflexive, and transitive.
If $B$ is an equivalence relation on $X$ and $x\in X$ then
the set $\lbrace y\in Y:(x,y)\in B\rbrace$ is called
the ``equivalence class of $x$ under the equivalence relation $B$''.
We also write $x\sim y$ if $(x,y)\in B$.

The equivalence classes of $B$ form a partition of $X$ as follows.
Clearly the equivalence classes of $B$ are nonempty sets and 
overlap $X$. Moreover
by, transitivity, if two equivalence classes of $B$ have a common element
then they are equal.
\subsection{Topology, topological space, open set, closed set, closure}
\label{A.3}
A collection, $\tau$, of subsets of
a set $X$ is called a ``topology on $X$'' if $\tau$ is closed under arbitrary
unions and finite intersections and if $X,\emptyset\in\tau$.
Any pair $(X,\tau)$ is called a ``topological space (over $X$)''.
The elements of $\tau$ are called the ``open'' sets of $(X,\tau)$.

The ``closed'' sets of $(X,\tau)$ are the complements of the open sets
$(X,\tau)$. If $M$ is a subset of $X$ then its ``closure'' $\bar{M}$
is defined as the intersection of all closed supsets of $M$.

If $(X,\tau)$ is a topological space and if $X'$ is a subset of $X$
then the 
``subspace topology'' $\tau'$ of $X'$ from $X$ is the collection
$\lbrace X'\cap M:M\in\tau\rbrace$ and the topological space $(X',\tau')$
is called a ``topological subspace'' of $(X,\tau)$.

Since the topology $\tau$
is always clear from the context we write $X$ instead of $(X,\tau)$.
For example the topology of $\Rd$ is obtained from the Euclidean norm
and the topology of $\Zd$ is discrete, i.e., every subset of $\Zd$ is open.\\
\subsection{ Continuous function, homeomorphism, cross section}
\label{A.4}
Let $(X,\tau)$ and $(X',\tau')$ be topological spaces. Then a function
$f:X\rightarrow X'$ is called ``continuous w.r.t. $(X,\tau)$ and 
$(X',\tau')$'' if 
for every $M\in\tau'$ the inverse image of $M$ under
$f$ belongs to $\tau$, i.e.,
$f^{-1}(M)\in\tau$. We denote the collection
of continuous functions by ${\cal C}(X,X')$. A function
$f\in{\cal C}(X,X')$ is called a ``homeomorphism'' and
$X,X'$ are called a ``homeomorphic'' if
$f$ is a bijection and if its inverse is continuous.
We denote the collection of those homeomorphisms by
$\Homeo(X,X')$ and we also define $\Homeo(X):=\Homeo(X,X)$.
The topological spaces $X$ and $X'$ are called ``homeomorphic'' if
$\Homeo(X,X')$ is nonempty.

If $f\in{\cal C}(Z,Z')$ with $Z'$ being a subspace of $X\times Y$
then we denote the two components of $f$ by
$f_1,f_2$, i.e., $f_1\in{\cal C}(Z,X)$ and $f_2\in{\cal C}(Z,Y)$
where $f(z)=(f_1(z),f_2(z))$.
If $f\in{\cal C}(X,Y)$ then the inverse image
$f^{-1}(\lbrace y\rbrace)$ is called
the ``fibre of $f$ over $y$''.
If $f\in {\cal C}(X,Y)$ then a function
$\sigma\in {\cal C}(Y,X)$ is called a ``cross section of $f$'' if
$f\circ\sigma=id_Y$. Note that a cross section is often called a ``section''.
\subsection{Product topology, Hausdorff space, 
compact space, \\path-connected space}
\label{A.5}
If $X$ and $Y$ are topological spaces then the product topology on
$X\times Y$ is defined such that sets $M\times N$ are open if
$M$ and $N$ are open and such that every open set of $X\times Y$
is a union of those $M\times N$. For example
the topology of ${\cal E}_d=\Td\times SO(3)$ is
the product topology where
the topology of $SO(3)$ is the subspace
topology from $\R^{3\times 3}$.

A topological space $X$ is called ``Hausdorff'' if for every pair of
distinct elements $x,x'$ of $X$ open sets $M,M'$ exists such that
$x\in M,x'\in M'$ and $M\cap M'=\emptyset$.
A topological space $X$ is called ``compact'' if for any union of $X$ by open
sets of $X$ already the union of finitely many of those open sets
equals $X$. 

If $X$ is a topological space and then a subset $A$ of $X$ 
is called ``compact'' if $A$ is, as a topological subspace of $X$, compact.

A topological space $X$ is called ``path-connected'' if for elements
$x,x'\in X$ a continuous function $f:[0,1]\rightarrow X$ exists such that
$f(0)=x$ and $f(1) = x'$. 
A subset $A$ of $X$ 
is called ``path-connected'' if $A$ is, as a topological subspace of $X$, 
path-connected. One has the following intermediate-value theorem:
If $X,Y$ are topological spaces such that 
$X$ is path-connected and if $g:X\rightarrow Y$ is a continuous function
then the range of $g$ is a path-connected subset of $Y$.
\subsection{Co-induced topology, identifying function}
\label{A.6}
Let $X$ be a topological space and let $p:X\rightarrow Y$ be a surjection where
$Y$ is a set. A natural topology on $Y$ is defined such that
a subset $B\subset Y$ is open iff the inverse image
$p^{-1}(B)\subset X$ is open. One calls
the topology on $Y$ ``co-induced by $p$'' \cite{wiki}. 
Using older terminology, one also says that the topology on $Y$
is the ``identification topology'' w.r.t. $p$ and that 
$p$ is ``identifying'' \cite{Du,Hu}.
Of course an identifying function is continuous (but not vice versa).
Time and again we will use co-induced topologies and we will often
use the following lemma to prove the continuity of a function:\\

\noindent (Continuity Lemma)\\
\noindent 
Let $X$ be a topological space and let $p:X\rightarrow Y$ be a surjection where
$Y$ is a set. Let the topology on $Y$ be co-induced by $p$. 
If $Z$ is a topological space and $f\in{\cal C}(X,Z)$
and $F:Y\rightarrow Z$ are functions such that 
$F\circ p=f$ then $F$ is continuous. If in addition $f$ is
an identification map then $F$ is an identification map, too. \\

\noindent Proof of the Continuity Lemma:
To see that $F$ is continuous we
need to show that the inverse image $F^{-1}(V)$ is open 
for all open subsets $V$ of $Z$.
In fact since the topology on $Y$ is co-induced by $p$ we get
$p^{-1}(F^{-1}(V))=(F\circ p)^{-1}(V)=f^{-1}(V)$, thus if $V$ is 
open, $f^{-1}(V)=p^{-1}(F^{-1}(V))$ is open.
Thus indeed $F$ is continuous. The second claim is shown in the same vein
(see also Section VI.3 in \cite{Du}).
\hfill $\Box$
\section{Various Proofs}
\label{11}
\setcounter{equation}{0}
This appendix contains those proofs which are
too long for the main text.
\subsection{Proof of Lemma \ref{L10.5}} 
\label{11.1}
\noindent{\em Proof of Lemma \ref{L10.5}a:} 
By Definitions \ref{D5.x} and \ref{D8.x} of $\unlhd$ and
$N(Iso(E,l;x),Iso(E',l';x'))$ it is clear
that the first claim follows from (\ref{eq:10.83a}).
To show (\ref{eq:10.83a}) we first prove the inclusion
(\ref{eq:11.200}) so let $\beta\in B(E,l,E',l';x,x')$, i.e., 
by Definition \ref{D8.x},
$\beta$ is a topological 
$SO(3)$-map from $(l(SO(3);x),l_{dec}[x])$ to $(l'(SO(3),x'),l_{dec}'[x'])$.
Clearly we can pick an 
$r_0\in SO(3)$ such that
\begin{eqnarray}
&&  \hspace{-1cm}
\beta(x) = l'(r_0^t;x') \; .
\label{eq:10.91}
\end{eqnarray}
By (\ref{eq:10.69}) and (\ref{eq:10.91}) we have, for all $r\in SO(3)$,
\begin{eqnarray}
&&  \hspace{-1cm}
\beta(l(r;x)) = \beta(l_{dec}[x](r;x)) =  l_{dec}'[x'](r;\beta(x)) =
l'(r;\beta(x)) =l'(rr_0^t;x') \; ,
\label{eq:10.91n}
\end{eqnarray}
where in the second equality we used that $\beta$ is a topological 
$SO(3)$-map from $(l(SO(3);x),l_{dec}[x])$ to $(l'(SO(3),x'),l_{dec}'[x'])$.
On the other hand if $r_1\in Iso(E,l;x)$ then, for
all $r\in SO(3)$,
\begin{eqnarray}
&&  \hspace{-1cm}
 \beta(l(r;x)) = \beta(l(r;l(r_1;x))) = \beta(l(rr_1 ;x))
= l'(rr_1 r_0^t;x') \; ,
\nonumber
\end{eqnarray}
where in the third equality we used (\ref{eq:10.91n})
whence, by using again 
(\ref{eq:10.91n}), $l'(rr_0^t;x')=l'(rr_1 r_0^t;x')$ so that 
$r_0 r_1 r_0^t\in Iso(E',l';x')$ which implies that
$r_0 Iso(E,l;x)r_0^t\subset Iso(E',l';x')$, i.e., 
$r_0\in N(Iso(E,l;x),Iso(E',l';x'))$. Thus 
$\hat{\beta}[E,l,E',l';x,x',r_0]$ is well defined and, by  
(\ref{eq:10.83}) and (\ref{eq:10.91n}),
$\beta=\hat{\beta}[E,l,E',l';x,x',r_0]$
whence $\beta$ belongs to the set on the rhs of (\ref{eq:10.83a}) so that
we have shown that 
\begin{eqnarray}
&&  \hspace{-1cm}
B(E,l,E',l';x,x') \subset \lbrace
\hat{\beta}[E,l,E',l';x,x',r_0]:r_0\in N(Iso(E,l;x),Iso(E',l';x'))\rbrace \; .
\label{eq:11.200}
\end{eqnarray}
To show the reverse inclusion 
let $r_0\in N(Iso(E,l;x),Iso(E',l';x'))$ so we have to
show that $\hat{\beta}[E,l,E',l';x,x',r_0]$ belongs to  $B(E,l,E',l';x,x')$.
We first show that $\hat{\beta}[E,l,E',l';x,x',r_0]$ is an $SO(3)$-map. In fact it
follows from (\ref{eq:10.69}) and (\ref{eq:10.83}) that
\begin{eqnarray}
&&  \hspace{-1cm}
\hat{\beta}[E,l,E',l';x,x',r_0](l_{dec}[x](r_0;l(r_1;x))) =
\hat{\beta}[E,l,E',l';x,x',r_0](l(r_0;l(r_1;x))) 
\nonumber\\
&&
=
\hat{\beta}[E,l,E',l';x,x',r_0](l(r_0r_1;x)) =
l'(r_0r_1r_0^t;x')
 \; ,
\nonumber\\
&&  \hspace{-1cm}
l_{dec}'[x'](r_0;\hat{\beta}[E,l,E',l';x,x',r_0](l(r_1;x))) =
l'(r_0;\hat{\beta}[E,l,E',l';x,x',r_0](l(r_1;x))) 
\nonumber\\
&&
= l'(r_0;l'(r_1r_0^t;x')) 
= l'(r_0r_1r_0^t;x') \; ,
\nonumber
\end{eqnarray}
whence, for every $y\in l(SO(3);x)$, we have
$\hat{\beta}[E,l,E',l';x,x',r_0](l_{dec}[x](r_0;y)) =\\
l_{dec}'[x'](r_0;\hat{\beta}[E,l,E',l';x,x',r_0](y))$ so that indeed
$\hat{\beta}[E,l,E',l';x,x',r_0]$ is an $SO(3)$-map.
To show that $\hat{\beta}[E,l,E',l';x,x',r_0]$ is continuous
we first note that the function 
$l_{dec}[x](\cdot;x):SO(3)\rightarrow l(SO(3);x)$, defined by
$(l_{dec}[x](\cdot;x))(r):=l_{dec}[x](r;x)=l(r;x)$, is continuous since
$l$ is continuous. Moreover
since $E$ is Hausdorff, its subspace $l(SO(3);x)$ is Hausdorff, too.
Thus $l_{dec}[x](\cdot;x)$ is a continuous function from the compact space
$SO(3)$ onto the Hausdorff space $l(SO(3);x)$ whence, by the Closed Map Lemma
\cite[Section XI.2]{Du}, $l_{dec}[x](\cdot;x)$ 
is a closed map so that it is an identification map, i.e.,
the topology on $l(SO(3);x)$ is co-induced by $l_{dec}[x](\cdot;x)$ 
(for the notions of
co-induced and identification map see Appendix \ref{A.6}).
On the other hand, by  (\ref{eq:10.69}) and (\ref{eq:10.83}),
\begin{eqnarray}
&&  \hspace{-1cm}
\hat{\beta}[E,l,E',l';x,x',r_0](l_{dec}[x](r;x)) =
\hat{\beta}[E,l,E',l';x,x',r_0](l(r;x)) 
\nonumber\\
&&
=l'(rr_0^t;x') =
l'(r;l'(r_0^t;x')) 
= l'(r;x'') = l_{dec}'[x'](r;x'') 
\; ,
\nonumber
\end{eqnarray}
i.e.,
\begin{eqnarray}
&&  \hspace{-1cm}
\hat{\beta}[E,l,E',l';x,x',r_0]\circ l_{dec}[x](\cdot;x) =
l_{dec}'[x'](\cdot;x'')  \; ,
\label{eq:11.110}
\end{eqnarray}
where $x'':=l'(r_0^t;x')$.
Since by the above argument, $l_{dec}[x](\cdot;x)$ and
$l_{dec}'[x'](\cdot;x'')$ are identification maps, it follows
from (\ref{eq:11.110}) and the Continuity Lemma in Appendix \ref{A.6}
that the surjection 
$\hat{\beta}[E,l,E',l';x,x',r_0]$ is an identification map (whence it is
continuous).
We conclude that $\hat{\beta}[E,l,E',l';x,x',r_0]$ is a continuous $SO(3)$-map
whence it is a topological $SO(3)$-map from 
$(l(SO(3);x),l_{dec}[x])$ to $(l'(SO(3),x'),l_{dec}'[x'])$, i.e.,
$\hat{\beta}[E,l,E',l';x,x',r_0]$ belongs to  $B(E,l,E',l';x,x')$ so that indeed
\begin{eqnarray}
&&  \hspace{-1cm}
B(E,l,E',l';x,x') \supset \lbrace
\hat{\beta}[E,l,E',l';x,x',r_0]:r_0\in N(Iso(E,l;x),Iso(E',l';x'))\rbrace \; ,
\nonumber
\end{eqnarray}
whence (\ref{eq:10.83a}) follows from  (\ref{eq:11.200}).
\hfill $\Box$

\noindent{\em Proof of Lemma \ref{L10.5}b:} 
Let $Iso(E,l;x)\unlhd Iso(E',l';x')$
and let $r_0\in N(Iso(E,l;x),Iso(E',l';x'))$. Let also 
$r_1,r_2\in SO(3)$. Then, by (\ref{eq:10.110ln}),
\begin{eqnarray}
&&  \hspace{-1cm}
Iso(E,l;l(r_1;x)) = r_1Iso(E,l;x)r_1^t \; , \quad
Iso(E',l';l'(r_2;x'))=r_2Iso(E',l';x')r_2^t \; ,
\label{eq:10.110l}
\end{eqnarray}
whence, since
$r_0\in N(Iso(E,l;x),Iso(E',l';x'))$,
\begin{eqnarray}
&&  \hspace{-1cm}
 r_2 r_0 Iso(E,l;x)r_0^tr_2^t \subset r_2Iso(E',l';x')r_2^t 
= Iso(E',l';l'(r_2;x'))
\; ,
\label{eq:10.116l}
\end{eqnarray}
so that, by (\ref{eq:10.110l}),
\begin{eqnarray}
&&  \hspace{-1cm}
r_2r_0 r_1^tIso(E,l;l(r_1;x))r_1 r_0^tr_2^t  = r_2r_0 Iso(E,l;x) r_0^tr_2^t
\subset  Iso(E',l';l'(r_2;x')) \; ,
\label{eq:10.117l}
\end{eqnarray}
which implies that $(r_2 r_0 r_1^t)\in 
N(Iso(E,l;l(r_1;x)),Iso(E',l';l'(r_2;x')))$, i.e.,
$Iso(E,l;l(r_1;x))\unlhd Iso(E',l';l'(r_2;x'))$. 
Thus $\hat{\beta}[E,l,E',l';y,y',r_2r_0r_1^t]$ is well defined where
$y:=l(r_1;x)$ and $y':=l'(r_2;x')$ and we compute, by (\ref{eq:10.83}),
\begin{eqnarray}
&&  \hspace{-1cm}
\hat{\beta}[E,l,E',l';y,y',r_2r_0r_1^t](l(rr_1;x))
=\hat{\beta}[E,l,E',l';y,y',r_2r_0r_1^t](l(r;l(r_1;x)))
\nonumber\\
&&
=\hat{\beta}[E,l,E',l';y,y',r_2r_0r_1^t](l(r;y))=l'(r(r_2r_0r_1^t)^t;y')
\nonumber\\
&&\quad
=l'(r r_1r_0^tr_2^t;l'(r_2;x'))
=l'(r r_1r_0^t;x')=\hat{\beta}[E,l,E',l';x,x',r_0](l(rr_1;x)) \; ,
\nonumber
\end{eqnarray}
i.e., $\hat{\beta}[E,l,E',l';y,y',r_2r_0r_1^t]=\hat{\beta}[E,l,E',l';x,x',r_0]$.

Choosing $r_1,r_2\in SO(3)$ such that
$r_2r_0r_1^t=I_{3\times 3}$ (e.g., $r_1:=r_0, r_2:=I_{3\times 3}$)
and since $(r_2 r_0 r_1^t)\in 
N(Iso(E,l;l(r_1;x)),Iso(E',l';l'(r_2;x')))$ one gets \\
$I_{3\times 3}\in N(Iso(E,l;y),Iso(E',l';y'))$ whence
$Iso(E,l;y)\subset Iso(E',l';y')$. 
\hfill $\Box$

\noindent{\em Proof of Lemma \ref{L10.5}c:} 
We first prove the first claim.\\
\noindent
``$\Rightarrow$'': 
Let $Iso(E,l;x),Iso(E',l';x')$ be conjugate, i.e., let
$r_0\in SO(3)$ exist such that \\
$r_0 Iso(E,l;x)r_0^t= Iso(E',l';x')$.
Thus $r_0\in N(Iso(E,l;x),Iso(E',l';x'))$ whence, by Lemma \ref{L10.5}a,
$\hat{\beta}[E,l,E',l';x,x',r_0]\in B(E,l,E',l';x,x',r_0)$. To show that 
$\hat{\beta}[E,l,E',l';x,x',r_0]$ is an isomorphism from 
$(l(SO(3);x),l_{dec}[x])$ to $(l'(SO(3),x'),l_{dec}'[x'])$ we only have
to show that 
$\hat{\beta}[E,l,E',l';x,x',r_0]$ is a
homeomorphism. From the proof of Lemma \ref{L10.5}a we know that
$\hat{\beta}[E,l,E',l';x,x',r_0]$ is an identification map whence,
$\hat{\beta}[E,l,E',l';x,x',r_0]$ is a homeomorphism iff it is one-one.
To show that $\hat{\beta}[E,l,E',l';x,x',r_0]$ is one-one let
$r_1,r_2\in SO(3)$ such that $\hat{\beta}[E,l,E',l';x,x',r_0](l(r_1;x))
= \hat{\beta}[E,l,E',l';x,x',r_0](l(r_2;x))$ whence, by (\ref{eq:10.83}),
\begin{eqnarray}
&&  \hspace{-1cm}
l'(r_1r_0^t;x') = \hat{\beta}[E,l,E',l';x,x',r_0](l(r_1;x))) 
= \hat{\beta}[E,l,E',l';x,x',r_0](l(r_2;x)))
= l'(r_2r_0^t;x') \; ,
\nonumber
\end{eqnarray}
so that $l'(r_0r_1^tr_2r_0^t;x')=x'$ which implies, 
by (\ref{eq:12.17bnncaa}), that $r_0r_1^tr_2r_0^t\in Iso(E',l';x')$.
Since $r_0 Iso(E,l;x)r_0^t= Iso(E',l';x')$ this implies that
$r_0r_1^tr_2r_0^t\in r_0 Iso(E,l;x)r_0^t$ whence
$r_1^tr_2\in Iso(E,l;x)$ so that $l(r_1^tr_2;x) =x$, i.e.,
$l(r_1;x)=l(r_2;x)$. Thus $\hat{\beta}[E,l,E',l';x,x',r_0]$ is one-one as was to be
shown.\\

\noindent ``$\Leftarrow$'': 
We first prove the useful formula (\ref{eq:10.83an}).
In fact if in addition $(E'',l'')$ is an
$SO(3)$-space, $E''$ is Hausdorff, $x''\in E''$ and if
$r_0\in N\biggl(Iso(E,l;x),Iso(E',l';x')\biggr)$ and 
$r_1\in N\biggl(Iso(E',l';x'),Iso(E'',l'';x'')\biggr)$ then, by
(\ref{eq:12.17dbt}),
$(r_1r_0) \in N\biggl(Iso(E,l;x),Iso(E'',l'';x'')\biggr)$ whence, 
by (\ref{eq:10.83}),
\begin{eqnarray}
&&  \hspace{-1cm}
\hat{\beta}[E,l,E'',l'';x,x'',r_1r_0]=\hat{\beta}[E',l',E'',l'';x'',x',r_1]\circ
\hat{\beta}[E,l,E',l';x,x',r_0] \; .
\label{eq:10.83an}
\end{eqnarray}
Let $(l(SO(3);x),l_{dec}[x])$, $(l'(SO(3),x'),l'_{dec}[x'])$ be isomorphic.
Thus an isomorphism, say $\beta$, exists from
$(l(SO(3);x),l_{dec}[x])$ to $(l'(SO(3),x'),l_{dec}'[x'])$
whence, by Lemma \ref{L10.5}a, 
we can pick an $r_0$ in $N(Iso(E,l;x),Iso(E',l';x'))$ 
such that $\beta=\hat{\beta}[E,l,E',l';x,x',r_0]$ and we have,
by (\ref{eq:12.17dbt}),
\begin{eqnarray}
&& r_0 Iso(E,l;x)r_0^t \subset Iso(E',l';x')\; .
\label{eq:7.4.51f}
\end{eqnarray}
On the other hand, by
(\ref{eq:10.83an}), $\hat{\beta}[E,l,E',l';x,x',r_0^t]$ is the inverse, say 
$\beta^{-1}$, of $\beta$.
Since $\beta^{-1}$ is an isomorphism from
$(l'(SO(3),x'),l'_{dec}[x'])$ to $(l(SO(3);x),l_{dec}[x])$,
the above argument, which used $\beta$ to give us
(\ref{eq:7.4.51f}), can now be repeated
for $\beta^{-1}=\hat{\beta}[E,l,E',l';x,x',r_0^t]$ 
giving us in analogy to (\ref{eq:7.4.51f})
\begin{eqnarray}
&& r_0^t Iso(E',l';x')r_0 \subset Iso(E,l;x)\; ,
\label{eq:7.4.51h}
\end{eqnarray}
whence $Iso(E',l';x')\subset r_0 Iso(E',l';x')r_0^t$ so that, by
(\ref{eq:7.4.51f}),\\
$r_0Iso(E,l;x)r_0^t \subset Iso(E',l';x')
\subset r_0 Iso(E,l;x) r_0^t$
which implies that $r_0 Iso(E,l;x)r_0^t =Iso(E',l';x')$,
i.e., $Iso(E,l;x)$ and \\
$Iso(E',l';x')$ are conjugate.
This completes the proof of the first claim.
At the same time we have proven the second claim, i.e., that, 
for every $r_0\in SO(3)$ such that 
$r_0 Iso(E,l;x)r_0^t= Iso(E',l';x')$, 
$\hat{\beta}[E,l,E',l';x,x',r_0]$ is an isomorphism.
To prove the third claim, let\\
$Iso(E,l;x)$ and $Iso(E',l';x')$ be conjugate, i.e., let
$r\in SO(3)$ exist such that 
$r Iso(E,l;x)r^t= Iso(E',l';x')$. Defining $y:=x,y':=l'(r^t;x')$ we
conclude from Remark 7 that
$Iso(E',l';y')= Iso(E',l';l'(r^t;x'))=
r^t Iso(E',l';x')r = r^tr Iso(E,l;x)r^tr=Iso(E,l;x)=Iso(E,l;y)$
as was to be shown.
\hfill $\Box$
\subsection{Proof of Theorem \ref{T10.6}} 
\label{11.2}
\noindent{\em Proof of Theorem \ref{T10.6}a:} 
Using the notation of Section \ref{10.2.5} we define
$f_x\in{\cal C}(\Td,l(SO(3);x))$ by $f_x(z):=f(z)$ and
$g_{x'}\in{\cal C}(\Td,l'(SO(3);x'))$ by $g_{x'}(z):=g(z)$.
It follows from Theorem \ref{TT2} that
$(g_{x'})'=\hat{\beta}[E,l,E',l';x,x',r_0]\circ (f_x)'$ whence, by
(\ref{eq:10.72}),
$(g')_{x'}=\hat{\beta}[E,l,E',l';x,x',r_0]\circ (f')_x$ so that
$g'(z)=\hat{\beta}[E,l,E',l';x,x',r_0](f'(z))$.
\hfill $\Box$

\noindent{\em Proof of Theorem \ref{T10.6}b:} 
Let $f$ be an invariant $(E,l)$-field of $(j,A)$. 
Then, by Theorem \ref{T10.6}a, $g$ is an invariant $(E',l')$-field of $(j,A)$.
Let $g$ be an invariant $(E',l')$-field of $(j,A)$. 
Also by, Lemma \ref{L10.5}c,
$\hat{\beta}[E,l,E',l';x,x',r_0]^{-1}$ is a topological $SO(3)$-map from 
$(l'(SO(3),x'),l_{dec}'[x'])$ to $(l(SO(3);x),l_{dec}[x])$.
Thus, by Theorem \ref{T10.6}a, we conclude that
$f$ is an invariant $(E,l)$-field of $(j,A)$.
\hfill $\Box$
\end{document}